\documentclass[letterpaper, onecolumn, notitlepage, longbibliography, floatfix, superscriptaddress, aps, prx, accepted=2023-10-10, 12pt]{quantumarticle}
\pdfoutput=1
\usepackage{geometry}
\usepackage[utf8]{inputenc}
\usepackage[usenames, dvipsnames]{xcolor}
\usepackage{dsfont}
\usepackage{amsmath}
\usepackage{amsfonts}
\usepackage{amssymb}
\usepackage{graphicx}
\usepackage[pdfencoding=auto, psdextra]{hyperref}
\hypersetup{linktocpage}
\hypersetup{colorlinks=true,citecolor=quantumpurple,linkcolor=quantumpurple, urlcolor=quantumpurple}
\usepackage{natbib}
\usepackage{environ}
\usepackage{enumitem}
\usepackage{pdfpages}
\usepackage{mathrsfs}
\usepackage{changepage}
\usepackage[caption=false]{subfig}
\usepackage{amsmath, amssymb}
\usepackage{makecell}
\usepackage{tikz-cd}

\NewEnviron{eqs}{%
\begin{equation}\begin{split}
    \BODY
\end{split}\end{equation}
}

\makeatletter 
\def\l@subsubsection#1#2{}
\makeatother

\newcommand{\ZZ}{\mathbb{Z}}
\newcommand{\ra}{\rangle}
\newcommand{\la}{\langle}

\definecolor{quantumpurple}{RGB}{82, 35, 124}

\definecolor{arpit}{RGB}{127,0,0}

\definecolor{tyler}{rgb}{1,.3,0}

\definecolor{NT}{rgb}{0.8,0,0.8}

\definecolor{dom}{rgb}{.8,.1,.3}

\definecolor{YC}{RGB}{34,139,34}

\newtheorem{definition}{Definition}
\newtheorem{proposition}{Proposition}
\newtheorem{lemma}{Lemma}
\newtheorem{corollary}{Corollary}

\title{\textcolor{quantumpurple}{Pauli topological subsystem codes from Abelian anyon theories}}

\begin{document}

\author{Tyler~D. Ellison}
\thanks[]{tyler.ellison@yale.edu}
\affiliation{Department of Physics, Yale University, New Haven, CT 06511, USA}

\author{Yu-An Chen}
\affiliation{Department of Physics, Condensed Matter Theory Center, Joint Quantum Institute, and Joint Center for Quantum Information and Computer Science, University of Maryland, College Park, MD 20742, USA}
 
\author{Arpit Dua}
\affiliation{Department of Physics and Institute for Quantum Information and Matter, California Institute of Technology, Pasadena, CA 91125, USA}
 
\author{Wilbur Shirley}
\affiliation{School of Natural Sciences, Institute for Advanced Study, Princeton, NJ 08540, USA}

\author{Nathanan Tantivasadakarn}
\affiliation{Walter Burke Institute for Theoretical Physics and Department of Physics, California Institute of Technology, Pasadena, CA 91125, USA}
\affiliation{Department of Physics, Harvard University, Cambridge, MA 02138, USA}

\author{Dominic~J. Williamson}
\affiliation{Centre for Engineered Quantum Systems, School of Physics, University of Sydney, Sydney, New South Wales 2006, Australia}

\date{October 10, 2023}

\maketitle

 \vspace{-.3cm}

\begin{abstract}
\noindent \footnotesize \textcolor{quantumpurple}{\textbf{\textsf{Abstract:}}} We construct Pauli topological subsystem codes characterized by arbitrary two-dimensional Abelian anyon theories -- this includes anyon theories with degenerate braiding relations and those without a gapped boundary to the vacuum. Our work both extends the classification of two-dimensional Pauli topological subsystem codes to systems of composite-dimensional qudits and establishes that the classification is at least as rich as that of Abelian anyon theories.~We exemplify the construction with topological subsystem codes defined on four-dimensional qudits based on the $\ZZ_4^{(1)}$ anyon theory with degenerate braiding relations and the chiral semion theory -- both of which cannot be captured by topological stabilizer codes. The construction proceeds by ``gauging out'' certain anyon types of a topological~stabilizer code. This amounts to defining a gauge group generated by the stabilizer group of the topological stabilizer code and a set of anyonic string operators for the anyon types that are gauged out.
The resulting topological subsystem code is characterized by an anyon theory containing a proper subset of the anyons of the topological stabilizer code.~We thereby show that every Abelian anyon theory is a subtheory of a stack of toric codes and a certain family of twisted quantum doubles that generalize the double semion anyon theory.~We further prove a number of general statements about the logical operators of translation invariant topological subsystem codes and define their associated anyon theories in terms of higher-form symmetries.
\end{abstract}

\maketitle

\clearpage

\tableofcontents

\section{Introduction}

Topological quantum error-correcting codes are integral to many of the leading approaches to scalable, fault-tolerant quantum computation~\cite{Bravyi1998boundary, dennis2002memory, Kitaev2003quantumdouble, Raussendorf2006oneway, Fowler2012surface}. This is a consequence of their~compatibility with
spatially local architectures -- requiring only local checks to protect against the local errors that arise from coupling to the environment.
Moreover, topological quantum error-correcting codes exhibit a number of other desirable properties, including high error thresholds~\cite{dennis2002memory, Tuckett2018Ultrahigh} and natural sets of protected logical gates~\cite{Bombin2010Twist,Fowler2012surface,Brown2017poking, Webster2020defects}. 

The beneficial properties of topological quantum error-correcting codes are fundamentally tied to the physical properties~of the underlying topological order. While the classification of topological phases of matter in two spatial dimensions is well developed~\cite{Levin2005stringnet}, the vast majority of work on two-dimensional topological quantum error-correcting codes has focused on the simplest of topological orders -- namely, those that are locally equivalent to decoupled copies of Kitaev's toric code (TC). This is, to a large extent, due to the effectiveness and simplicity of the Pauli stabilizer formalism, with which the error-correcting properties of the TC can be readily analyzed~\cite{gottesman1997stabilizer,dennis2002memory,Fowler2012surface,Chubb2021statistical}. 

This makes the classification of topological quantum error-correcting codes described within the Pauli stabilizer formalism a problem of both fundamental and practical interest. 
Such error-correcting codes encompass both stabilizer codes and subsystem codes -- wherein the quantum information is stored in only a subsystem of the code space~\cite{Poulin2005stabilizer, Nielsen2007algebraic, Bombin2009fermions, Bombin2010subsystem, Bombin2012universal, Bombin2014structure}. As for the classification of stabilizer codes, it was shown in Ref.~\cite{Bombin2012universal} that, under a technical assumption, all translation invariant (TI) Pauli topological stabilizer codes on systems of qubits are locally equivalent to copies of the TC. The technical assumption was later removed in Ref.~\cite{Haah2018classification}, and it was proven that, more generally, TI Pauli topological stabilizer codes on qudits of prime dimension $p$ are equivalent to copies of the $\mathbb{Z}_p$ TC. 
More recently, new families of TI Pauli topological stabilizer codes were introduced on composite-dimensional qudits, which go beyond copies of the TC, and, in fact, realize all  {(bosonic)}\footnote{ {Throughout the text, we implicitly consider bosonic anyon theories, i.e., where the operator algebra of the underlying degrees of freedom exhibit bosonic commutation relations, such as the case for qudits.}} Abelian anyon theories that admit a gapped boundary to the vacuum~\cite{Ellison2022Pauli}. This demonstrates that the classification of Pauli topological stabilizer codes is significantly richer for the case of composite-dimensional qudits. 

The classification of Pauli topological subsystem codes \cite{Bombin2014structure} has received less attention, despite subsystem codes offering clear benefits over stabilizer codes, such as low-weight checks, which can improve the efficiency of error detection, and the measurement of gauge operators, which can provide additional information for decoding, resulting in, for example, single-shot error correction~\cite{Bravyi2011local, Suchara2011subsystem, Paetznick2013universal, Anderson2014Conversion, Bombin2015Gaugecolorcodes, Bravyi2013Subsystem, Vuillot2019code, Bombin2007branyons, Bombin2015Gaugecolorcodes, Brown2016gaugecolorcode, Brown2020nonClifford}. 
In Refs.~\cite{Bombin2009fermions, Bombin2010subsystem, Bombin2014structure, Bombin2012universal}, it was found that Pauli topological subsystem codes built from qubits can be characterized by anyon theories that are distinct from those of Pauli topological stabilizer codes. In particular, Pauli topological subsystem codes can be characterized by non-modular anyon theories (i.e., those with degenerate braiding relations) as well as chiral anyon theories (i.e., those with no possible gapped boundary to the vacuum). Therefore, Pauli topological subsystem codes promise to capture an even wider class of Abelian anyon theories than stabilizer codes. This naturally leads us to the question: 
what new classes of Pauli topological subsystem codes are made possible by considering systems of composite-dimensional qudits -- furthermore, do they capture all possible Abelian anyon theories?

In this work, we establish that the classification of Pauli topological subsystem codes is indeed at least as rich as the classification of Abelian anyon theories. We do so by~constructing a Pauli topological subsystem code on composite-dimensional qudits for every Abelian anyon theory,\footnote{Here, and throughout the text, we use ``anyon theory'' to broadly refer to braided fusion categories, as opposed to modular tensor categories in particular.} 
including non-modular 
anyon theories and chiral anyon theories. To construct the codes, we leverage the Pauli topological stabilizer models of Ref.~\cite{Ellison2022Pauli} and a process we refer to as gauging out anyon types (see also Ref.~\cite{Bombin2012universal}). At the level of the anyon theory, gauging out a set of anyon types removes all of the anyon types that have nontrivial braiding relations with the gauged out anyon types. This yields Pauli topological subsystem codes, whose anyon theories are proper subtheories of the parent stabilizer codes' anyon theories. 
Therefore, our work lays the foundation for
a complete classification of Pauli topological subsystem codes in two dimensions. Combined with Ref.~\cite{Ellison2022Pauli}, it initiates the investigation of topological quantum error correction with arbitrary Abelian anyon theories -- all within the Pauli stabilizer formalism. 

The paper is organized as follows. 
In Section~\ref{sec:Background}, we establish basic definitions required to discuss Pauli operators on composite-dimensional qudits, and recount the definition of subsystem codes. 
In the following section, Section~\ref{sec: topological subsystem codes}, we define Pauli topological subsystem codes for qudits of arbitrary finite dimension and argue that they can be characterized by Abelian anyon theories. 
In Section~\ref{sec: Z41 parafermion}, we introduce an example of a Pauli topological subsystem code based on the $\mathbb{Z}_4^{(1)}$ Abelian anyon theory, which has degenerate braiding relations. 
In Section~\ref{sec: chiral semion}, we introduce another example of a Pauli topological subsystem code, this one based on the chiral semion anyon theory, which does not admit a gapped boundary condition. 
In Section~\ref{sec: general}, we present our general construction of Pauli topological subsystem codes based on arbitrary Abelian anyon theories. 
Subsequently, in Section~\ref{sec: more examples}, we present further examples of Pauli topological subsystem codes, including one based on a $\mathbb{Z}_3\times\mathbb{Z}_9$ Abelian anyon theory, which has not appeared in the literature before, to the best of our knowledge. 
Finally, in Section~\ref{sec: discussion}, we summarize our findings and discuss the connection between the classification of topological subsystem codes and topological phases of matter. 

The results in the main text are supplemented by a number of appendices. 
We prove a correctability condition and a cleaning lemma for Pauli topological subsystem codes, in Appendix~\ref{app: cleaning lemma}.
In Appendix~\ref{app: 1form}, we describe the concept of a 1-form symmetry with an associated anyon theory. 
In Appendix~\ref{app: cellular}, we review the concepts from cellular homology that are used in our description of 1-form symmetries. 
In Appendix~\ref{app: fluxes}, we describe the concept of a flux in a topological subsystem code, following Ref.~\cite{Bombin2014structure}.
In Appendix~\ref{app: detectable=opaque}, we provide a proof of the technical statement that detectable anyon types are opaque. 
In Appendix~\ref{app: bare logicals and nonlocal stabilizers}, we then show that nontrivial bare logical operators and nonlocal stabilizers in Pauli topological subsystem codes are given by string operators that move anyon types around non-contractible paths. 
The final two appendices, Appendix~\ref{app: TQD review} and \ref{app: WW}, cover the background material used in our general construction of topological subsystem codes.

\section{Background}
\label{sec:Background}

In this section, we define the algebra of Pauli operators on composite-dimensional qudits and review the definition of subsystem codes. We encourage readers that are familiar with these concepts to proceed to Section~\ref{sec: topological subsystem codes} for a definition of topological subsystem codes and a discussion of the anyon theories that characterize them.

\subsection{Primer on composite-dimensional qudits} \label{sec: composite qudits}

Composite-dimensional qudits are essential to the construction of the topological subsystem codes introduced in this work. Therefore, it is important that
we generalize the usual notion of Pauli operators to composite-dimensional qudits. For a qudit of dimension $N$, we label the computational basis states by $\alpha \in \ZZ_N$. The Pauli $X$ and Pauli $Z$ operators on the $N$-dimensional qudit can then be represented in the computational basis as:
\begin{align}
    X \equiv \sum_{\alpha \in \ZZ_N} |\alpha+1\rangle \langle \alpha |, \quad Z \equiv \sum_{\alpha \in \ZZ_N} \omega^\alpha |\alpha \rangle \langle \alpha |,
\end{align}
where the addition is computed modulo $N$, and $\omega = e^{2 \pi i / N}$. We further define a Pauli $Y$ operator as:
\begin{align} \label{eq: Y def}
    Y \equiv 
    \begin{cases} X^\dagger Z^\dagger & \text{if } N \text{ is odd}, \\
    \sqrt{\omega} X^\dagger Z^\dagger & \text{if } N \text{ is even}.
    \end{cases}
\end{align}
For $N=2$, this reproduces the familiar Pauli $Y$ operator, given by $Y=iXZ$.
It can then be checked that the 
Pauli operators satisfy the relations:
\begin{align}
X^N = Y^N = Z^N = 1.
\end{align}
Note that the phase $\sqrt{\omega}$ in Eq.~\eqref{eq: Y def} ensures that $Y^N=1$ for even-dimensional qudits. The commutation relations amongst the Pauli operators are given by:
\begin{align}
X Y = \omega Y X, \qquad Y Z = \omega Z Y,\qquad Z X = \omega X Z.
\end{align}
For an odd-dimensional qudit, the full operator algebra is generated by the Pauli $X$ and Pauli $Z$ operators, while for an even-dimensional qudit, we need to include the Pauli $Y$ operator or the phase $\sqrt{\omega}$ in the generating set. 

For systems of more than one qudit, we index the single-site Pauli operators by their corresponding sites. The single-site Pauli operators at different sites are taken to commute with one another, i.e., for any sites $j$ and $k$ with $j \neq k$, we have:
\begin{align}
    X_j Y_k = Y_k X_j, \qquad Y_j Z_k = Z_k Y_j,\qquad Z_j X_k = X_k Z_j.
\end{align}
We use the term ``Pauli operator'' to refer to any product of finitely-many single-site Pauli operators, and we define the Pauli group $\mathcal{P}$ to be the group of Pauli operators. We say the support of a Pauli operator is the set of sites on which the Pauli operator acts non-identically, and we define the weight of a Pauli operator $P$ to be the number of qudits in the support of $P$. Finally, we say that a Pauli operator is local (or geometrically local) if its support can be contained within a constant-sized region.

\subsection{Review of subsystem codes} \label{sec: subsystem review}

The first step in defining a subsystem code is to specify a stabilizer group $\mathcal{S}$. Recall that a stabilizer group is a group of mutually commuting Pauli operators with the property that the only element of $\mathcal{S}$ proportional to the identity is the identity itself. The stabilizer group defines the code space $\mathcal{H}_C$, which is, by convention, the mutual $+1$ eigenspace of the stabilizers:
\begin{align} \label{eq: code space def}
\mathcal{H}_C \equiv \{|\psi\rangle : S|\psi \rangle = |\psi \rangle, \, \forall S \in \mathcal{S} \}.
\end{align}
The Hilbert space $\mathcal{H}$ then decomposes into the direct sum:
\begin{align} \label{eq: code space factorizes}
\mathcal{H} = \mathcal{H}_C \oplus \mathcal{H}_C^\perp,
\end{align}
where $\mathcal{H}_C^\perp$ is the orthogonal complement of $\mathcal{H}_C$. 

For a subsystem code, quantum information is only stored in a subsystem of the code space, known as the logical subsystem. This is to say that the code space further factorizes as a tensor product of $\mathcal{H}_G$ and $\mathcal{H}_L$:
\begin{align} \label{eq: Hilbert space factorization}
\mathcal{H} =(\mathcal{H}_G \otimes \mathcal{H}_L) \oplus \mathcal{H}_C^\perp,
\end{align} 
where $\mathcal{H}_G$ is the gauge subsystem, and $\mathcal{H}_L$ is the logical subsystem. Therefore, the second step in defining a subsystem code is to specify the subsystems $\mathcal{H}_G$ and $\mathcal{H}_L$, as in Eq.~\eqref{eq: Hilbert space factorization}. This is accomplished by defining a group of Pauli operators $\mathcal{G}$, referred to as the gauge group. The only requirement of the gauge group is that its center, denoted by $\mathcal{Z}(\mathcal{G})$, is equivalent to the stabilizer group up to roots of unity: 
\begin{align} \label{eq: gauge group condition}
    \mathcal{Z}(\mathcal{G}) \propto \mathcal{S}.
\end{align}
This guarantees that the gauge operators, i.e., the elements of $\mathcal{G}$, preserve the code space. 
The action of the gauge operators within the code space defines an algebra of Pauli $X$ and Pauli $Z$ operators on the gauge subsystem, which induces the factorization in Eq.~\eqref{eq: Hilbert space factorization}~\cite{Zanardi2004Tensor}. In other words, the group $\mathcal{G}/\mathcal{S}$ is isomorphic to the group of Pauli operators on the gauge subsystem.
We note that it is common to define the gauge group by a choice of generators, referred to as the gauge generators.

There are a few additional comments that we would like to make about this structure: 
\begin{enumerate} [label={(\roman*)}]
    \item According to the condition in Eq.~\eqref{eq: gauge group condition}, the gauge group is almost sufficient to determine the structure of the subsystem code, since it specifies the stabilizer group up to a choice of phases. Therefore, a subsystem code can alternatively be defined by first specifying the gauge group, then choosing a stabilizer group consistent with Eq.~\eqref{eq: gauge group condition}. This is the~approach used for the examples in this work.
    \item If the gauge group $\mathcal{G}$ is proportional to the stabilizer group $\mathcal{S}$, then the gauge subsystem $\mathcal{H}_G$ is trivial. This means that the logical subsystem and the code space are equivalent, and the subsystem code is equivalent to a stabilizer code defined by $\mathcal{S}$. On the other hand, if the gauge group generates the full operator algebra of the code space, then $\mathcal{H}_L$ is trivial. This is exemplified by the subsystem code in Section~\ref{sec: 1 psi subsystem code}, based on the honeycomb model of Ref.~\cite{kitaev2006anyons}.
    \item For systems of composite-dimensional qudits, the dimension of the logical subsystem may differ from the dimensions of the physical qudits. Depending on the choice of $\mathcal{S}$ and $\mathcal{G}$, the dimension of the logical subsystem can be any factor of the dimension of the full Hilbert space. For example, the logical subsystem of the $\ZZ_4^{(1)}$ subsystem code in Section~\ref{sec: Z41 parafermion} is two-dimensional, despite being defined on four-dimensional physical qudits. 
    \item  {The gauge group needs to include all of the requisite roots of unity to generate a representation of the Pauli group. Throughout the text, we implicitly assume that the gauge group includes all $U(1)$-valued phases.} 
\end{enumerate}

Next, we discuss the logical operators of subsystem codes. Similar to stabilizer codes, we refer to any Pauli operator that preserves the code space as a logical operator. These operators generate the group $\mathcal{Z}_\mathcal{P}(\mathcal{S})$, where $\mathcal{Z}_\mathcal{P}(\mathcal{S})$ denotes the centralizer of the stabilizer group $\mathcal{S}$ over the Pauli group $\mathcal{P}$.\footnote{In other words, $\mathcal{Z}_\mathcal{P}(\mathcal{S})$ is the subgroup of the Pauli operators in $\mathcal{P}$ that commute with every stabilizer of $\mathcal{S}$.} We say a logical operator is nontrivial if it acts non-identically on the logical subsystem. The nontrivial logical operators are given by the set $\mathcal{Z}_\mathcal{P}(\mathcal{S}) - \mathcal{G}$, where the gauge operators have been subtracted, since by construction, they act as the identity on $\mathcal{H}_L$. Up to roots of unity, the group $\mathcal{Z}_\mathcal{P}(\mathcal{S})/\mathcal{G}$ is isomorphic to the group of Pauli operators on the logical subsystem. Thus, we say that the elements of $\mathcal{Z}_\mathcal{P}(\mathcal{S})$ represent Pauli operators on the logical subsystem.

It is convenient to further distinguish between logical operators that act nontrivially on the gauge subsystem and those that are fully supported on the logical subsystem. We define the bare logical operators to be the subgroup of logical operators that act as the identity on the gauge subsystem. The group of bare logical operators $\mathcal{L}$ is given explicitly by:
\begin{align} \label{eq: bare logical def}
    \mathcal{L} \equiv \mathcal{Z}_\mathcal{P}(\mathcal{G}),
\end{align} 
i.e., it is the group of Pauli operators that commute with the gauge group.
Up to roots of unity, the group $\mathcal{L}/\mathcal{S}$ is isomorphic to the group of Pauli operators on the logical subsystem. Thus, the Pauli operators on $\mathcal{H}_L$ can be represented by bare logical operators. We note that the elements of $\mathcal{Z}_{\mathcal{P}}(\mathcal{S})$ are sometimes referred to as dressed logical operators, since they differ from bare logical operators by ``dressing'' them with products of gauge operators.

At this point, we have defined all of the essential structures of subsystem codes. This might motivate one to ask:
what have we gained by sacrificing a portion of the code~space to the gauge subsystem?
In fact, the additional structure leads to two notable advantages over stabilizer codes. 
The first is that the stabilizer syndrome -- i.e., the set of measurement outcomes of the stabilizers -- can be inferred from measurements of the gauge generators.\footnote{Note that, although the gauge operators may be non-Hermitian for systems of qudits, the eigenspaces of each gauge operator are mutually orthogonal. Therefore, we can effectively measure a gauge operator $G$ by instead measuring the sum of projectors: $\sum_{\alpha} \alpha \Pi_{\lambda_\alpha}$, where $\alpha \in \ZZ$ indexes the eigenvalues $\lambda_\alpha$ of $G$, and $\Pi_{\lambda_\alpha}$ is a projector onto the $\lambda_\alpha$ eigenspace.} 
This can simplify the detection of errors, if, for example, the gauge generators have lower weights than the generators of the stabilizer group. This is exemplified by the subsystem toric code of Ref.~\cite{Bravyi2013Subsystem}, where the stabilizer syndrome can be deduced from measurements of three-body gauge operators -- as opposed to the six-body measurements required to measure the stabilizers directly.
We emphasize that the order in which gauge generators are measured requires special care, however, since they do not commute with one another in general.\footnote{See the appendix of Ref.~\cite{Suchara2011subsystem} for sufficient condition for inferring the stabilizer syndrome from the measurements of the gauge generators.}

The second noteworthy advantage of subsystem codes is the ability to gauge fix~\cite{Poulin2005stabilizer}.
Formally, we gauge fix an Abelian subgroup $\mathcal{F}$ of the gauge group $\mathcal{G}$ by defining a new subsystem code, whose gauge group is equal to $\mathcal{Z}_\mathcal{G}(\mathcal{F})$.
We thus keep all of the elements in $\mathcal{G}$ that commute with $\mathcal{F}$. Since $\mathcal{F}$ is Abelian, it belongs to the center of $\mathcal{Z}_\mathcal{G}(\mathcal{F})$. This means that, up to roots of unity, $\mathcal{F}$ is in the stabilizer group of the new subsystem code. Consequently, gauge fixing amounts to adding $\mathcal{F}$ to the stabilizer group and removing any of the gauge operators that fail to commute with $\mathcal{F}$. 
This is a natural procedure to consider, since in the process of determining the stabilizer syndrome, we make consecutive measurements of a commuting set of gauge operators. The effect of measuring an Abelian subgroup of gauge operators $\mathcal{F}$ is the same as gauge fixing $\mathcal{F}$, up to multiplying the gauge operators by roots of unity depending on the measurement outcomes. 

One benefit of gauge fixing is that it provides a general framework for understanding code deformation and lattice surgery~\cite{Vuillot2019code}. In particular, gauge fixing can be used to switch between codes with complementary fault-tolerant gate sets -- thereby enabling universal fault-tolerant quantum computation by appropriately switching between the gauge fixed codes~\cite{Paetznick2013universal, Anderson2014Conversion, Bombin2015Gaugecolorcodes, Brown2020nonClifford}. 
Moreover, by scheduling the measurements of gauge operators carefully, the gauge fixed codes can lead to improved error thresholds \cite{Higgott2021subsystem} and dynamically generated logical qudits~\cite{Hastings2021dynamically, Gidney2021faulttolerant, Haah2022boundarieshoneycomb, Paetznick2022performance, Gidney2022benchmarking}.

 {As described below, another feature of subsystem codes is that they are characterized by a wider range of anyon theories than stabilizer codes. In this work, we construct such subsystem codes using a general process we call ``gauging out''. Abstractly, gauging out maps a subsystem code to another subsystem code using a group of Pauli operators $\mathcal{F}$, which needs not be Abelian. More specifically, given a gauge group $\mathcal{G}$ and a group of Pauli operators $\mathcal{F}$ to be gauged out, we define a new gauge group $\mathcal{G}'$ generated by $\mathcal{G}$ and $\mathcal{F}$, i.e., $\mathcal{G}' \equiv \langle \mathcal{G}, \mathcal{F} \rangle$. In contrast to gauge fixing, $\mathcal{F}$ does not need to be a subgroup of $\mathcal{G}$, nor does it need to be Abelian. Note that $\mathcal{Z}(\mathcal{G}') \subset \mathcal{Z}(\mathcal{G})$, so up to phases, the stabilizer group of the new subsystem code is contained in the stabilizer group of the original subsystem code. In all of the examples in this text, $\mathcal{G}$ is Abelian, so we use gauging out to construct a subsystem code from a stabilizer code.}



\section{Topological subsystem codes} \label{sec: topological subsystem codes}

The focus of this work is on a particular class of subsystem codes -- the topological subsystem codes. These feature gauge groups that can be generated by  {(geometrically)} local\footnote{ {In this text, we use local to mean geometrically local, i.e., the support of the gauge generators can be contained within a constant-sized disk.}} gauge operators and logical subsystems that are robust to local errors. Such properties are desirable for quantum error correcting codes, given that many of the leading quantum computing platforms are limited to local connectivity and suffer from local errors. 
In this section, we begin by defining topological subsystem codes, following Refs.~\cite{Bombin2010subsystem, Bombin2012universal, Bombin2014structure}. Similar to Refs.~\cite{Bombin2010subsystem, Bombin2012universal, Bombin2014structure}, we consider topological subsystem codes that are translation invariant (TI) and defined on two-dimensional lattices. In the second part of this section, we argue that topological subsystem codes can be characterized by Abelian anyon theories. We note that the discussion on Abelian anyon theories agrees with that of Ref.~\cite{Bombin2014structure} for system of qubits. In contrast to Refs.~\cite{Bombin2010subsystem, Bombin2012universal, Bombin2014structure}, throughout this section, we make no assumptions about the dimensions of the physical qudits, see Section~\ref{sec: composite qudits}.

\subsection{Definition of topological subsystem codes} \label{sec: definition of topological subsystem codes}

We now define TI topological subsystem codes in two-dimensions. We point out that the definition can be generalized straightforwardly to higher dimensions. However, we restrict to two-dimensional topological subsystem codes, since these are naturally characterized by Abelian anyon theories. It is also worth noting that the definition of TI topological subsystem codes below is a direct generalization of the definition of TI topological stabilizer codes, given in Ref.~\cite{Haah2018classification}. In particular, we recover the definition in Ref.~\cite{Haah2018classification} if the gauge group is proportional to the stabilizer group.
\begin{definition}[Translation invariant topological subsystem code] \label{def: topological subsystem code}
A two-dimensional translation invariant topological subsystem code is a subsystem code defined on a two-dimensional lattice with the following three properties:
\begin{enumerate}[label={(\roman*)}]
\item \textbf{Translation invariant}: For any gauge operator $G$, every translate of $G$ belongs to the gauge group. 
\item \textbf{Local}: The gauge group $\mathcal{G}$ admits a set of generators whose supports have linear size less than some constant-sized length $\ell_G$.
\item  \textbf{Topological}: On an infinite plane, the stabilizer group $\mathcal{S}$ admits a set of generators whose supports have linear size less than a constant-sized length $\ell_S$, and $\mathcal{Z}_\mathcal{P}(\mathcal{S}) \propto \mathcal{G}$.\footnote{Here,  {$\mathcal{Z}_\mathcal{P}(\mathcal{S})$} is the centralizer of  {$\mathcal{S}$} over the Pauli group $\mathcal{P}$, i.e. it is the group of Pauli operators that commute with  {$\mathcal{S}$}.}
\end{enumerate}
\end{definition} 

Before discussing properties (ii) and (iii) of TI topological subsystem codes, we would like to emphasize that translation invariance is not as restrictive as it might seem. The quantum error correcting properties of topological subsystem codes and the discussion of anyon theories in the next section hold even after conjugating the gauge group by an arbitrary constant-depth Clifford circuit -- which may explicitly break the translation symmetry. Therefore, many of our results also apply to subsystem codes that differ~from TI topological subsystem codes by a constant-depth Clifford circuit. In Section~\ref{sec: discussion}, we~describe how twist defects and boundaries can be further introduced to topological subsystem codes. These may break the translation invariance, but they preserve locality and, in some cases, the topological property.
Intuitively, TI topological subsystem codes capture the universal bulk properties of topological subsystem codes, far from the defects and boundaries.

We now turn to properties (ii) and (iii) in the definition above. First, the local property ensures that the stabilizer syndrome can be inferred from local measurements of the gauge generators.
It is important to note that, although the gauge group is required to have a set of local generators, the stabilizer group itself does not need to admit a set of local generators (except on an infinite plane). As illustrated by the example in Section~\ref{sec: Z41 parafermion}, topological subsystem codes may indeed have stabilizers that cannot be generated by local stabilizers. We refer to such stabilizers as nonlocal stabilizers, and note that their existence is a key difference between topological subsystem codes and topological stabilizer codes.
Second, the topological property of topological subsystem codes tells us that there are no logical operators or nonlocal stabilizers on an infinite plane. This agrees with our intuition that the logical operators of topological subsystem codes should be supported on topologically nontrivial regions -- such as a path that wraps around a non-contractible loop of the torus. 

The topological property also tells us about the errors that can be detected and corrected by topological subsystem codes. 
To unpack the topological property,
we define the subgroup of locally generated stabilizers, denoted by $\tilde{\mathcal{S}}$. In particular, every element of $\tilde{\mathcal{S}}$ can be generated by geometrically local stabilizers. As noted above, the subgroup $\tilde{\mathcal{S}}$ may differ from the full stabilizer group $\mathcal{S}$, for topological subsystem codes.
We define a code space $\mathcal{H}_{\tilde{C}}$ associated to $\tilde{\mathcal{S}}$ as:
\begin{align} \label{eq: local code space}
\mathcal{H}_{\tilde{C}} \equiv \{|\psi \rangle : S |\psi \rangle = |\psi \rangle, \, \forall S \in \tilde{\mathcal{S}} \}.
\end{align}
Notably, the nonlocal stabilizers, i.e., the elements of $\mathcal{S} - \tilde{\mathcal{S}}$, are left unfixed in $\mathcal{H}_{\tilde{C}}$. 

With this, we can specify the set of correctable errors for a TI topological subsystem code on an $L \times L$ torus. 
We let $\mathsf{E}$ be the set of Pauli operators supported on regions whose linear size is less than $L/2 - \ell_S$. We prove in Appendix~\ref{app: cleaning lemma} that, for any $E_1,E_2 \in \mathsf{E}$, the following condition is satisfied for some gauge operator $G$:
\begin{align} \label{eq: topological subsystem code ii}
{\Pi}_{\tilde{C}} E_1^\dagger E_2 {\Pi}_{\tilde{C}} \propto G \Pi_{\tilde{C}},
\end{align}
where $\Pi_{\tilde{C}}$ is the projector onto the code space $\mathcal{H}_{\tilde{C}}$. 
We point out that this property is stronger than the correctability condition of Refs.~\cite{Nielsen2007algebraic, Poulin2005stabilizer},
since here, 
we have replaced the projector onto the code space $\mathcal{H}_C$ with a projector onto $\mathcal{H}_{\tilde{C}}$. As such, the condition in Eq.~\eqref{eq: topological subsystem code ii} guarantees that Pauli operators supported entirely within a region of linear size $L-\ell_S$ create errors 
that can be detected and corrected using the measurement outcomes of the local stabilizers (see Appendix~\ref{app: cleaning lemma}).

The correctability condition in Eq.~\eqref{eq: topological subsystem code ii} also implies that the nontrivial bare logical operators and nonlocal stabilizers cannot be fully supported in regions whose linear size is less than $L - \ell_S$. In Appendix~\ref{app: cleaning lemma}, we prove a cleaning lemma for topological subsystem codes, which shows that the nontrivial bare logical operators and nonlocal stabilizers on a torus can be redefined by elements of $\tilde{\mathcal{S}}$ so as to be supported on the non-contractible region pictured in Fig.~\ref{fig: cleaning}. Further, in Appendix~\ref{app: bare logicals and nonlocal stabilizers}, we argue that the Pauli operators on the logical subsystem can be represented by string operators wrapped around non-contractible paths of the torus.

\begin{figure}[t]
\centering
\includegraphics[width=.4\textwidth]{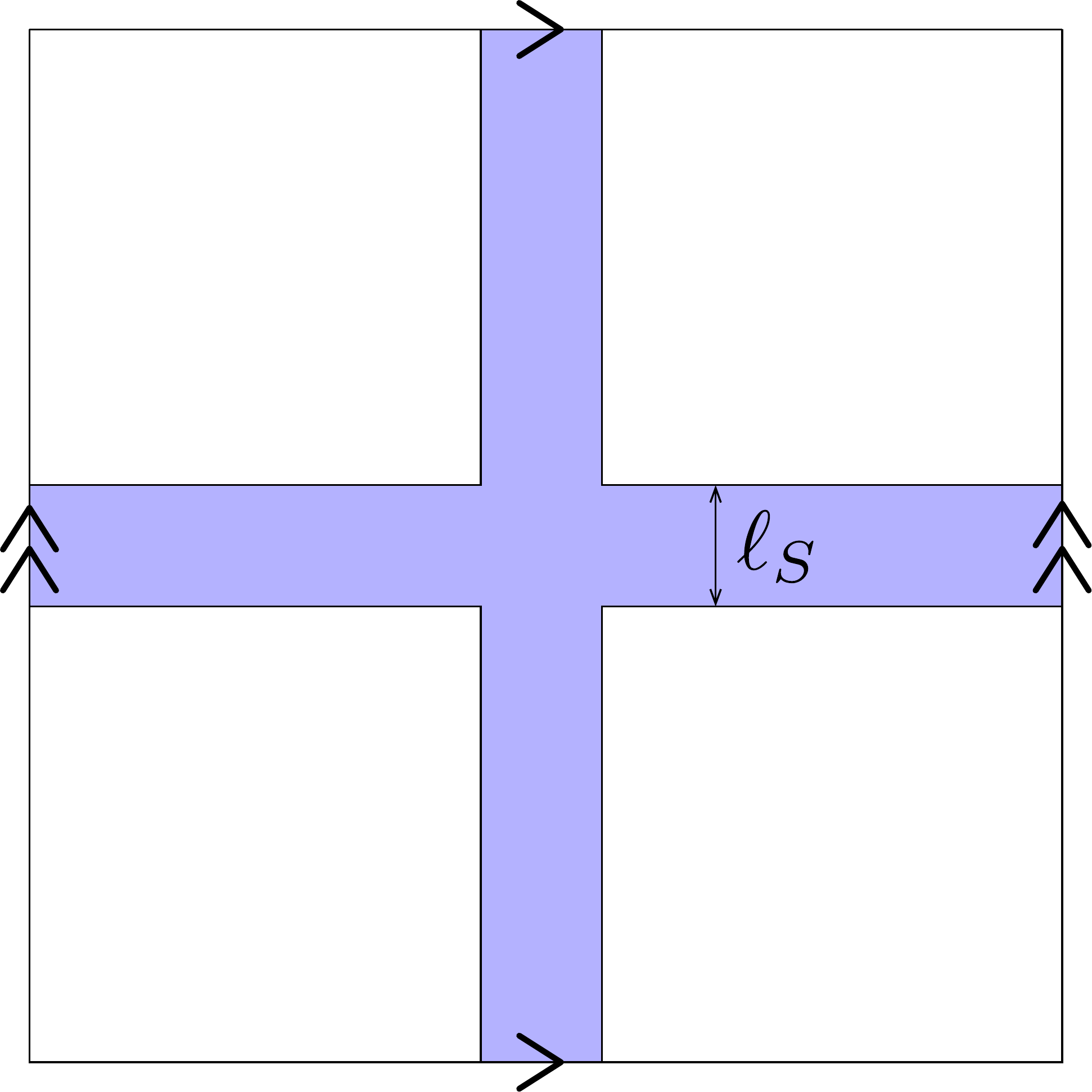}
\caption{The cleaning lemma in Appendix~\ref{app: cleaning lemma} shows that, for any nontrivial bare logical operator or nonlocal stabilizer $T$ on a torus, there is a locally generated stabilizer $S \in \tilde{\mathcal{S}}$ such that the operator $TS$ is fully supported on a region of intersecting strips (blue) with width less than $\ell_S$.}
\label{fig: cleaning}
\end{figure}

Moving forward, we find it instructive to note that a topological subsystem code defines a family of quasi-local Hamiltonians, the terms of which are the gauge operators weighted by coefficients whose strengths are bounded by a function of the size of the support. More explicitly, the quasi-local Hamiltonians are of the form:
\begin{align} \label{eq: subsystem Hamiltonian}
H_{\{J_G\}} = -\sum_{G \in \mathcal{G}} J_G G + \text{h.c.},
\end{align}
for some choice of coefficients $\{J_G\}$, whose strengths are bounded as:
\begin{align}
    |J_G| \leq J e^{-\mu  l(G)}.
\end{align}
Here, $J,\mu > 0$ are a choice of constants that are independent of the system size, and $l(G)$ is the linear size of the support of $G$. This is to say that the strength of the interaction $J_G$ decays exponentially with the range of the interaction $l(G)$.
By varying the coefficients $\{J_G\}$ of the Hamiltonian in Eq.~\eqref{eq: subsystem Hamiltonian}, we obtain the full parameter space of quasi-local Hamiltonians associated to the topological subsystem code. In general, the elements of $\mathcal{G}$ do not commute with one another, so the Hamiltonians might not be exactly-solvable. Therefore, unlike topological stabilizer codes, which correspond to a single exactly-solvable local Hamiltonian, topological subsystem codes generally correspond to a parameter space of frustrated Hamiltonians -- with distinct phases of matter and phase transitions.  {Note that we consider quasi-local Hamiltonians here, since those give a physically meaningful set of Hamiltonians. More generally, one could consider nonlocal or $k$-local Hamiltonians associated to subsystem codes.}

The elements of the bare logical group $\mathcal{L}$ can be interpreted as symmetries, or conserved quantities, common to every Hamiltonian in the parameter space. This is because, by definition [Eq.~\eqref{eq: bare logical def}], the elements of $\mathcal{L}$ commute with the elements of $\mathcal{G}$ and thus commute with the Hamiltonians. 
The nonlocal conserved quantities associated to nontrivial bare logical operators then guarantee that there are degenerate eigenspaces in which quantum information can be stored -- since the nontrivial bare logical operators come in non-commuting pairs. In contrast to topological stabilizer codes, the quantum information is not necessarily encoded in the ground-state subspace.
For the topological subsystem codes defined in this work, many of the conserved quantities of $\mathcal{L}$ are reminiscent of loops of anyon string operators in topological orders. This motivates characterizing topological subsystem codes by anyon theories, as elaborated on in the next section.

We also note that gauge fixing, as described at the end of Section~\ref{sec: subsystem review}, admits a natural interpretation in terms of the family of Hamiltonians $H_{\{J_G\}}$. To see this, we consider gauge fixing an Abelian subgroup $\mathcal{F}$ of the gauge group $\mathcal{G}$. 
After gauge fixing, the system is left in a definite eigenstate of the operators in $\mathcal{F}$. For example, we can consider the case in which the system is in a mutual $+1$ eigenstate of the gauge fixed operators. This amounts to taking the coefficient $J_G$ to infinity for any $G \in \mathcal{F}$.\footnote{This can be generalized to other eigenspaces by adding appropriate roots of unity to the coefficients.} We can then study the resulting Hamiltonian using perturbation theory. Intuitively, the terms of the effective Hamiltonian commute with the elements of $\mathcal{F}$ and generate a group proportional to $\mathcal{Z}_\mathcal{P}(\mathcal{F})$, which is precisely the gauge group after gauge fixing. We caution that, in general, the subsystem code obtained from gauge fixing need not satisfy the properties of a topological subsystem code in Definition~\ref{def: topological subsystem code}.\footnote{For instance, the topological subsystem code may have a nonlocal stabilizer on a torus. If the nonlocal stabilizer does not belong to $\mathcal{F}$, then it becomes a nonlocal gauge operator after gauge fixing.}

\subsection{Anyon theories of topological subsystem codes} \label{sec: anyon theories for subsystem codes}

In what follows, we characterize two-dimensional topological subsystem codes by Abelian anyon theories. 
This is a subtle task,
given that topological subsystem codes correspond to a parameter space of Hamiltonians
 $H_{\{J_G\}}$,
which may exhibit different topological orders based on the parameters
$\{J_G\}$. The main goal of this section is to clarify the sense in which a topological subsystem code can be characterized by a single anyon theory. After reviewing the standard description of anyon theories in terms of the local excitations of a gapped Hamiltonian, we formulate the anyon theories of topological subsystem codes more abstractly -- in terms of certain homomorphisms from the gauge group to $U(1)$, which we refer to as gauge twists. We find that, heuristically,
the anyon theory of a topological subsystem code is given by a subset of anyons common to all of the gapped Hamiltonians $H_{\{J_G\}}$ in the parameter space. In Appendix~\ref{app: 1form}, we give an alternative description of the anyon theory by considering the $1$-form symmetries generated by the conserved quantities of $\mathcal{L}$.
 
At an informal level, anyons are local excitations of a fixed gapped Hamiltonian.\footnote{We refer to Refs.~\cite{Doplicher1971observablesI, Doplicher1974observablesII, Cha2020Charges} for a more formal definition of anyons.} 
The first step in defining an anyon theory is to then organize the anyons into superselection sectors, where two anyons belong to the same superselection sector, if one can be constructed from the other by a local operator~\cite{kitaev2006anyons}.
It is often assumed that the superselection sectors can be labeled by a finite set of anyon types $\{1, a, b, c, \ldots\}$, 
where $1$ labels the superselection sector containing the trivial excitation (i.e., no excitation). After defining the anyon types, we can specify the fusion rules. Two or more anyons at the same location can be fused together by considering the superselection sector of the composite excitation. We denote the fusion product of the anyon types $a$ and $b$ as $a \times b$ or simply $ab$. For Abelian anyon theories, which are the focus of this work, the anyon types generate an Abelian group under fusion. 

The remaining data needed to specify an anyon theory are the so-called $F$- and $R$-symbols. The $F$-symbol is determined by the associativity of fusion, and the $R$-symbol encodes the braiding relations of the anyons and the exchange statistics -- i.e., the phase obtained by swapping two anyon types of the same type~\cite{kitaev2006anyons, Kawagoe2020microscopic}. For Abelian anyon theories, the $F$- and $R$-symbols are fully determined by the exchange statistics~\cite{Wang2020abelian}.
Thus, Abelian anyon theories are characterized by just two pieces of data: \textbf{(i)} an Abelian group $A$, specifying the anyon types and their fusion rules, and \textbf{(ii)} a function $\theta$ from $A$ to $U(1)$, encoding the $U(1)$-valued exchange statistics of the anyons. An Abelian anyon theory $\mathcal{A}$ is hence defined by a pair $(A,\theta)$. We give a general parameterization of this data in Section~\ref{sec: general}.

\begin{figure}[t] 
\centering
\raisebox{1cm}{\hbox{\includegraphics[scale=.3]{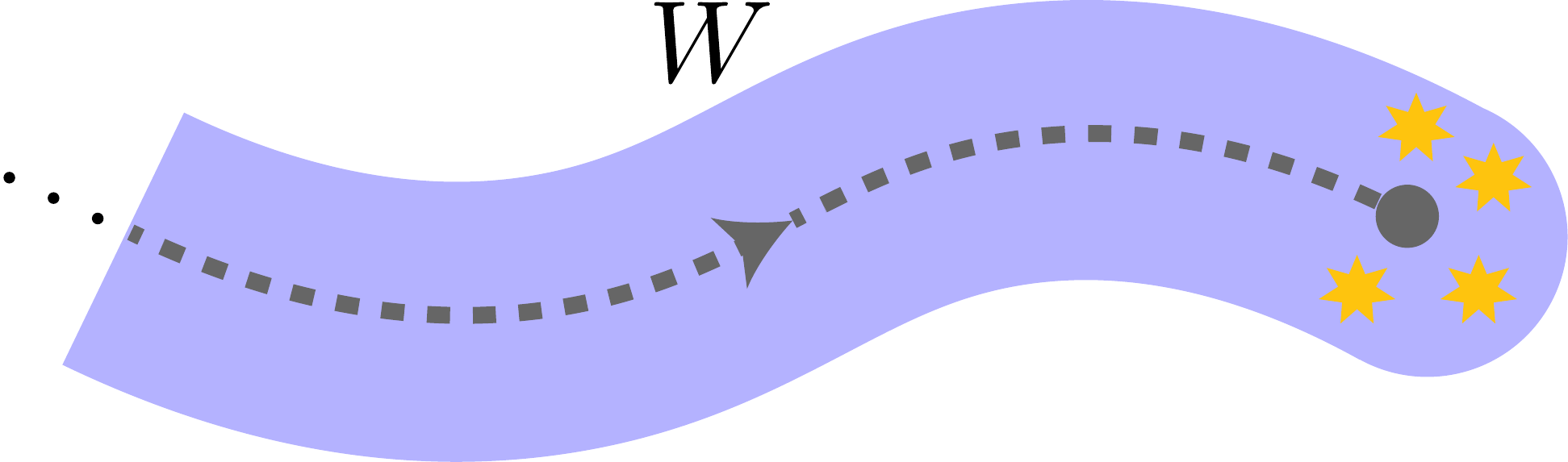}}}
\caption{At a heuristic level, any gauge twist $\Gamma$ that cannot be written as $\Phi_P$ in Eq.~\eqref{eq: trivial gauge functional} is defined by the commutation relations of some string operator $W$ (blue) supported on a semi-infinite path (gray), such that, for any gauge operator $G$, $\Gamma(G) = WGW^\dagger G^\dagger$. The string operator $W$ commutes with the gauge generators along its length and fails to commute with gauge generators near its endpoint (yellow). }
\label{fig: anyondef}
\end{figure}

Given that topological subsystem codes correspond to a family of Hamiltonians $H_{\{J_G\}}$, 
the definition of anyon types in terms of local excitations is, in general, ambiguous. 
We instead define the anyon types of topological subsystem codes more abstractly, by defining the concept of a gauge twist.
\begin{definition}[Gauge twist] \label{def: gauge twist}
Letting $\mathcal{G}$ be the gauge group of a TI Pauli topological subsystem code, a gauge twist is a group homomorphism $\Gamma: \mathcal{G} \to U(1)$, such that $\Gamma(G)=1$ for all but finitely many $G$ in some set of independent local generators of $\mathcal{G}$. 
\end{definition}
We note that, given that every gauge twist $\Gamma$ is a homomorphism, if a gauge operator $G$ has order $q$, then $\Gamma(G)$ satisfies $\Gamma(G)^q=1$. Consequently, $\Gamma(G)=\omega$, for some $q$th root of unity $\omega \in U(1)$. We also note that gauge twists are local on an infinite plane, in the sense that, the set of local gauge generators for which $\Gamma(G) \neq 1$ can be contained within a constant-sized region.

A special class of gauge twists are those that can be reproduced from the commutation relations of a Pauli operator $P$. For any Pauli operator $P$, we define the gauge twist $\Phi_P$ as the commutator with $P$, i.e., for any gauge operator $G$:
\begin{align} \label{eq: trivial gauge functional}
    \Phi_P(G) \equiv PGP^\dagger G^\dagger.
\end{align}
The function $\Phi_P$ is a group homomorphism, since for any $G,G'\in \mathcal{G}$, it satisfies:
\begin{align}
    \Phi_P(G)\Phi_P(G') = \left(PGP^\dagger G^\dagger \right) \left( P G'P^\dagger G'^\dagger \right) = P GG' P^\dagger G'^\dagger G^\dagger = \Phi_P(GG'),
\end{align}
In the second equality, we have used that $G^\dagger$ commutes with the phase $\Phi_P(G')$.
Furthermore, since Pauli operators are, by definition,\footnote{See Section~\ref{sec: composite qudits}} supported on finitely many sites, every Pauli operator $P$ fails to commute with only finitely many gauge generators, for any set of local independent gauge generators. Therefore, $\Phi_P$ is indeed a gauge twist. 

Importantly, there are topological subsystem codes featuring gauge twists that cannot be written in the form $\Phi_P$, for any Pauli operator $P$. 
We find it useful to think of these gauge twists as instead arising from the commutation relations of semi-infinite products of Pauli operators supported along a path that extends to spatial infinity, as shown in Fig.~\ref{fig: anyondef}. 
That is, we think of $\Gamma$ as being defined as $\Gamma(G) = WGW^\dagger G^\dagger$, for any gauge operator $G$, where the semi-infinite string operator $W$ only fails to commute with finitely many gauge generators localized at its endpoint. This captures the intuition that a single anyon can be created at the endpoint of a semi-infinite string operator. We caution that this interpretation of gauge twists should be taken loosely, since infinite products of Pauli operators are not generally well-defined (see Refs.~\cite{Naaijkens2017Infinite} and~\cite{Witten2021Thermodynamic}, for example).

Nonetheless, this leads us to a definition of the anyon types of a topological subsystem code. We define the anyon types of a TI Pauli topological subsystem to be equivalence classes of gauge twists on an infinite plane.
\begin{definition}[Anyon types] \label{def: anyon types subsystem}
The anyon type of a TI Pauli topological subsystem code are the equivalence classes of gauge twists under the following equivalence relation: two gauge twists $\Gamma$ and $\hat{\Gamma}$ are equivalent, or represent the same anyon type, if there exists a Pauli operator $P$, such that $\Gamma(G) = \Phi_P(G)\hat{\Gamma}(G)$, for all $G$ in the gauge group.
\end{definition}
In other words, two gauge twists belong to the same equivalence class if they differ only by the commutation relations of some Pauli operator. We write the equivalence class represented by a gauge twist $\Gamma$ as $[\Gamma]$. We define the trivial anyon type to be the equivalence class $[\Gamma^1]$, where $\Gamma^1$ is the gauge twist satisfying $\Gamma^1(G) = 1$, for every gauge operator $G$. The nontrivial anyon types are then represented by gauge twists that cannot be written as $\Phi_P$ as in Eq.~\eqref{eq: trivial gauge functional}. Intuitively, we think of the nontrivial anyon types as corresponding to semi-infinite string operators $W$ whose commutation relations with the gauge operators cannot be reproduced by a Pauli operator.
In what follows, we assume that TI topological subsystem codes have finitely many nontrivial anyon types. We note that, in Ref.~\cite{Bombin2014structure}, it was proven that there are finitely many anyon types in TI topological subsystem codes built from qubits.\footnote{The proof relies on the fact that one can always find a non-redundant TI set of local generators of the gauge group. For systems of composite-dimensional qudits, however, redundancy may be unavoidable. For example, the stabilizer group of the DS stabilizer code in Ref.~\cite{Ellison2022Pauli} does not admit a set of local generators that are simultaneously TI and non-redundant.}

To gain intuition for this definition of anyon types, it is helpful to consider the case in which the gauge group $\mathcal{G}$ is proportional to the stabilizer group $\mathcal{S}$, since we recover the usual anyon theory of a topological stabilizer code. In the case that $\mathcal{G}$ is proportional to $\mathcal{S}$, the Hamiltonians defined in Eq.~\eqref{eq: subsystem Hamiltonian} are Pauli stabilizer Hamiltonians of the form:
\begin{align}
    H_{\{J_S\}} = - \sum_{S \in \mathcal{S}} J_S S + \text{h.c.},
\end{align}
for some set of coefficients $\{J_S\}$, whose strength decays exponentially with the linear size of the support of $S$. Here, we assume that the coefficients $J_S$ are nonzero.
To see that a gauge twist $\Gamma$ corresponds to a local excitation, we define a twisted stabilizer Hamiltonian by multiplying each stabilizer term $S$ by $\Gamma(S)$:
\begin{align} \label{eq: gauge twisted stabilizer Hamiltonian}
    {H}^{\Gamma}_{\{J_S\}} = - \sum_{S \in \mathcal{S}} J_S \Gamma(S)S + \text{h.c.}.
\end{align}
Relative to the unmodified Hamiltonian, the ground state eigenvalue of $S$ is twisted by the phase $\Gamma(S)$. Therefore, the ground states of the Hamiltonians in Eq.~\eqref{eq: gauge twisted stabilizer Hamiltonian} correspond to excitations of the unmodified Hamiltonians $H_{\{J_S\}}$. We say that the stabilizer term $S$ is violated if $\Gamma(S)\neq 1$. Furthermore, on an infinite plane, the excitations are local, because the subset of violated stabilizers in $\mathsf{S}_{loc}$
can be contained in a constant-sized region. 
We see that, in this case, two gauge twists $\Gamma$ and $\hat{\Gamma}$ are equivalent if they correspond to the same local excitation of the stabilizer Hamiltonians up to the action of a Pauli operator $P$. 

For more general topological subsystem codes, where the gauge group might not be proportional to the stabilizer group, the gauge twists need not correspond to local excitations. This is because the Hamiltonians in Eq.~\eqref{eq: subsystem Hamiltonian} are typically frustrated, and hence, the ground states are not necessarily eigenstates of the gauge operators. 
We find it convenient to, nonetheless, adopt the language from the case of topological stabilizer codes and say that $\Gamma$ violates a gauge operator $G$ if $\Gamma(G) \neq 1$. 

To emphasize the distinction between the anyon types of topological subsystem codes and those of topological stabilizer codes,
we find it instructive to divide the anyon types of topological subsystem codes into three disjoint sets -- based on the collection of gauge operators violated by the representative gauge twists. \\

\vspace{-.2cm}

\begin{adjustwidth}{.4cm}{.4cm}
\noindent \textbf{Trivial:} The first family consists of the sole anyon type that can be represented by a gauge twist that does not violate any gauge operators. By definition, the only anyon type in this set is the trivial anyon type. \\
\end{adjustwidth}
\vspace{-.2cm}
\begin{adjustwidth}{.4cm}{.4cm}
\noindent \textbf{Detectable:} The second family of anyon types are those for which every representative gauge twist violates at least one stabilizer. We call these anyon types detectable, since they can be detected by measuring stabilizers. \\
\end{adjustwidth}
\vspace{-.2cm}
\begin{adjustwidth}{.4cm}{.4cm}
\noindent \textbf{Undetectable:} The last family of anyon types can be represented by a gauge twist that only violates Pauli operators in $\mathcal{G}-\mathcal{S}$. We refer to these anyon types as undetectable anyon types. 
This is motivated by the fact 
that the stabilizers are unable to distinguish the undetectable anyon types from the trivial anyon type. This family of anyon types is unique to topological subsystem codes, since, for topological stabilizer codes, the gauge twists either do not violate any stabilizers or violate at least one stabilizer. \\
\end{adjustwidth}
\vspace{-.2cm}

Having defined the anyon types of topological subsytem codes, we next discuss their fusion rules and exchange statistics in terms of gauge twists.
The fusion of two anyon types $a$ and $b$ is given by multiplying representative gauge twists for the anyon types. 
To make this explicit, let $a$ and $b$ be the equivalence classes of gauge twists $[\Gamma^a]$ and $[\Gamma^b]$, respectively. We define the fusion of $a$ and $b$ to be the anyon type $[\Gamma^a \cdot \Gamma^b]$, where the gauge twist $\Gamma^a \cdot \Gamma^b$ is given by its action on an arbitrary gauge operator $G$:
\begin{align} \label{eq: gauge twist product}
    \Gamma^a \cdot \Gamma^b (G) \equiv \Gamma^a(G)\Gamma^b(G).
\end{align}
Explicitly, the fusion of $a$ and $b$ is:
\begin{align} \label{eq: fusion def}
    a \times b = [\Gamma^a] \times [\Gamma^b] = [\Gamma^a \cdot \Gamma^b].
\end{align}
It can be checked,\footnote{Explicitly, we let $P_a$ and $P_b$ be the Pauli operators such that, for any gauge operator $G$, we have $\Gamma^a(G) = \Phi_{P_a}(G) \hat{\Gamma}^a(G)$ and $\Gamma^b(G) = \Phi_{P_b}(G) \hat{\Gamma}^b(G)$. This implies that the composition $\Gamma^{a} \cdot \Gamma^b (G)$ can be written as $\Phi_{P_a}(G) \Phi_{P_b}(G) \hat{\Gamma}^a(G) \hat{\Gamma}^b(G)$. Using that $P_b$ commutes with $\Phi_{P_a}(G)$, we find that $\Gamma^{a} \cdot \Gamma^b (G)$ is equal to $(P_bP_a G P_a^\dagger P_b^\dagger G ) \hat{\Gamma}^a(G) \hat{\Gamma}^b(G) = \Phi_{P_bP_a}(G)\hat{\Gamma}^a \cdot \hat{\Gamma}^b(G)$. Hence, $\Gamma^{a}\cdot \Gamma^b$ and $\hat{\Gamma}^{a} \cdot \hat{\Gamma}^b$ represent the same anyon type.} that this definition is well-defined. That is, if two other gauge twists $\hat{\Gamma}^{a}$ and $\hat{\Gamma}^{b}$ represent $a$ and $b$, respectively, then the composition $\hat{\Gamma}^a \cdot \hat{\Gamma}^b$ belongs to the same equivalence class as $\Gamma^{a} \cdot \Gamma^b$.

The equivalence classes of gauge twists, furthermore, form an Abelian group under the operation in Eq.~\eqref{eq: fusion def}. The identity is the trivial anyon type $[\Gamma^1]$, and 
the inverse of an anyon type $[\Gamma^a]$ is $[\Gamma^{a^{-1}}]$, where the gauge twist $\Gamma^{a^{-1}}$ acts on an arbitrary gauge operator $G$ as:
\begin{align}
    \Gamma^{a^{-1}}(G) \equiv \Gamma^a(G)^*.
\end{align}
Since the product of gauge twists in Eq.~\eqref{eq: gauge twist product} is commutative, the anyon types of a topological subsystem code generate an Abelian group under fusion. We note that, while the detectable anyon types can fuse into undetectable anyon types, the undetectable anyon types cannot fuse into detectable anyon types. Thus, the undetectable anyon types and the trivial anyon type form a subgroup under fusion.

\begin{figure}[t]
\centering
\includegraphics[scale=.3]{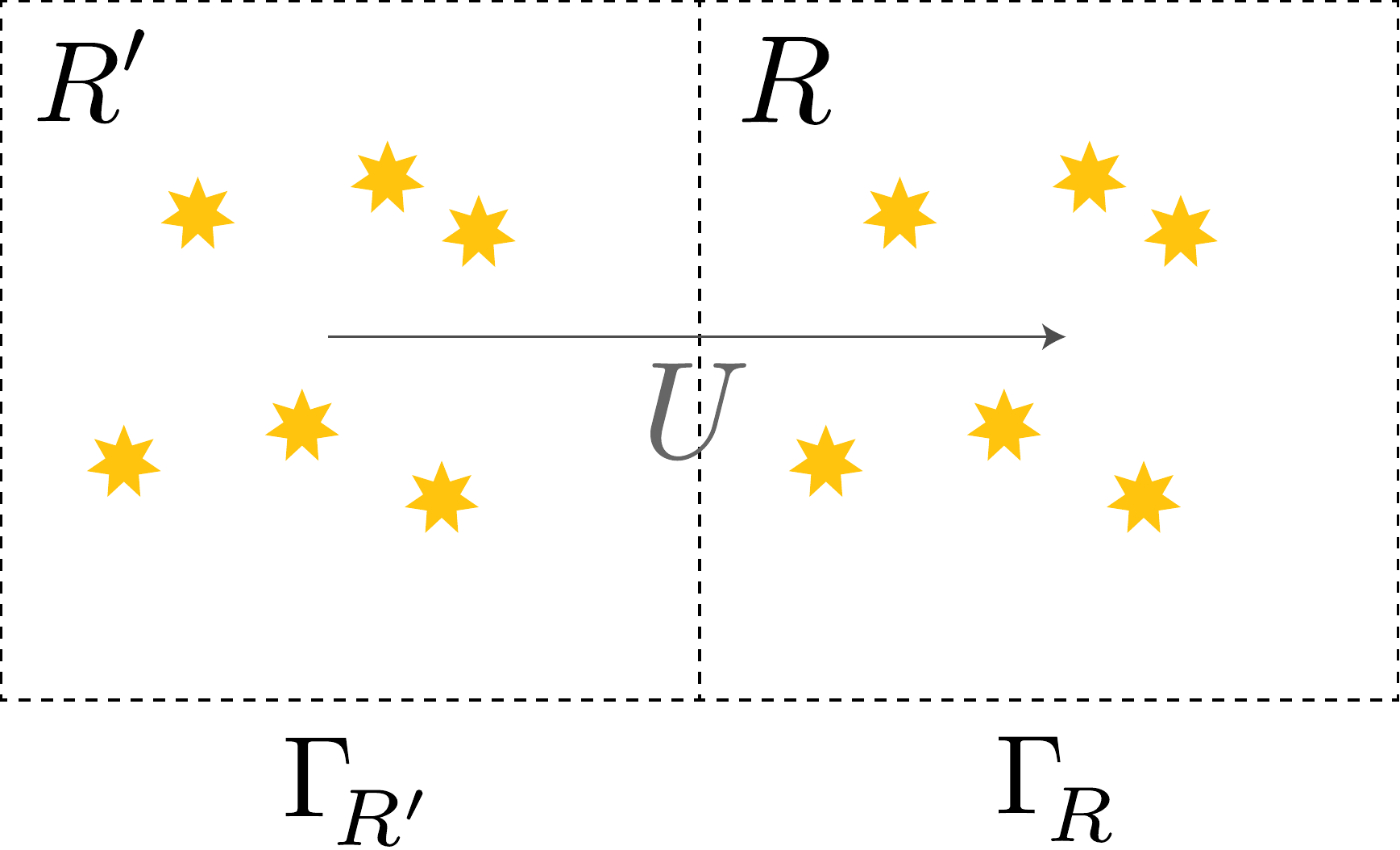}
\caption{If a gauge twist $\Gamma_{R'}$ only violates gauge generators (yellow, left) in a region $R'$, then the action of the translation $U$ on $\Gamma_{R'}$ produces a gauge twist ${}^U\Gamma_{R'}$ that only violates gauge generators (yellow, right) in the translated region $R$.}
\label{fig: violationtranslation}
\end{figure}

The remaining data needed to characterize the Abelian anyon theory are the exchange statistics. 
To clarify the sense in which the anyon types can be moved around the system and exchanged, we say a gauge twist $\Gamma_R$ is localized to a region $R$ if $\Gamma_R$ violates only the gauge generators of $\mathsf{G}_{loc}$ supported on $R$, where $\mathsf{G}_{loc}$ denotes some independent set of local gauge generators. We further say that a gauge twist $\Gamma_R^a$ represents an anyon type $a$ at $R$, if $\Gamma^a_R$ represents $a$ and is localized to $R$.

We next define string operators that move the anyon types, by considering the action of the translation symmetry on the gauge twists. We let $U$ represent a translation symmetry operator, which acts on an arbitrary gauge operator $G$ as $UGU^\dagger$. The induced action of the translation symmetry $U$ on a gauge twist $\Gamma$ is $U: \Gamma \mapsto {}^U\Gamma$, where ${}^U\Gamma$ is defined as:
\begin{align}
    {}^U\Gamma(G) \equiv \Gamma(U G U^\dagger).
\end{align}
In words, we translate the gauge operator $G$ and then apply the gauge twist $\Gamma$.
Notice that, if $\Gamma$ only violates gauge generators supported on $R$, then ${}^U\Gamma$ only violates gauge generators supported on the translated region $R'$ (Fig.~\ref{fig: violationtranslation}).

Importantly, the gauge twist ${}^U\Gamma$ may represent a different anyon type than $\Gamma$. In other words, the translation symmetry may act nontrivially on the anyon types. Since~(by assumption) there are finitely many anyon types, after a sufficient amount of coarse graining, i.e., blocking together unit cells and redefining the unit translation, the anyon types transform trivially under the coarse grained translation symmetry. We assume that the system has been coarse grained so that, if $\Gamma^a_R$ represents an anyon type $a$ at a region $R$, then ${}^U\Gamma^a_R$ represents $a$ at the translated region, for any translation symmetry operator $U$, i.e.:
\begin{align}
    [\Gamma^a_R] = [{}^U\Gamma^a_R].
\end{align}

\begin{figure}[t]
\centering
\includegraphics[scale=.3]{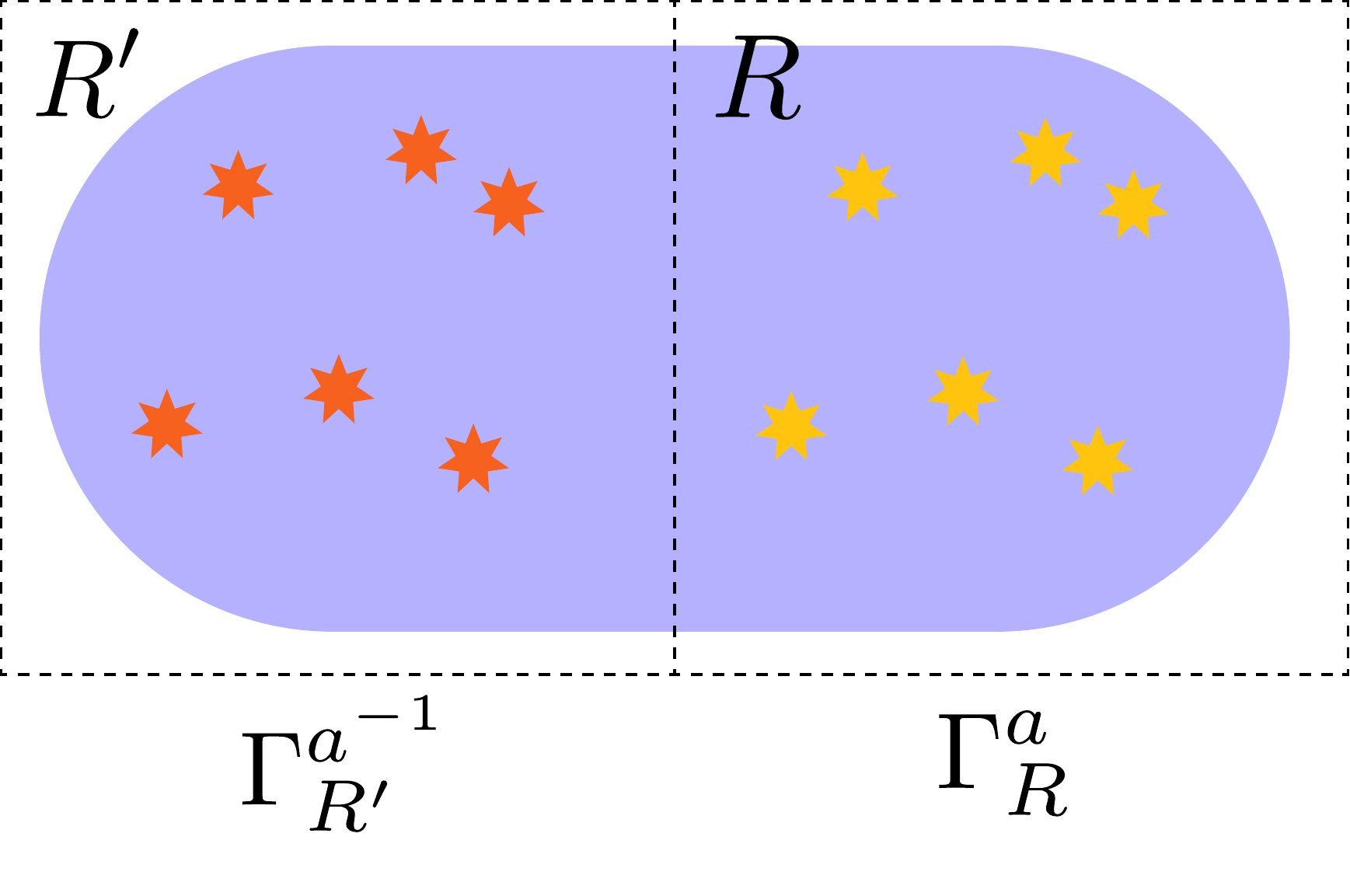}
\caption{$\Gamma^a_R$ violates gauge generators (yellow) within $R$, while $\Gamma^{a^{-1}}_{R'}$ violates gauge generators (orange) in the region $R'$, which is a translate of $R$ by a unit translation. The gauge twist $\Gamma^a_R \cdot \Gamma^{a^{-1}}_{R'}$ represents the trivial anyon type, by construction, and defines a short string operator $W_{RR'}^a$ that moves the anyon type $a$ from $R'$ to $R$.}
\label{fig: shortstringdef}
\end{figure}

\begin{figure*}[t]
\centering
\subfloat[\label{fig: buildingstring1}]{\includegraphics[width=.4\textwidth]{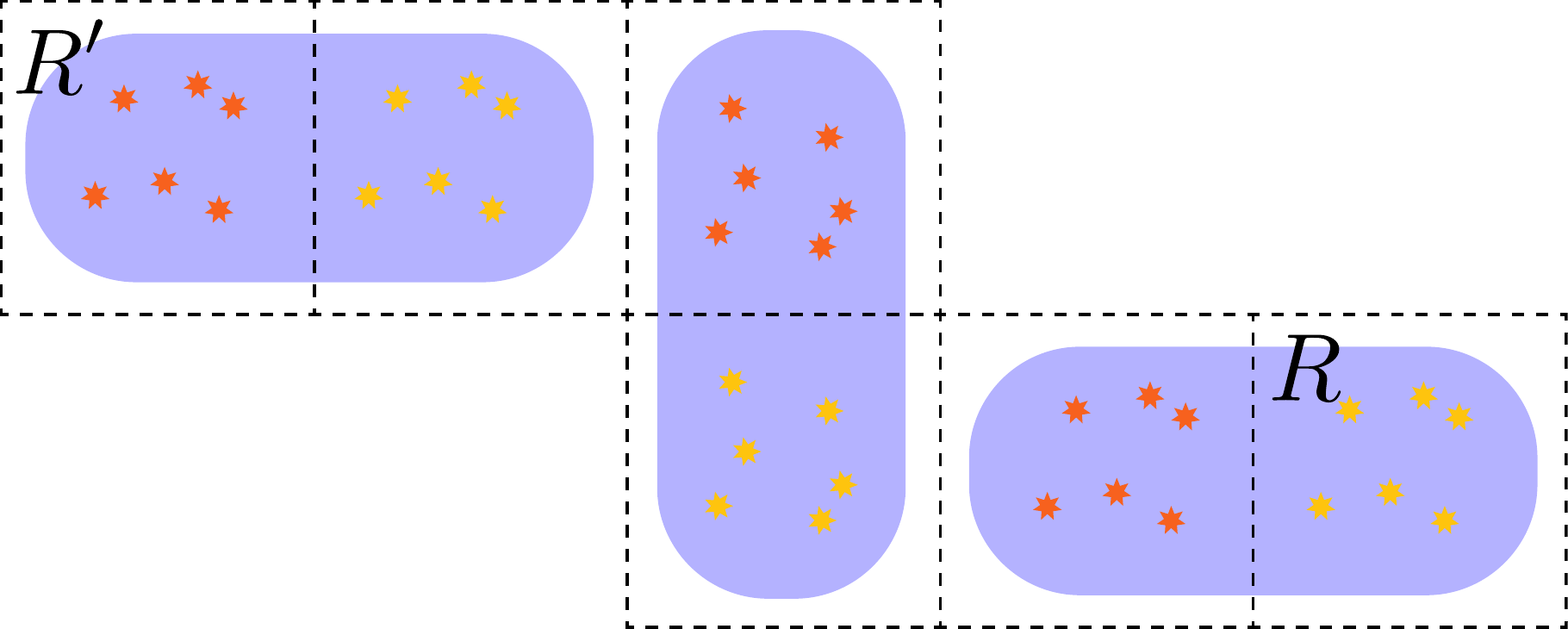}} \qquad
\subfloat[\label{fig: buildingstring2}]{\includegraphics[width=.4\textwidth]{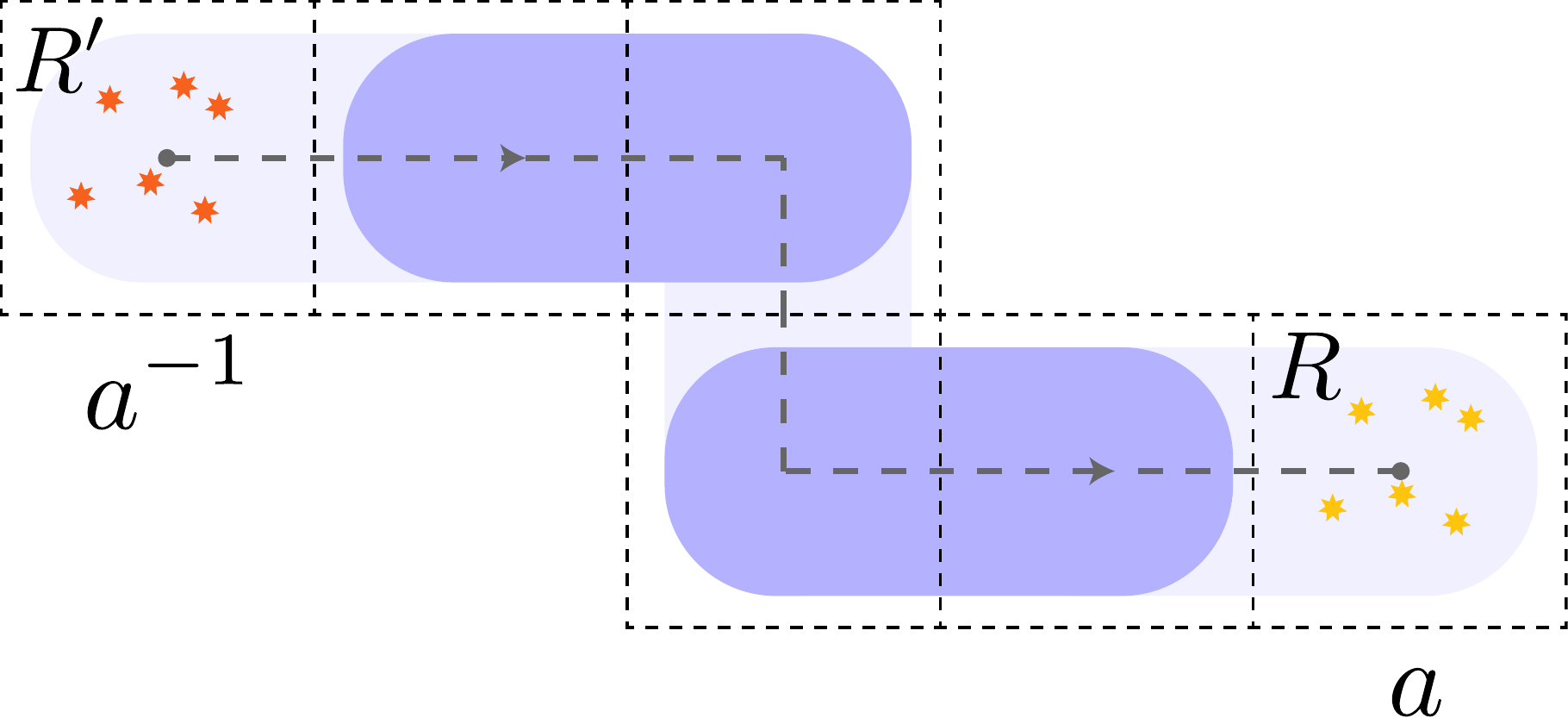}}
\caption{To create a string operator that moves the anyon type $a$ from $R'$ to another region $R$, we do the following. (a) We apply Pauli operators (blue) along a path from $R'$ to $R$ to create pairs of $a$ and $a^{-1}$ anyon types (corresponding to the yellow and orange violated gauge operators, respectively). In particular, the Pauli operators are the short string operators shown in Fig.~\ref{fig: shortstringdef}. (b) We annihilate pairs of $a$ and $a^{-1}$ anyon types with short string operators (blue), leaving behind an $a^{-1}$ anyon type at $R'$ and an $a$ anyon type at $R$. The product of the short string operators gives a string operator that moves $a$ from $R'$ to $R$.}
\label{fig: buildingstring}
\end{figure*}

The translation symmetry action on the gauge twists allows us to define string operators that move the anyon types between two regions.
To see this, we let $\Gamma^a_R$ be a gauge twist representing $a$ at $R$ and  $\Gamma_{R'}^{a^{-1}} = {}^{U^\dagger}\Gamma_R^{a^{-1}}$ be a gauge twist representing $a^{-1}$ at a region $R'$ that is shifted by a unit translation, as in Fig.~\ref{fig: shortstringdef}. By definition, the gauge twist $\Gamma^a_R \cdot \Gamma_{R'}^{a^{-1}}$ represents the trivial anyon type. This implies that there is a Pauli operator $W^a_{RR'}$ such that for any gauge operator $G$, we have:
\begin{align}
    \Gamma^a_R\cdot \Gamma_{R'}^{a^{-1}}(G) = W^a_{RR'} G (W^a_{RR'})^\dagger G^\dagger.
\end{align}
That is, the commutation relations of $W^a_{RR'}$ with the gauge operators reproduces the gauge twist $\Gamma_R^a$ on $R$ and the gauge twist $\Gamma_{R'}^{a^{-1}}$ on $R'$. We say that $W^a_{RR'}$ is a short string operator that moves the anyon type $a$ from $R'$ to $R$. Alternatively, we say that $W^a_{RR'}$ creates an anyon type $a^{-1}$ at $R'$ and an anyon type $a$ at $R$. By gluing together short string operators, we can then create longer string operators along arbitrary paths, as illustrated in Fig.~\ref{fig: buildingstring}. 

We would like to make a few comments about the string operators defined above: \textbf{(i)} If the gauge group is proportional to the stabilizer group, then the string operators create local excitations at their endpoints. More generally, the string operators commute with the local gauge generators along their length and fail to commute with local gauge generators at their endpoints. However, since the Hamiltonians $H_{\{J_G\}}$ are generally frustrated, the string operators might not create local excitations at their endpoints. \textbf{(ii)} Up to Pauli operators localized at the endpoints, the string operators that move the trivial anyon type are products of stabilizers, as shown in Appendix~\ref{app: bare logicals and nonlocal stabilizers}. Further, the string operators that move detectable anyon types correspond to detectable errors, since they fail to commute with stabilizers at their endpoints. Lastly, up to Pauli operators localized to the endpoints, the string operators that move undetectable anyon types are undetectable errors, in that they preserve the code space \cite{Poulin2005stabilizer}.

\begin{figure}[t] 
\centering
\includegraphics[scale = .35]{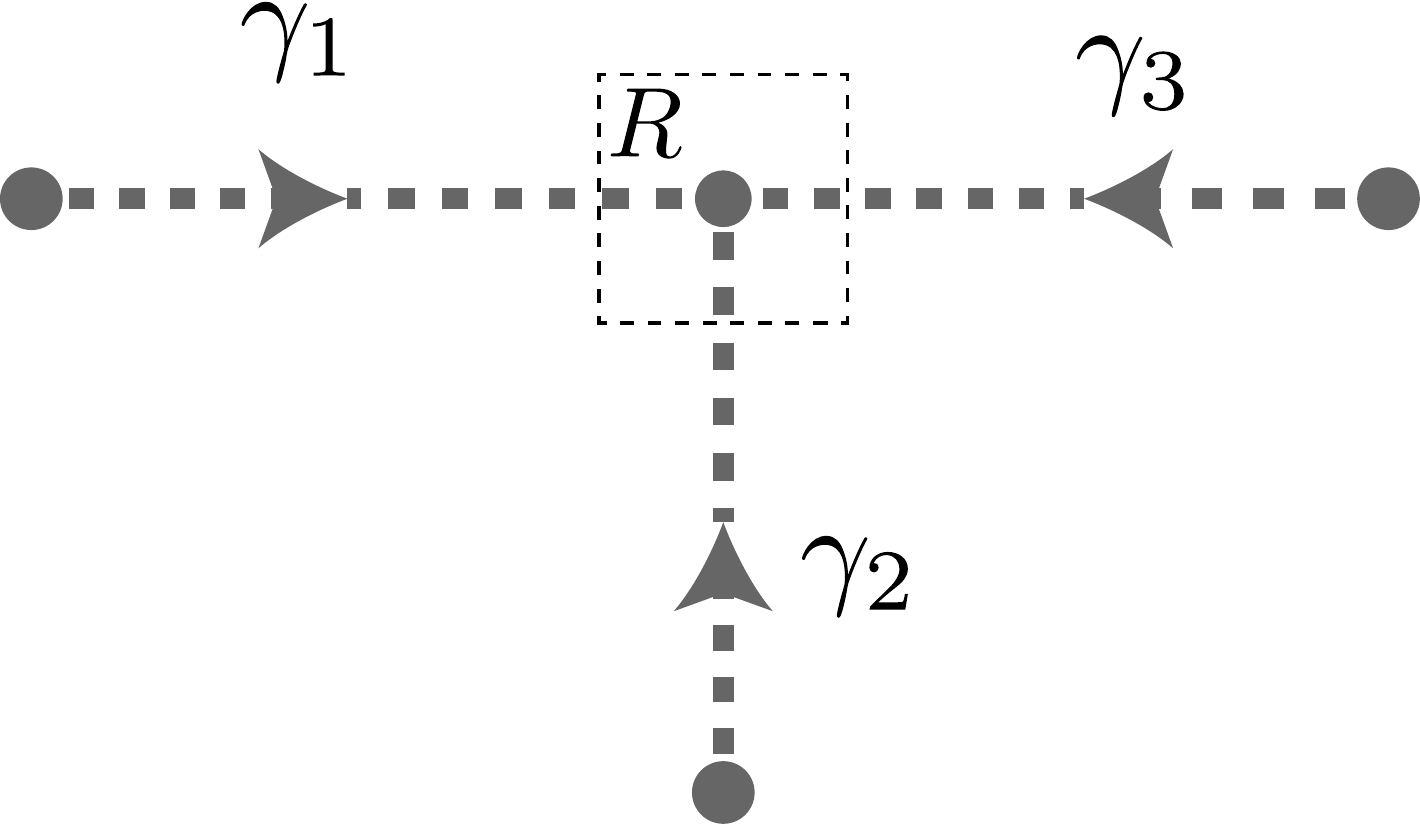}
\caption{The exchange statistics of an anyon type $a$ are computed from the commutation relations of the three string operators $W^a_{\gamma_1}$, $W^a_{\gamma_2}$, and $W^a_{\gamma_3}$. These move the anyon type $a$ along paths $\gamma_1$, $\gamma_2$, and $\gamma_3$ (gray) oriented towards the same point in a region $R$.}
\label{fig: statistics}
\end{figure}

With this, we are prepared to define the exchange statistics of the anyon types. In fact, the exchange statistics can be computed using the well-established methods of Refs.~\cite{Levin2003fermions, Kawagoe2020microscopic}. For completeness, we review the details here. For an anyon type $a$, we consider string operators $W^a_{\gamma_1}$, $W^a_{\gamma_2}$, and $W^a_{\gamma_3}$, defined along paths\footnote{The paths are not completely arbitrary. We require that the endpoints at the tails of $\gamma_1$, $\gamma_2$, and $\gamma_3$ are far from $R$ and that they do not wrap around a tail of any other path.} $\gamma_1$, $\gamma_2$, and $\gamma_3$, which move the anyon type $a$ to a constant-sized region $R$. We assume the paths are ordered counter-clockwise around $R$, as shown in Fig.~\ref{fig: statistics}. Importantly, we require that all three string operators have the exact same commutation relations with the gauge operators in $R$. This is always possible by redefining the string operators by Pauli operators localized near $R$, since this does not affect the anyon type. The exchange statistics $\theta(a)$ of the anyon type $a$ is then given by:
\begin{align} \label{eq: statistics formula}
W^a_{\gamma_1}\left(W^a_{\gamma_2}\right)^\dagger W^a_{\gamma_3} = \theta(a) W^a_{\gamma_3}\left(W^a_{\gamma_2}\right)^\dagger W^a_{\gamma_1}.
\end{align}
Heuristically, the string operators $\left(W^a_{\gamma_2}\right)^\dagger W^a_{\gamma_3}$ on the left hand side of Eq.~\eqref{eq: statistics formula} move an $a$ anyon type from the tail of $\gamma_3$ to the tail of $\gamma_2$, then $W^a_{\gamma_1}$ moves another $a$  anyon type from the tail of $\gamma_1$ to the head of $\gamma_1$. This is compared to the right hand side, in which, the $a$ anyon type at the tail of $\gamma_1$ is first moved to the tail of $\gamma_2$, then the $a$ anyon type at the tail of $\gamma_3$ is moved to the head of $\gamma_3$. These two processes differ by an exchange of $a$ anyon types -- thus, they differ by the phase $\theta(a)$. The pattern of string operators in Eq.~\eqref{eq: statistics formula} is chosen so that the dynamical phases cancel (i.e., the phases obtained by moving anyons along a path). We refer to Fig.~5 of Ref.~\cite{Bombin2012universal} for a proof that the formula for the exchange statistics is well defined.

\begin{figure}[t] 
\centering
\includegraphics[scale=.35]{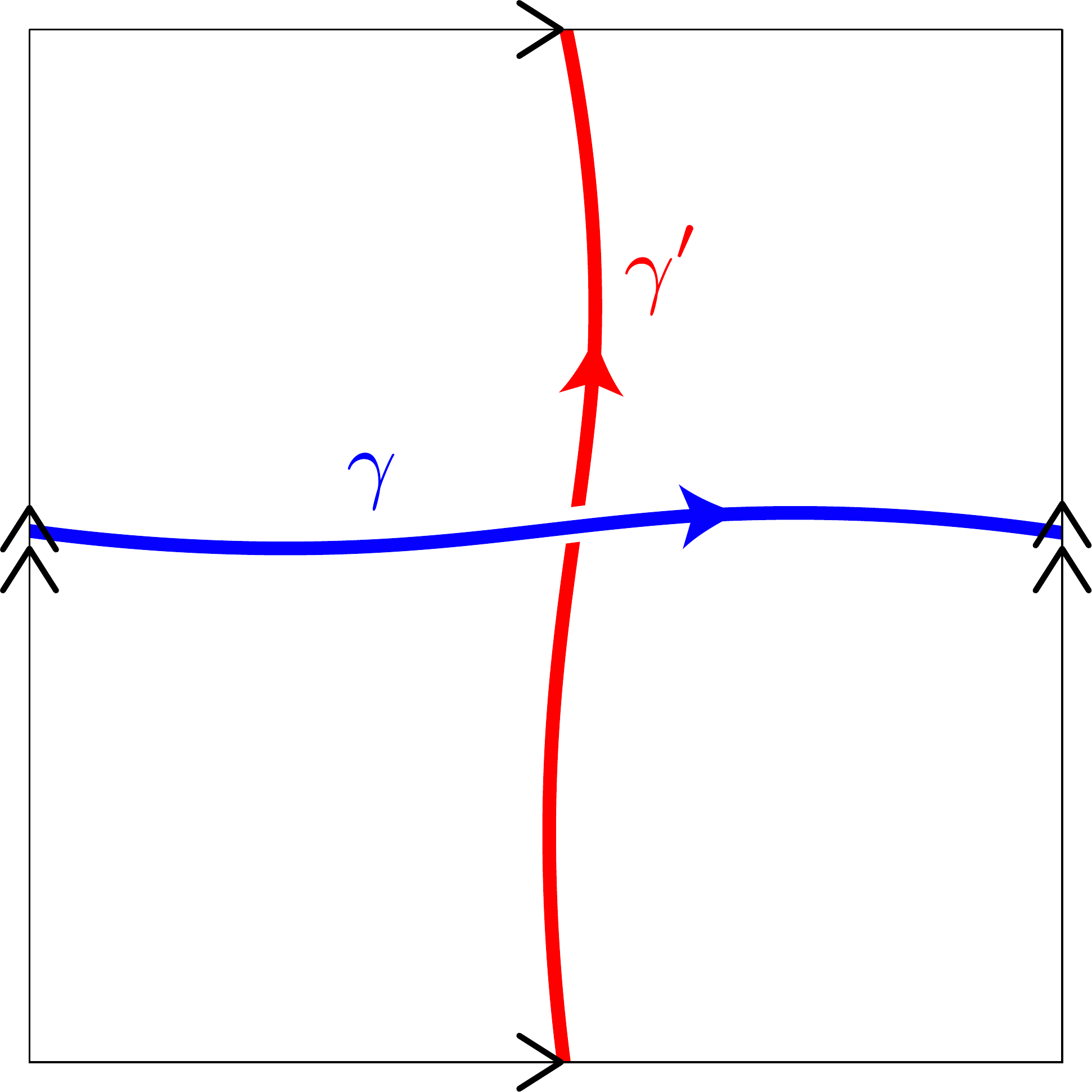}
\caption{The braiding relations of two anyon types can be computed from the commutation relations of string operators supported along the noncontractible paths $\gamma$ (blue) and $\gamma'$ (red) that wrap around the torus and intersect at a single point.}
\label{fig: braiding}
\end{figure}

For Abelian anyon theories, the $U(1)$-valued phase $B_\theta(a,b)$, obtained from a full braid of the anyon types $a$ and $b$, is completely determined by the exchange statistics of $a$, $b$, and their fusion product $ab$, according to the formula \cite{kitaev2006anyons, Ellison2022Pauli}:
\begin{align} \label{eq: braiding identity}
B_\theta(a,b) \equiv \frac{\theta(ab)}{\theta(a)\theta(b)}.
\end{align}
However, we find it valuable to describe how the braiding relation $B_\theta(a,b)$ can be computed directly.
We consider string operators supported along the paths $\gamma$ and $\gamma'$, as shown in Fig.~\ref{fig: braiding}. In fact, any choice of sufficiently long paths with an odd number of intersections and the same orientations as $\gamma$ and $\gamma'$ suffice. 
The phase $B_\theta(a,b)$ is then obtained from the commutation relations of $W^a_\gamma$ and $W^b_{\gamma'}$:
\begin{align} \label{eq: braiding formula}
W^b_{\gamma'}W^a_\gamma = B_\theta(a,b) W^a_\gamma W^b_{\gamma'}.
\end{align}
Importantly, the dynamical phases cancel, since the anyon types are moved along the same paths on either side of Eq.~\eqref{eq: braiding formula}. We also note that, by physical arguments, the braiding relations for any anyon types $a$, $b$, and $c$ are required to satisfy:
\begin{align} \label{eq: braiding relations relations}
B_\theta(a,b)=B_\theta(b,a), \quad \text{and} \quad B_\theta(a,bc) = B_\theta(a,b)B_\theta(a,c).
\end{align}

With this, we have defined the full set of data for the Abelian anyon theory of a topological subsystem code -- namely, the anyon types, their fusion rules, and their exchange statistics.
Before turning to a general construction of topological subsystem codes based on arbitrary Abelian anyon theories, 
we would like to emphasize two ways in which the anyon theories of topological subsystem codes go beyond those captured by topological stabilizer codes.
First, there are convincing arguments to say that topological stabilizer codes capture only the Abelian anyon theories that admit gapped boundaries \cite{kitaev2006anyons, Kapustin2020thermal}. Topological subsystem codes, in contrast, can be characterized by anyon theories that do not admit gapped boundaries, such as the chiral semion subsystem code in Section~\ref{sec: chiral semion}. 

Second, in Appendix~\ref{app: detectable=opaque}, we argue that the anyon theories of topological stabilizer codes must be modular, meaning that, for every anyon type $a$, there exists some anyon type $b$ that braids nontrivially with $a$.
As described below, topological subsystem codes can in fact be characterized by anyon theories that are non-modular -- wherein there exists an anyon type $a$ with braiding relations $B_\theta(a,b)=1$, for every anyon type $b$. The anyon type $a$ with trivial braiding relations is known as a transparent anyon type.\footnote{We remark that every transparent anyon type must be either a boson or a fermion. This follows from the fact that a transparent anyon type $a$ braids trivially with itself. Then, since a full braid is equivalent to two exchanges, we find:
$B_\theta(a,a) = \theta(a)^2 =1$.
Thus, the transparent anyon type $a$ has exchange statistics $\theta(a) = \pm 1$, meaning that it is either a boson or a fermion.}
For contrast, we refer to anyon types that have at least one nontrivial braiding relation as opaque anyon types. An example of a topological subsystem code characterized by a non-modular anyon theory is given in Section~\ref{sec: Z41 parafermion}. We note that Ref.~\cite{Bombin2014structure} further introduced the concept of a flux in topological subsystem codes, which includes objects that braid nontrivially with the transparent anyon types. For completeness, we discuss fluxes in Appendix~\ref{app: fluxes}.

\begin{figure}[t] 
\centering
\includegraphics[scale=.4]{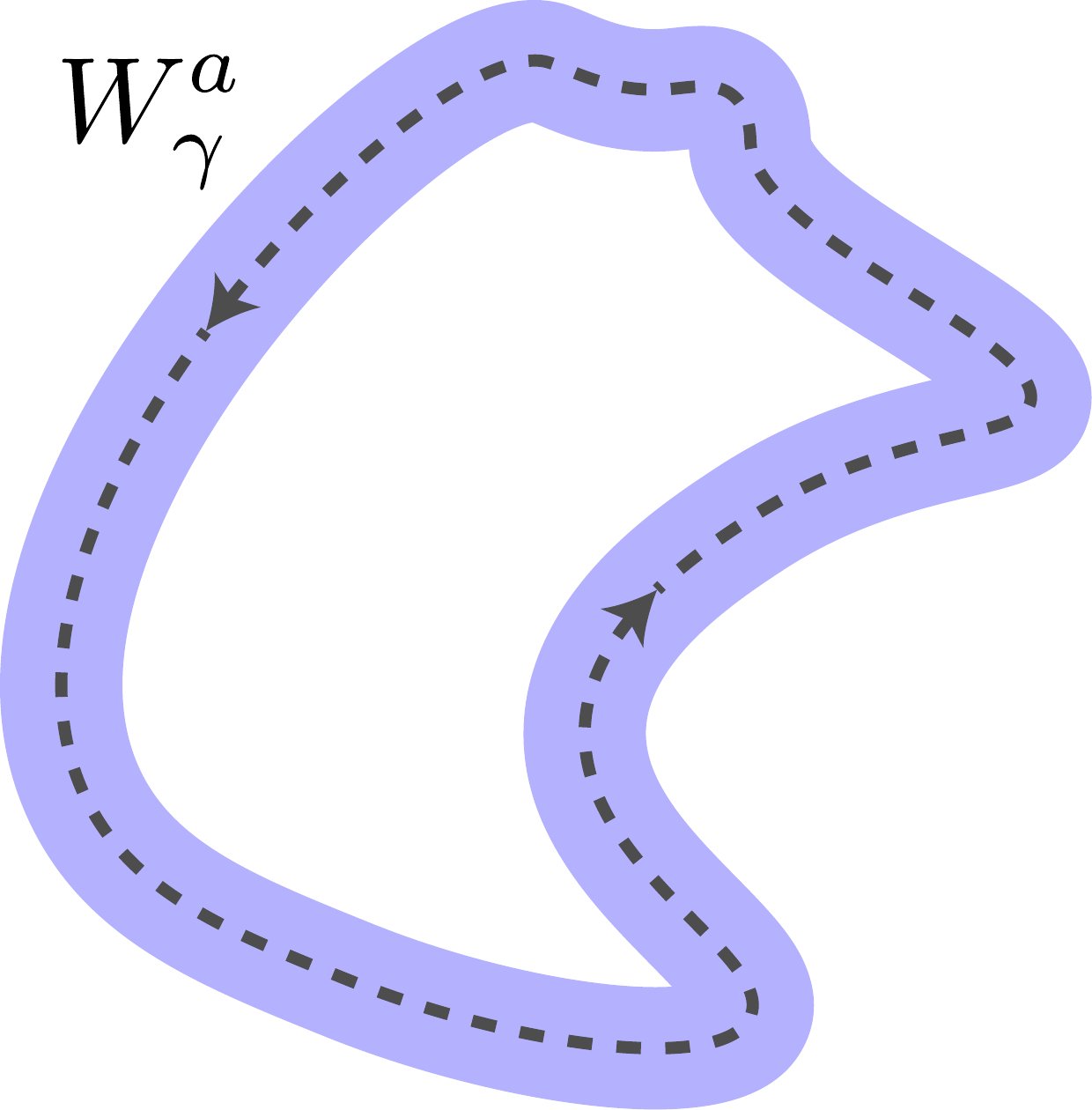}
\caption{By moving an anyon type $a$ along a closed path $\gamma$ (gray), we can create a string operator $W^a_\gamma$ (blue) that commutes with all of the gauge operators. If $\gamma$ is contractible, then $W^a_\gamma$ is stabilizer. If $\gamma$ is non-contractible, then $W^a_\gamma$ can be either a nontrivial bare logical operator or a nonlocal stabilizer, depending on whether $a$ is opaque or transparent, respectively.}
\label{fig: loopmovement}
\end{figure}

To clarify the differences between transparent anyon types and opaque anyon types in topological subsystem codes, 
we consider the string operators formed by moving anyon types along closed paths.
That is, we imagine moving a nontrivial anyon type $a$ along a closed path $\gamma$ using a string operator $W^a_{\gamma}$, as depicted in Fig.~\ref{fig: loopmovement}. The string operator $W^a_{\gamma}$ is supported along the loop formed by $\gamma$ and commutes with all of the gauge operators. Therefore, if $\gamma$ is contractible, the string operator $W^a_{\gamma}$ defines an element of the stabilizer group -- it cannot be a logical operator, because of property (iii) of Definition~\ref{def: topological subsystem code}. 

Furthermore, if the topological subsystem code is defined on a torus, then we can also consider a path $\gamma$ which wraps around a non-contractible loop. In this case, the string operator $W^a_{\gamma}$ may be a nontrivial bare logical operator or a nonlocal stabilizer -- depending on whether the anyon type is opaque or transparent, respectively. We note that the loops of string operators described here generate an anyonic $1$-form symmetry, defined more precisely in Appendix~\ref{app: 1form}. 

The nature of the string operators formed by moving nontrivial anyon types around non-contractible paths, i.e., whether they are nontrivial bare logical operators or nonlocal stabilizers, can be determined by the formula for the braiding relations in Eq.~\eqref{eq: braiding formula}. In particular, the string operators obtained by moving opaque anyon types around non-contractible paths must be nontrivial bare logical operators. This is because every opaque anyon type $a$ braids nontrivially with some other anyon type $b$. The string operators for $a$ and $b$ supported along paths with an odd intersection number (such as in Fig.~\ref{fig: braiding}) generate non-commuting operators on the logical subsystem. Transparent anyon types, on the other hand, yield nonlocal stabilizers when moved around non-contractible loops. This is because
the corresponding string operators commute with all other string operators created by moving anyon types around closed paths. Furthermore, in Appendix~\ref{app: bare logicals and nonlocal stabilizers}, we argue that every nontrivial bare logical operator can be generated by moving nontrivial anyon types around non-contractible loops, up to stabilizers. Therefore, the string operators of transparent anyon types commute with all of the logical operators, so they cannot be logical operators themselves. 
We note that, as the terminology suggests, the detectable anyon types of a topological subsystem code are opaque, while the undetectable anyon types are transparent (see Appendix~\ref{app: detectable=opaque}).

We conclude this section by commenting on the anyon theory of a topological subsystem code in terms of the family of Hamiltonians $H_{\{J_G\}}$, defined in Eq.~\eqref{eq: subsystem Hamiltonian}. To simplify the discussion, we assume that the Hamiltonians $H_{\{J_G\}}$ have no local degeneracies, as is the case for a generic set of coefficients $\{J_G\}$. 
Then, for any Hamiltonian in the parameter space, the string operators for opaque anyon types create gapped excitations at their endpoints. This is because the eigenstates of an arbitrary $H_{\{J_G\}}$ are simultaneously eigenstates of the stabilizers. The endpoints of a string operator representing an opaque anyon type fail to commute with some subset of the generators of $\tilde{S}$ at their endpoints, thus creating local excitations.\footnote{We have used here that every opaque anyon type has $B_\theta(a,b) \neq 1$ for some anyon type $b$. The open string operators for $a$ fail to commute with the $b$ string operators encircling the endpoint. Thus, the open string operators for $a$ fail to commute with some stabilizers.} 
These can be interpreted as anyonic excitations when the Hamiltonian is gapped. The string operators of transparent anyon types, on the other hand, are guaranteed to create anyonic excitations for the gapped Hamiltonians only if the corresponding anyon types do not have bosonic exchange statistics. 
This follows from the fact that the algebra of open string operators for anyon types with nontrivial statistics requires a Hilbert space of dimension greater than one.
Indeed, if a state $|\psi\rangle$ is invariant under the open string operators $W^a_{\gamma_1}$, $W^a_{\gamma_2}$, and $W^a_{\gamma_3}$ in Eq.~\eqref{eq: statistics formula}, then we have:
\begin{align}
|\psi\rangle = W^a_{\gamma_1} \left( W^a_{\gamma_2} \right)^\dagger W^a_{\gamma_3} |
\psi\rangle = \theta(a)W^a_{\gamma_3} \left(W^a_{\gamma_2}\right)^\dagger 
W^a_{\gamma_1} |\psi \rangle = \theta(a)|\psi \rangle.
\label{eq: T-junction for statistic}
\end{align}
Therefore, the phase $\theta(a)$ must be $1$, implying that $a$ is a boson. If $a$ has nontrivial exchange statistics, then $|\psi\rangle$ cannot be invariant under the open string operators. Thus, the string operators create local excitations at their endpoints.
We see that, modulo 
the transparent anyon types with bosonic exchange statistics, the anyon theory of a topological subsystem code captures a set of anyonic excitations common to the gapped Hamiltonians of the parameter space.  
We comment further on the gapped topological phases of matter exhibited by the family of Hamiltonians $H_{\{J_G\}}$, in Section~\ref{sec: discussion}.

\section{${\ZZ_4^{(1)}}$ subsystem code} \label{sec: Z41 parafermion}

As a first example of a topological subsystem code, we introduce the $\ZZ_4^{(1)}$ subsystem code. 
The $\ZZ_4^{(1)}$ subsystem code is named after the $\ZZ_4^{(1)}$ anyon theory, where we have adopted the nomenclature of Ref.~\cite{Bonderson2012interferometry}.\footnote{See also the beginning of Section~\ref{sec: z41 construction} for an explanation of the notation.}
The anyon types of this theory generate a $\ZZ_4$ group under fusion and are labeled as $\{1, s, s^2, s^3\}$. The exchange statistics of the $\ZZ_4^{(1)}$ anyon types~are:
\begin{align} \label{eq: z41 statistics}
\theta(1) = 1, \quad \theta(s) = i, \quad \theta(s^2) =1, \quad \theta(s^3) = i.
\end{align}  
Notably, the generator $s$ is a semion, which is to say that interchanging two $s$ anyon types yields a phase of $\theta(s) = i$.

The braiding relations of the anyon types can then be computed from the exchange statistics using the identity in Eq.~\eqref{eq: braiding identity}. We point out that the braiding relations of the $s^2$ anyon type are:
\begin{align}
B_\theta(s^2,1) = B_\theta(s^2,s) = B_\theta(s^2,s^2) = B_\theta(s^2,s^3) = 1.
\end{align}
This tells us that $s^2$ is a boson with trivial braiding relations. Therefore, it is a transparent anyon type, and the $\ZZ_4^{(1)}$ anyon theory is non-modular. As a result, the subsystem code defined below is an example of a topological subsystem code characterized by a non-modular anyon theory -- this is distinct from topological stabilizer codes, which only capture modular anyon theories (Appendix~\ref{app: bare logicals and nonlocal stabilizers}). 

We note that the $\ZZ_4^{(1)}$ subsystem code is based on a generalization of Kitaev's honeycomb model given in Ref.~\cite{Barkeshli2015generalized}. Here, we reformulate the model as a topological subsystem code and derive it by starting from a $\ZZ_4$ TC.
We give the details of the subsystem code in Section~\ref{sec: z41 definition} and describe its construction from a $\ZZ_4$ TC in Section~\ref{sec: z41 construction}.

\subsection{Definition of the subsystem code} \label{sec: z41 definition}

The $\ZZ_4^{(1)}$ subsystem code is defined on a hexagonal lattice with a four-dimensional qudit at each vertex. We assume, for simplicity, that the system has periodic boundary conditions. As described in Section~\ref{sec: composite qudits}, the basis states for a single four-dimensional qudit can be labeled by elements of $\ZZ_4$ as $|\alpha \rangle$, for $\alpha \in \ZZ_4$. The operator algebra at the vertex $j$ is then generated by the (generalized) Pauli $X$ and Pauli $Z$ operators:
\begin{align}
X_j = \sum_{\alpha \in \ZZ_4} | \alpha +1 \rangle \langle \alpha |, \quad Z_j = \sum_{\alpha \in \ZZ_4} i^{\alpha} |\alpha \rangle \langle \alpha|,
\end{align}
where the addition is computed modulo~$4$. We also find it convenient to introduce a Pauli $Y$ operator at each site $j$, defined by:
\begin{align}
Y_j \equiv \sqrt{i} X_j^\dagger Z_j^\dagger.
\end{align}
The Pauli operators at site $j$ satisfy the relations:
\begin{align}
X_j^4 = Z_j^4 = Y_j^4 = 1, \qquad  X Y = i Y X, \qquad Y Z = i Z Y,\qquad Z X = i X Z,
\end{align}
while the Pauli operators at sites $j$ and $k$ with $j \neq k$ are mutually commuting:
\begin{align}
    X_j Y_k = Y_k X_j, \qquad Y_j Z_k = Z_k Y_j,\qquad Z_j X_k = X_k Z_j.
\end{align}

\begin{figure}[t] 
\centering
\includegraphics[width=.7\textwidth]{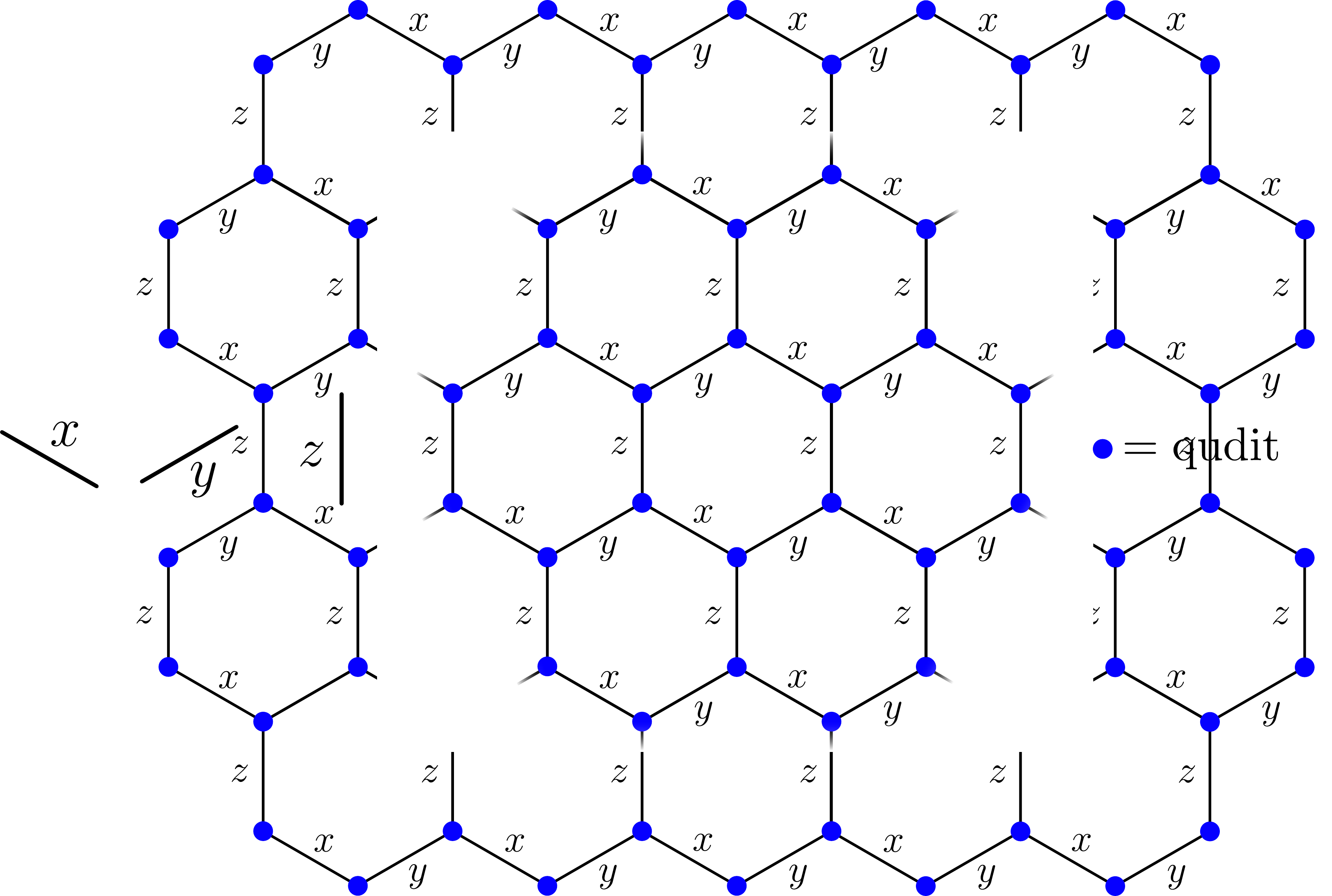}
\caption{The Hilbert space of the $\ZZ_4^{(1)}$ subsystem code is built from a four-dimensional qudit (blue) at each vertex of a hexagonal lattice. We refer to the negative slope, positive slope, and vertical~edges as $x$-, $y$-, and $z$-edges, respectively.}
\label{fig: xyzedges}
\end{figure}

To define the $\ZZ_4^{(1)}$ subsystem code, we first specify the gauge group $\mathcal{G}$. To do so, we label each edge of the hexagonal lattice with an $x$, $y$, or $z$ as shown in Fig.~\ref{fig: xyzedges}. We then define a gauge generator for each edge, depending on the label $x$, $y$, or $z$. The gauge generator for an $x$-edge is product of Pauli $X$ operators on the neighboring vertices. Likewise, the gauge generator for a $y$-edge is a product of Pauli $Y$ operators, and for a $z$-edge, it is a product of Pauli $Z$ operators. More explicitly, the gauge group $\mathcal{G}$ is the group generated by the following gauge operators:
\begin{align} \label{eq: z41 gauge group}
\mathcal{G} \equiv \left\langle 
~~
\raisebox{-0.2cm}{\hbox{\includegraphics[scale=.35]{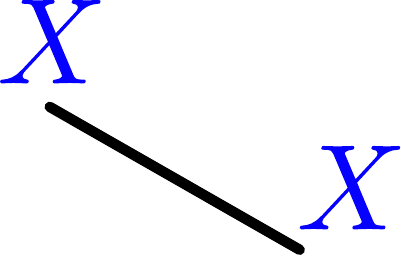}}}
, \quad
\raisebox{-0.4cm}{\hbox{\includegraphics[scale=.35]{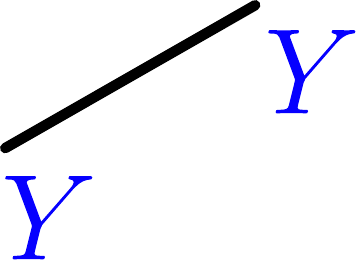}}}
, \quad
\raisebox{-0.6cm}{\hbox{\includegraphics[scale=.35]{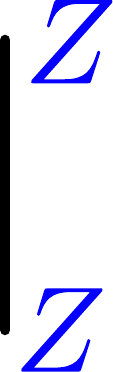}}}
~~\right\rangle.
\end{align}
Note that, here, the generating set implicitly includes an operator for each edge. 

Up to a choice of phases, the stabilizer group $\mathcal{S}$ is given by the center of $\mathcal{G}$. We choose the stabilizer group to be the group of Pauli operators generated by the plaquette stabilizers $S_p$:
\begin{align} \label{eq: z41 local stabilizer}
S_p \equiv \vcenter{\hbox{\includegraphics[scale=.35]{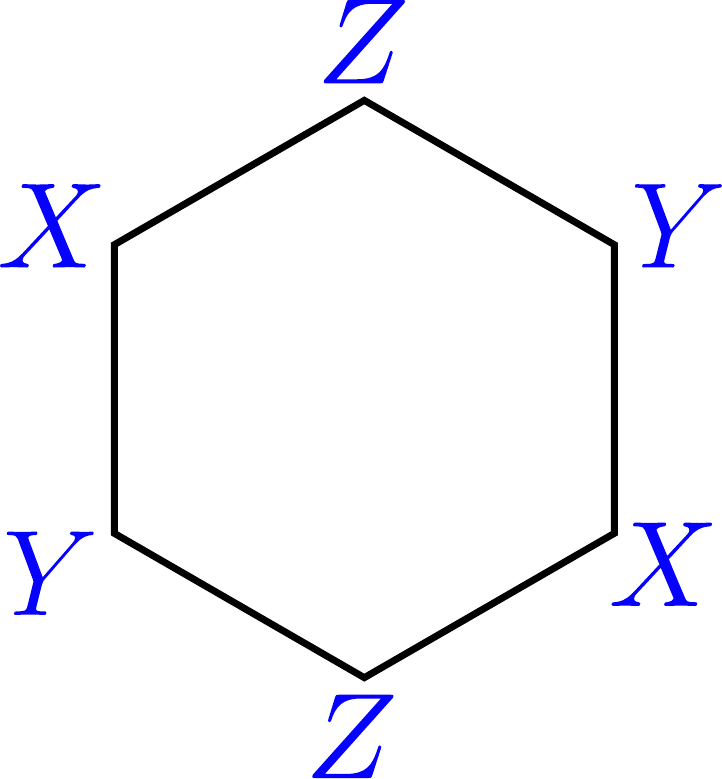}}},
\end{align}
and the nonlocal stabilizers:
\begin{align}
S_\gamma \equiv~~
\vcenter{\hbox{\includegraphics[scale=0.28]{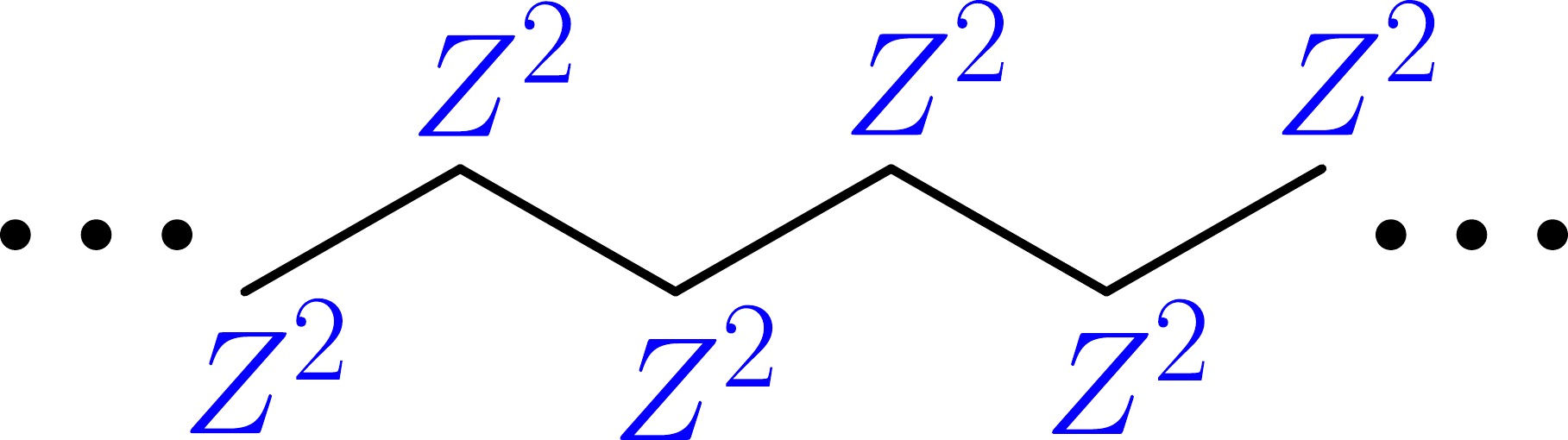}}}
~~, \quad S_{\gamma'} \equiv
\vcenter{\hbox{\includegraphics[scale=0.26]{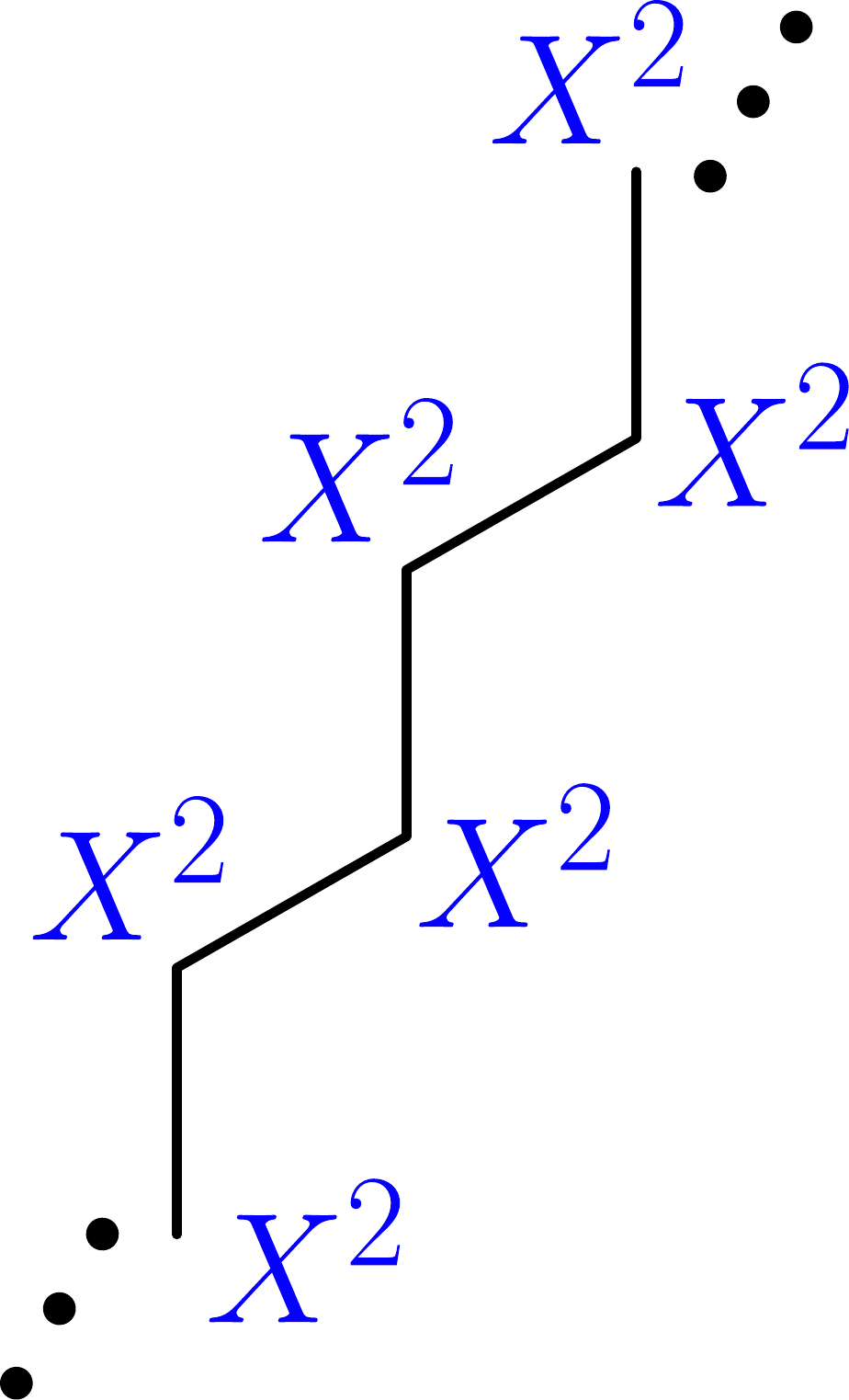}}}.
\end{align}
These string operators wrap around the non-contractible paths $\gamma$ and $\gamma'$ shown in Fig.~\ref{fig: braiding}. We see that, although the gauge group is generated by local operators, the stabilizer group includes nonlocal  {stabilizer generators}. 
This differs from topological stabilizer codes, which are required to have a stabilizer group that admits a set of local generators \cite{Haah2018classification}.

Lastly, the bare logical operators are fully determined by the gauge group and the choice of stabilizer group. The bare logical group $\mathcal{L}$ is generated by the stabilizer group and the two nontrivial bare logical operators:
\begin{align} \label{eq: z41 logicals}
L_\gamma \equiv
\vcenter{\hbox{\includegraphics[scale=0.26]{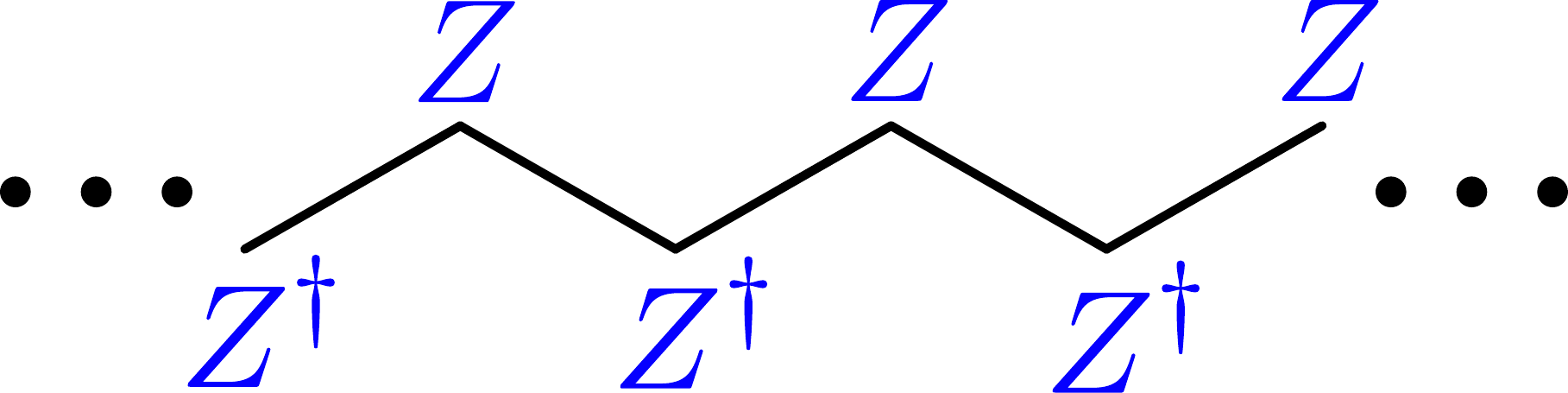}}}, \quad L_{\gamma'} \equiv
\vcenter{\hbox{\includegraphics[scale=0.25]{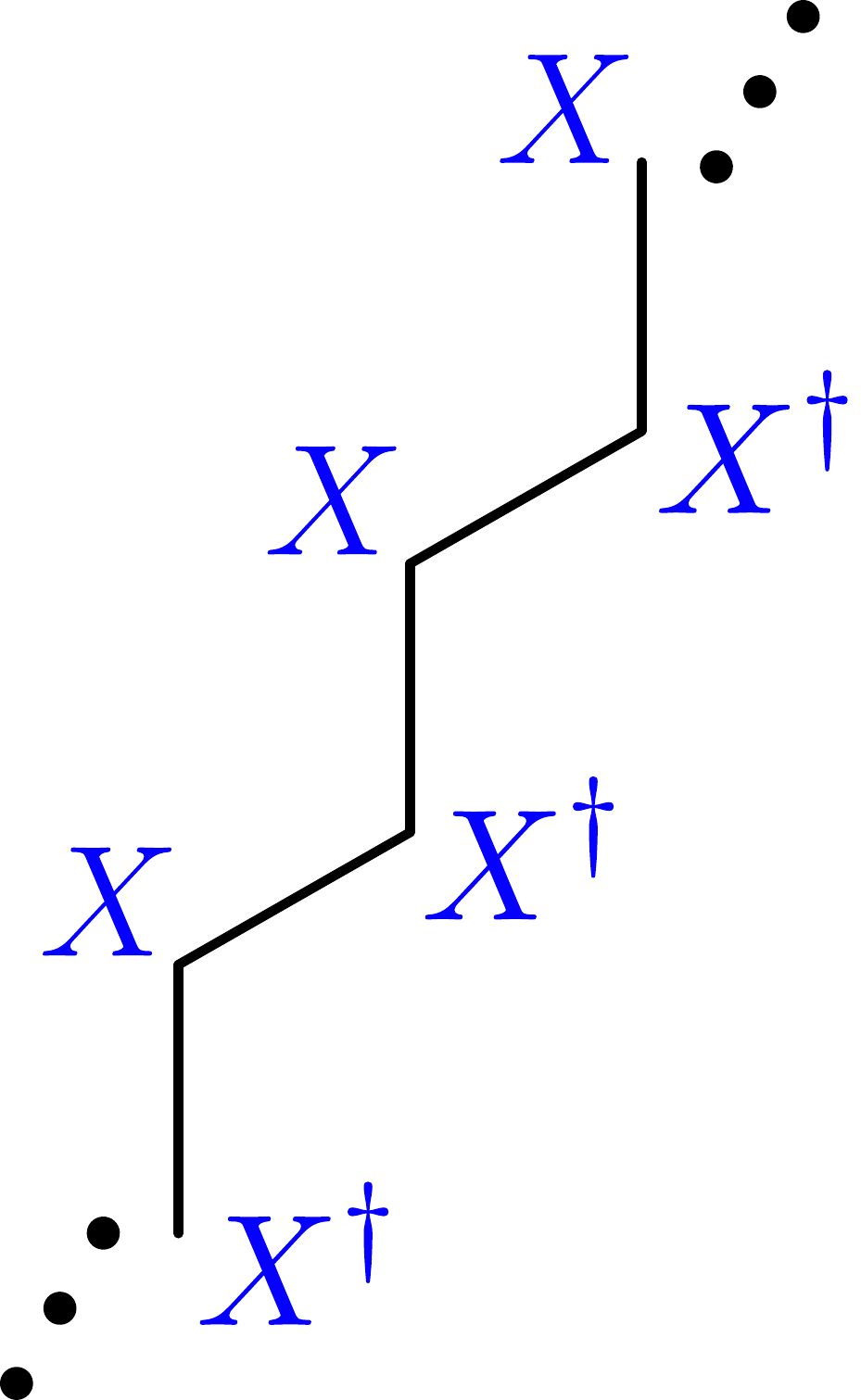}}}.
\end{align}
In fact, the bare logical group can be generated by the local stabilizers $S_p$ in Eq.~\eqref{eq: z41 local stabilizer} and the nontrivial bare logical operators $L_\gamma$ and $L_{\gamma'}$ in Eq.~\eqref{eq: z41 logicals}, since the nonlocal stabilizers $S_\gamma$ and $S_{\gamma'}$ can be generated from $L_\gamma$ and $L_{\gamma'}$. The logical operators satisfy the commutation relations:
\begin{align} \label{eq: z41 logical commutations}
L_{\gamma'} L_\gamma = - L_\gamma L_{\gamma'},
\end{align}
implying that the logical subsystem consists of a single qubit. The operators $L_{\gamma}$ and $L_{\gamma'}$ act as the logical Pauli $X$ and Pauli $Z$ operators on the logical qubit.

Before discussing the anyon theory of the subsystem code, let us make sure that we have accounted for all of the gauge operators, stabilizers, and nontrivial bare logical operators. We do this by checking that the dimension of the code space satisfies a certain consistency condition. On the one hand, the definition of the code space in Eq~\eqref{eq: code space def} tells us that each stabilizer imposes a constraint on the full Hilbert space, implying that the dimension of $\mathcal{H}_C$ is:\footnote{We refer to Ref.~\cite{Gheorghiu2014standard} for a proof of Eq.~\eqref{eq: stabilizer constraints dim} for composite-dimensional qudits.}
\begin{align} \label{eq: stabilizer constraints dim}
\text{dim}(\mathcal{H}_C) = \text{dim}(\mathcal{H}) / |\mathcal{S}|,
\end{align}
where $|\mathcal{S}|$ is the order of the stabilizer group.
On the other hand, from the factorization of the code space in Eq.~\eqref{eq: Hilbert space factorization}, we have:
\begin{align} \label{eq: code space factorization dim}
\text{dim}(\mathcal{H}_C) = \text{dim}(\mathcal{H}_G \otimes \mathcal{H}_L) = \text{dim}(\mathcal{H}_G)\text{dim}(\mathcal{H}_L).
\end{align}
Together, Eqs.~\eqref{eq: stabilizer constraints dim} and \eqref{eq: code space factorization dim} give us the consistency condition:
\begin{align} \label{eq: consistency dim no log}
\text{dim}(\mathcal{H})/|\mathcal{S}| = \text{dim}(\mathcal{H}_G) \text{dim}(\mathcal{H}_L).
\end{align}
To proceed, we find it convenient to take the logarithm (base $4$) of both sides:
\begin{align} \label{eq: consistency dim log}
\log_4\left[\text{dim}(\mathcal{H})\right] = \log_4|\mathcal{S}| + \log_4\left[ \text{dim}(\mathcal{H}_G)\right] + \log_4 \left[ \text{dim}(\mathcal{H}_L) \right].
\end{align} 
This allows us to count the dimensions in terms of the number of four-dimensional qudits. 
We further introduce the notation:
\begin{align} \label{eq: qudit counting notation}
\mathbf{N}_Q \equiv \log_4\left[\text{dim}(\mathcal{H})\right], \quad \mathbf{N}_S \equiv \log_4|\mathcal{S}|, \quad \mathbf{N}_G \equiv \log_4\left[ \text{dim}(\mathcal{H}_G)\right], \quad \mathbf{N}_L \equiv \log_4 \left[ \text{dim}(\mathcal{H}_L) \right].
\end{align}
Here, $\mathbf{N}_Q$ is the total number of qudits, while $\mathbf{N}_S$, $\mathbf{N}_G$, and $\mathbf{N}_L$ correspond to the number of stabilized qudits, gauge qudits, and logical qudits, respectively. With this notation, the consistency condition in Eq.~\eqref{eq: consistency dim log} can be written succinctly as:
\begin{align} \label{eq: counting qubits z41}
\mathbf{N}_Q = \mathbf{N}_S + \mathbf{N}_G + \mathbf{N}_L.
\end{align}

Let us now count $\mathbf{N}_Q$, $\mathbf{N}_S$, $\mathbf{N}_G$, and $\mathbf{N}_L$ to verify Eq.~\eqref{eq: counting qubits z41} for the $\ZZ_4^{(1)}$ subsystem code. We assume that the subsystem code is on a torus. We let $\mathbf{P}$ be the total number of plaquettes in the system. Then, $\mathbf{N}_Q$ is $2\mathbf{P}$, since there are two four-dimensional qudits for each plaquette. There is one order-four stabilizer generator $S_p$ for every plaquette. However, their product over all plaquettes is the identity, meaning that there are only $\mathbf{P}-1$ independent local stabilizer generators. Including the two order-two nonlocal stabilizers, we find that $\mathbf{N}_S$ is $\mathbf{P}$.  Next, there are three order-four gauge generators for every plaquette. Since the product of all gauge generators is the identity, however, there are only $3\mathbf{P} -1$ independent gauge generators. The gauge operators also generate the stabilizers, so we need to subtract $\mathbf{P}$ from $3\mathbf{P}-1$. The number of gauge qudits is then:
\begin{align}
\mathbf{N}_G = [(3\mathbf{P}-1)-\mathbf{P}]/2 = \mathbf{P} - \frac{1}{2}.
\end{align}
We divide by $2$ because the gauge generators form both the Pauli $Z$ and Pauli $X$ operators for the gauge qudit. Finally, the number of logical qudits is $\mathbf{N}_L = \frac{1}{2}$, i.e., there is a single logical qubit. All together, we have:
\begin{align}
\mathbf{N}_S + \mathbf{N}_G + \mathbf{N}_L = \mathbf{P} + (\mathbf{P} - \frac{1}{2}) + \frac{1}{2} = 2 \mathbf{P} = \mathbf{N}_Q.
\end{align} 
Therefore, the counting of the stabilizers, gauge operators, and logical operators is self-consistent.

We now show that the anyon theory of the subsystem code above is indeed the $\ZZ_4^{(1)}$ anyon theory described at the beginning of this section. As shown in Appendix~\ref{app: bare logicals and nonlocal stabilizers}, all of the nontrivial bare logical operators and nonlocal stabilizers of a topological subsystem code are generated by moving anyon types along non-contractible paths. Thus, truncations of the string operator $L_\gamma$ (or $L_{\gamma'}$) can be used to reveal the anyon types of the subsystem code. To see this, let us imagine truncating $L_\gamma$ on an infinite-sized system, so that one endpoint is at spatial infinity. A possible truncation of $L_\gamma$ is:
\begin{align} \label{eq: truncated logicals z41}
\vcenter{\hbox{\includegraphics[scale=0.28]{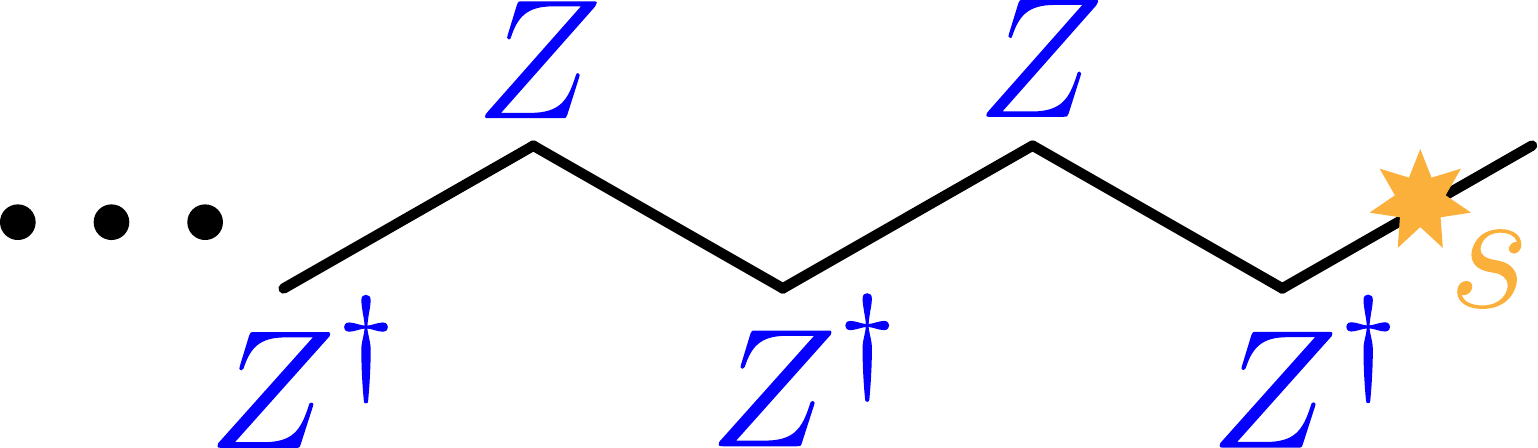}}}
~.
\end{align}
This truncated string operator commutes with all of the gauge operators away from the endpoint, but fails to commute with the gauge generators associated  {with} the $y$-edges at the endpoint (labeled by $s$). Heuristically, the commutation relations of the semi-infinite string operator with the gauge operators defines a gauge twist (see Section~\ref{sec: anyon theories for subsystem codes}). We claim that the commutation relations cannot be reproduced by the commutation relations of Pauli operators (with constant-sized support). This follows from the fact that the string operators that move trivial anyon types around a closed path are necessarily products of local stabilizers along the path (see Proposition~\ref{prop: anyon strings trivial}). Hence, the open string operator in Eq.~\eqref{eq: truncated logicals z41} corresponds to a nontrivial anyon type. Moreover, it is a detectable anyon type, as argued below. With some foresight, we have called this anyon type $s$.

\begin{figure}[t] 
\centering
\includegraphics[scale=.26]{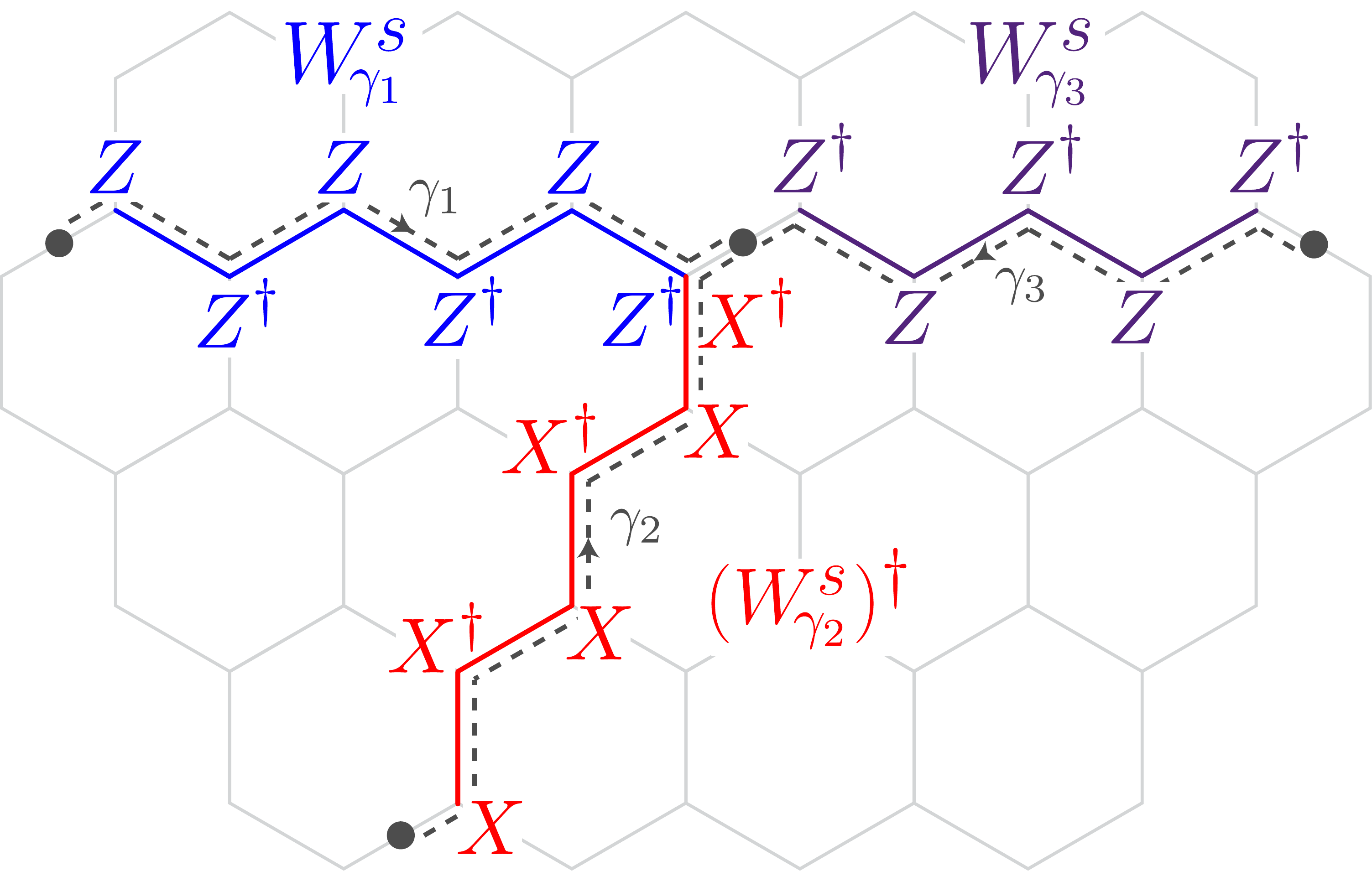}
\caption{The string operators $W^s_{\gamma_1}$, $W^s_{\gamma_2}$, and $W^s_{\gamma_3}$ (blue, red, purple, respectively) move the anyon type $s$ along the paths $\gamma_1$, $\gamma_2$, and $\gamma_3$ (gray). The string operators have the exact same commutation relations with the gauge operators supported  near the common endpoint of $\gamma_1$, $\gamma_2$, and $\gamma_3$. The exchange statistics of $s$ is computed using Eq.~\eqref{eq: statistics formula}, which shows that $s$ is a semion, i.e., $\theta(s) = i$.}
\label{fig: Z41 statistics}
\end{figure}

We are now able to compute the fusion rules and exchange statistics of the anyon type $s$ following the discussion in Section~\ref{sec: anyon theories for subsystem codes}. The fusion rules can be seen from the powers of $L_\gamma$. The fourth power of $L_\gamma$ is the identity, so the anyon type is at most order four. The string operators $L_\gamma$, $L_\gamma^2$, and $L_\gamma^3$ are either nontrivial bare logical operators or nonlocal stabilizers. Therefore, $s$, $s^2$, and $s^3$ are nontrivial anyon types. We see that $s$ generates the group $\ZZ_4$ under fusion. Next, the exchange statistics of the anyon types $\{1, s, s^2, s^3\}$ can be computed using Eq.~\eqref{eq: statistics formula}. The calculation of $\theta(s)$ is shown in Fig.~\ref{fig: Z41 statistics}.
We find the exchange statistics:
\begin{align}
\theta(1) = 1, \quad \theta(s) = i, \quad \theta(s^2) = 1, \quad \theta(s^3) = -i.
\end{align}
These are precisely the exchange statistics of the anyon types in the $\ZZ_4^{(1)}$ anyon theory, as claimed. Using either Eq.~\eqref{eq: braiding identity} or Eq.~\eqref{eq: braiding formula}, the nontrivial braiding relations are given by:
\begin{align}
B_\theta(s,s) = B_\theta(s,s^3) = B_\theta(s^3,s^3) = -1.
\end{align}
This tells us that $s$ and $s^3$ are opaque anyon types, while $s^2$ is transparent. The anyon types $s$ and $s^3$ are detectable, since the corresponding open string operators fail to commute with a loops of $s$ or $s^3$ string operators wrapped around the endpoint.

The $\ZZ_4^{(1)}$ subsystem code defines a family of Hamiltonians according to Eq.~\eqref{eq: subsystem Hamiltonian}, which for a particular choice of coefficients takes the form:
\begin{align} \label{eq: z41 subsystem Hamiltonians}
H_{\{J_e,J_p\}} = - \sum_{e \in x\text{-edges}} J_e \, \vcenter{\hbox{\includegraphics[scale=0.35]{Figures/xedge2.pdf}}} -  \sum_{e \in y\text{-edges}} J_e  \,  \vcenter{\hbox{\includegraphics[scale=0.35]{Figures/yedge2.pdf}}} -  \sum_{e \in z\text{-edges}} J_e \,  \vcenter{\hbox{\includegraphics[scale=0.35]{Figures/zedge2.pdf}}} - \sum_p J_p \vcenter{\hbox{\includegraphics[scale=.35]{Figures/Sp2.pdf}}}.
\end{align}
Given that the anyon types $s$ and $s^3$ are detectable, i.e., the endpoints of the corresponding string operators fail to commute with stabilizers, they correspond to gapped anyonic excitations common to the Hamiltonians in Eq.~\eqref{eq: z41 subsystem Hamiltonians} (assuming $J_p \neq 0$ for each plaquette $p$). The $s^2$ anyon types, on the other hand, are transparent bosons, so they do not enforce any anyonic excitations, as discussed at the end of Section~\ref{sec: anyon theories for subsystem codes}. Therefore, the gapped Hamiltonians in the parameter space must have a semion -- which may have order two. Note that (besides the $J_p$ term) this Hamiltonian is precisely the generalized honeycomb model introduced in Ref.~\cite{Barkeshli2015generalized}.

We point out that, although the most natural string operator for $s^2$ is the square of the string operator in Eq.~\eqref{eq: truncated logicals z41}:
\begin{align} \label{eq: simple s2 string}
    \vcenter{\hbox{\includegraphics[scale=0.28]{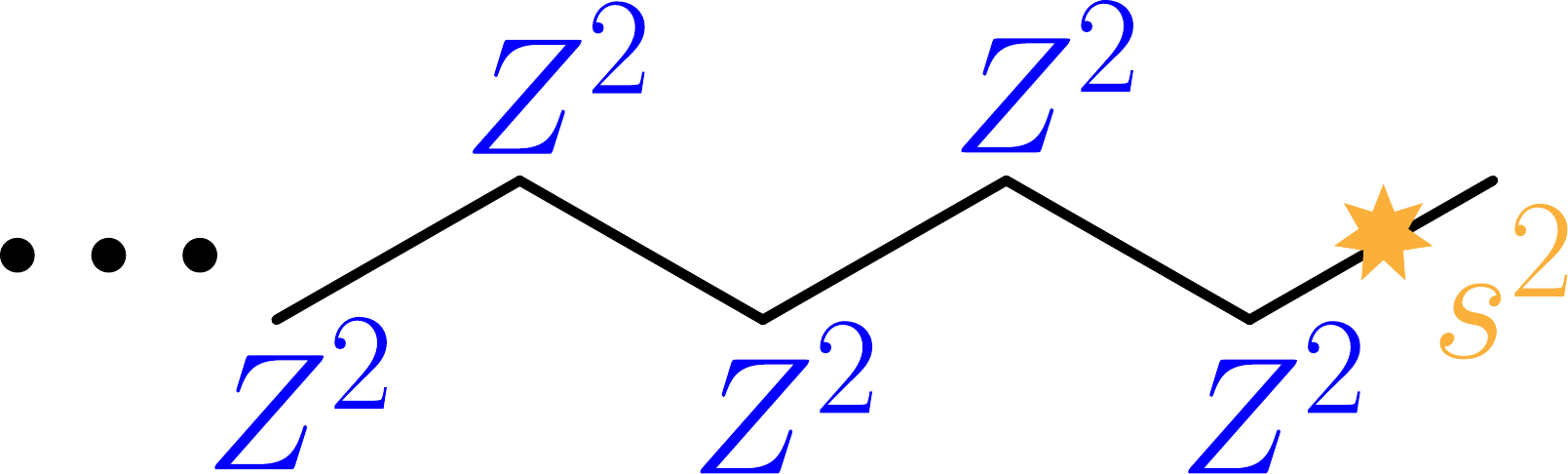}}},
\end{align}
this choice of string operator fails to commute with stabilizers. To find a string operator for $s^2$ that is manifestly undetectable, we can modify the string operator in Eq.~\eqref{eq: simple s2 string} at its endpoint by:
\begin{align}\label{eq: short s^2 string}
    \vcenter{\hbox{\includegraphics[scale=0.26]{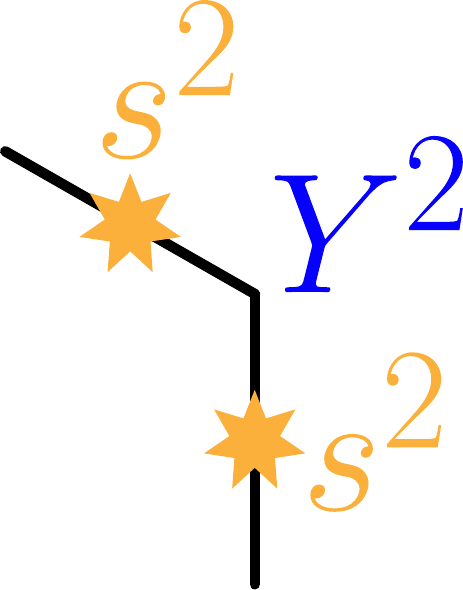}}}.
\end{align}
This gives us the string operator:
\begin{align}\label{eq: three s^2 long string}
    \vcenter{\hbox{\includegraphics[scale=0.28]{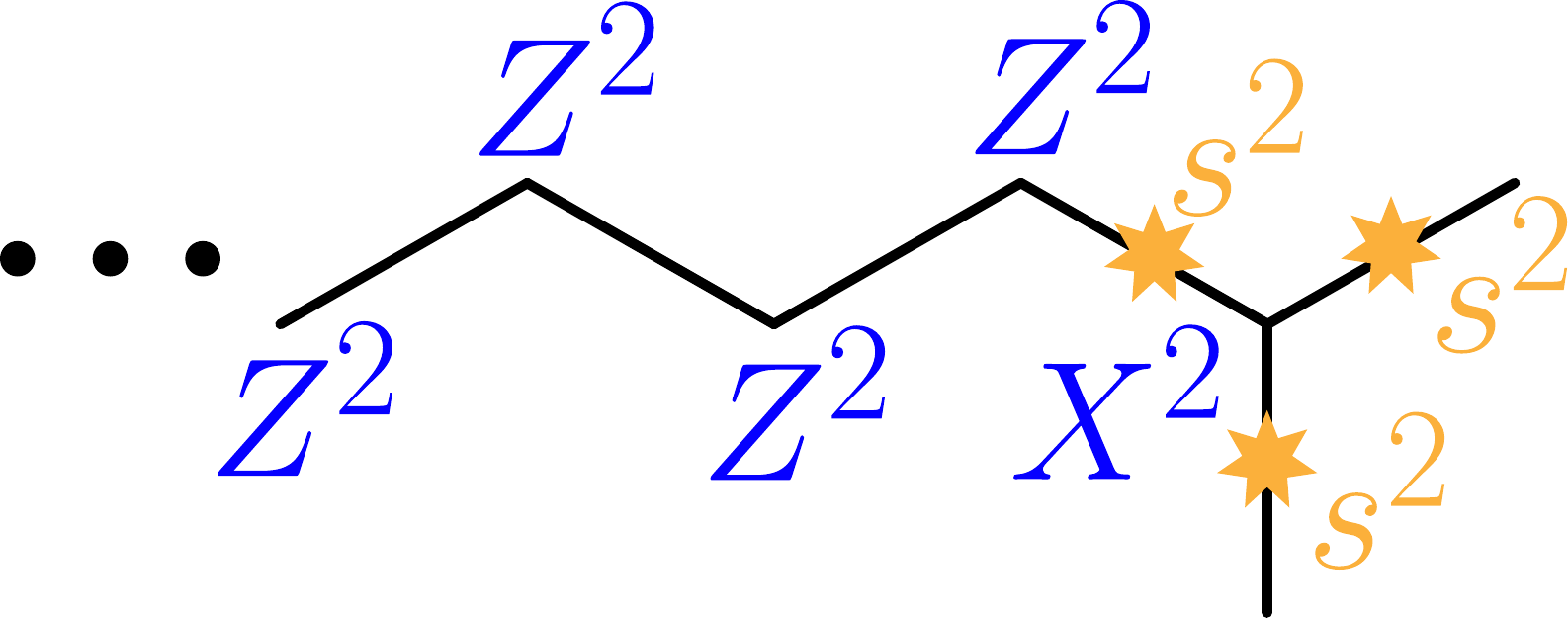}}},
\end{align}
which commutes will all stabilizers and therefore the anyon type $s^2$ is undetectable. Note that, although there are three $s^2$ anyon types shown in Eq.~\ref{eq: three s^2 long string}, it is equivalent to a single $s^2$ anyon type, since two $s^2$ anyons are created by the local operator in Eq.~\eqref{eq: short s^2 string}.

\subsection{Construction of the subsystem code} \label{sec: z41 construction}

In this section, we show that the $\ZZ_4^{(1)}$ subsystem code can be constructed from a $\ZZ_4$ TC. This follows from the simple observation that the $em$ excitations of the $\ZZ_4$ TC generate the $\ZZ_4^{(1)}$ anyon theory as a subgroup. In particular, the $em$ excitation generates the anyon types: $\{1, em, e^2m^2, e^3m^3\}$, which have the exchange statistics \cite{Kitaev2003quantumdouble}:
\begin{align}
\theta(1) = 1, \quad \theta(em) = i, \quad \theta(e^2m^2) =1, \quad \theta(e^3m^3) = i.
\end{align}
These match the exchange statistics of the $\ZZ_4^{(1)}$ anyon types in Eq.~\eqref{eq: z41 statistics}. Moreover, the anyon type $e^2m^2$ is transparent in the subtheory, since it has trivial braiding relations with the anyon types in $\{1, em, e^2m^2, e^3m^3\}$. We note that, in general, the notation $\ZZ_N^{(p)}$ of Ref.~\cite{Bonderson2012interferometry} can be interpreted as the anyon theory generated by the $e^pm$ anyon type of a $\ZZ_N$ TC.

\begin{figure}[tb] 
\centering
\hspace{3cm}\includegraphics[scale=0.25]{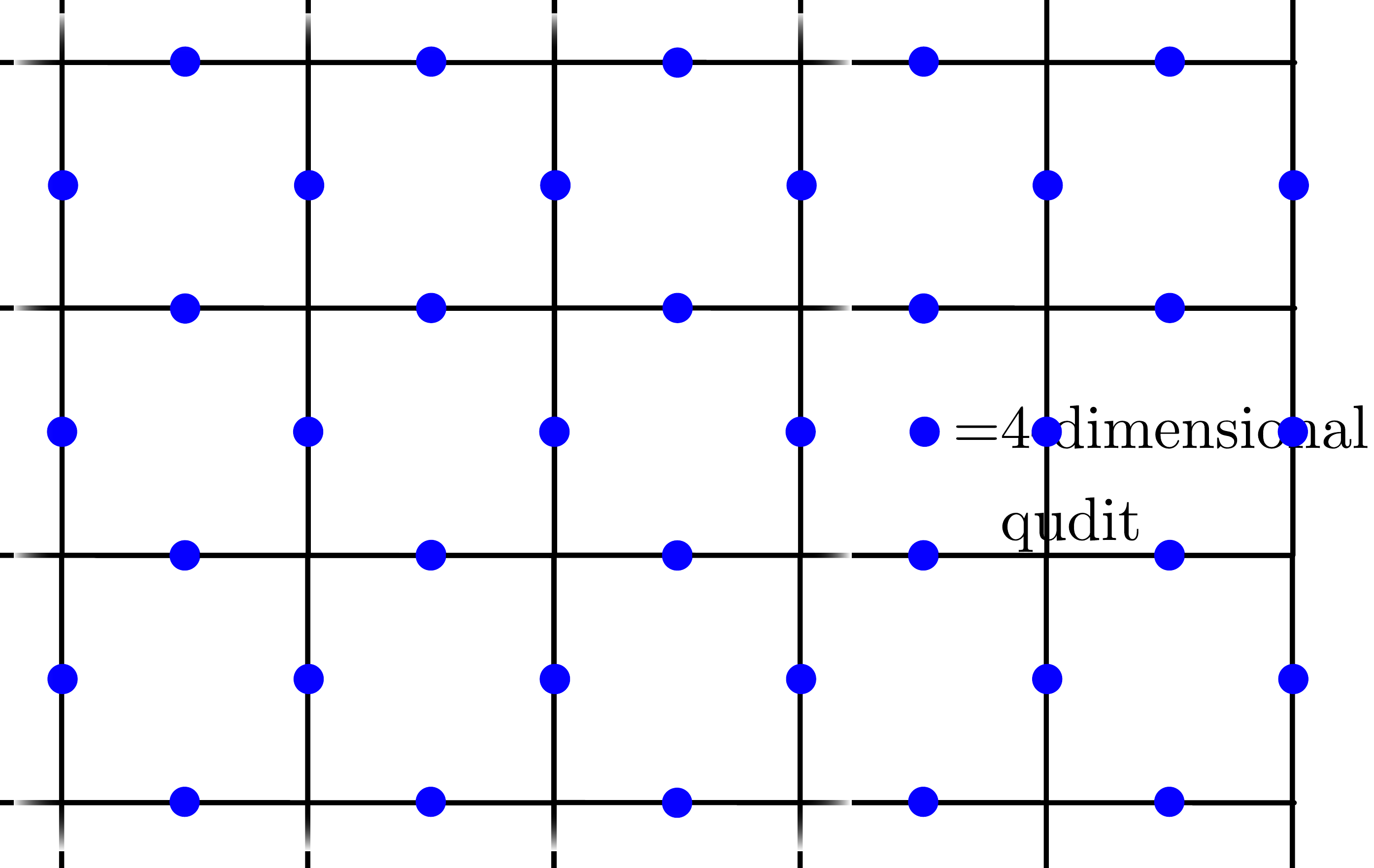}
\caption{The $\ZZ_4$ TC, used to build the $\ZZ_4^{(1)}$ subsystem code, is defined on a square lattice with a four-dimensional qudit on each edge (blue).}
\label{fig: Z4dof}
\end{figure}

Therefore, our construction of the $\ZZ_4^{(1)}$ subsystem code starts with a $\ZZ_4$ TC. The $\ZZ_4$ TC is defined on a square lattice with a four-dimensional qudit at each edge (Fig.~\ref{fig: Z4dof}). Explicitly, the Hamiltonian of the $\ZZ_4$ TC is:
\begin{align}
H_\text{TC} \equiv - \sum_v A^\text{TC}_v - \sum_p B^\text{TC}_p,
\end{align}
where the vertex term $A^\text{TC}_v$ and plaquette term $B^\text{TC}_p$ are graphically represented as:
\begin{align}
A^\text{TC}_v \equiv \vcenter{\hbox{\includegraphics[scale=0.25]{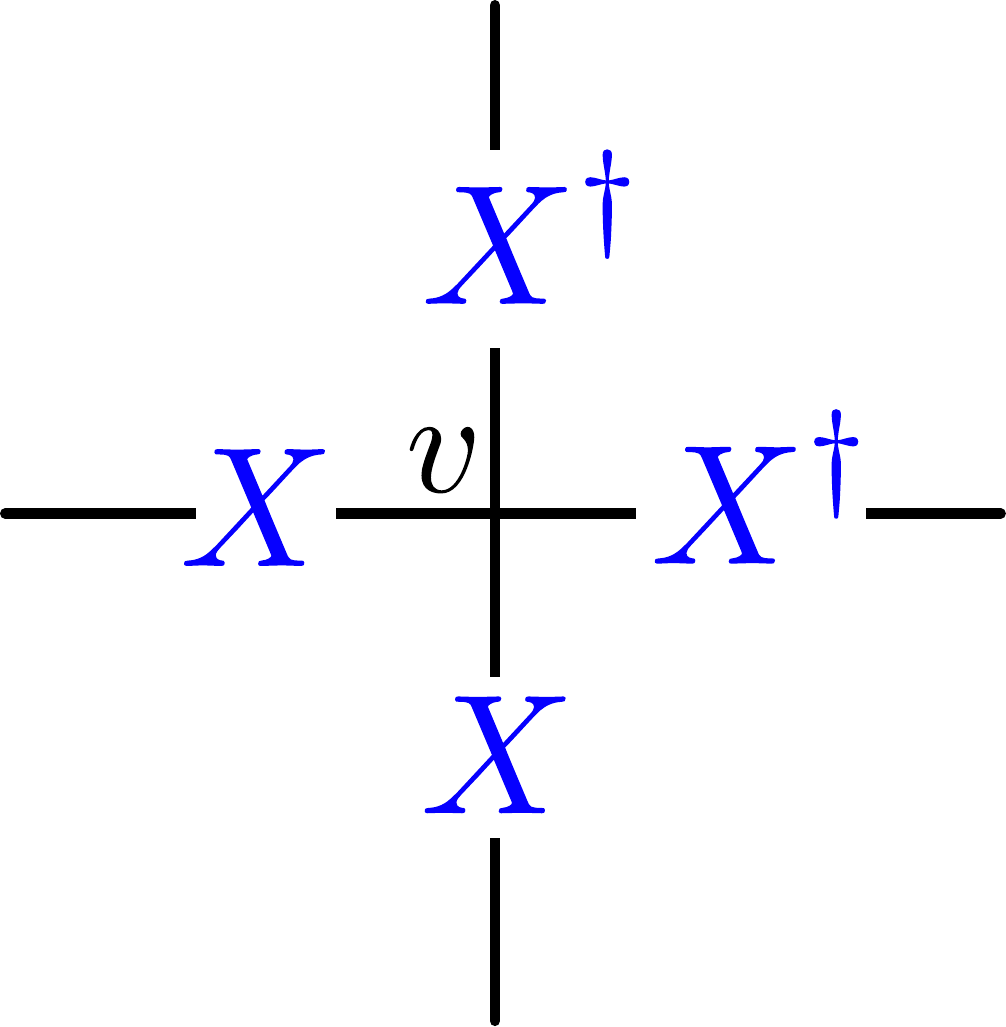}}}, \quad B^\text{TC}_p \equiv \vcenter{\hbox{\includegraphics[scale=0.25]{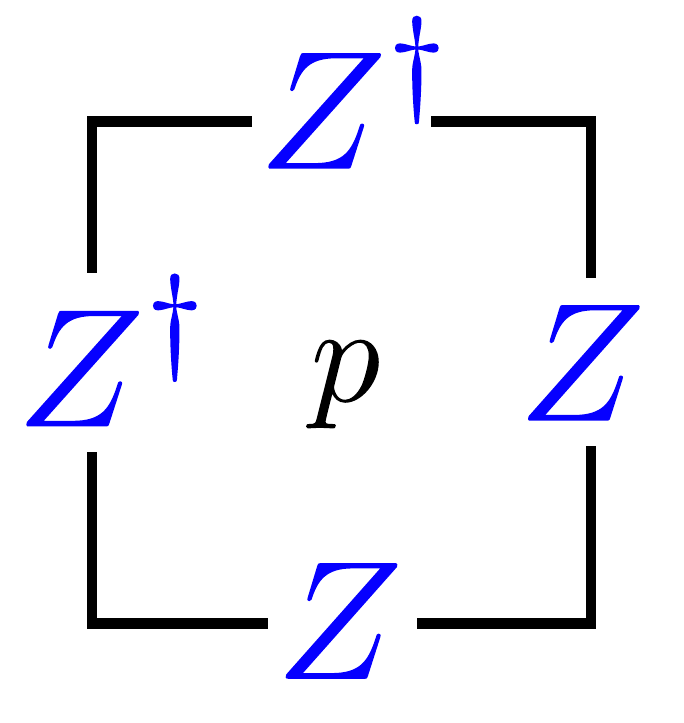}}}.
\end{align}
This model has sixteen types of anyonic excitations, tabulated below:
\begin{align} \nonumber
\vcenter{\hbox{\includegraphics[width=.35\textwidth]{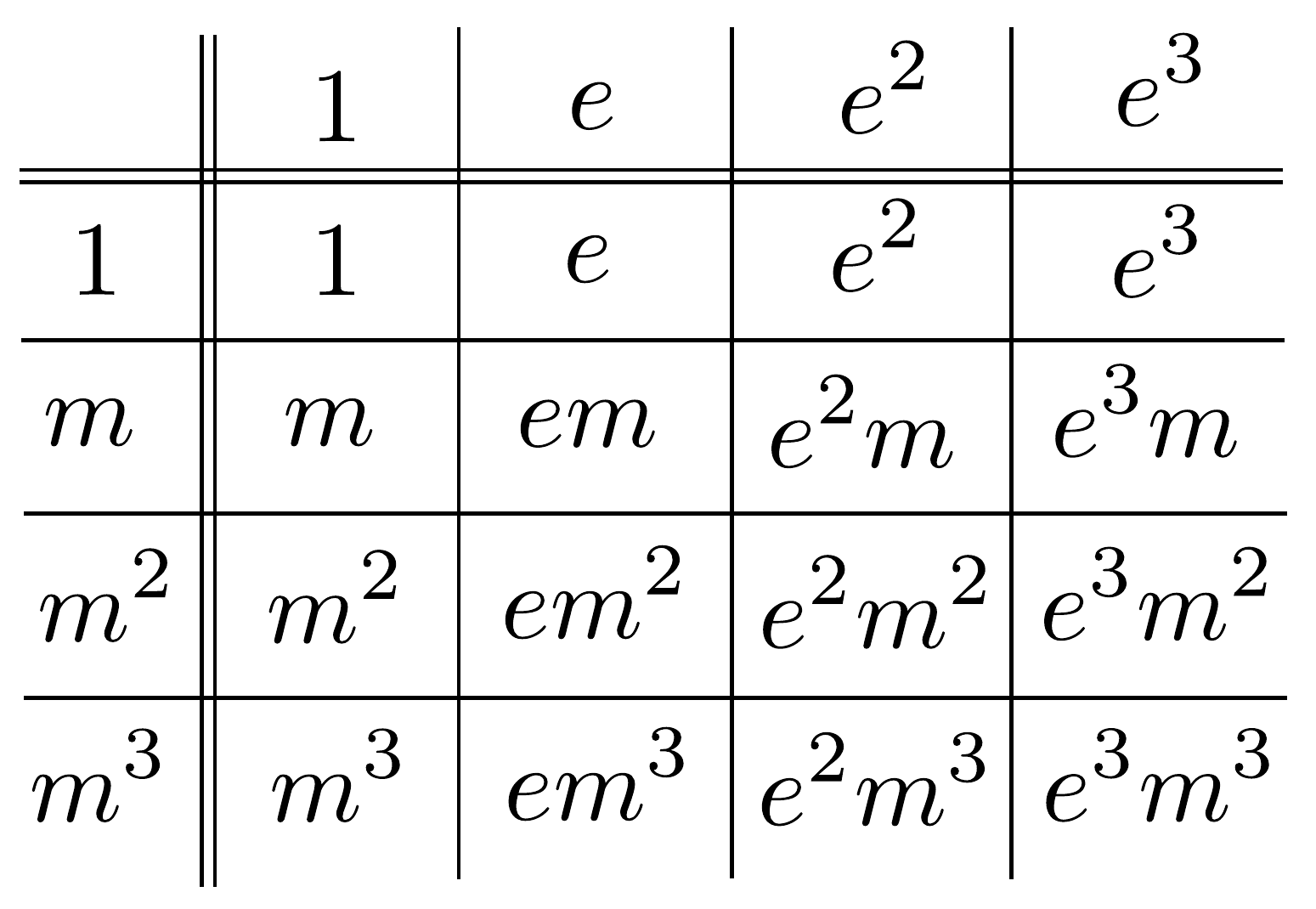}}}.
\end{align}
The string operators for the anyon types are generated by the short string operators for $e$ and $m$ anyon types:
\begin{align}
    W_e^e \equiv \vcenter{\hbox{\includegraphics[scale=0.25]{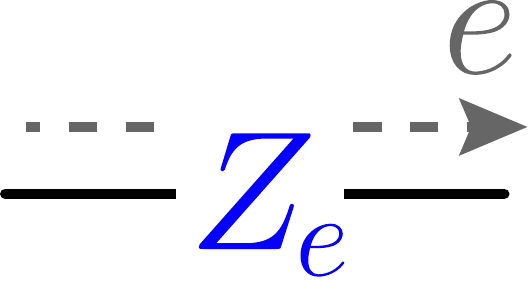}}}
    ,  \, \,
    \vcenter{\hbox{\includegraphics[scale=0.25]{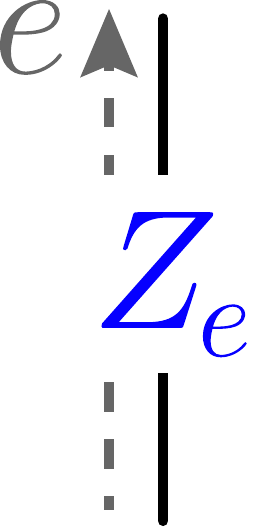}}}
    ~, \quad
   W_e^m \equiv \vcenter{\hbox{\includegraphics[scale=0.25]{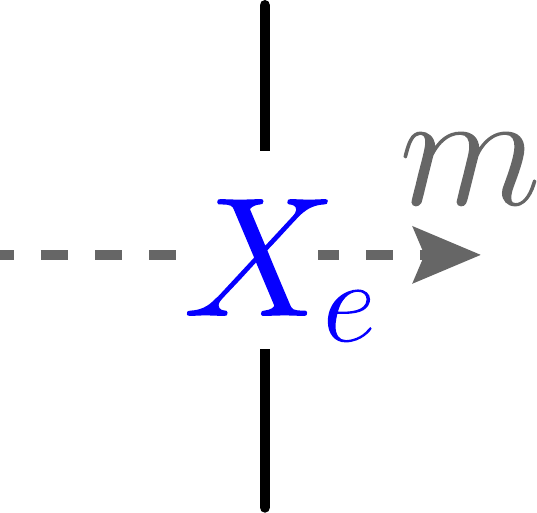}}}
    , \, \, 
    \vcenter{\hbox{\includegraphics[scale=0.25]{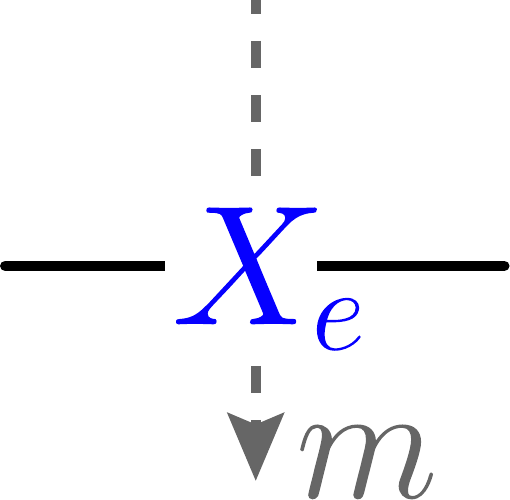}}}
    ~,
\label{eq: short string of Z4 e and m}
\end{align}
where the single Pauli $Z$ and $X$ move $e$ and $m$ anyons according to the convention above. Note that the horizontal $Z$ can be interpreted as either moving $e$ from left to right or moving $e^3$ from right to left. That is to say, an anyon type traveling along a path is equivalent to its inverse anyon type traveling in the opposite direction. The Hermitian conjugate thus moves the anyon types in the opposite direction.

To build a topological subsystem code characterized by $\{1,em,e^2m^2,e^3m^3\}$, we leverage the braiding relations of the $\ZZ_4$ TC anyon types -- in particular, the braiding relations of the anyon type $e^3m$. We recall that the exchange statistics of a general $\ZZ_4$ TC anyon type $e^pm^q$ is given by:
\begin{align}
\theta(e^pm^q) = i^{pq},
\end{align}
where $p$ and $q$ are valued in $\ZZ_4$. Note that the $e^3m$ anyon type is an antisemion, meaning that $\theta(em^3)=-i$. According to Eq.~\eqref{eq: braiding identity}, the braiding relations between two general $\ZZ_4$ anyon types $e^pm^q$ and $e^rm^s$ are then:
\begin{align}
B_\theta(e^pm^q,e^rm^s) = i^{ps + qr},
\end{align}
for $p,q,r,s$ in $\ZZ_4$.
Importantly, the antisemion $e^3m$ has trivial braiding relations with the anyon types in the subtheory $\{1, em, e^2m^2, e^3m^3\}$ and nontrivial braiding relations with all of the other anyon types. 

The braiding relations of the antisemion suggest a means of constructing the $\ZZ_4^{(1)}$ subsystem code from the $\ZZ_4$ TC. As described in Section~\ref{sec: anyon theories for subsystem codes}, the anyon types of a topological subsystem code generate bare logical operators when they are moved along closed paths. This means that the corresponding string operators commute with all of the gauge operators. Therefore, if we construct a gauge group that includes the open string operators for the $e^3m$ antisemion, we can exclude the anyon types outside of the set $\{1, em, e^2m^2, e^3m^3\}$. This is because any string operator for an anyon type outside of this set fails to commute with the $e^3m$ string operators, due to the formula for the braiding relations in Eq.~\eqref{eq: braiding formula}.

More specifically, we define a gauge group $\mathcal{G}'$, generated by the terms of the $\ZZ_4$ TC Hamiltonian as well as the short string operators for the $e^3m$ anyon types: 
\begin{align}
\mathcal{G}' \equiv \left \langle 
\vcenter{\hbox{\includegraphics[scale=0.25]{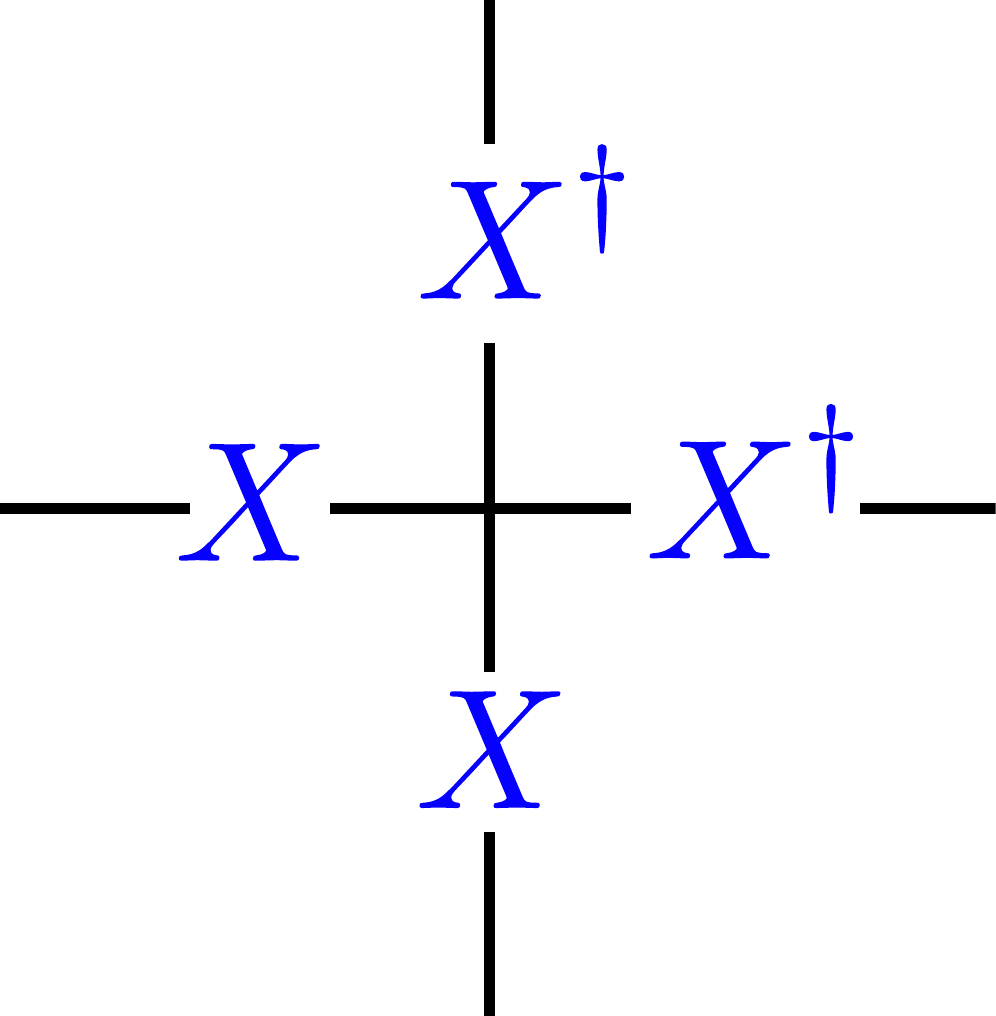}}}, \quad
\vcenter{\hbox{\includegraphics[scale=0.25]{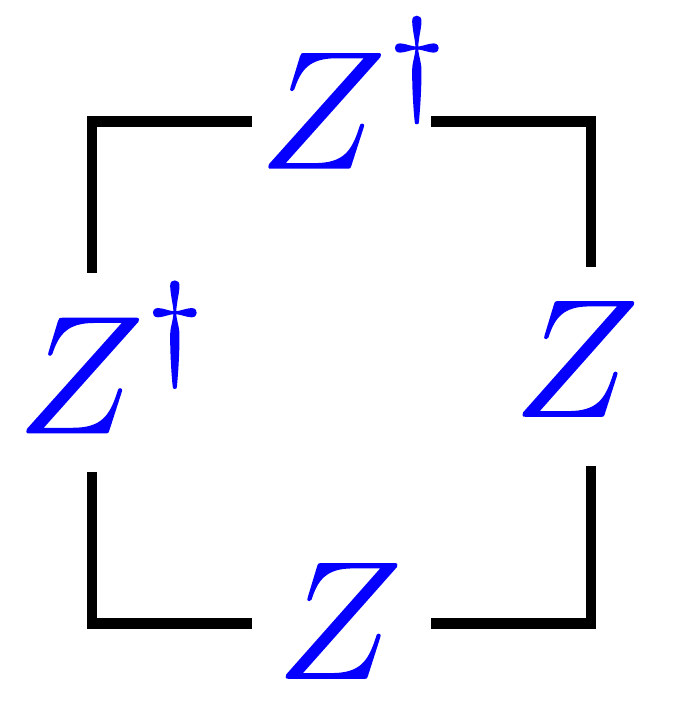}}}, \quad
\vcenter{\hbox{\includegraphics[scale=0.25]{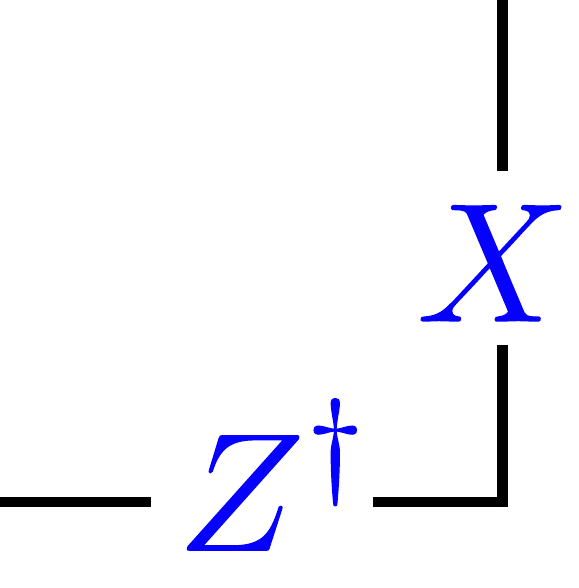}}}, \quad
\vcenter{\hbox{\includegraphics[scale=0.25]{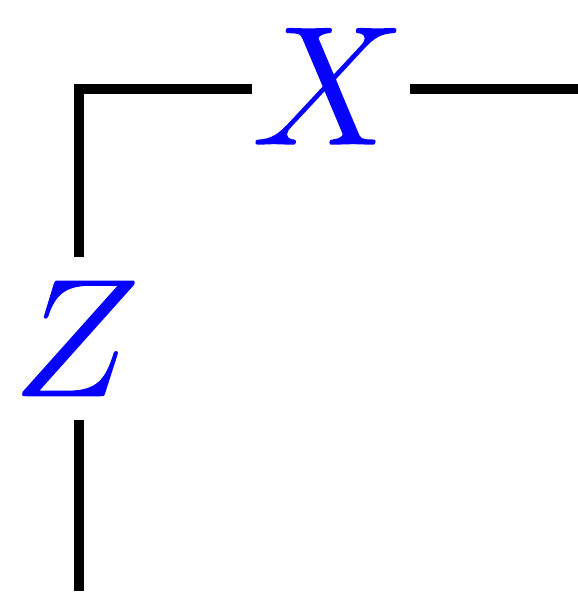}}}
\right \rangle.
\end{align}
The generators of $\mathcal{G}'$ can be simplified by noticing that the vertex term is generated by the short string operators and the plaquette term. Further, the weight of the generators can be reduced by multiplying the plaquette term by short string operators. This gives us:
\begin{align} \label{eq: z41 gauge group construction}
\mathcal{G}' \equiv \left \langle 
\vcenter{\hbox{\includegraphics[scale=0.23]{Figures/e3mh.pdf}}}, \quad
\vcenter{\hbox{\includegraphics[scale=0.23]{Figures/e3mv_2.pdf}}}, \quad
\vcenter{\hbox{\includegraphics[scale=0.23]{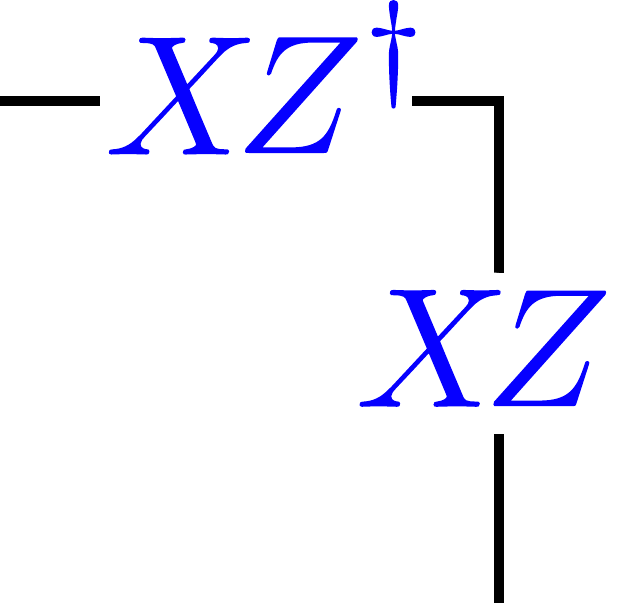}}}
\right \rangle,
\end{align}
whose stabilizer group  {$\mathcal{S}' \propto \mathcal{Z}(\mathcal{G}')$} is generated by string operators along closed paths of the form:
\begin{align} \label{eq: z41 stabilizer group construction}
    \vcenter{\hbox{\includegraphics[scale=0.23]{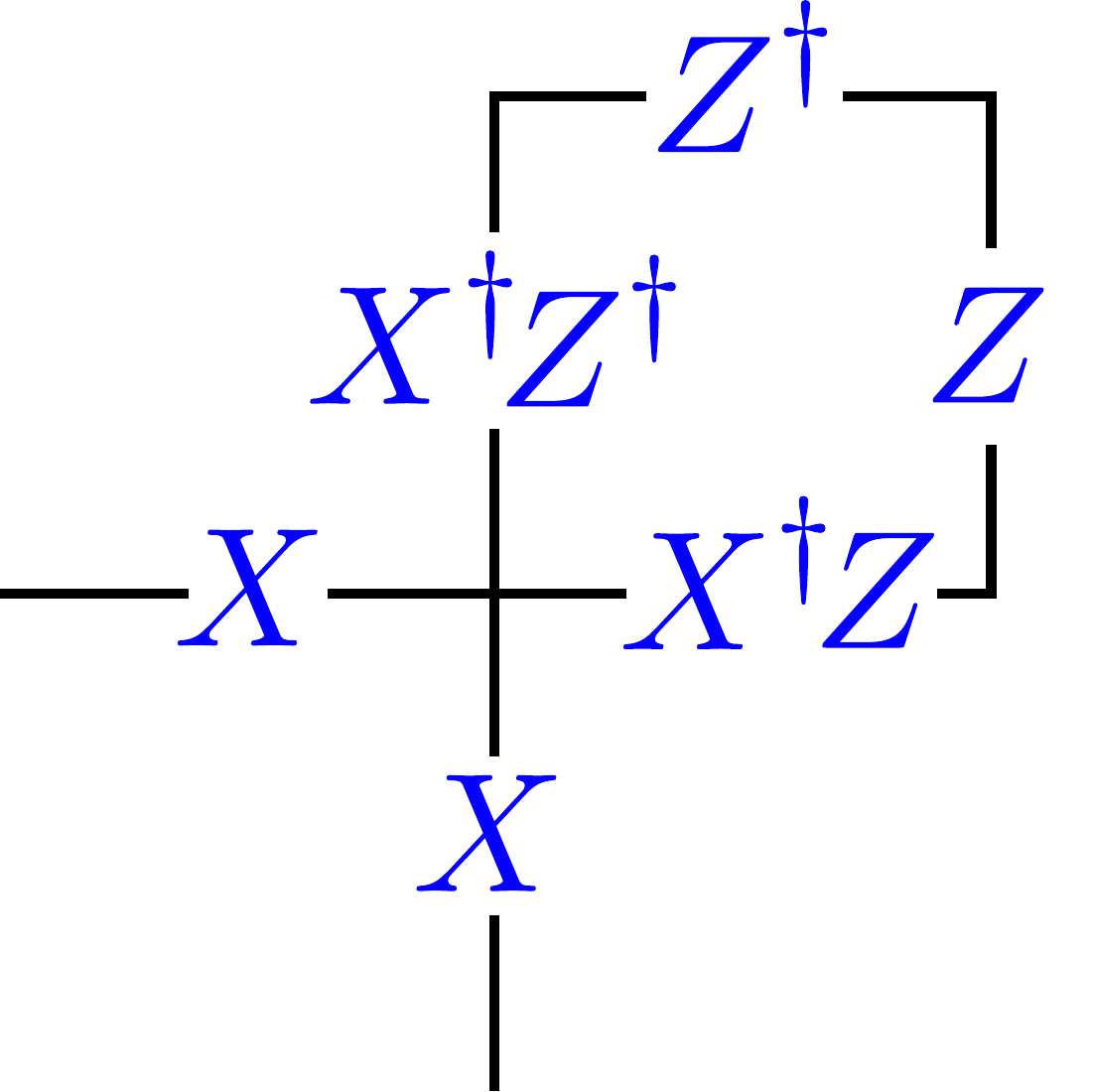}}}, \qquad \vcenter{\hbox{\includegraphics[scale=0.23]{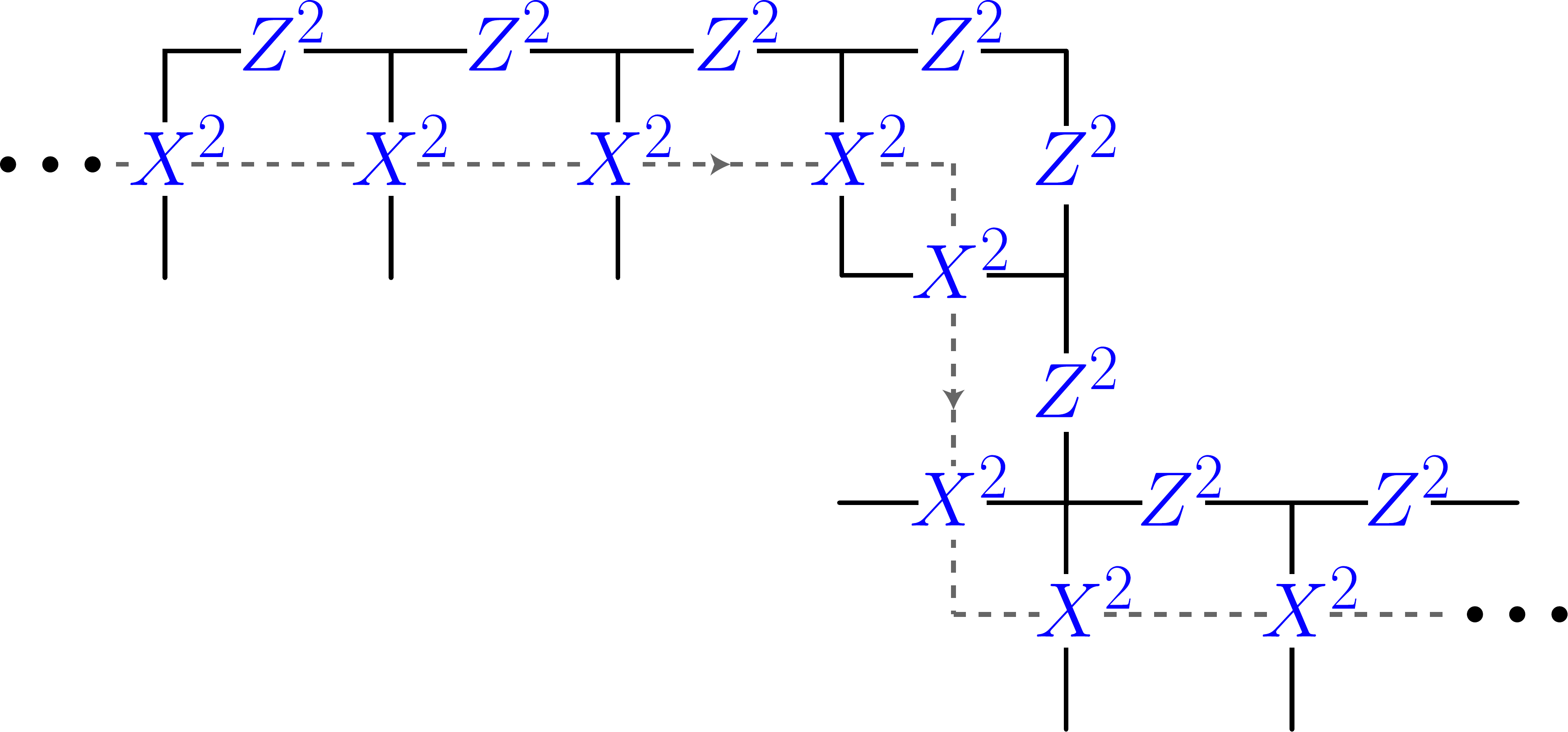}}}
\end{align}
Here, we have assumed that the system is defined on a torus, so the second string operator may wrap around a non-contractible path.

The gauge group $\mathcal{G}'$ implicitly defines the group of bare logical operators $\mathcal{L}$. 
In particular, $\mathcal{L}$ is generated by loops of $em$ string operators, such as:
\begin{align}
    \vcenter{\hbox{\includegraphics[scale=0.23]{Figures/emloop.pdf}}},
    \qquad
    \vcenter{\hbox{\includegraphics[scale=0.23]{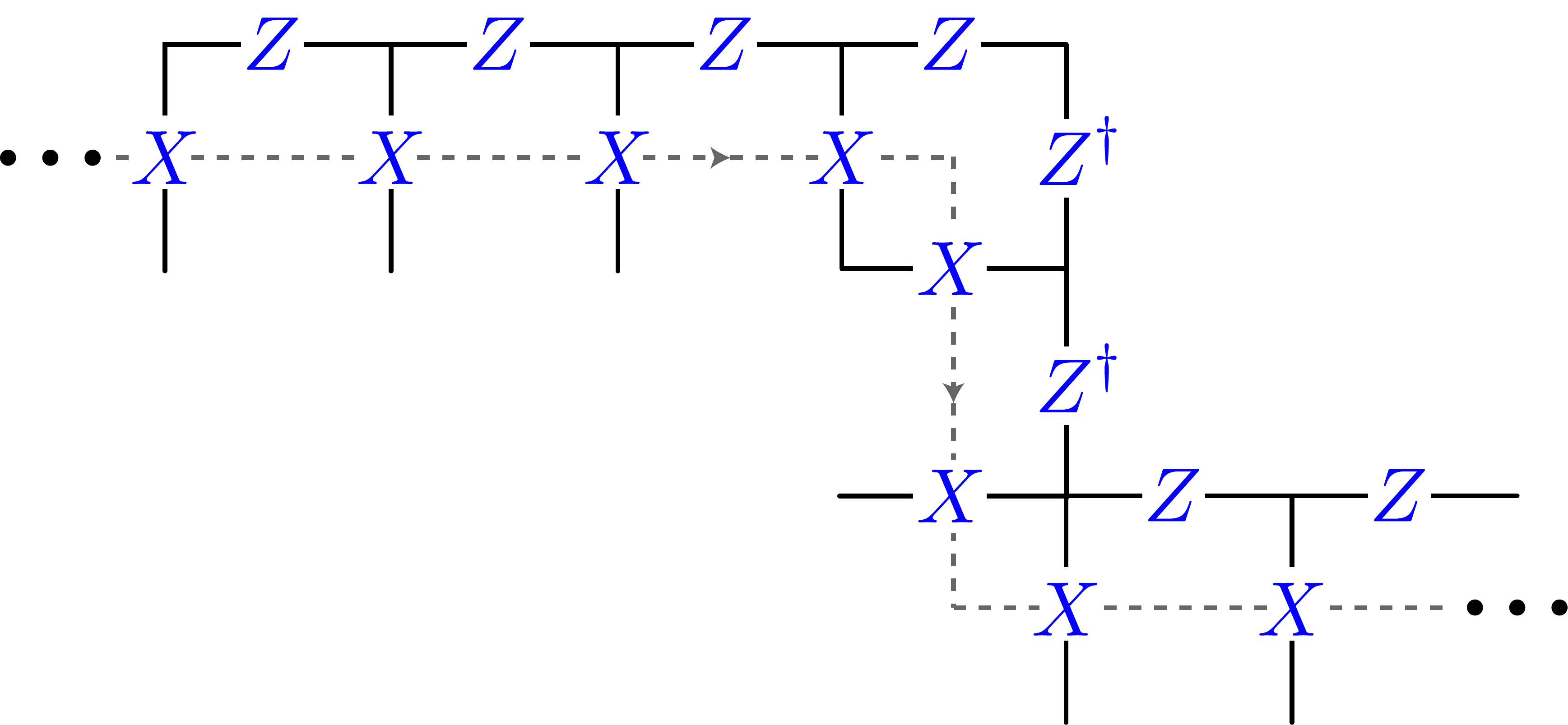}}},
\end{align}
where the small loop of $em$ string operators is one of the generators of the stabilizer group in Eq.~\eqref{eq: z41 stabilizer group construction}, and the $em$ string operators along non-contractible loops give nontrivial bare logical operators.
We thus see that the topological subsystem code is characterized by the $\ZZ_4^{(1)}$ subtheory $\{1, em, e^2m^2, e^3m^3 \}$ of the $\ZZ_4$ TC. 

We refer to this procedure of introducing gauge operators to the $\ZZ_4$ TC as ``gauging out'' the $e^3m$ anyon types of the $\ZZ_4$ TC.\footnote{This should be distinguished from the concept of ``gauging a $1$-form symmetry'', in which a system with a $1$-form symmetry is mapped to a system with a $0$-form symmetry (in two dimensions). Notably, the $1$-form symmetry associated to $e^3m$ is anomalous, as captured by the exchange statistics. This implies that the $1$-form symmetry corresponding to $e^3m$ cannot be gauged in this sense. Here, we instead use the word ``gauging'' in the phrase ``gauging out'' to refer to the gauge subsystem $\mathcal{H}_G$.} This has two effects on the anyon theory. First, the anyon types generated by $e^3m$ become undetectable, because they are created by string operators built from gauge operators. The $e^2m^2$ anyon type, in particular, is undetectable -- i.e, the string operator for an $e^2m^2$ anyon type can be chosen such that it commutes with all of the stabilizers of the topological subsystem code at its endpoints. Second, we remove the anyon types that have nontrivial braiding relations with $e^3m$. For example, $e$ and $m$ have nontrivial braiding relations with $e^3m$. Consequently, the $e$ and $m$ anyons are not anyon types of the topological subsystem code, since their string operators fail to commute with the gauge operators, namely the short $e^3m$ string operators.

 {Although the effects are similar, we would like to pause the construction to emphasize the distinction between gauging out and condensation. To this end, we let $\mathcal{F}$ be the group of open string operators for some anyon type. In the case above, $\mathcal{F}$ is the group generated by the short string operators for $e^3m$. The process of gauging out takes a stabilizer group $\mathcal{S}$ (e.g. the $\ZZ_4$ TC) to a gauge group $\mathcal{G} = \langle \mathcal{S}, \mathcal{F} \rangle$ (up to phases). Importantly, we do not require $\mathcal{F}$ to be Abelian. In fact, $\mathcal{F}$ is only Abelian if the corresponding anyons are bosons, which is not the case for the $e^3m$ anyon types. If $\mathcal{F}$ is Abelian, then we can further gauge fix $\mathcal{F}$. This has the effect of condensing the anyon types. Therefore, gauging out can be interpreted as the first step of condensation, where the second step can only be carried out if the string operators are mutually commuting, i.e., the anyon types are bosons.}  


\begin{figure}[t] 
\centering
\includegraphics[scale=0.28]{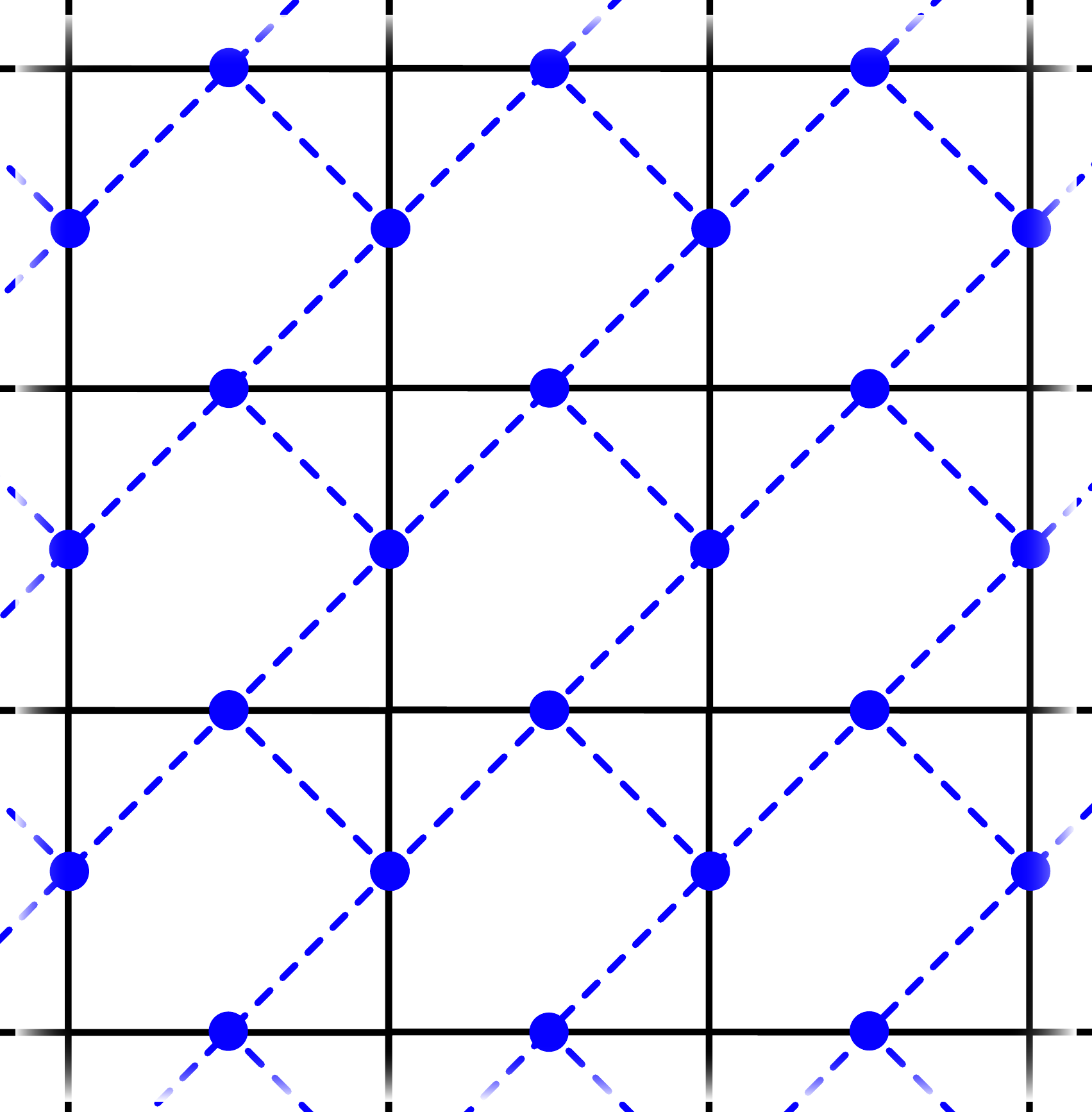}
\caption{The four-dimensional qudits (blue) on the edges of a square lattice define a hexagonal lattice (dashed blue lines) with a four-dimensional qudit at each vertex.}
\label{fig: squaretohexagon}
\end{figure}

Finally, to relate the topological subsystem code defined by $\mathcal{G}'$ in Eq.~\eqref{eq: z41 gauge group construction}
to the $\ZZ_4^{(1)}$ subsystem code defined by Eq.~\eqref{eq: z41 gauge group}, we apply a Clifford transformation. We define the single site Clifford unitary $F$ according to its action on the Pauli operators via conjugation:
\begin{align} \label{eq: F Clifford def}
F: X \mapsto Z, \quad Y \mapsto -\sqrt{i} Z^\dagger X, \quad Z \mapsto X^\dagger.
\end{align}
We then apply the transformation in Eq.~\eqref{eq: F Clifford def} to all of the horizontal edges. This maps the gauge group $\mathcal{G}'$ to:
\begin{align}
\mathcal{G} = \left \langle \,
\vcenter{\hbox{\includegraphics[scale=0.25]{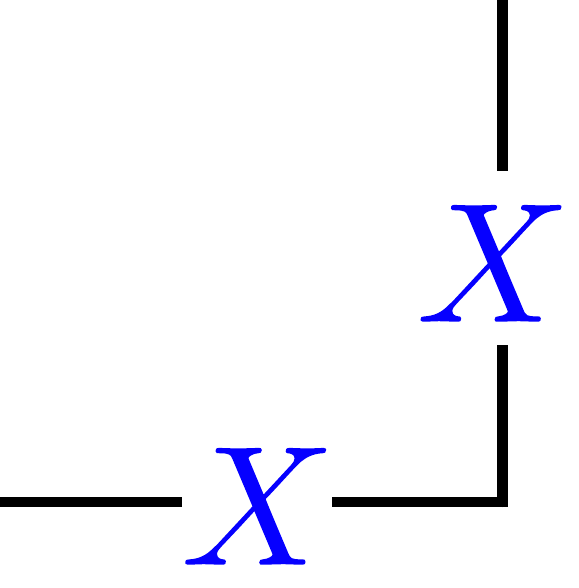}}}, \quad
\vcenter{\hbox{\includegraphics[scale=0.25]{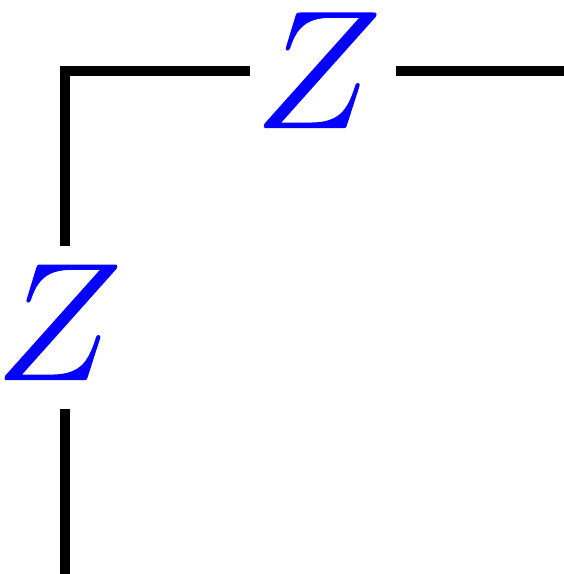}}}, \quad
\vcenter{\hbox{\includegraphics[scale=0.25]{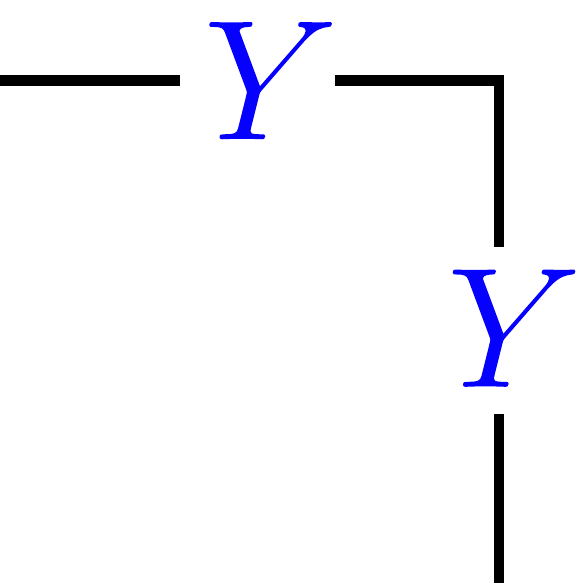}}}
 \right \rangle.
\end{align}
By redefining the system on a hexagonal lattice with four-dimensional qudits on the vertices, as in Fig.~\ref{fig: squaretohexagon}, we see that $\mathcal{G}$ is precisely the gauge group in Eq.~\eqref{eq: z41 gauge group}. Thus, we have derived the $\ZZ_4^{(1)}$ subsystem code from a $\ZZ_4$ TC by gauging out the $e^3m$ antisemions.

\section{Chiral semion subsystem code} \label{sec: chiral semion} 

Our next example is a topological subsystem code characterized by the chiral semion anyon theory. We label the anyon types of the chiral semion anyon theory by $\{1,s\}$. These form a $\ZZ_2$ group under fusion and the nontrivial anyon type $s$ has semionic exchange statistics (i.e., $\theta(s) = i$). As the name suggests, this anyon theory is chiral. Formally, this means that the chiral central charge $c_-$ is nonzero. We recall that the chiral central charge of a modular Abelian anyon theory can be computed modulo $8$ using the formula:
\begin{align} \label{eq: chiral anyons}
e^{2\pi i c_-/8} = \frac{1}{\sqrt{|A|}} \sum_a \theta(a),
\end{align}
where the sum is over all anyon types and $|A|$ is the total number of anyon types. From this, we see that the chiral central charge of the chiral semion theory is ${c_-= 1 \text{ mod }8}$. The widely held belief \cite{kitaev2006anyons, Kapustin2020thermal} is that anyon theories with a nonzero chiral central charge cannot be modeled by commuting projector Hamiltonians. This suggests, in particular, that the chiral semion theory cannot be modeled by the Hamiltonian of a topological stabilizer code. 
Thus, the subsystem code described below is distinct from any topological stabilizer code in that it is characterized by a chiral anyon theory.

\subsection{Definition of the subsystem code}

\begin{figure}[tb] 
\centering
\hspace{3cm}\includegraphics[scale=0.25]{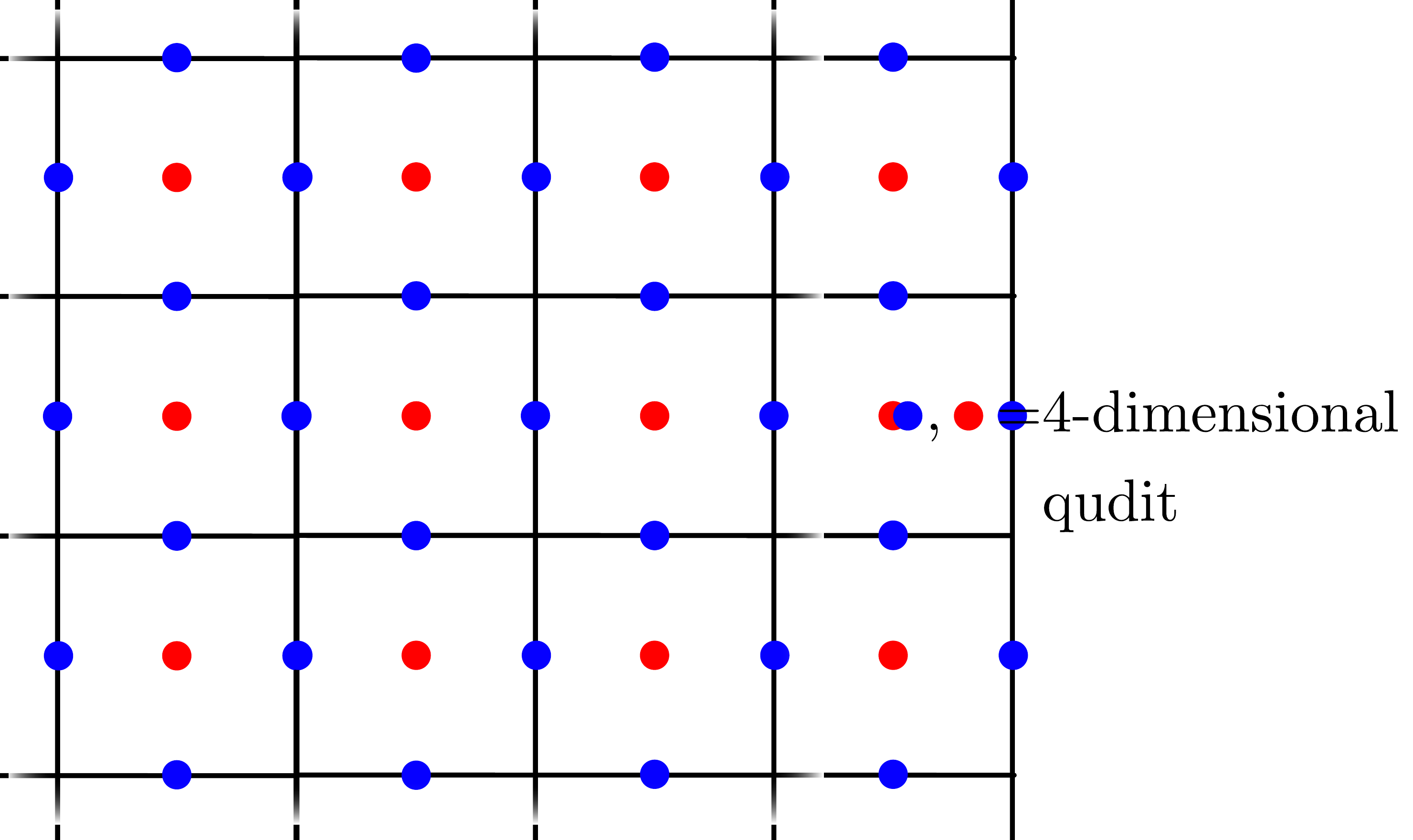}
\caption{The Hilbert space of the chiral semion subsystem code is composed of a four-dimensional qudit on each edge (blue) and a four-dimensional qudit (red) on each plaquette.}
\label{fig: modifieddofZ4}
\end{figure}

The chiral semion subsystem code is defined on a square lattice with a four-dimensional qudit at each edge and each plaquette, as in Fig.~\ref{fig: modifieddofZ4}. The gauge group $\mathcal{G}$ is given by:
\begin{align}
\mathcal{G} \equiv \left \langle 
\vcenter{\hbox{\includegraphics[scale=0.23]{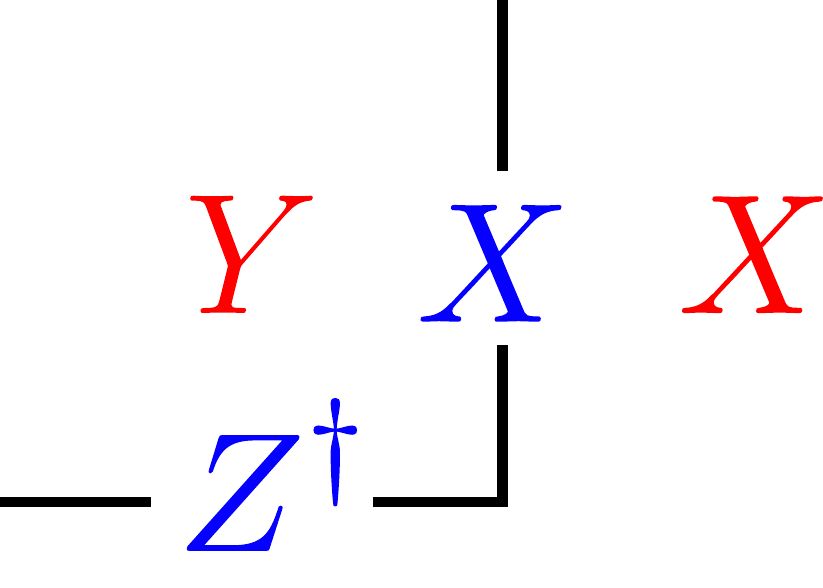}}}, \quad
\vcenter{\hbox{\includegraphics[scale=0.23]{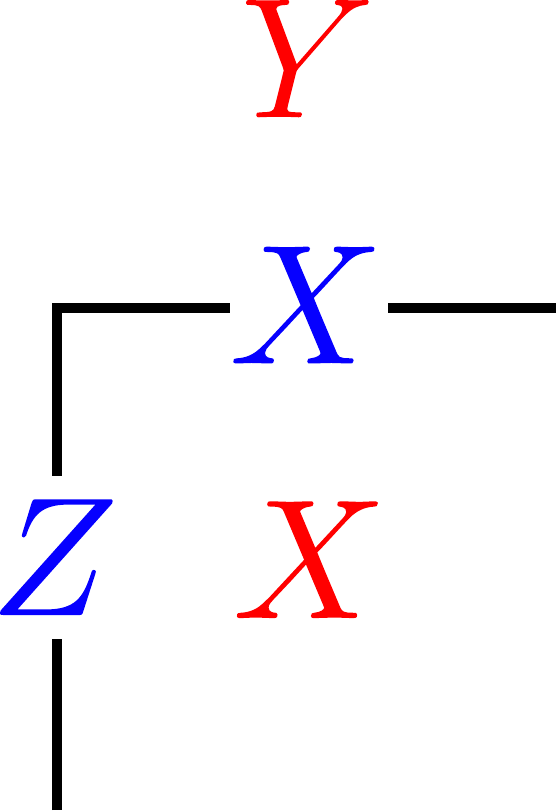}}}, \quad
\vcenter{\hbox{\includegraphics[scale=0.23]{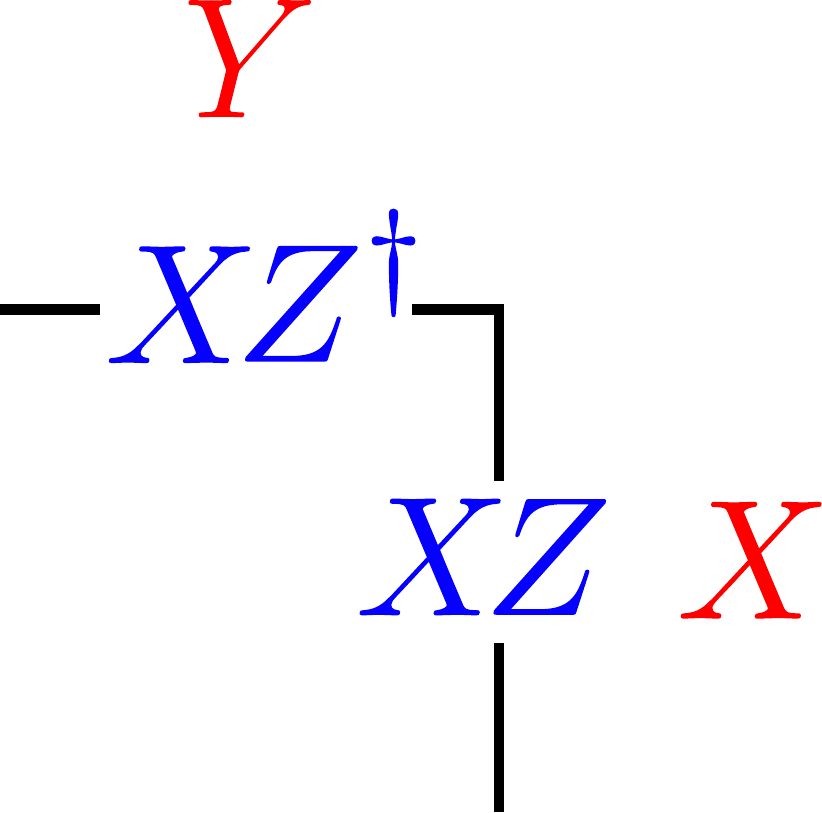}}}, \quad
\textcolor{blue}{Z^2} \right \rangle.
\end{align}
This defines the stabilizer group up to a factor of $i$. We take the stabilizer group to be:
\begin{align}
\mathcal{S} \equiv \left \langle
i \times
\raisebox{-1.4cm}{\hbox{\includegraphics[scale=0.23]{Figures/emloop.pdf}}}
, \quad
\vcenter{\hbox{\includegraphics[scale=0.23]{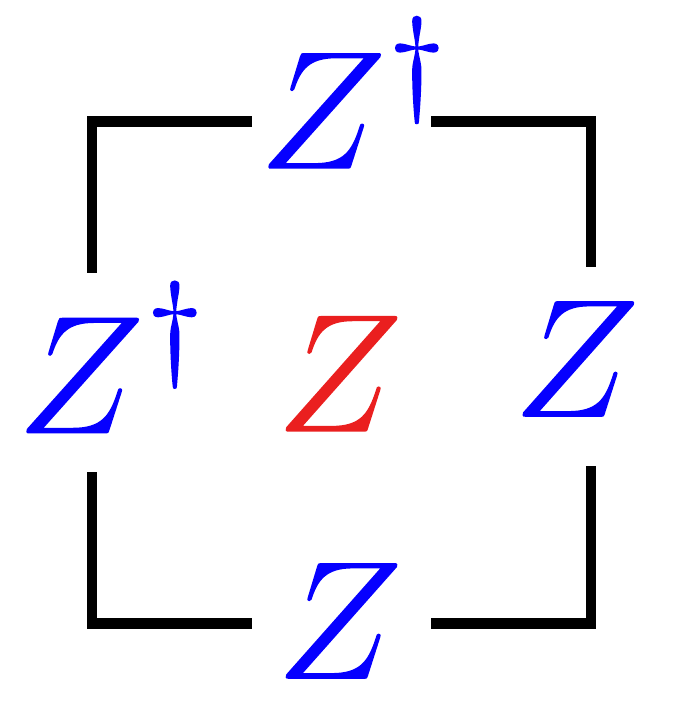}}}, \quad
\vcenter{\hbox{\includegraphics[scale=0.23]{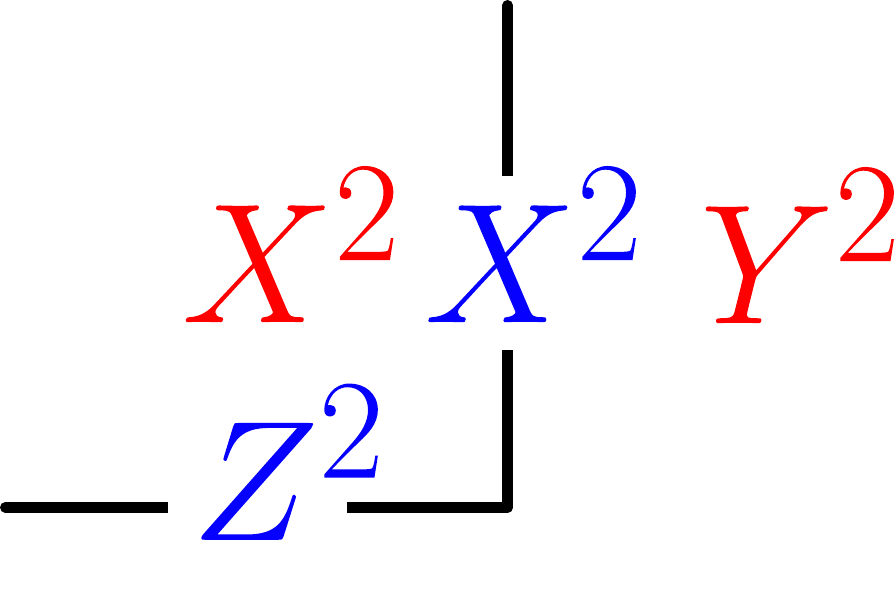}}}, \quad
\vcenter{\hbox{\includegraphics[scale=0.23]{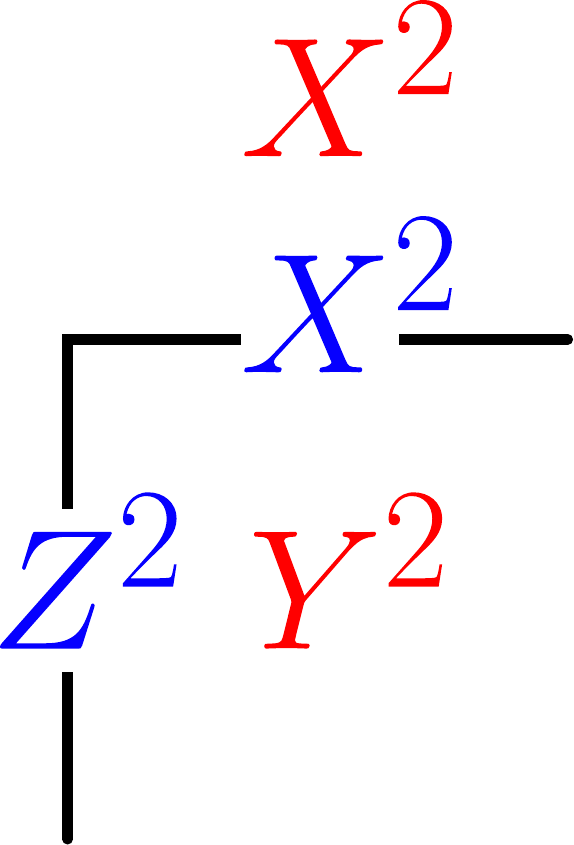}}}
\right \rangle.
\end{align}
Notice that the vertex term (on the far left) has a factor of $i$. Given the relations between the stabilizers, this is necessary to ensure that the only element proportional to the identity is the identity itself.
Unlike the $\ZZ_4^{(1)}$ subsystem code in the previous section, the stabilizer group of the chiral semion subsystem code admits a set of local generators on a torus. We attribute this to the fact that the chiral semion anyon theory is modular,  {since nonlocal stabilizer generators necessarily correspond to string operators of transparent anyon types (see Appendix~\ref{app: bare logicals and nonlocal stabilizers} for more details)}. 

Lastly, assuming that the subsystem code is defined on a torus, the bare logical group is generated by the stabilizer group and the following two nontrivial bare logical operators, which are supported along the non-contractible paths $\gamma$ and $\gamma'$ in Fig.~\ref{fig: braiding}:
\begin{align} \label{eq: chiral semion logicals}
L_\gamma \equiv \vcenter{\hbox{\includegraphics[scale=0.23]{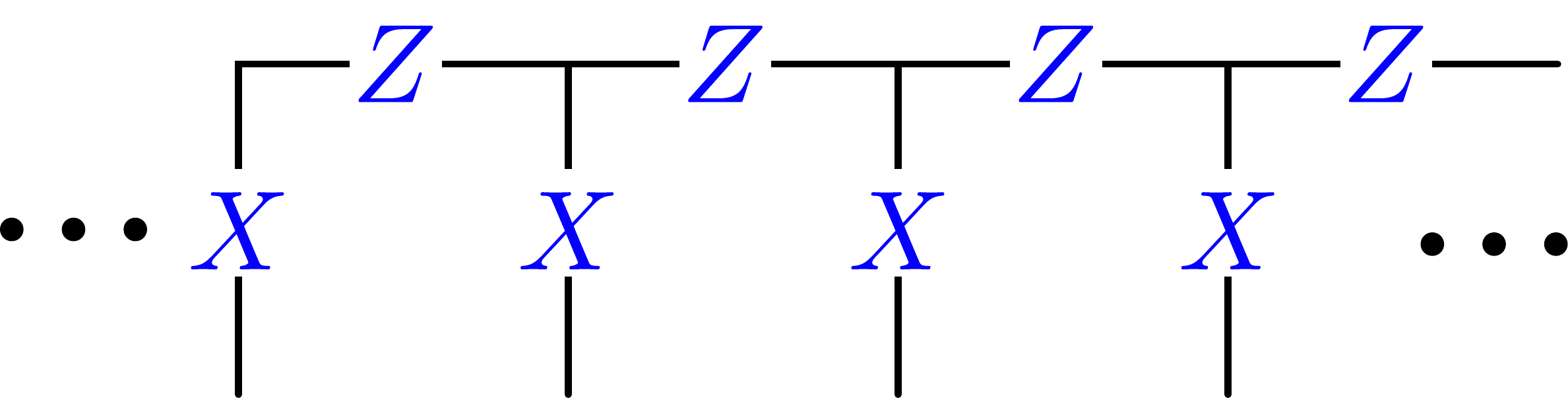}}}~,
\qquad 
L_{\gamma'} \equiv \vcenter{\hbox{\includegraphics[scale=0.23]{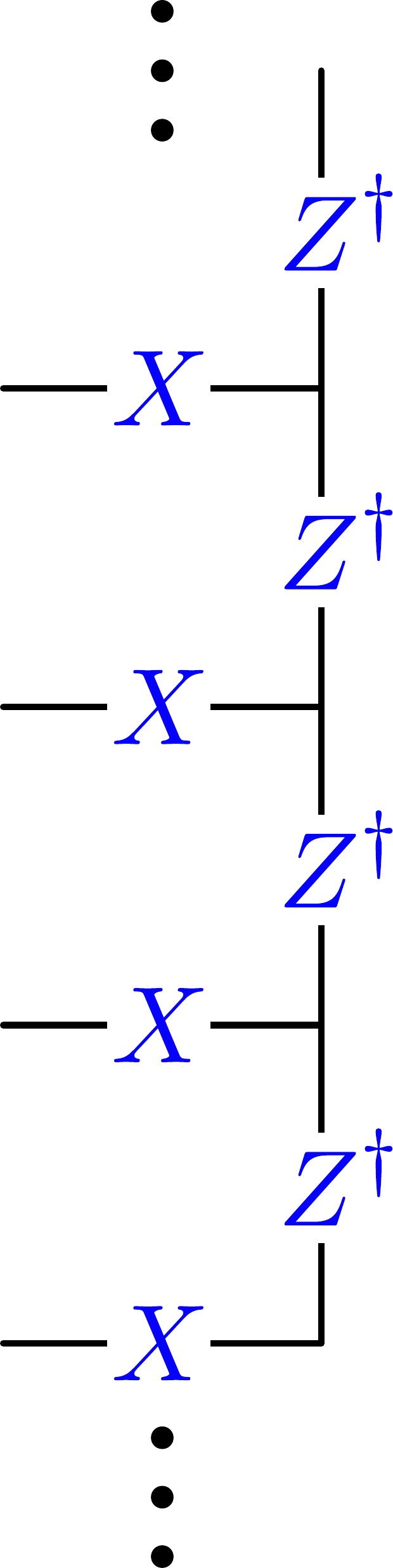}}}.
\end{align}
These string operators obey the commutation relations:
\begin{align} \label{eq: chiral semion logical commutation}
L_{\gamma'} L_\gamma = - L_\gamma L_{\gamma'}.
\end{align}
Therefore, the chiral semion subsystem code encodes a single qubit when defined on a torus. We defer the counting of the number of stabilized qudits, gauge qudits, and logical qubits to the general construction in Section~\ref{sec: general}.

The anyon theory of the subsystem code can be determined straightforwardly from the nontrivial bare logical operators in Eq.~\eqref{eq: chiral semion logicals}. A possible truncation of $L_\gamma$ is given by:
\begin{align}
\vcenter{\hbox{\includegraphics[scale=0.25]{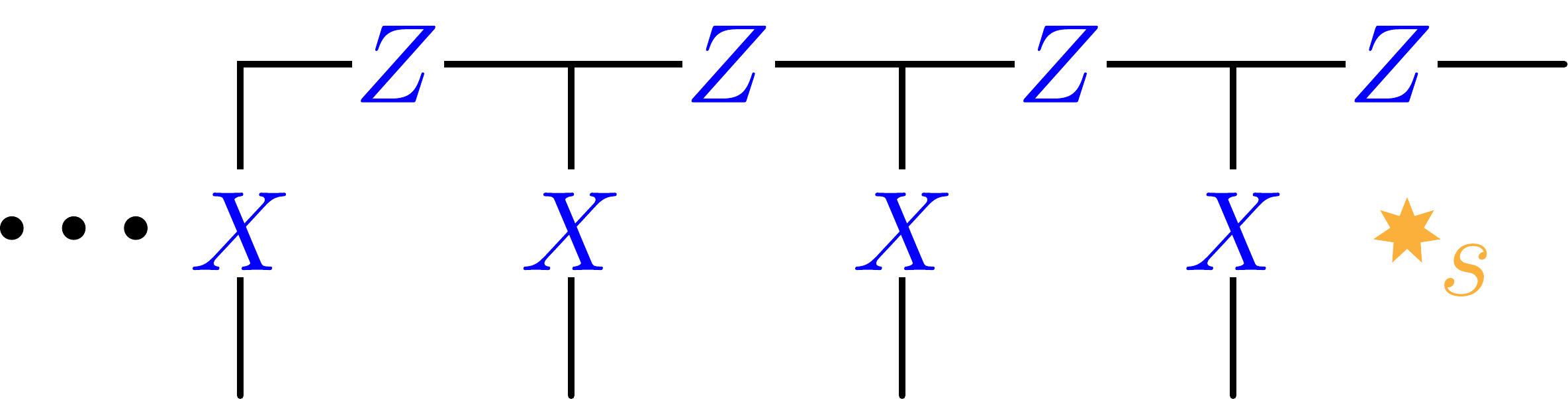}}} ~. 
\end{align}
This fails to commute with stabilizers at the endpoint. Moreover, the commutation relations with the stabilizers cannot be reproduced by a Pauli operator localized near the endpoint, since $L_\gamma$ is a nontrivial bare logical operator. This follows from Proposition~\ref{prop: bare logicals and nonlocal stabilizers} in Appendix~\ref{app: bare logicals and nonlocal stabilizers}. Therefore, the endpoint of the truncation of $L_\gamma$ is a detectable anyon type. We suggestively call the anyon type $s$. 

\begin{figure}[tb] 
\centering
\includegraphics[width=.6\textwidth]{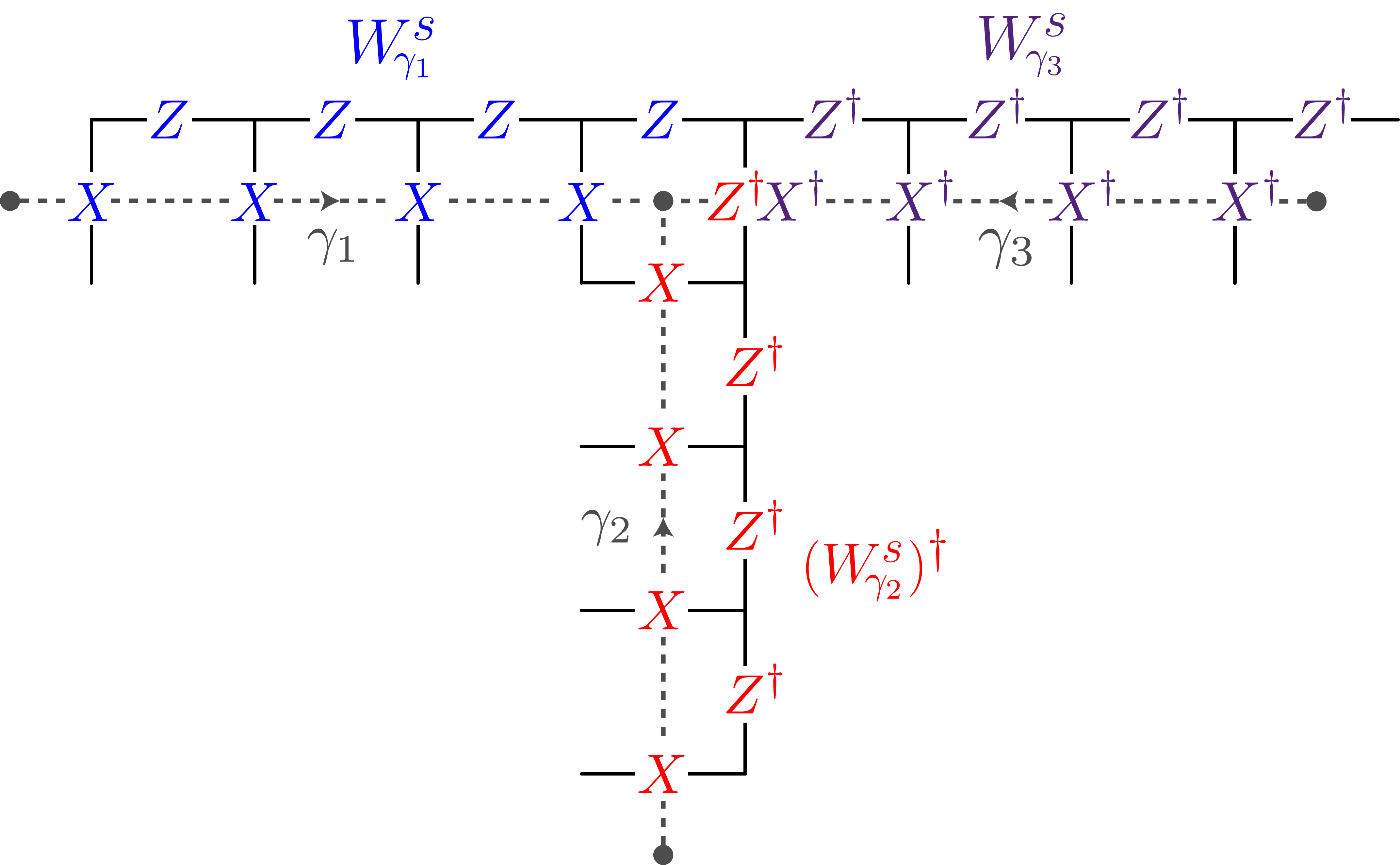}
\caption{The exchange statistics of the anyon type $s$ can be computed from the commutation relations of the string operators $W^s_{\gamma_1}$, $W^s_{\gamma_2}$, and $W^s_{\gamma_3}$ (blue, red, and purple, respectively), defined along the paths $\gamma_1$, $\gamma_2$, and $\gamma_3$ (gray). Importantly, the string operators have the same commutation relations with the gauge operators supported at the common endpoint. The exchange statistics of $s$ then follow from Eq.~\eqref{eq: statistics formula}.
}
\label{fig: semionstatistics}
\end{figure}

We next determine the fusion rules and exchange statistics for $s$. Although the logical operators are order four, the anyon type $s$ has order two under fusion, i.e., $s \times s =1$. This is because the truncated string operator squares to a product of stabilizers, shown explicitly in the next section. The braiding relations for $s$ follow from Eq.~\eqref{eq: chiral semion logical commutation}, which tells us that $B_\theta(s,s)=-1$. Further, the exchange statistics of $s$ can be computed using the formula in Eq.~\eqref{eq: statistics formula}. The calculation of the exchange statistics of $s$ is shown in Fig.~\ref{fig: semionstatistics}. We find that $s$ is indeed a semion: $\theta(s) = i$. 

\subsection{Construction of the subsystem code} \label{sec: chiral semion construction}

\begin{figure}[tb] 
\centering
\includegraphics[width=.7\textwidth]{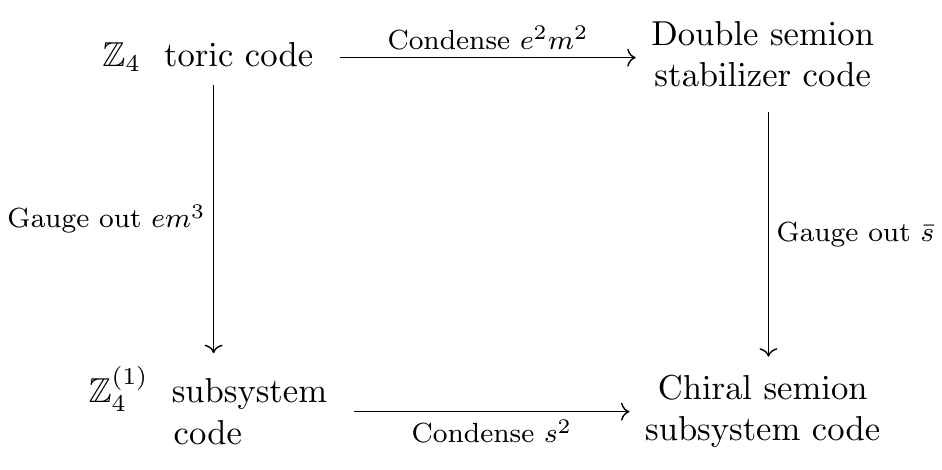}
\caption{We construct the chiral semion subsystem code by gauging out the antisemion $\bar{s}$ in a DS stabilizer code.~Alternatively, the chiral semion subsystem code can be constructed from the $\ZZ_4^{(1)}$ subsystem code by condensing the transparent boson $s^2$. The condensation of $s^2$ can be made explicit by first adding a trivial subsystem code to the plaquettes, with gauge generators that are~single-site $X$ and $Z$. The $s^2$ boson is then condensed by measuring $s^2$ short string operators, such as the edge terms in Eq.~\eqref{eq: DS Hamiltonian terms} redefined on a hexagonal lattice according to the mapping in Fig.~\ref{fig: squaretohexagon}.}
\label{fig: chiral semion construction}
\end{figure}

We now describe how the chiral semion subsystem code can be constructed from a double semion (DS) stabilizer code \cite{Ellison2022Pauli}. The construction is motivated by the fact that the  anyon theory of the DS stabilizer code contains the chiral semion theory as a subtheory. To make this explicit, we recall that the anyon types of the DS stabilizer code form a $\ZZ_2 \times \ZZ_2$ group under fusion, with elements labeled by $\{1, s, \bar{s}, s \bar{s}\}$. The anyon type $s$ is a semion, $\bar{s}$ is an antisemion, and $s \bar{s}$ is a boson, i.e.:
\begin{align} \label{eq: DS statistics}
\theta(1) = 1, \quad \theta(s) = i, \quad \theta(\bar{s}) = -i , \quad \theta(s \bar{s}) = 1.
\end{align}
This implies, in particular, that the semions and antisemions have the following braiding relations:
\begin{align}
B_\theta(s,s) = -1, \quad B_\theta(\bar{s},\bar{s}) = -1, \quad B_\theta(s,\bar{s})=1.
\end{align}
Since the semion $s$ braids trivially with the antisemion, the subgroups $\{1, s\}$ and $\{1,\bar{s}\}$ are independent. Hence, the anyon theory generated by $s$ is precisely the chiral semion theory. Consequently, we build the chiral semion subsystem code by gauging out the antisemion $\bar{s}$ in a DS stabilizer code. This mirrors the construction of the $\ZZ_4^{(1)}$ subsystem code in Section~\ref{sec: z41 construction}, where we gauged out an antisemion in a $\ZZ_4$ TC.
We note that an alternative method for constructing the chiral semion subsystem code is to in fact gauge the $1$-form symmetry associated to the $s^2$ boson in the $\ZZ_4^{(1)}$ subsystem code (Fig.~\ref{fig: chiral semion construction}).

We start by describing a DS stabilizer code, which is a slight modification of the DS stabilizer code introduced in Ref.~\cite{Ellison2022Pauli}.\footnote{Specifically, the DS model in Ref.~\cite{Ellison2022Pauli} is constructed by condensing the $e^2m^2$ anyon types of the $\ZZ_4$ TC. The DS stabilizer code used here, in contrast, is constructed from the $\ZZ_4$ TC by gauging the $1$-form symmetry associated to $e^2m^2$. 
The key difference is that the latter has gauge field degrees of freedom on the plaquettes.} 
Here, the DS stabilizer code is built using four-dimensional qudits on the edges and plaquettes of a square lattice (Fig.~\ref{fig: modifieddofZ4}). We denote the Pauli $X$ and Pauli $Z$ operators on a plaquette $p$ by $X_p$ and $Z_p$.
The Hamiltonian $H_\text{DS}$ of the DS stabilizer code is then a sum of four types of terms, corresponding to vertices~$v$, edges~$e$, and plaquettes~$p$:
\begin{align} \label{eq: DS stabilizer Hamiltonian}
H_\text{DS} = -\sum_v A_v - \sum_p B_p - \sum_e C_e - \sum_p Z^2_p + \text{h.c.}
\end{align}
The vertex terms $A_v$, plaquette terms $B_p$, and edge terms $C_e$ are graphically represented as: 
\begin{equation} \label{eq: DS Hamiltonian terms}
   A_v \equiv i \times \raisebox{-1.4cm}{\hbox{\includegraphics[scale=0.23]{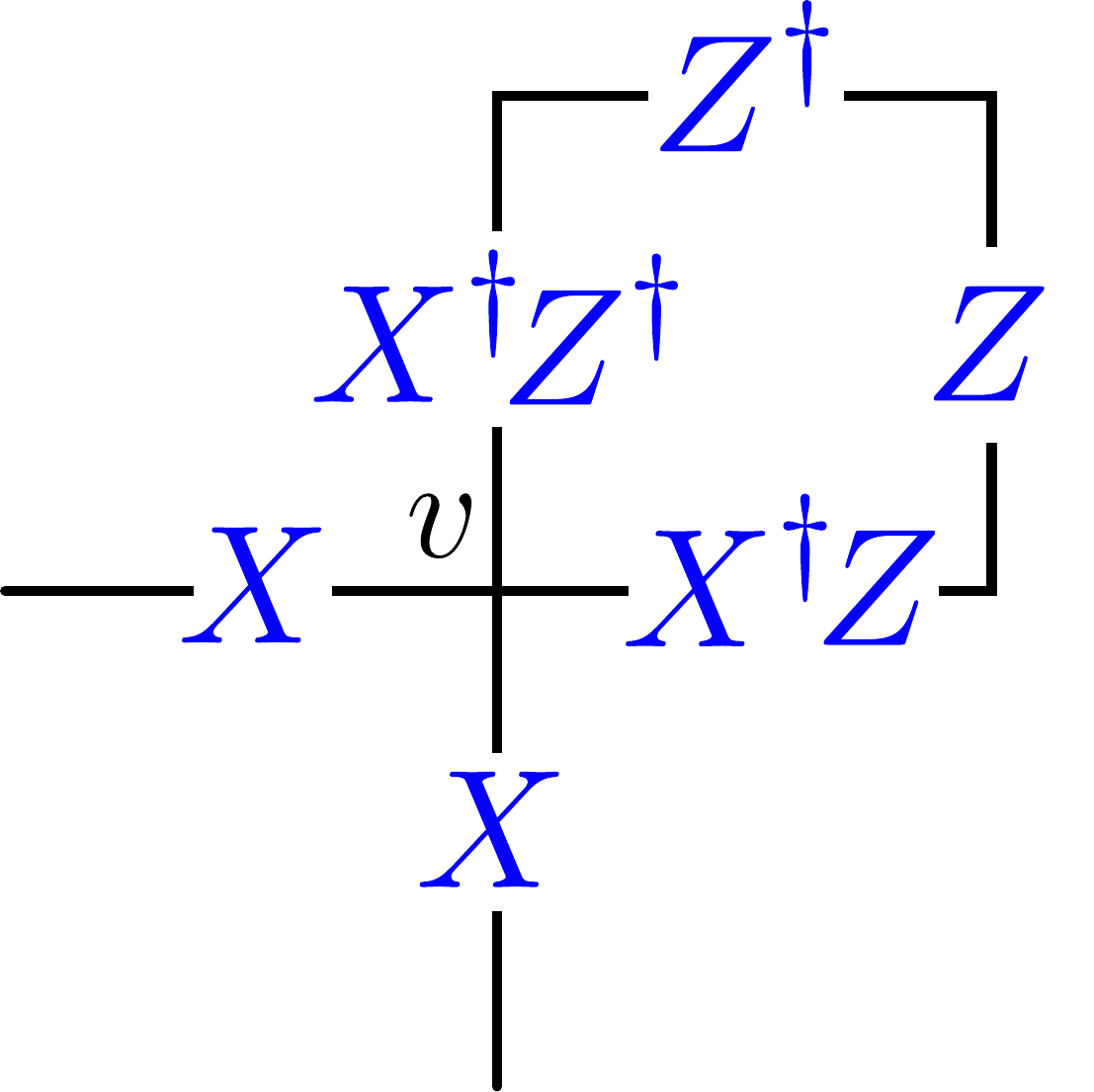}}},\quad 
   B_p \equiv  \vcenter{\hbox{\includegraphics[scale=0.23]{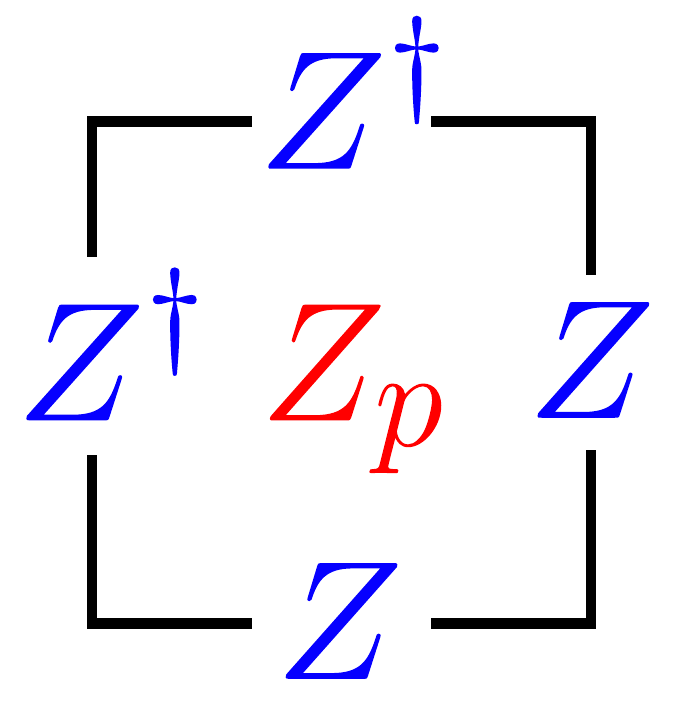}}}, \quad 
   C_e \equiv  \vcenter{\hbox{\includegraphics[scale=0.23]{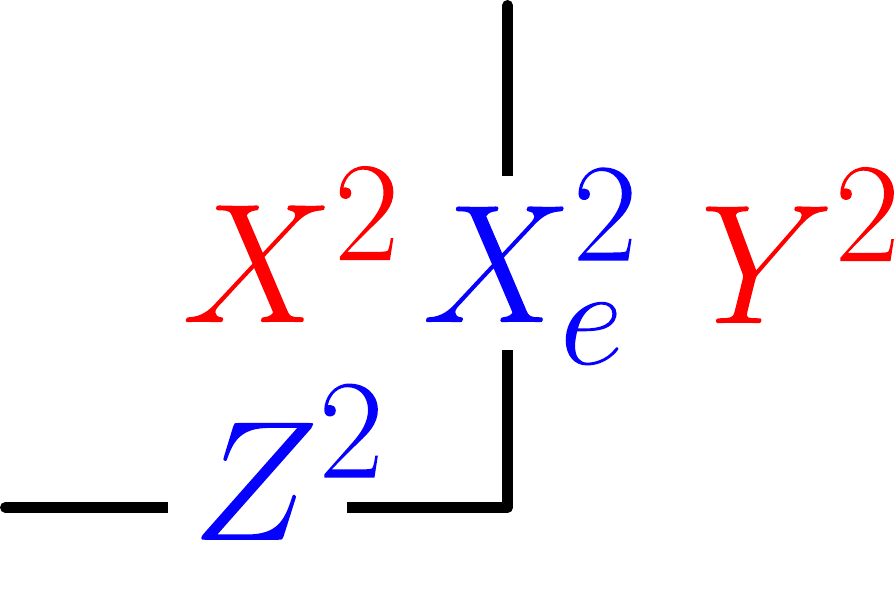}}}, \,\,
   \vcenter{\hbox{\includegraphics[scale=0.215]{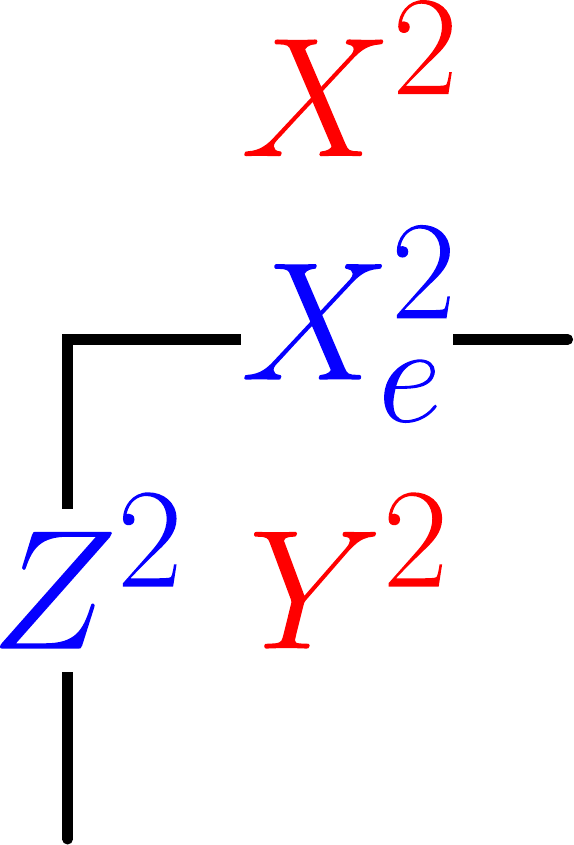}}}.
\end{equation}
The stabilizer group $\mathcal{S}_\text{DS}$ is generated by the Hamiltonian terms:
\begin{align} \label{eq: DS stabilizer group}
\mathcal{S}_\text{DS} \equiv \la \{A_v\}, \{B_p\}, \{{C}_e\}, Z^2_p \ra.
\end{align}

It can be checked that the anyon theory of the DS stabilizer code is indeed given by the anyon types $\{1, s, \bar{s}, s\bar{s}\}$, described above. The anyons can be created by string operators along oriented paths, formed by gluing together the following operators from head-to-tail according to their orientation:
\begin{eqs} \label{eq: DSshortstrings}
    W^s_e \equiv \vcenter{\hbox{\includegraphics[scale=0.23]{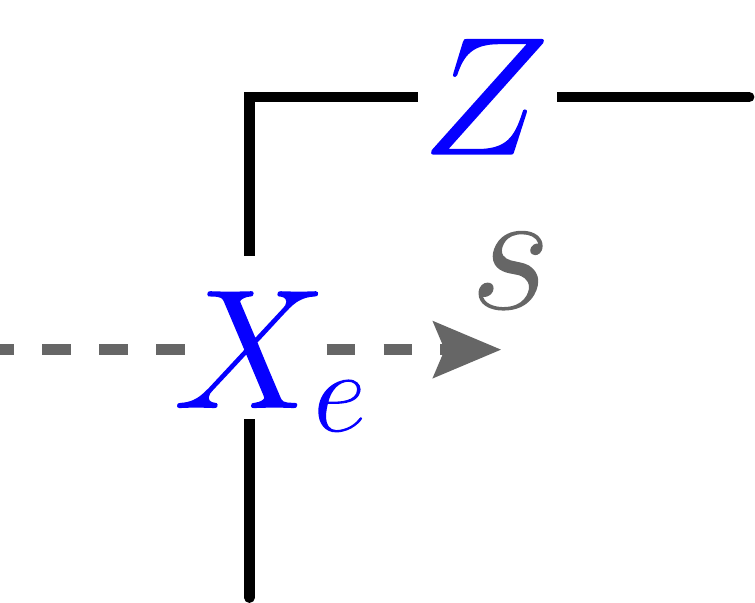}}},
    \,\,
    \vcenter{\hbox{\includegraphics[scale=0.23]{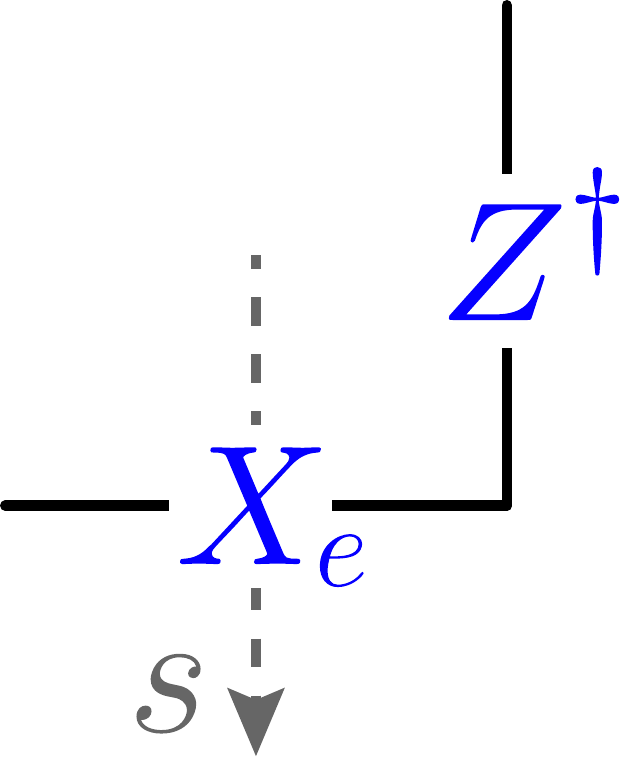}}}, \quad 
    W^{\bar{s}}_e \equiv  \vcenter{\hbox{\includegraphics[scale=0.23]{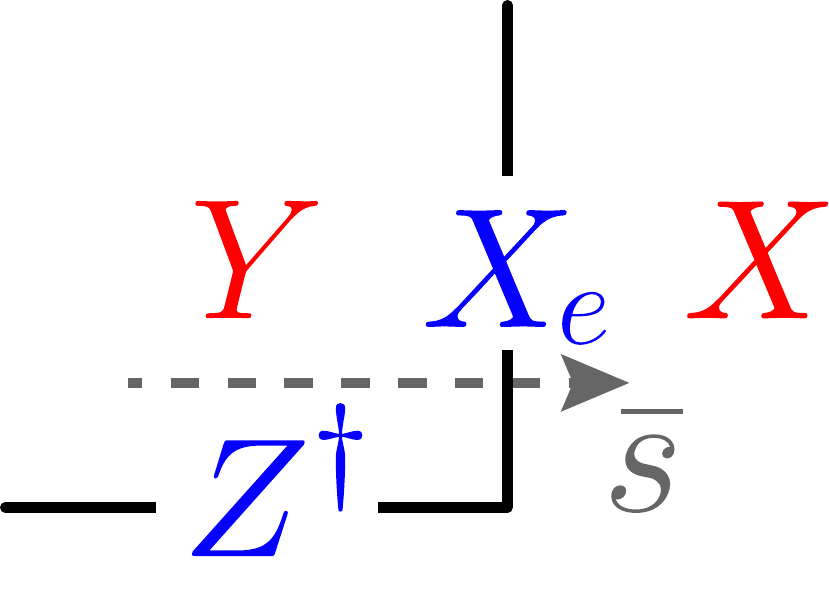}}},
    \,\,
    \vcenter{\hbox{\includegraphics[scale=0.23]{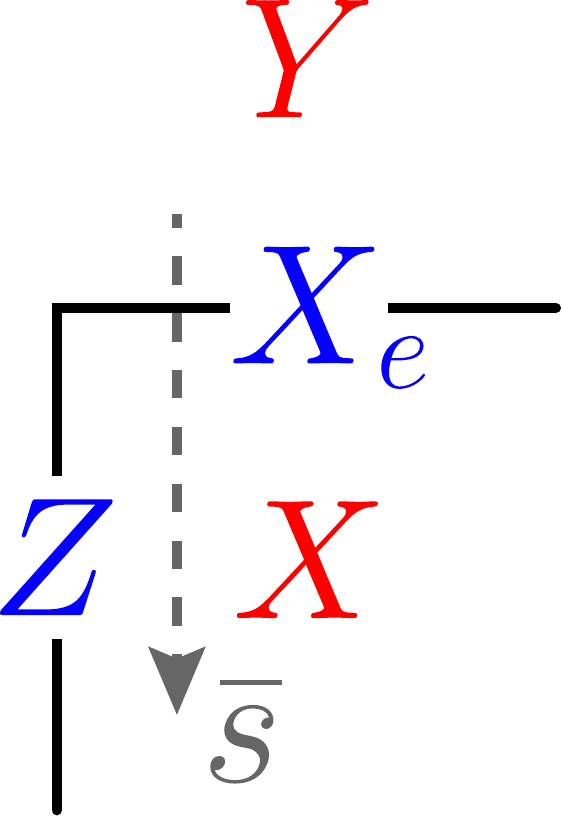}}}, \quad
    W^{s \bar{s}}_e \equiv \vcenter{\hbox{\includegraphics[scale=0.23]{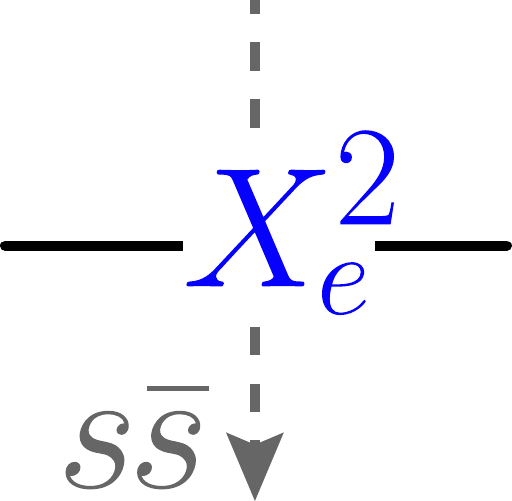}}}, \,\,
    \vcenter{\hbox{\includegraphics[scale=0.23]{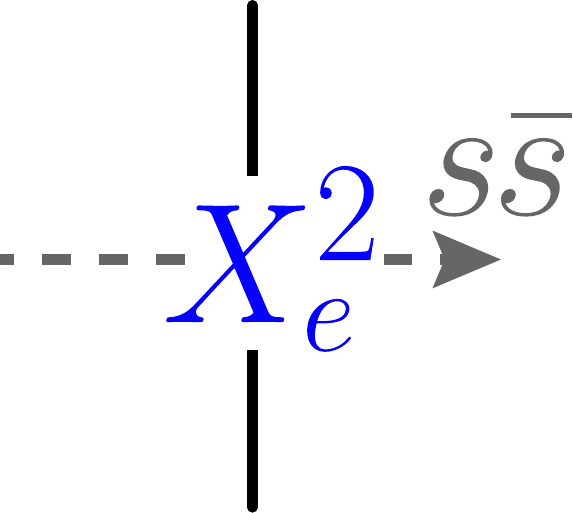}}}.
\end{eqs}
The short string operators that move the anyon types in the opposite direction are given by $(W^s_e)^\dagger$, $(W^{\bar{s}}_e)^\dagger$, and $(W^{s \bar{s}}_e)^\dagger$.
We note that, although the product of short string operators $W^s_e W^{\bar{s}}_e$ differs from $W^{s \bar{s}}_e$, the long string operators $\prod_{e \in \bar{\gamma}} W^{s \bar{s}}_e$ and $\prod_{e \in \bar{\gamma}} (W^s_e W^{\bar{s}}_e)$ along a path $\bar{\gamma}$ in the dual lattice are equivalent, up to stabilizers along $\bar{\gamma}$ and local operators at the endpoints. Thus, they correspond to the same anyon type. We choose the expression $W^{s \bar{s}}_e = X_e^2$ for simplicity.
In this sense, the string operators reproduce the fusion rules of the DS anyon types. Furthermore, using Eq.~\eqref{eq: statistics formula}, one can show that the exchange statistics of the anyon types match those in Eq.~\eqref{eq: DS statistics}. The short string operators in Eq.~\eqref{eq: DSshortstrings}, in fact, account for all of the anyon types of the DS stabilizer code.\footnote{This can be verified by computing the dimension of the logical subspace on a torus. Using the~notation in Section~\ref{sec: z41 definition}, 
there are $\mathbf{N}_Q=3\mathbf{P}$ qudits in total. As for the stabilized qudits, each stabilizer $Z_p^2$ and $C_e$ term gives an order $2$ constraint, so there are $\frac{3}{2}\mathbf{P}$ stabilized qudits from these terms. Each $B_p$ term gives an order $4$ constraint, but the product of all $B_p^2$ is equivalent to a product of $Z_p^2$. Therefore, $\mathbf{P}- \frac{1}{2}$ stabilized qudits are from the $B_p$ terms. Lastly, $A_v$ may seem to give an order $4$ constraint, but $A_v^2$ can be generated from $B_p$ and $C_e$. Thus, each $A_v$ gives only one order $2$ constraint. Moreover, the product of all $A_v$ is proportional to the identity, so there are $\frac{1}{2}\mathbf{P} - \frac{1}{2}$ stabilized qudits from $A_v$. This gives us $\mathbf{N}_S=3 \mathbf{P}-1$. According to Eq.~\eqref{eq: consistency dim log}, the number of logical qudits is then: $\mathbf{N}_L = \mathbf{N}_Q - \mathbf{N}_S = 1$. This means that the logical subspace is four dimensional.
The string operators of the anyon types $s$ and $\bar{s}$ along non-contractible paths already account for a four-dimensional logical subspace, so there are no other anyon types in the stabilizer code.}
It is important to observe that the logical operators in Eq.~\eqref{eq: chiral semion logicals} are products of the semion string operators $W^s_e$.

We gauge out the antisemion by defining a topological subsystem code whose gauge group is generated by the stabilizers of the DS stabilizer code and the short string operators of the $\bar{s}$ anyon type in Eq.~\eqref{eq: DSshortstrings}. The intuition is that, since the antisemions braid nontrivially with themselves, the short string operators fail to commute with antisemion string operators, in agreement with Eq.~\eqref{eq: braiding formula}. As a consequence, the antisemion string operators cannot represent anyon types in the topological subsystem code, because they fail to commute with gauge operators along the length of the string. The gauge group resulting from gauging out the antisemions is:
\begin{align}
\mathcal{G} = \left \langle 
\vcenter{\hbox{\includegraphics[scale=0.23]{Figures/emloop.pdf}}}, \quad
\vcenter{\hbox{\includegraphics[scale=0.23]{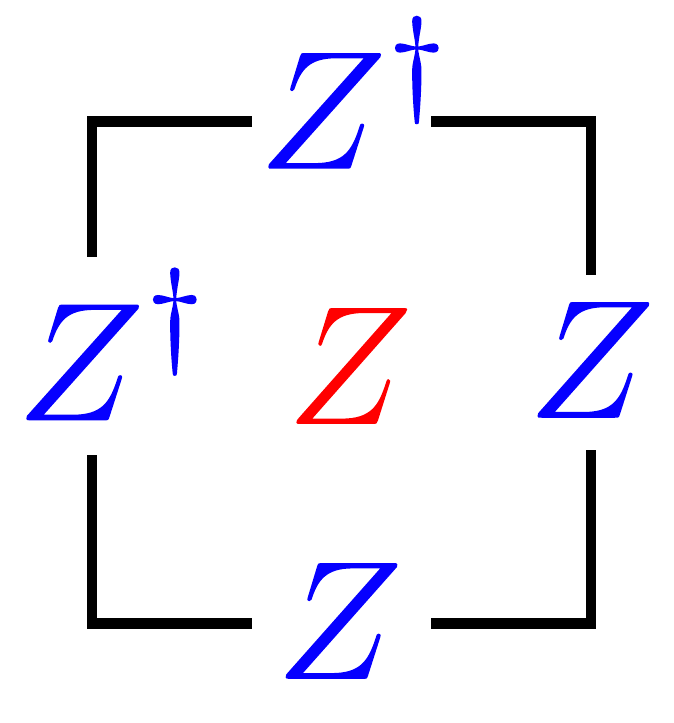}}}, \quad
\vcenter{\hbox{\includegraphics[scale=0.23]{Figures/sbarh}}}, \quad
\vcenter{\hbox{\includegraphics[scale=0.23]{Figures/sbarv}}}, \quad
\textcolor{blue}{Z^2}
 \right \rangle.
\end{align}
We have not included the edge terms $C_e$ as generators, since they are generated by the antisemion short string operators.
Furthermore, the vertex term can be generated by the other gauge generator, and the plaquette term can be simplified by multiplying with the antisemion short string operators.
This gives us the gauge group introduced the previous section: 
\begin{align}
\mathcal{G} \equiv \left \langle 
\vcenter{\hbox{\includegraphics[scale=0.23]{Figures/sbarh}}}, \quad
\vcenter{\hbox{\includegraphics[scale=0.23]{Figures/sbarv}}}, \quad
\vcenter{\hbox{\includegraphics[scale=0.23]{Figures/semionthirdgenerator}}}, \quad
\textcolor{blue}{Z^2}
\right \rangle.
\end{align}
Note that one could subsequently apply the transformation in Eq.~\eqref{eq: F Clifford def} to the horizontal edges and rewrite the model on a hexagonal lattice. This makes it apparent that the chiral semion subsystem code can be constructed from the $\ZZ_4^{(1)}$ subsystem code by gauging the $1$-form symmetry associated to the closed string operators of the transparent boson.  

\section{Subsystem codes from general Abelian anyon theories} \label{sec: general}

We now generalize the constructions of topological subsystem codes in Sections~\ref{sec: Z41 parafermion} and \ref{sec: chiral semion} to arbitrary Abelian anyon theories. In particular, we show that, given any Abelian anyon theory $\mathcal{A}$, we can construct a topological subsystem code characterized by $\mathcal{A}$.\footnote{Note that, in some cases, the anyon theory $\mathcal{A}$ is already equivalent to the anyon theory of a topological stabilizer code. This is the case if $\mathcal{A}$ is modular and can be realized in a system that admits a gapped boundary (to the vacuum), as shown in Ref.~\cite{Ellison2022Pauli}. For such an anyon theory, we can define a topological subsystem code whose gauge group is simply the stabilizer group of the associated topological stabilizer code in Ref.~\cite{Ellison2022Pauli}. This gives a topological subsystem code characterized by $\mathcal{A}$, albeit with a trivial gauge subsystem $\mathcal{H}_G$.}~Similar~to the constructions in Sections~\ref{sec: Z41 parafermion} and \ref{sec: chiral semion}, the first step is to identify a topological stabilizer code whose anyon theory includes the anyon types of $\mathcal{A}$. The second step is to then ``gauge out'' a subset of anyon types such that the remaining anyon types are exactly those of $\mathcal{A}$. We discuss the procedure for gauging out anyon types in more detail in Section~\ref{sec: Gauging out anyon types}. 

In Section~\ref{sec: General Abelian anyon theories}, we give a complete parameterization of the data that characterizes Abelian anyon theories. In Section~\ref{sec: Construction of twisted quantum doubles}, we construct a topological stabilizer code based on the data of a chosen Abelian anyon theory $\mathcal{A}$, such that the anyon theory of the topological stabilizer code includes $\mathcal{A}$. Finally, in Sections~\ref{sec: Gauging out anyon types}, we gauge out particular anyons of the topological stabilizer code to obtain a topological subsystem code characterized by $\mathcal{A}$.
By comparing the total number of qudits to the number of stabilized qudits, gauge qudits, and logical qudits, we verify that the topological subsystem code is indeed described by the anyon theory $\mathcal{A}$.
In the subsequent section, Section~\ref{sec: more examples}, we give further examples of our construction for simple Abelian anyon theories.

\subsection{General Abelian anyon theories}\label{sec: General Abelian anyon theories}

As described in Section~\ref{sec: anyon theories for subsystem codes}, an Abelian anyon theory $\mathcal{A}$ is fully determined by two pieces of data: \textbf{(i)} a finite Abelian group $A$, which specifies the anyon types and their fusion rules, and \textbf{(ii)} a function $\theta:A \to U(1)$, which encodes the exchange statistics. Equivalently, an Abelian anyon theory can be specified by a set of generators for the group $A$ along with their exchange statistics and mutual braiding relations. The statistics of every other anyon type can then be recovered from the statistics and braiding relations of the generators using the identity in Eq.~\eqref{eq: braiding identity}. This leads to the following parameterization of an arbitrary Abelian anyon theory, which we use throughout this section.
\begin{itemize}
    \item The finite Abelian group $A$ takes the general form $A = \prod_i^M \ZZ_{N_i}$, and without loss of generality, we assume that each $N_i$ is a power of a prime. We define a generating set of anyon types $\{a_i\}_{i=1}^M$ such that $a_i$ generates the factor $\ZZ_{N_i}$.
    \item The statistics of the generators take the form:\footnote{See Appendix~F of Ref.~\cite{Hsin2019Comments}.}
    \begin{align}
        \theta(a_i) = e^{\frac{2 \pi i}{N_i}t_i},
    \end{align}
    where $t_i$ is half-integer valued if $N_i$ is even, and integer valued if $N_i$ is odd:
    \begin{eqs}
    t_i \in
    \begin{cases}
        \{0, \pm \frac{1}{2}, \pm 1, \ldots\}  & \text{if } N_i \text{ is even}, \\
        \{0, \pm 1, \pm 2, \ldots \} & \text{if }N_i \text{ is odd}.
    \end{cases}
\end{eqs}
    \item The braiding relations between the generators $a_i$ and $a_j$ with $i \neq j$ take the form:
    \begin{align}
        B_\theta(a_i,a_j) = e^{\frac{2 \pi i}{N_{ij}}p_{ij}},
    \end{align}
    where $N_{ij}$ is the greatest common divisor of $N_i$ and $N_j$, and $\{p_{ij}\}_{i \neq j}$ is a set of integers satisfying $p_{ij}=p_{ji}$.
\end{itemize}
This means that a general Abelian anyon theory can instead be specified by the following three pieces of data \textbf{(i)} a finite Abelian group $A = \prod_{i=1}^M \ZZ_{N_i}$, \textbf{(ii)} a set of half integers $\{t_i\}_{i=1}^M$, 
which specify the statistics of the generators, and \textbf{(iii)} a set of integers $\{p_{ij}\}_{i \neq j}$ satisfying $p_{ij} = p_{ji}$, 
which determine the braiding relations of the generators. We find it convenient to further define $p_{ii}$ to be:\footnote{Here, we have used the floor function $\lfloor \cdot \rfloor$, which satisfies $\lfloor t_i \rfloor = t_i$, if $t_i$ is an integer, and $\lfloor t_i \rfloor = t_i -1/2$, if it is a half-integer.}
\begin{align}
    p_{ii} \equiv \lfloor t_i \rfloor.
\end{align}
Note that we recover the $\ZZ_4^{(1)}$ anyon theory with $A=\ZZ_4$ and  $t=1$, while the chiral semion anyon theory is obtained by $A=\ZZ_2$ and $t=1/2$. For these examples, there is only a single generator, so we do not need to specify $p_{ij}$.

\subsection{Construction of twisted quantum doubles}\label{sec: Construction of twisted quantum doubles}

Given an Abelian anyon theory $\mathcal{A}$ specified by the data $A=\prod_{i=1}^M \ZZ_{N_i}$, $\{t_i\}_{i=1}^M$, and $\{p_{ij}\}_{i \neq j}$, our first objective is to find a topological stabilizer code that includes $\mathcal{A}$ as a subtheory. In Ref.~\cite{Ellison2022Pauli}, it was shown that every modular Abelian anyon theory that admits a gapped boundary (to the vacuum) can be realized by a Pauli stabilizer Hamiltonian. This class of Abelian anyon theories coincides with those known as the Abelian twisted quantum doubles (TQDs)~\cite{HW13, Kapustin2011boundary, Kaidi2021higher}.   
This implies that it is sufficient to look for an Abelian TQD that includes $\mathcal{A}$ as a subtheory, since then the corresponding topological stabilizer code, referred to as the TQD stabilizer code in Ref.~\cite{Ellison2022Pauli}, is characterized by an anyon theory containing $\mathcal{A}$. To supplement the discussion below, we provide a review of Abelian TQDs in Appendix~\ref{app: TQD review}.

An Abelian TQD that includes the anyon theory $\mathcal{A}$ as a subtheory can be motivated by constructing a quasi-two-dimensional Walker-Wang model based on $\mathcal{A}$, as described in Appendix~\ref{app: WW}. To streamline the presentation, however, we simply state the associated Abelian TQD here. 
The particular TQD decomposes into a stack of $M$ decoupled TQDs, where the TQD in the $i$th layer is characterized by the group $\ZZ_{N_i}$ and an integer:\footnote{We note that the anyon theory for $n_i = N_i/2$ is that of a $U(1)_{N_i} \times U(1)_{-N_i}$ Chern-Simons theory.}
\begin{eqs}
    n_i \equiv
    \begin{cases}
        0  & \text{if } t_i = 0~(\text{mod }1), \\
        \frac{N_i}{2} & \text{if } t_i = \frac{1}{2}~(\text{mod }1).
    \end{cases}
\end{eqs}
We take the integers $n_{ij}$, defined in Appendix~\ref{app: TQD review}, to be trivial (i.e., $n_{ij}=0$, for all $i,j$).
As described in Appendix~\ref{app: TQD review}, the anyon types of this TQD are generated by the sets of anyon types $\{c_i\}_{i=1}^M$ and $\{\varphi_i\}_{i=1}^M$, which can be interpreted as the gauge charges and choices of elementary fluxes, respectively. The generators satisfy the fusion rules:
\begin{align}
    c_i^{N_i}=1, \quad  \varphi_i^{N_i} = 1,
\end{align}
and their exchange statistics and braiding relations are given by (for $i \neq j$):
\begin{align} \label{eq: charge flux TQD stack}
    \theta(c_i) = 1, \quad \theta (\varphi_i) = e^{\frac{ 2 \pi i n_i }{N_i^2}}, \quad B_\theta(\varphi_i, c_i) = e^{\frac{ 2 \pi i}{N_i}}, \quad B_\theta(\varphi_i, c_j) = 1 .
\end{align}
Note that the identity in Eq.~\eqref{eq: braiding identity} can be used in conjuction with the data above to recover the exchange statistics for an arbitrary composite of gauge charges and elementary fluxes.

To see that this anyon theory contains $\mathcal{A}$ as a subtheory, we consider the following anyon type $a_i$, defined for each layer $i$:
\begin{align} \label{eq: a generators def}
    a_{i} = {\varphi}_{i}  \prod_{j=1}^{i} c_j^{\frac{p_{ji}N_{j}}{N_{ij}}}.
\end{align}
First of all, we notice that the anyon types $\{a_i\}_{i=1}^M$ generate the group $A = \prod_{i=1}^M \ZZ_{N_i}$. This is because each anyon type $a_i$ has order $N_i$, which follows from the fact that $N_iN_j/N_{ij}$ is a multiple of both $N_i$ and $N_j$. Second, using the data given in Eq.~\eqref{eq: charge flux TQD stack} and the identity in Eq.~\eqref{eq: braiding identity}, we compute the exchange statistics and braiding relations of $a_i$ and $a_j$ with $i \neq j$ to be:
\begin{eqs}
    \theta (a_i) = e^{\frac{2 \pi i }{N_i} t_i}, \quad B_\theta(a_i, a_j) = e^{\frac{ 2 \pi i }{N_{ij}} p_{ij}}.
\end{eqs}
Therefore, the anyon theory generated by $\{a_i\}_{i=1}^M$ is precisely the anyon theory $\mathcal{A}$ defined by $A$, $\{t_i\}_{i=1}^M$, and~$\{p_{ij}\}_{i \neq j}$!

For the construction below, we find it convenient to also introduce the anyon types $\{\overline{\varphi}_i\}_{i=1}^M$ of the Abelian TQD, which can be interpreted as an alternative choice of elementary flux for each layer $i$. The anyon type $\overline{\varphi}_i$ has the fusion rules:
\begin{align}
    \overline{\varphi}_i^{N_i} = 1, \quad \varphi_i \times \overline{\varphi}_i = c_i^{\frac{2n_i}{N_i}},
\end{align}
and the exchange statistics and braiding relations of $\overline{\varphi}_i$ are:
\begin{eqs} \label{eq: barphi braiding}
    \theta (\overline{\varphi}_i) = \theta ({\varphi}_i)^*, \quad B(\varphi_i, \overline{\varphi}_j)=1, \quad B_\theta(\overline{\varphi}_i, c_j) = B_\theta({\varphi}_i, c_j)^*.
\end{eqs}
For example, if we take $\mathcal{A}$ to be the chiral semion anyon theory, then $c$ is the boson $s\bar{s}$, $\varphi$ is the semion $s$, and $\overline{\varphi}$ is the antisemion $\bar{s}$.

At this point, we could write down an explicit TQD stabilizer code following Ref.~\cite{Ellison2022Pauli}. However, for the construction of topological subsystem codes, we find it convenient to use a TQD stabilizer code that differs from the TQD stabilizer code of Ref.~\cite{Ellison2022Pauli}. The important difference is that (if there are layers with $n_i = N_i/2$) the stabilizer codes in this work have ancillary qudits, relative to those of Ref.~\cite{Ellison2022Pauli}. More technically, the difference is that the TQD stabilizer codes in Ref.~\cite{Ellison2022Pauli} are constructed by condensing certain anyon types, whereas the TQD stabilizer codes here are obtained by gauging the corresponding $1$-form symmetry.  {Specifically, to gauge a $\ZZ_N$ $1$-form symmetry represented by the string operators for an anyon type $a$, we first add ancillary degrees of freedom that transform under a $\ZZ_N$ $0$-form symmetry. We then condense the anyon type $a$ bound to a $0$-form symmetry charge. The difference between condensing $a$ and gauging the associated $1$-form symmetry is highlighted in Section~V of Ref.~\cite{Ellison2022Pauli}.} In the present text, the ancillary qudits are valuable in the next step of the construction, where we define short string operators to gauge out certain anyon types.

\begin{figure}[t] 
\centering
\hspace{3cm}\includegraphics[scale=0.26]{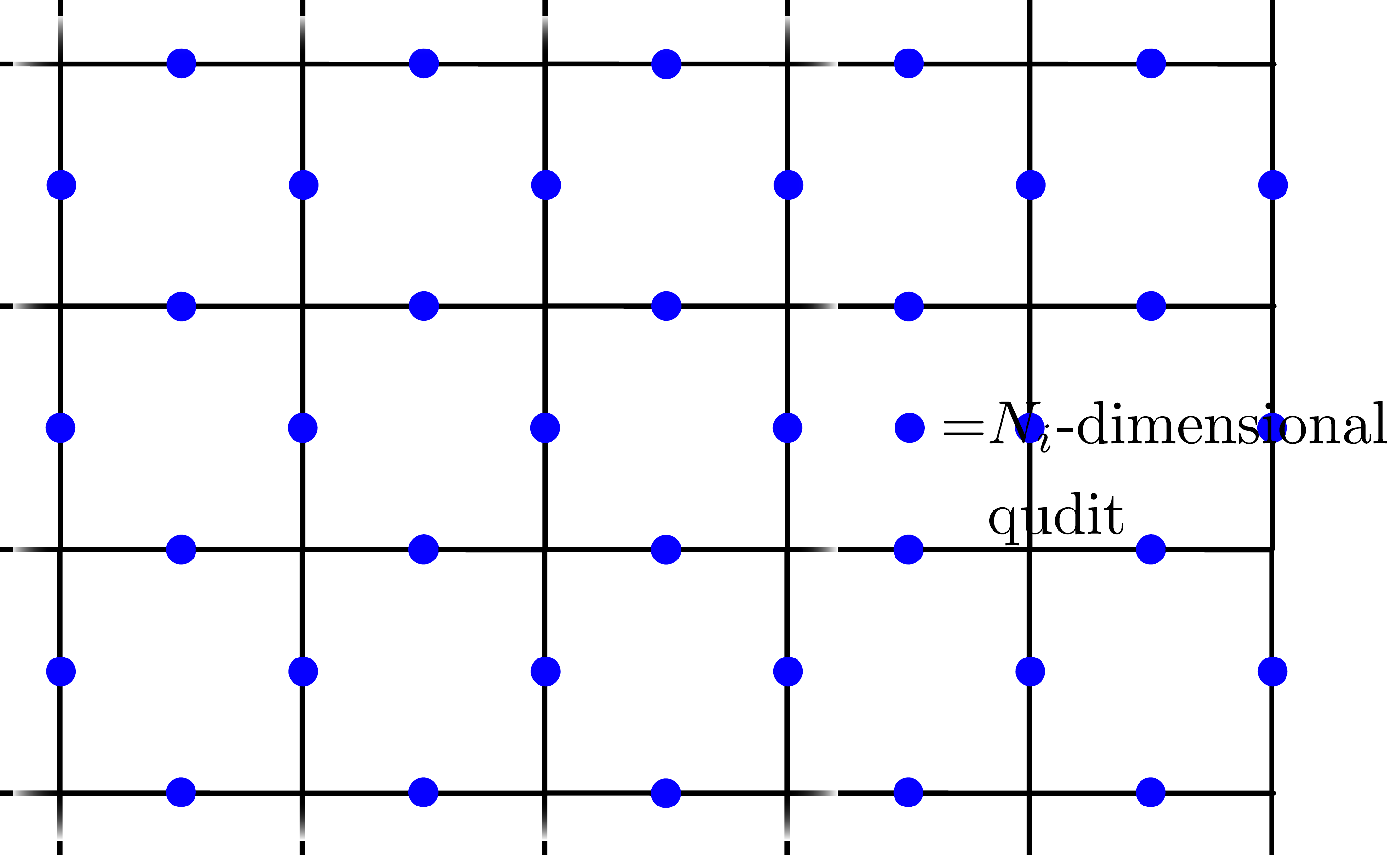}
\caption{If $n_i = 0$, the TQD stabilizer code in the $i$th layer is defined on a square lattice with an $N_i$-dimensional qudit (blue) at each edge.}
\label{fig: ni 0 dof}
\end{figure}

To write the general form for the TQD stabilizer code with $M$ layers, we first focus on the individual layers.
If the $i$th layer is characterized by $n_i=0$, then the TQD stabilizer code for the $i$th layer is precisely a $\ZZ_{N_i}$ TC. We define the $\ZZ_{N_i}$ TC on a square lattice with an $N_i$-dimensional qudit on each edge,
as shown in Fig.~\ref{fig: ni 0 dof}. Explicitly, the corresponding Hamiltonian for the $\ZZ_{N_i}$ TC is:
\begin{align} \label{eq: 0 TC}
H^{(i)}_0 \equiv - \sum_v A_{v,i}^\text{TC} - \sum_p B_{p,i}^\text{TC} + \text{h.c.},
\end{align}
with the vertex term $A_{v,i}^\text{TC}$ and plaquette term $B_{p,i}^\text{TC}$ pictured below:
\begin{align}
A_{v,i}^\text{TC} \equiv  \vcenter{\hbox{\includegraphics[scale=0.23]{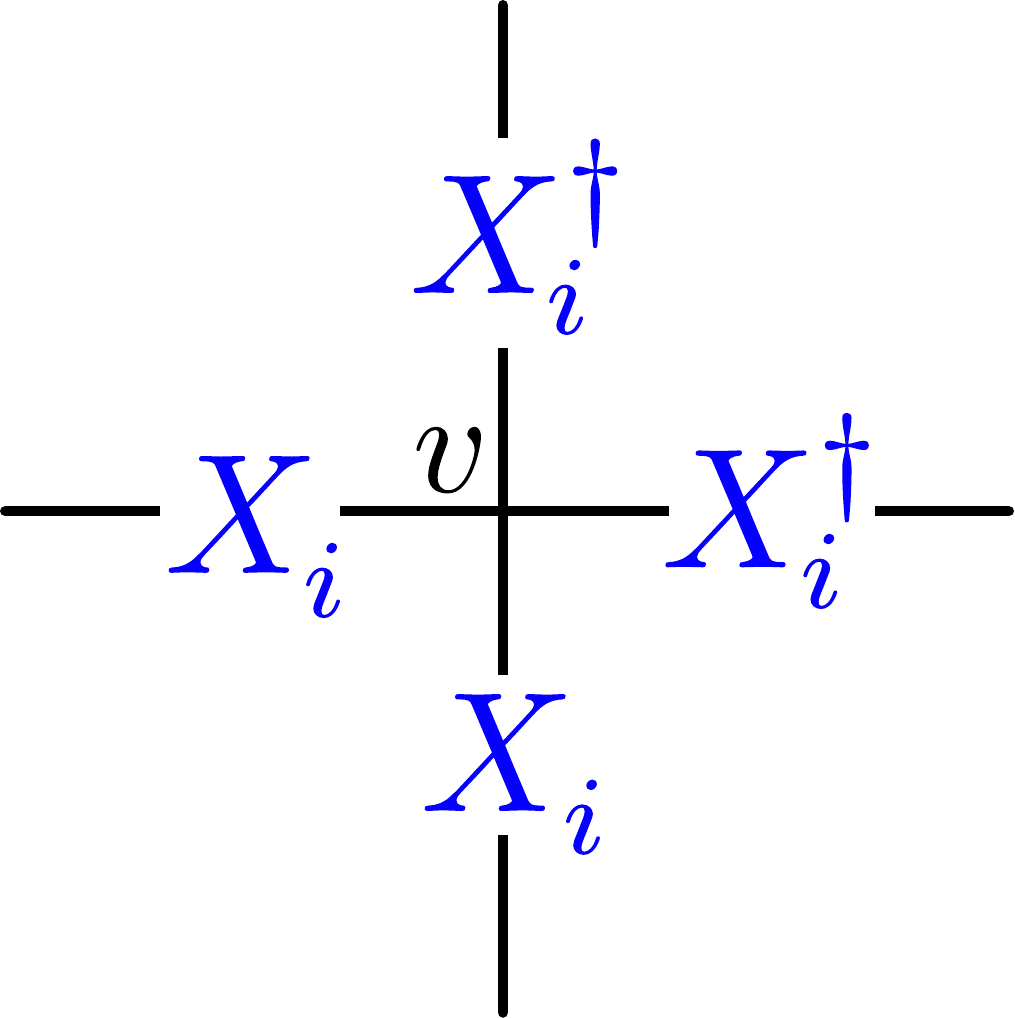}}}, \quad B_{p,i}^\text{TC} \equiv \vcenter{\hbox{\includegraphics[scale=0.23]{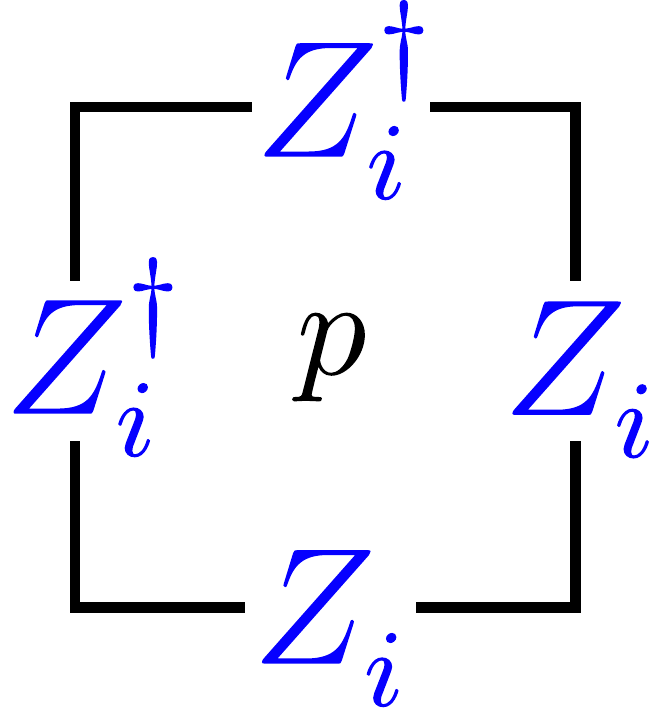}}}\,.
\end{align}
Here, we have labeled the Pauli operators by the layer $i$. These can be interpreted as string operators created by moving an $m$ or $e$ anyon type counterclockwise around a loop, respectively. The $e$ and $m$ anyon types play the role of the charge $c_i$ and elementary flux $\varphi_i$ of the $i$th layer (with $n_i=0$). The anyon type $m^{-1}$ then corresponds to the choice of elementary flux $\overline{\varphi}_i$. The short string operators that move the $c_i$ and $\varphi_i$ anyon types are:
\begin{align}
    W_e^{c_i} \equiv \vcenter{\hbox{\includegraphics[scale=0.28]{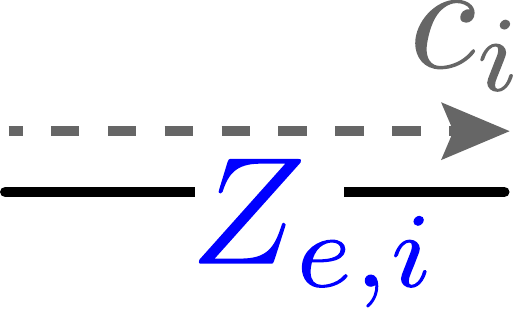}}}
    ~, \,\,
    \vcenter{\hbox{\includegraphics[scale=0.28]{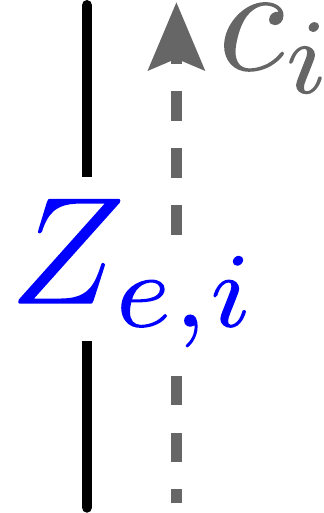}}}
    ~, \qquad
   W_e^{\varphi_i} \equiv \vcenter{\hbox{\includegraphics[scale=0.28]{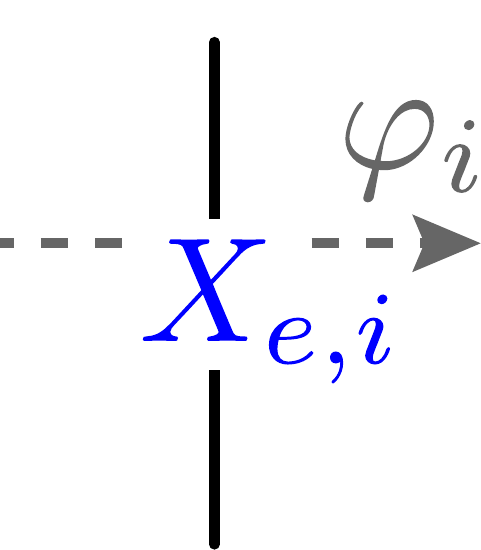}}}
    ~, \,\,
    \vcenter{\hbox{\includegraphics[scale=0.28]{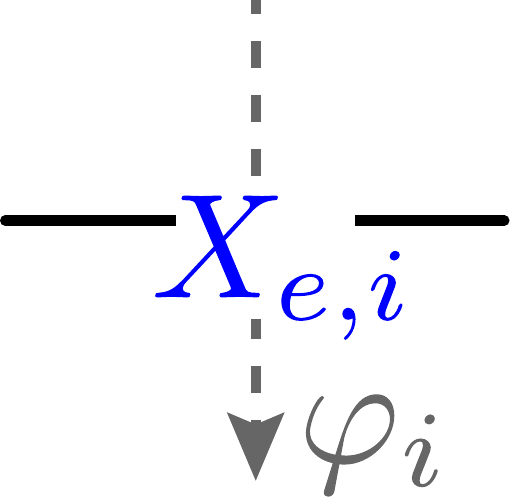}}},
\label{eq: short string of Z_Ni e and m}
\end{align}
where the subscript $e$ labels an edge. 

\begin{figure}[t] 
\centering
\hspace{3.5cm}\includegraphics[scale=0.26]{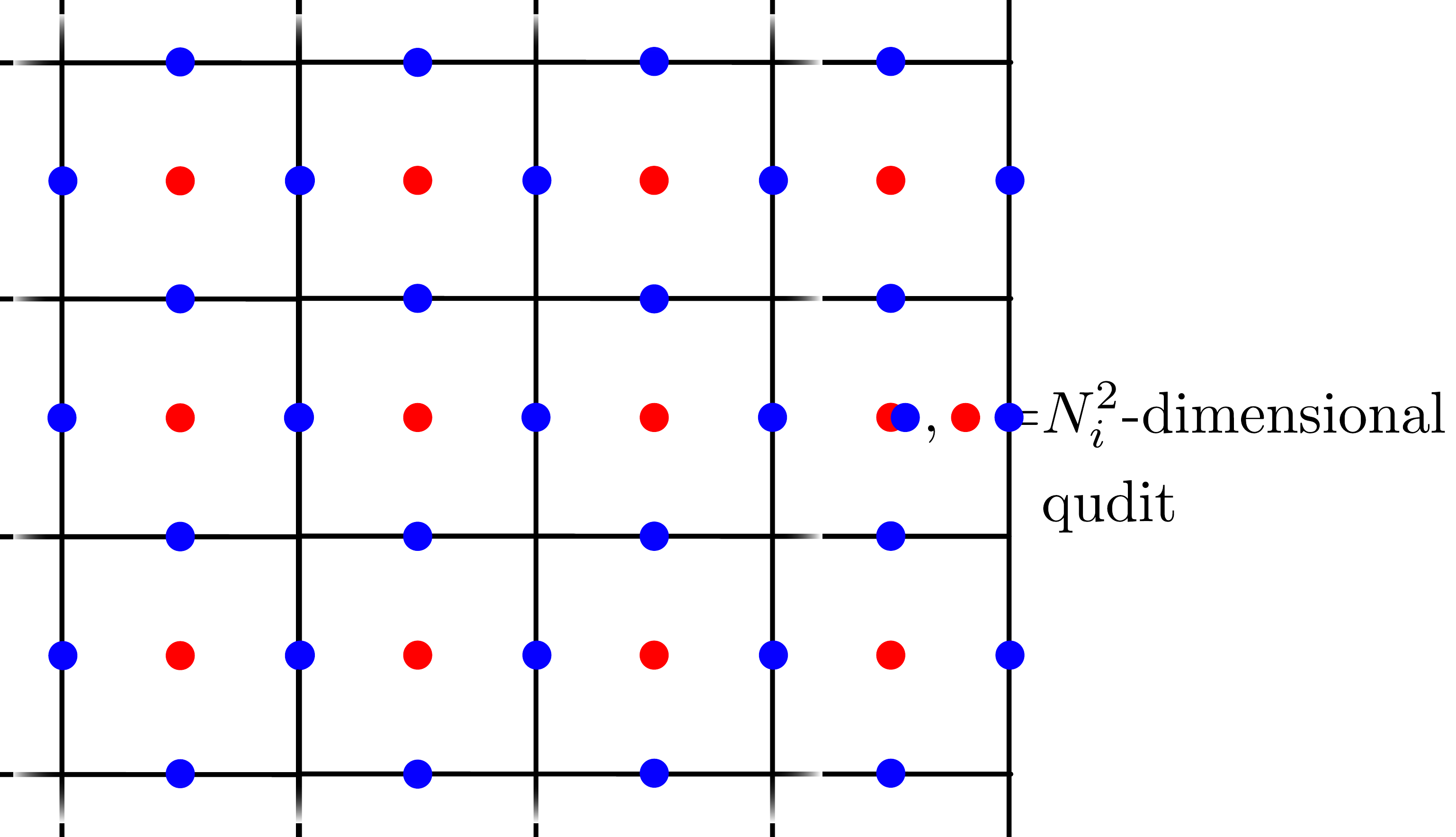}
\caption{If $n_i=N_i/2$, the TQD stabilizer code in the $i$th layer is defined on a square lattice with an  $N_i^2$-dimensional qudit on each edge (blue) and on each plaquette (red).}
\label{fig: ni Ni/2 dof}
\end{figure}

Next, if the $i$th layer is instead characterized by $n_i=N_i/2$, then the TQD stabilizer code in the $i$th layer is a generalization of the DS stabilizer code in Eq.~\eqref{eq: DS stabilizer Hamiltonian}. In particular, we define the TQD stabilizer code on a square lattice with an $N_i^2$-dimensional qudit at each edge and plaquette, as illustrated in Fig.~\ref{fig: ni Ni/2 dof}. The Hamiltonian is a sum of vertex terms $A_v$, plaquette terms $B_p$, edge terms $C_e$, and Pauli operators $Z_{p,i}^{N_i}$ defined on the plaquette qudits labeled by a plaquette $p$ and layer $i$:
\begin{align} \label{eq: Ni/2 TQD}
H^{(i)}_{N_i/2} \equiv - \sum_v A_{v,i} - \sum_p B_{p,i}
- \sum_e C_{e,i} - \sum_p Z_{p,i}^{N_i} + \text{h.c.}
\end{align}
Here, the terms $A_{v,i}$, $B_{p,i}$, and $C_{e,i}$ are pictorially represented as:
\begin{eqs}
&A_{v,i}\equiv \vcenter{\hbox{\includegraphics[scale=0.23]{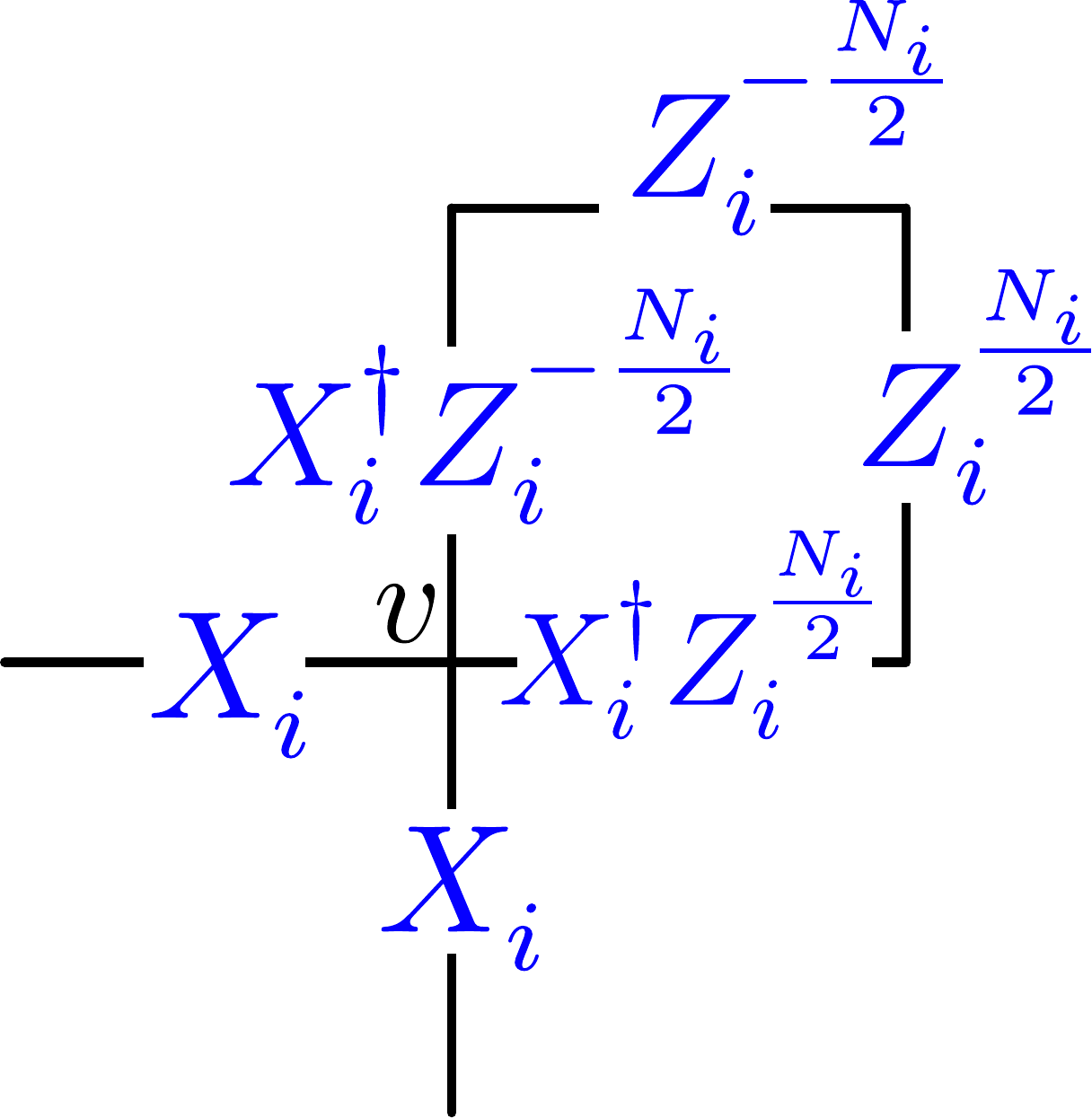}}}~,\quad
B_{p,i} \equiv \vcenter{\hbox{\includegraphics[scale=0.23]{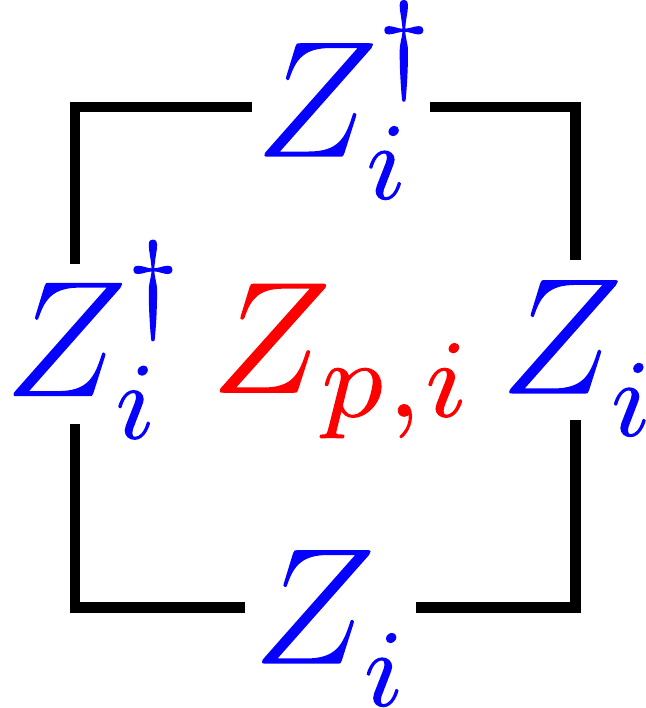}}}~, \quad \\
&C_{e,i}\equiv  \vcenter{\hbox{\includegraphics[scale=0.26]{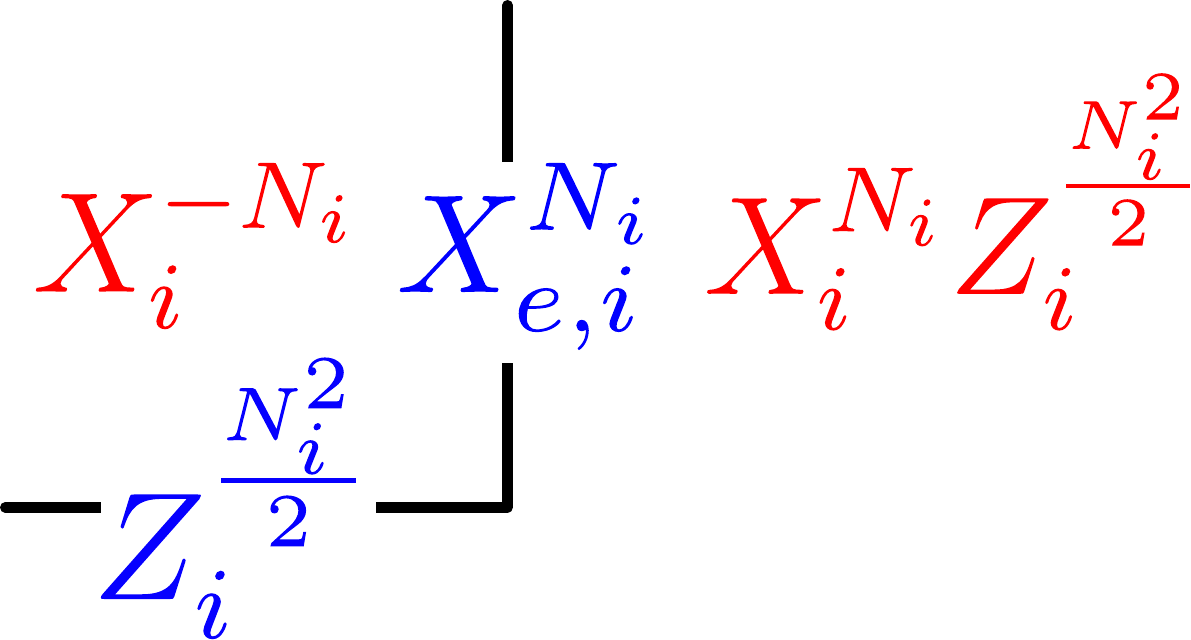}}}~, \quad
\raisebox{-0.5cm}{\hbox{\includegraphics[scale=0.25]{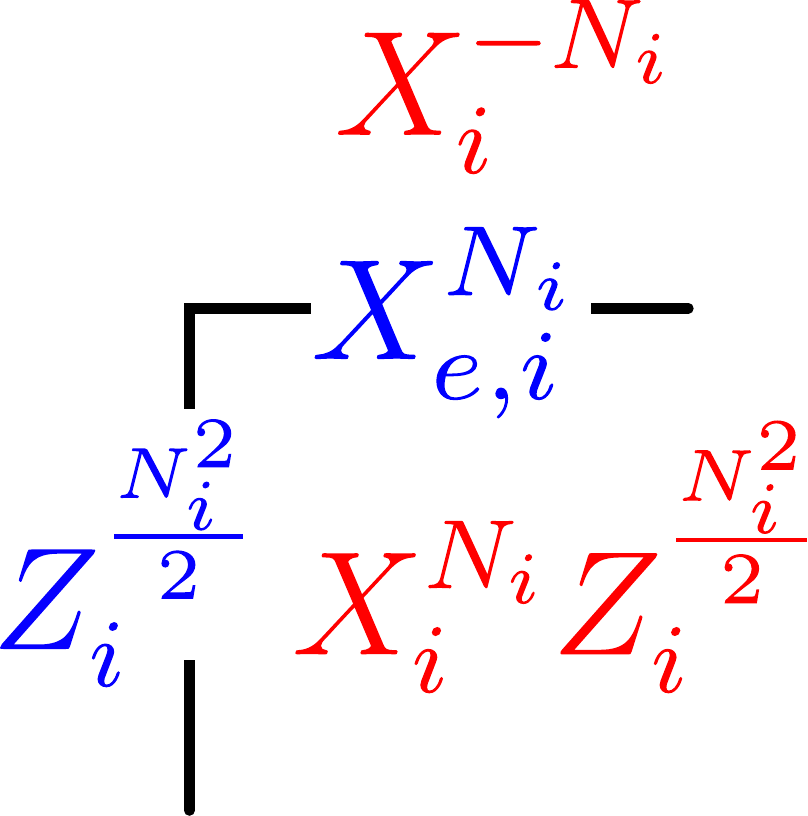}}}~.
\end{eqs}
This model is constructed from a $\ZZ_{N_i^2}$ TC, where the vertex term can be interpreted as a counterclockwise loop of string operator for an $e^{N_i/2}m$ anyon type. In the $\ZZ_{N_i^2}$ TC, the edge terms correspond to short string operators for $e^{N^2_i/2}m^{N_i}$ anyon types bound to $\ZZ_{N_i}$ charges on the plaquettes. Here, the $\ZZ_{N_i}$ charges are the $Z_p^{\pm N_i}$ operators on the plaquettes. These are a consequence of gauging the $1$-form symmetry corresponding to the boson $e^{N^2_i/2}m^{N_i}$ and do not appear in the models of Ref.~\cite{Ellison2022Pauli}. 

The short string operators for the anyon types $\varphi_i = e^{\frac{N_i}{2}} m$ and $\overline{\varphi}_i = e^{\frac{N_i}{2}} m^{-1}$ can be written as:
\begin{eqs} \label{eq: phi and barphi short strings}
    W^{\varphi_i}_e &\equiv~ \vcenter{\hbox{\includegraphics[scale=0.28]{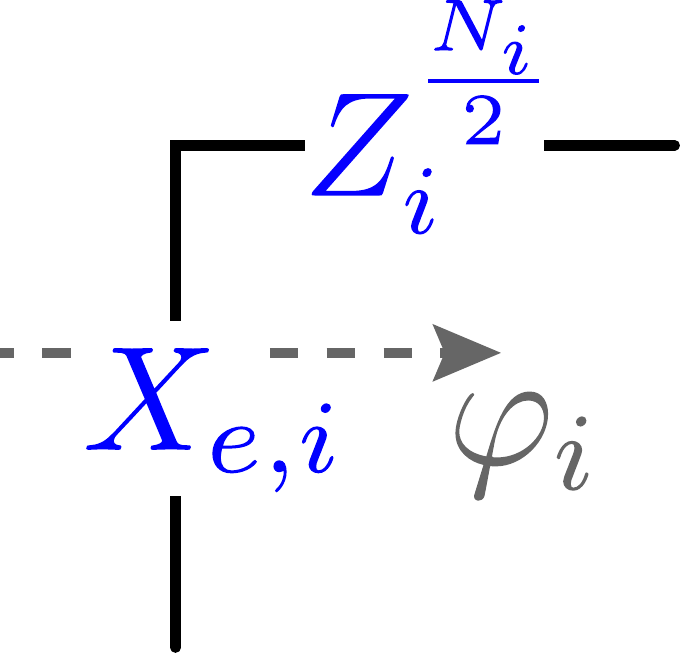}}}~,
    \quad
    \vcenter{\hbox{\includegraphics[scale=0.28]{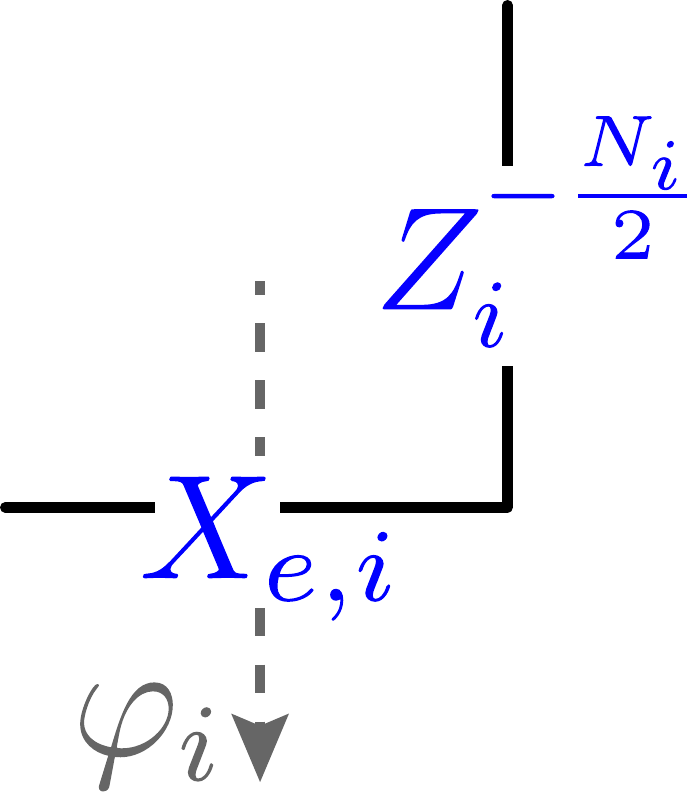}}}~, \quad 
    W^{\overline{\varphi}_i}_e \equiv~ \vcenter{\hbox{\includegraphics[scale=0.28]{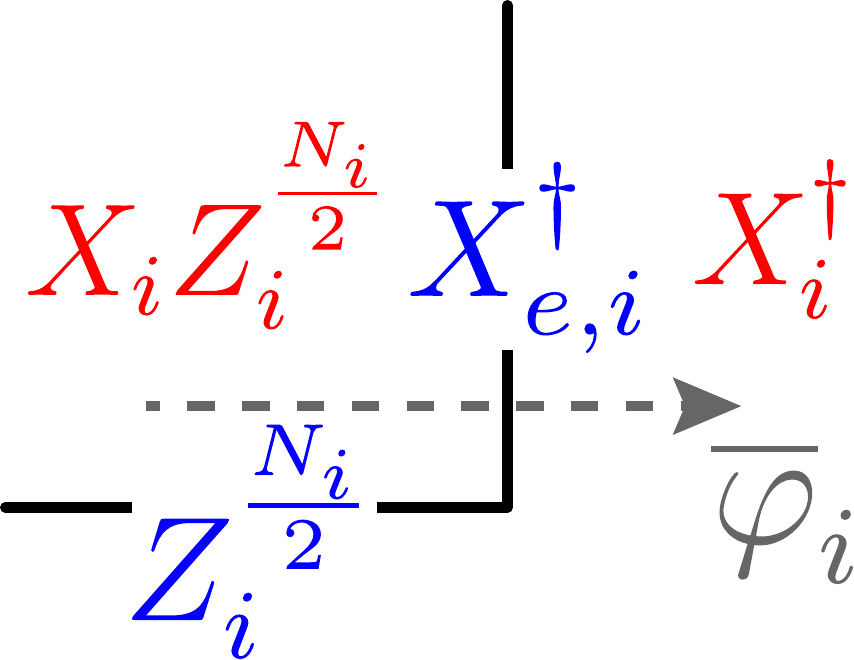}}}~,
    \quad
    \vcenter{\hbox{\includegraphics[scale=0.28]{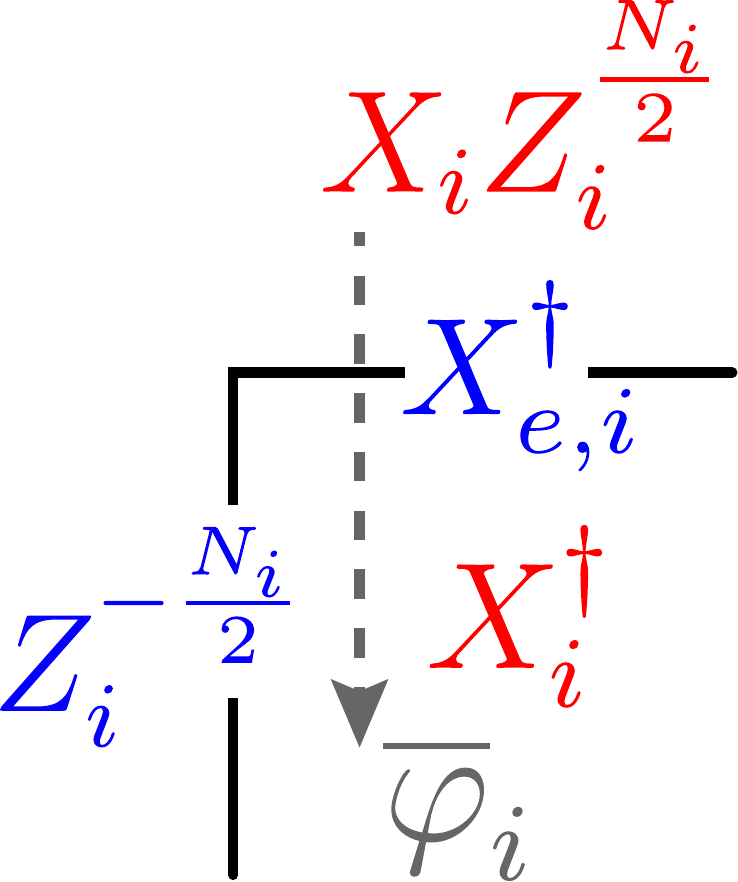}}}~.
\end{eqs}
Note that for later convenience, the short string operators for $\overline{\varphi}_i$ have been chosen so that they commute with $A_{v,i}$, $B_{p,i}$, and $C_{e,i}$. 
Using Eq.~\eqref{eq: statistics formula}, we can verify that the exchange statistics of $\varphi_i$ and $\overline{\varphi}_i$ are $\theta(\varphi_i) = \theta(\overline{\varphi}_i)^* = e^{\pi i/N_i}$. Furthermore, since $W^{\varphi_i}_e$ and $W^{\overline{\varphi}_i}_e$ commute, $\varphi_i$ and $\overline{\varphi}_i$ have trivial braiding relations, as expected from Eq.~\eqref{eq: barphi braiding}. The gauge charge $c_i$ is given by the product $\varphi_i \times \overline{\varphi}_i$, and its short string operators can be written as:
\begin{eqs} 
    W^{c_i}_e &\equiv~ \vcenter{\hbox{\includegraphics[scale=0.28]{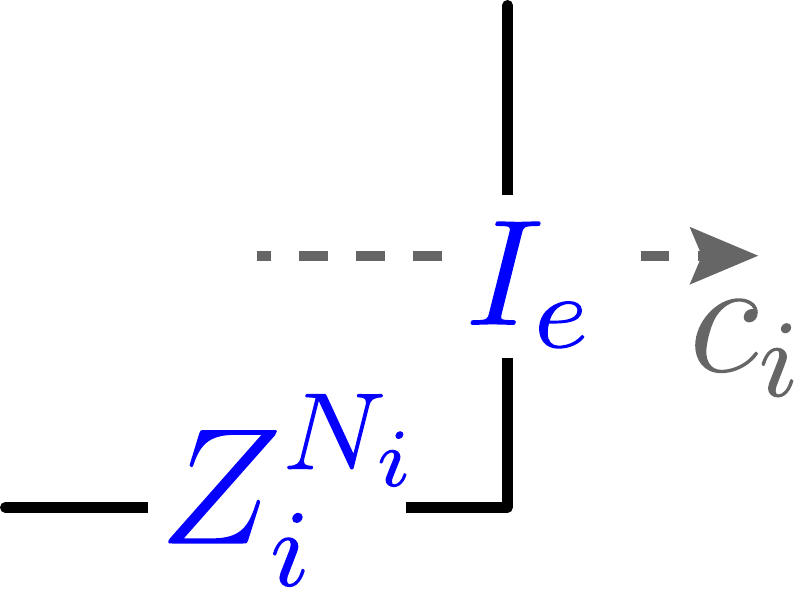}}}~,
    \quad
    \vcenter{\hbox{\includegraphics[scale=0.28]{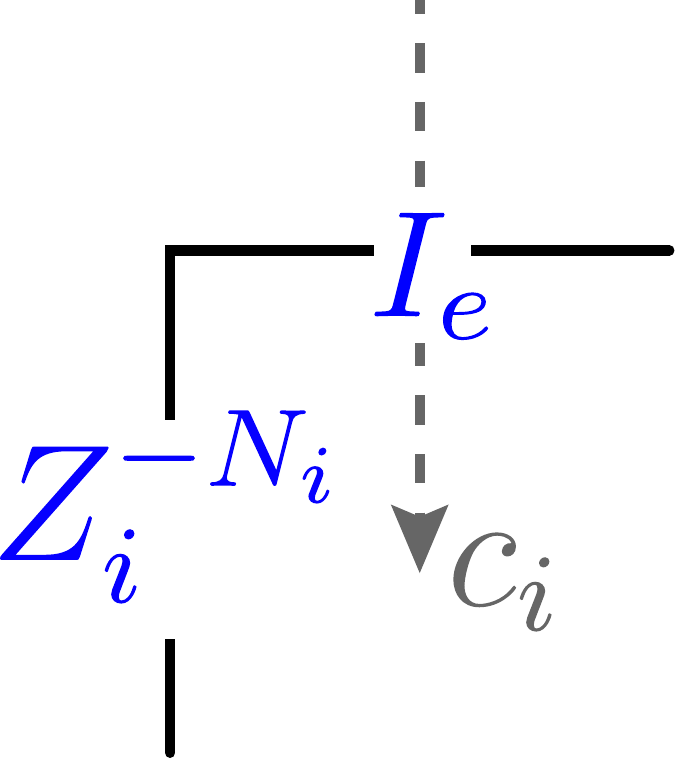}}}~,
\end{eqs}
where $I_e$ denotes the identity on the edge $e$.
Note that long string operators formed from products of $W^{c_i}_e$ and $W^{\varphi_i}_e W^{\overline{\varphi}_i}_e$ differ only by stabilizers along the string operator and local operators near the endpoints. 

Putting Eqs.~\eqref{eq: 0 TC} and \eqref{eq: Ni/2 TQD} together, we obtain a TQD stabilizer code characterized by an anyon theory that contains $\mathcal{A}$ as a subtheory. We write the corresponding Hamiltonian of the TQD stabilizer code as:
\begin{align}
H_{\text{TQD}} \equiv \sum_{i = 1}^M H^{(i)}_{n_i}.
\label{eq: total TQD}
\end{align}

\subsection{Gauging out anyon types}\label{sec: Gauging out anyon types}

In the next step of the construction, we gauge out a set of anyon types, denoted by $\bar{\mathcal{A}}$. At the level of the anyon theory, gauging out $\bar{\mathcal{A}}$ amounts to removing all of the anyon types that braid nontrivially with some anyon type in $\bar{\mathcal{A}}$.  {As mentioned in Section~\ref{sec: z41 construction},} this should be distinguished from condensing an anyon type, since, in contrast to anyon condensation, \textbf{(i)} we are free to gauge out anyon types that do not have bosonic statistics, and \textbf{(ii)} we do not identify the gauged out anyon types with the trivial anyon type. 

At an operational level, we gauge out the anyon types in $\bar{\mathcal{A}}$ by defining a new topological subsystem code, whose gauge group is generated by the stabilizers of the TQD stabilizer code and a certain set of short string operators for the anyon types in $\bar{\mathcal{A}}$. In other words, we append short string operators for the anyon types in $\bar{\mathcal{A}}$ to the (Abelian) gauge group of the TQD stabilizer code. We note that the short string operators need to be chosen so that they commute with the string operators for the anyon types in $\mathcal{A}$ along closed paths.
This motivates the introduction of ancillary plaquette degrees of freedom and the peculiar choice of short string operator for $\overline{\varphi}_i$ in Eq.~\eqref{eq: phi and barphi short strings} -- i.e., the extra structure ensures that the short string operators commute with the string operators for the $a_i$ anyon types. 

Adding the short string operators of the anyon types in $\bar{\mathcal{A}}$ to the gauge group of the TQD stabilizer code has two notable effects. The first is that the short string operators of the anyon types in $\bar{\mathcal{A}}$ fail to commute with the stabilizers that correspond to loops of string operators of anyon types that braid nontrivially with the anyon types of $\bar{\mathcal{A}}$. Consequently, any such stabilizer is not included in the stabilizer group of the topological subsystem code after gauging out the anyon types. We argue below that the anyon types in $\bar{\mathcal{A}}$ in fact braid nontrivially with all of the anyon types of the TQD other than those in $\mathcal{A}$. In terms of the anyonic $1$-form symmetries defined in Appendix~\ref{app: 1form}, the short string operators employed in this step break the anyonic $1$-form symmetries associated to all of the anyon types outside of $\mathcal{A}$.
The second effect is that the logical operators of the TQD stabilizer code that correspond to moving the anyon types in $\bar{\mathcal{A}}$ along non-contractible paths become products of gauge operators in the resulting topological subsystem code. If the anyon type is transparent in $\bar{\mathcal{A}}$, then the logical operator is converted into a nonlocal stabilizer. Hence, the corresponding anyon type of the TQD becomes a transparent anyon type in the topological subsystem code. 

More concretely, for the $\ZZ_4^{(1)}$ subsystem code in Section~\ref{sec: Z41 parafermion}, the anyon types of $\bar{\mathcal{A}}$ are generated by the $e^3m$ anyons of the $\ZZ_4$ TC. Thus, the gauge group of the subsystem code is generated by the stabilizer group of the $\ZZ_4$ TC and short string operators for the $e^3m$ anyon types. For the chiral semion subsystem code in Section~\ref{sec: chiral semion}, $\bar{\mathcal{A}}$ is taken to be the subtheory generated by the antisemion of the DS stabilizer code. Therefore, the gauge group obtained from gauging out $\bar{\mathcal{A}}$ is generated by the stabilizers of the DS stabilizer code and a set of short string operators for the antisemions.

We now specify the anyon theory $\bar{\mathcal{A}}$ that needs to be gauged out to build a topological subsystem code characterized by $\mathcal{A}$. We take $\bar{\mathcal{A}}$ to be the anyon theory generated by the anyon types $\{\overline{a}_i\}_{i=1}^M$, where $\overline{a}_i$ in the $i$th layer is defined as:
\begin{align} \label{eq: anyons to be gauged out}
    \overline{a}_i = \overline{\varphi}_i \prod_{j=i}^M c_j^{\frac{p_{ij} N_j}{N_{ij}} }.
\end{align}
We claim that the only anyon types that remain after gauging out the anyon types of $\bar{\mathcal{A}}$ are those of $\mathcal{A}$, generated by the anyon types $\{a_i\}_{i=1}^M$ with $a_i$ given in Eq.~\eqref{eq: a generators def}:
\begin{align}
a_{i} = {\varphi}_{i}  \prod_{j=1}^{i} c_j^{\frac{p_{ji}N_{j}}{N_{ij}}}.
\label{eq: remaining anyon}
\end{align}
It can be checked using Eqs.~\eqref{eq: charge flux TQD stack} and \eqref{eq: barphi braiding} that $a_i$ and $\overline{a}_j$ indeed have trivial braiding relations, for any choice of $i$ and $j$. Note that gauging out $\overline{a}_i$ has the same effect as gauging out $(\overline{a}_i)^{-1}$. In Sections~\ref{sec: z41 construction} and \ref{sec: chiral semion construction}, we gauged out $(\overline{a}_i)^{-1}$ to simplify the notation.

We argue that all of the anyon types outside of $\mathcal{A}$ have nontrivial braiding relations with the anyon types in $\bar{\mathcal{A}}$. This is because if an anyon type $b$ is composed of gauge charges $c_i$, it must braid non-trivially with some $\overline{a}_i$. If $b$ instead contains a single elementary flux $\varphi_i$ and braids trivially with all $\overline{a}_i$, then it must be the anyon type $a_i$. This is because, if $b$ is not equal to $a_i$, then $b^{-1} a_i$ is a product of gauge charges that braids trivially with all of the $\bar{A}$ anyon types -- a contradiction. Similarly, if $b$ contains multiple elementary fluxes and braids trivially with $\overline{a}_i$ it must be a product of anyon types in $\mathcal{A}$. The uniqueness of the product of $\mathcal{A}$ anyon types holds for the same reason as above. 

We next turn to the lattice construction of the topological subsystem code. We define the gauge group $\mathcal{G}$ after gauging out the anyon types $\bar{\mathcal{A}}$ to be the group generated by stabilizers of the TQD in Eq.~\eqref{eq: total TQD} and certain short string operator for the anyon types $\{\overline{a}_i\}_{i=1}^M$. In particular, we take the short string operators for $\overline{a}_i$ to be:
\begin{eqs}
    W_e^{\overline{a}_i} = W_e^{\overline{\varphi}_i} \prod_{j=i}^M (W_e^{c_j})^{\frac{p_{ij} N_j}{N_{ij}} }.
\end{eqs}
These can be represented pictorially as:
\begin{align} \label{eq: barai short strings}
    W_e^{\overline{a}_i} = \vcenter{\hbox{\includegraphics[scale=0.25]{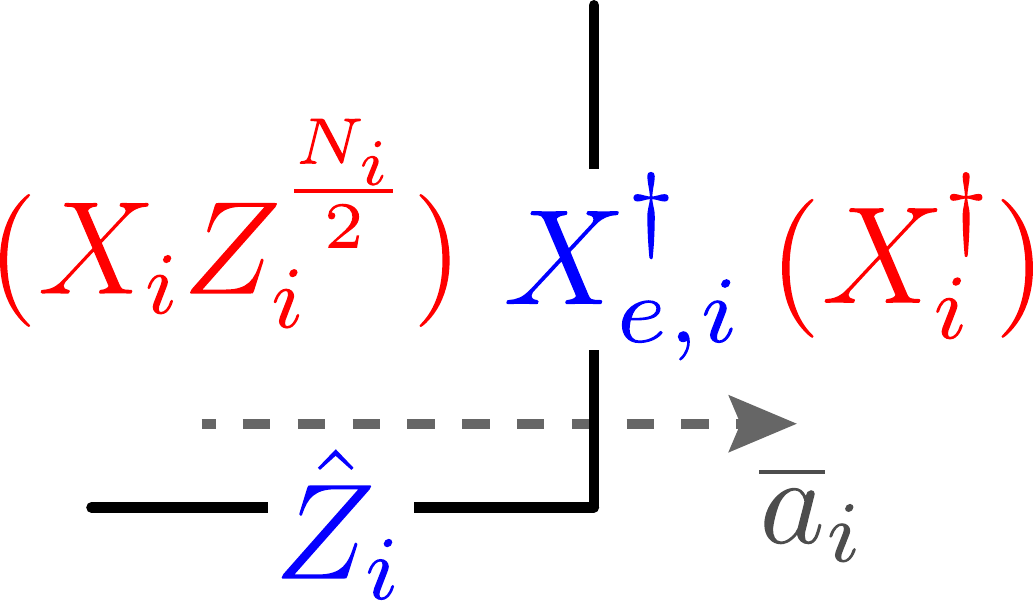}}}, \,\,\, \vcenter{\hbox{\includegraphics[scale=0.25]{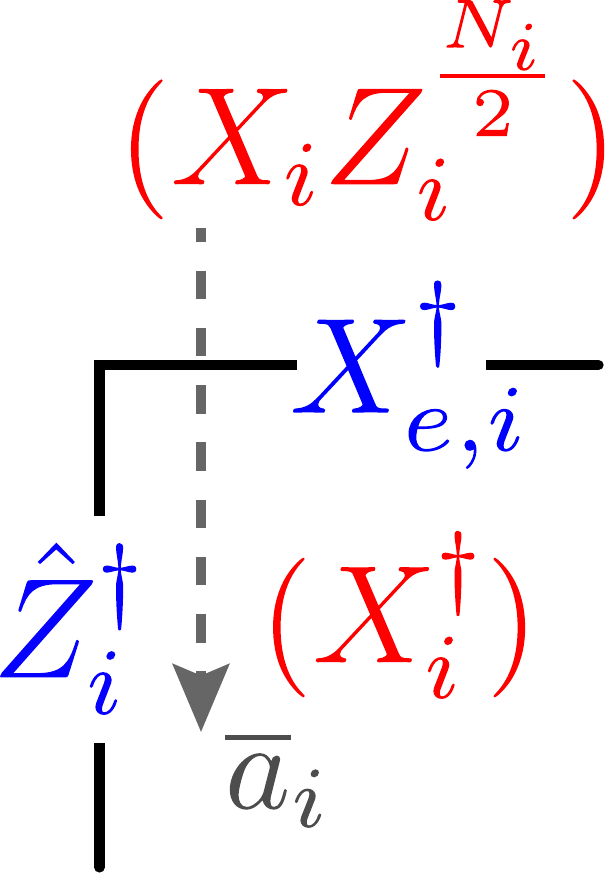}}},
\end{align}
where the plaquette Pauli operators in the parentheses should be ignored if $n_i = 0$, and the operator $\hat{Z}_i$ is given by:
\begin{align}
\hat{Z}_i \equiv Z_i^{n_i}
\prod_{j : \, j\geq i,~n_j =0} Z_j^{\frac{p_{ij} N_j}{N_{ij}}}
\prod_{j : \, j\geq i,~n_j = N_j/2} Z_j^{\frac{p_{ij} N^2_j}{N_{ij}}}.
\end{align}

To make the gauge group $\mathcal{G}$ of the topological subsystem code explicit, we separate the generators of the gauge group by layers. We define $\mathcal{G}_i$ to be the group of gauge operators associated to the $i$th layer, and represent it graphically as:
\begin{align}
\mathcal{G}_i \equiv \left \langle 
\vcenter{\hbox{\includegraphics[scale=0.23]{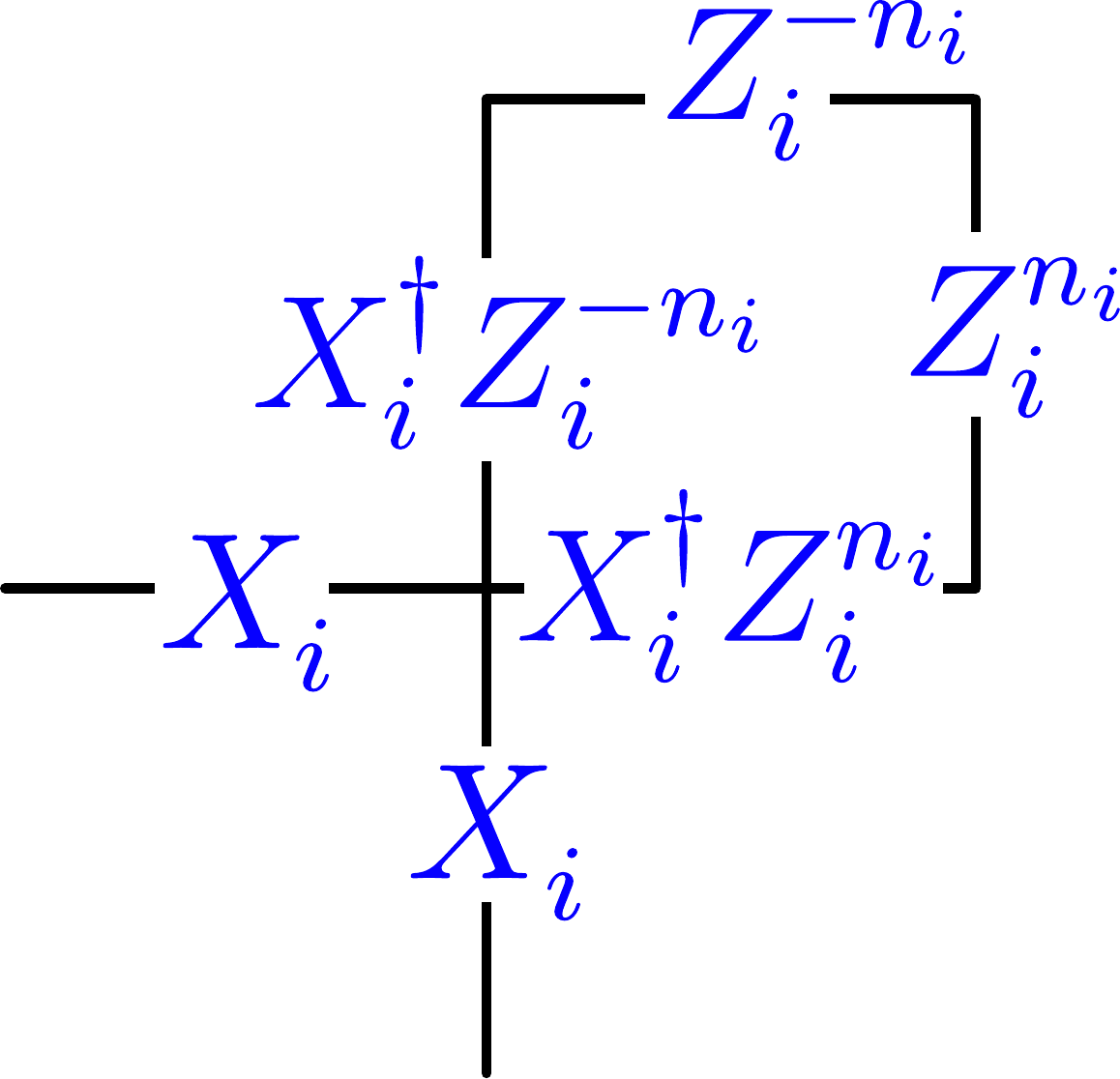}}}~, \quad
\vcenter{\hbox{\includegraphics[scale=0.23]{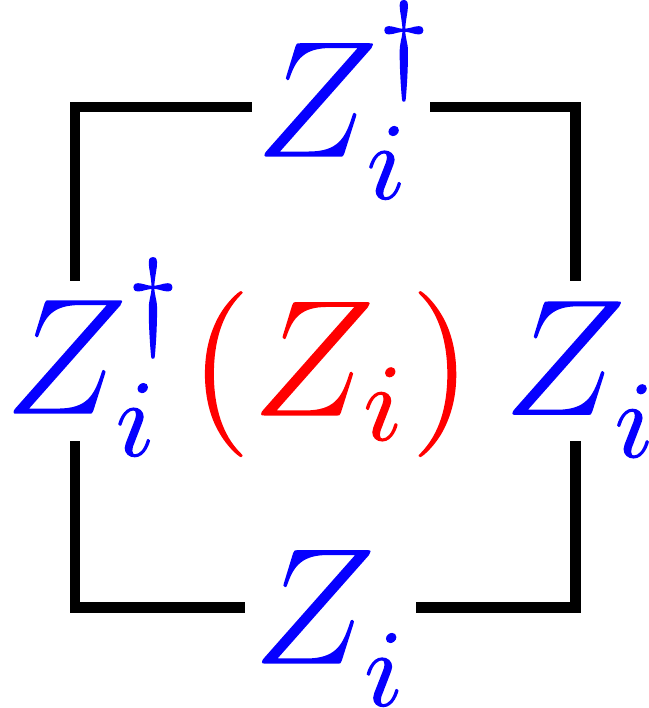}}}~, \quad
\vcenter{\hbox{\includegraphics[scale=0.25]{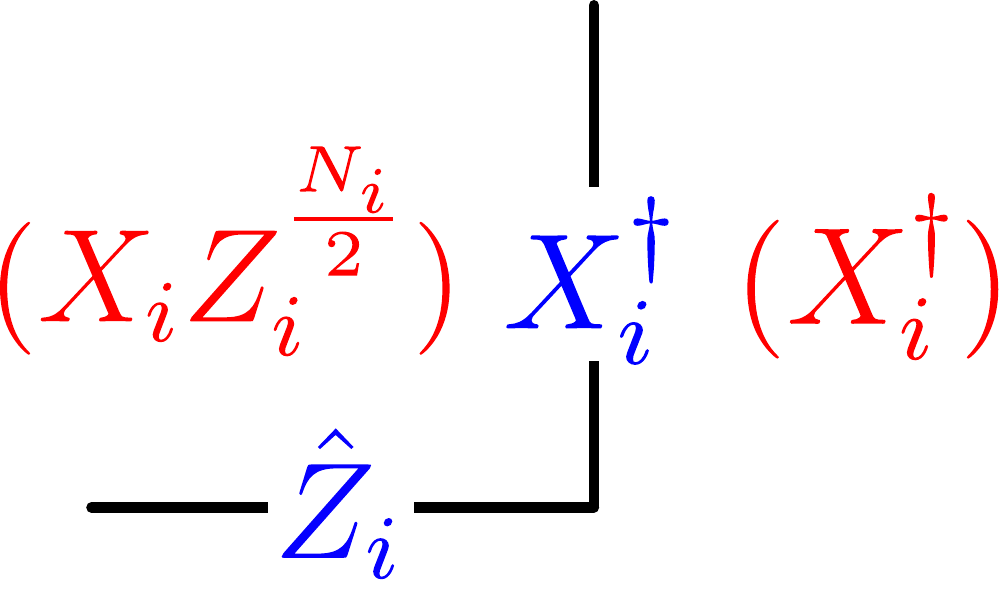}}}~, \quad
\vcenter{\hbox{\includegraphics[scale=0.25]{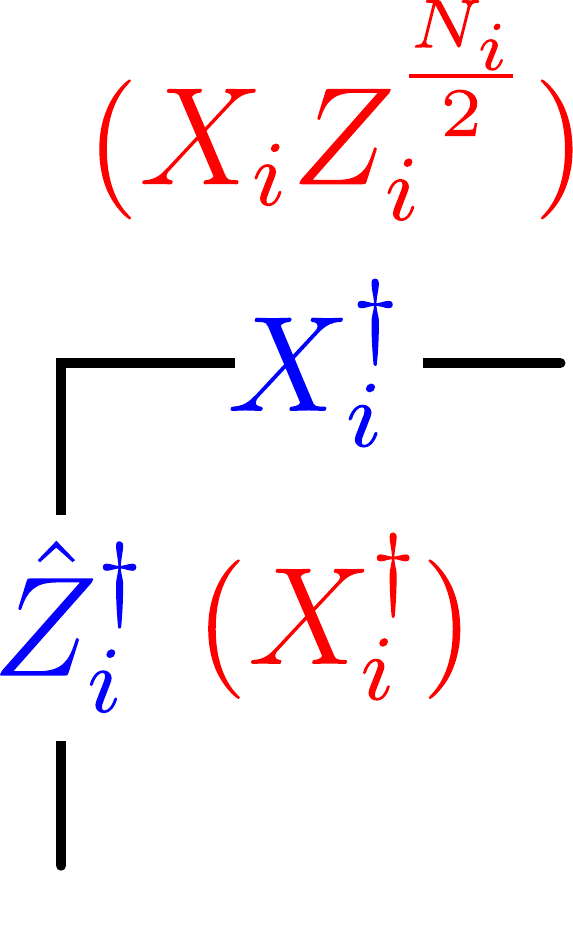}}}~, \quad
\textcolor{blue}{(Z_i^{N_i})}
 \right \rangle.
\end{align}
Similar to the expression for the string operators in Eq.~\eqref{eq: barai short strings}, the plaquette Pauli operators in parentheses are meant to be ignored if $n_i = 0$. The generating set implicitly includes the terms for each vertex, plaquette, and edge. It also includes the requisite root of unity $\omega$ or $\sqrt{\omega}$ for odd- or even-dimensional qudits, respectively (see Section~\ref{sec: composite qudits}). Note that we have not included the edge stabilizers $C_{e,i}$ of the TQD stabilizer code, since these can generated by $W_e^{\overline{a}_i}$ and the plaquette Pauli operators {$Z_i^{N_i}$}. We can further simplify $\mathcal{G}_i$, due to the fact that the vertex term is redundant. The group of gauge operators in the $i$th layer is more simply given by:
\begin{align}
\mathcal{G}_i = \left \langle 
\vcenter{\hbox{\includegraphics[scale=0.25]{Figures/Bpgi.pdf}}}~, \quad
\vcenter{\hbox{\includegraphics[scale=0.25]{Figures/gaugehatphih2.pdf}}}~, \quad
\vcenter{\hbox{\includegraphics[scale=0.25]{Figures/gaugehatphiv2.pdf}}}~, \quad
\textcolor{blue}{(Z_i^{N_i})}
 \right \rangle.
\end{align}
The full gauge group $\mathcal{G}$, obtained from gauging out the anyon types in $\bar{\mathcal{A}}$, is then:
\begin{align}
    \mathcal{G} = \left \langle \{\mathcal{G}_i\}_{i=1}^M \right \rangle.
\end{align}

Similar to the gauge group, we specify the stabilizer group $\mathcal{S}$ by separating the generators by layers. We denote the group of locally generated stabilizers associated to the $i$th layer by $\tilde{\mathcal{S}}_i$.
For $n_i=0$, the subgroup $\tilde{\mathcal{S}}_i$ is:
\begin{align} \label{eq: stabilizer ni0}
\tilde{\mathcal{S}}_i \equiv \left \langle 
\vcenter{\hbox{\includegraphics[scale=0.23]{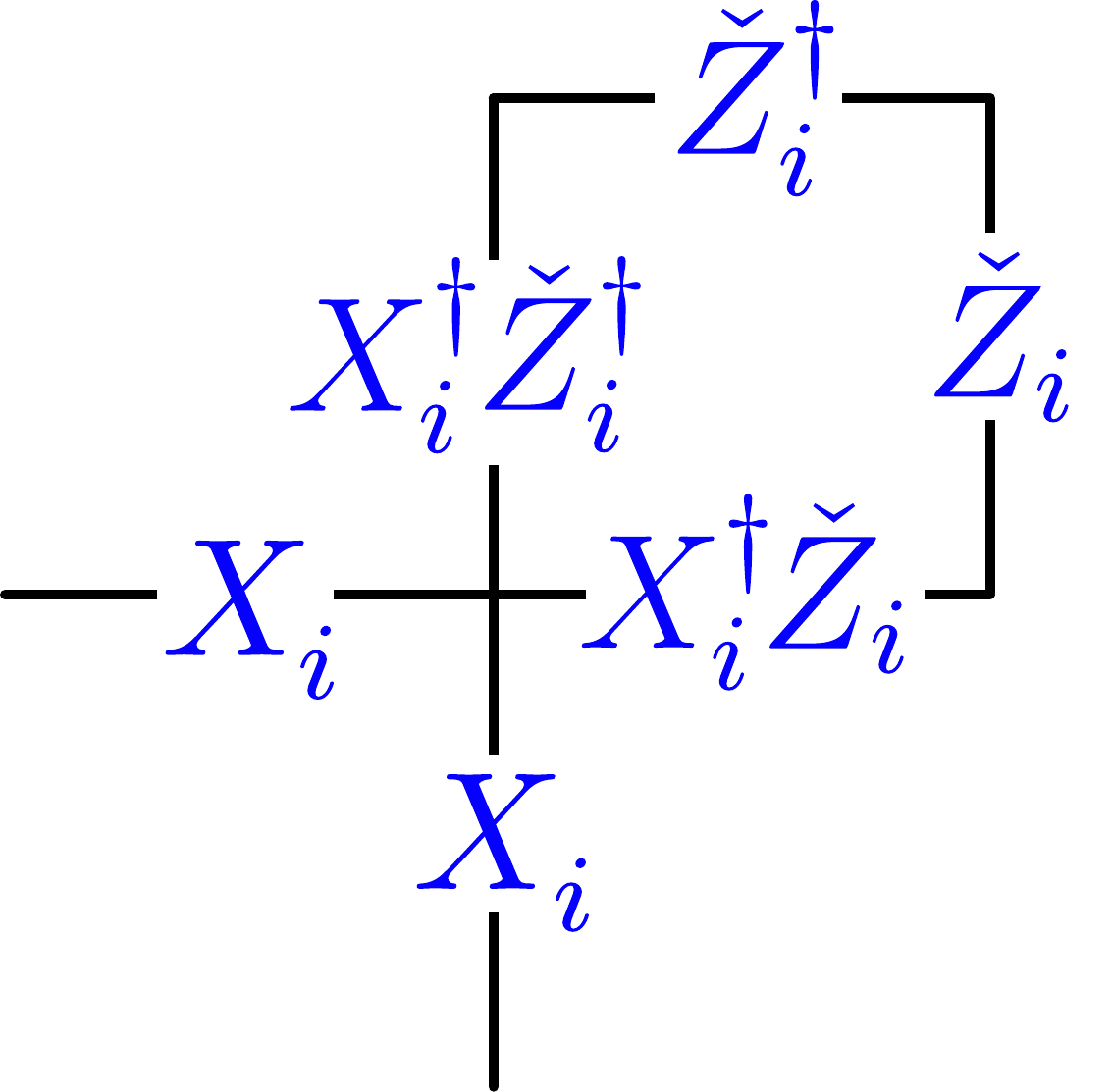}}}
\right \rangle,
\end{align}
while for $n_i=N_i/2$, it is:
\begin{align} \label{eq: stabilizer niNi2}
\tilde{\mathcal{S}}_i = \left \langle 
\vcenter{\hbox{\includegraphics[scale=0.23]{Figures/AvZcheck.pdf}}}~, \quad
\vcenter{\hbox{\includegraphics[scale=0.23]{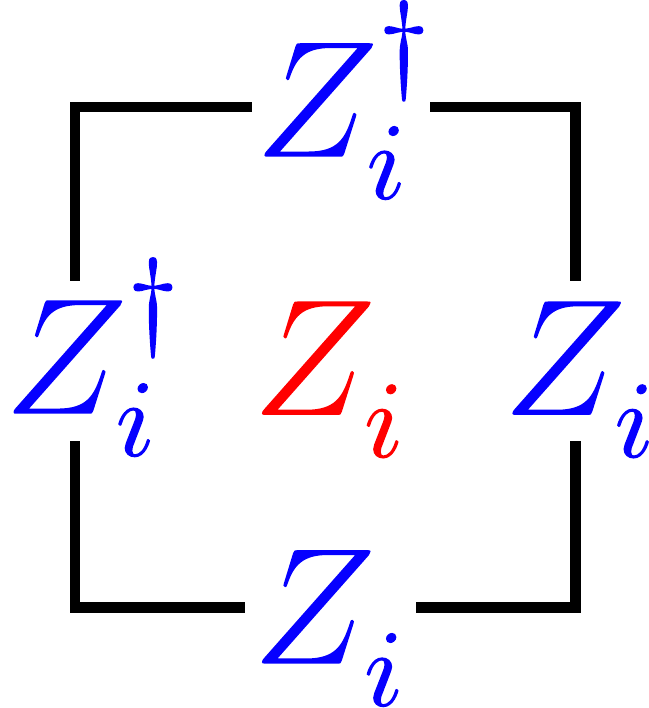}}}~, \quad
\vcenter{\hbox{\includegraphics[scale=0.26]{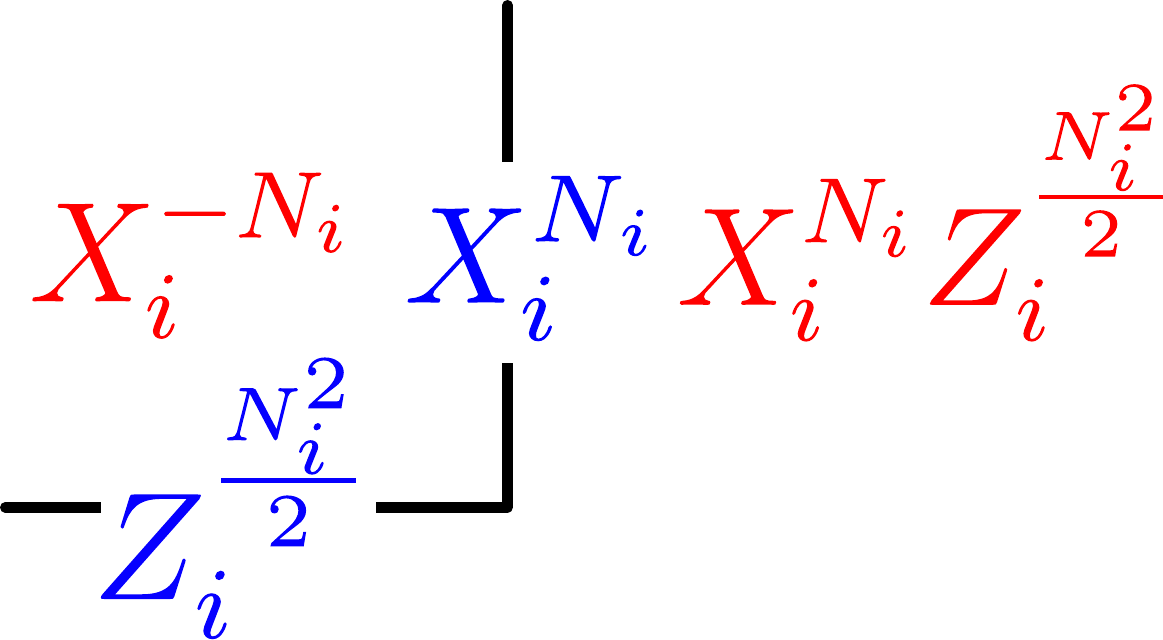}}}~, \quad
\vcenter{\hbox{\includegraphics[scale=0.25]{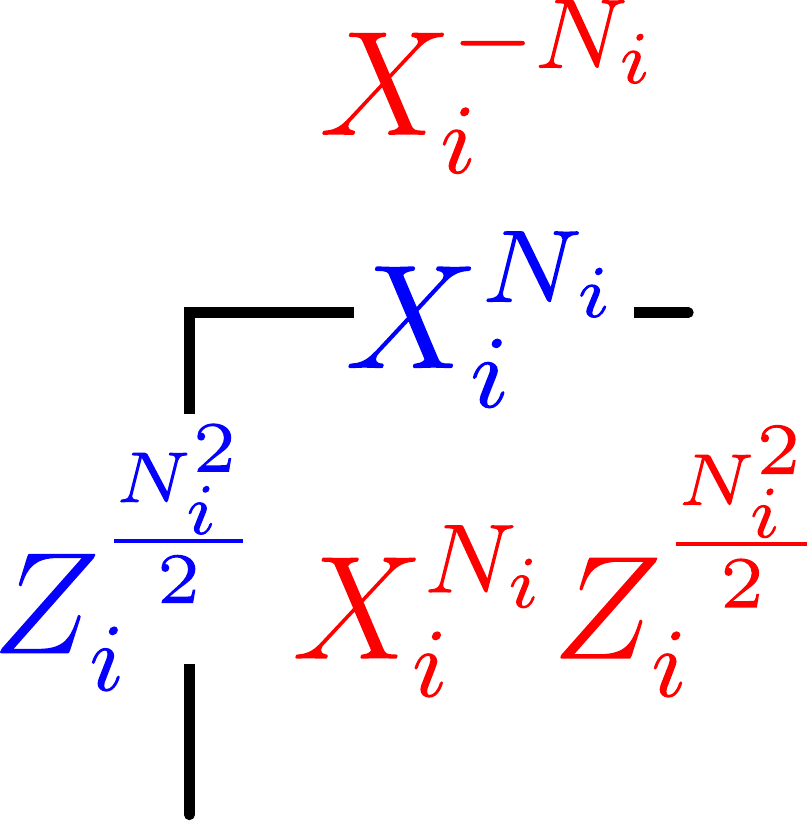}}}~
\right \rangle.
\end{align}
Here, we have used the operator $\check{Z}_i$, which is defined as:
\begin{align}
\check{Z}_i \equiv Z_i^{n_i}
\prod_{j : \, j\leq i,~n_j =0} Z_j^{\frac{p_{ji}N_j}{N_{ij}}} \prod_{j : \, j\leq i,~ n_j = N_j/2} Z_j^{\frac{p_{ji}N_j^2}{N_{ij}}}.
\end{align}
The full group of locally generated stabilizers $\tilde{\mathcal{S}}$ is:
\begin{align}
    \tilde{\mathcal{S}} = \left \langle \{\tilde{\mathcal{S}}_i\}_{i=1}^M \right \rangle.
\end{align}
It is important to note that, for $n_i=N_i/2$, our particular choice of short string operators for $\overline{a}_i$ ensure that the stabilizer group contains the plaquette terms and edge terms of the TQD stabilizer code. This is necessary so that the
vertex terms (on the far left of Eq.~\eqref{eq: stabilizer niNi2}) only generate an order $N_i$ constraint, due to the relations with the other stabilizer generators.

If the topological subsystem code is defined on a torus, there are nonlocal stabilizers or nontrivial bare logical operators. These can be represented by powers of string operators of the form:
\begin{align} \label{eq: nonlocal strings niNi2}
\vcenter{\hbox{\includegraphics[scale=.25]{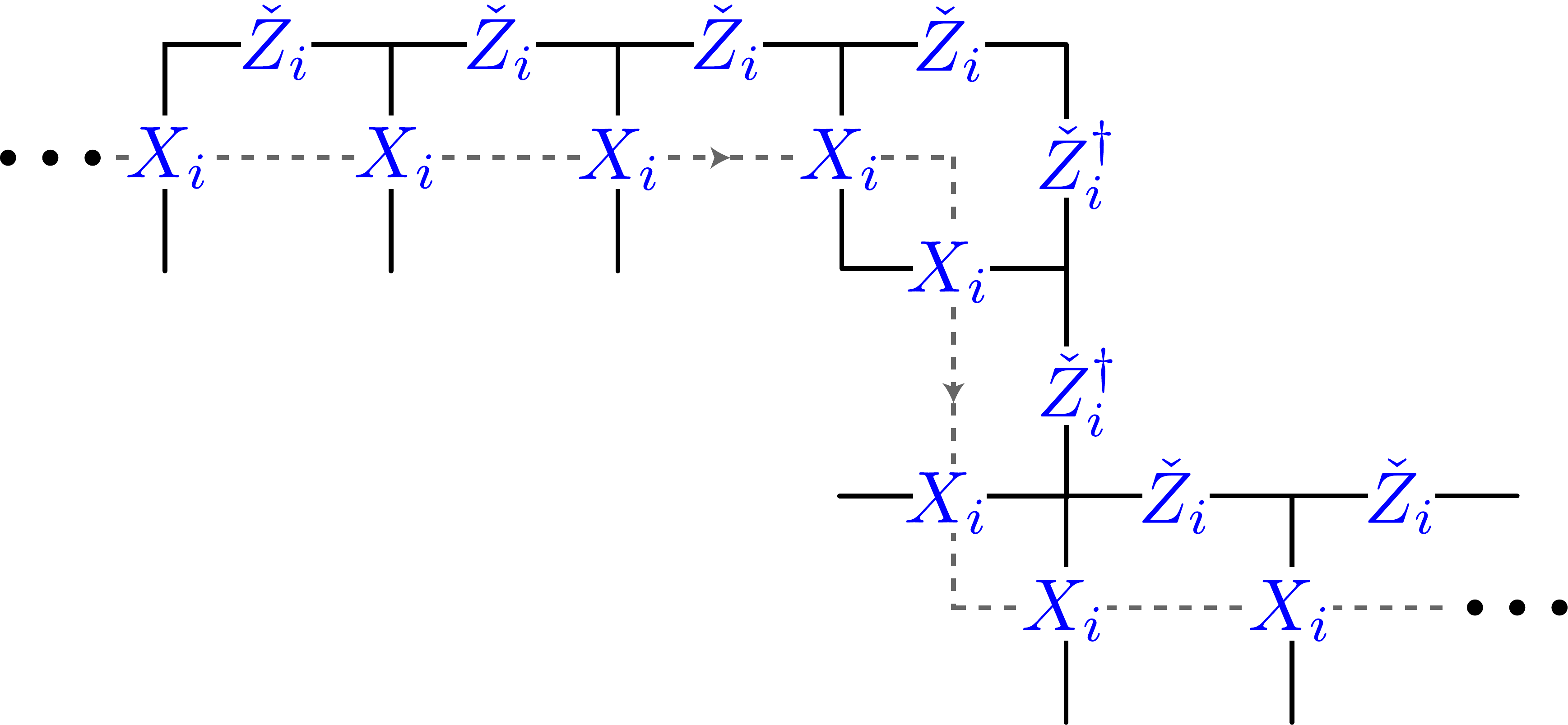}}},
\end{align} 
where the string operator implicitly wraps around a non-contractible path of the torus.
Importantly, the vertex terms of the stabilizer group and the nonlocal stabilizers or nontrivial bare logical operators are generated by loops of string operators for the anyon type $a_i$. This is made explicit by defining the following short string operators:
\begin{eqs}
    W^{a_i}_e &\equiv~ \vcenter{\hbox{\includegraphics[scale=0.28]{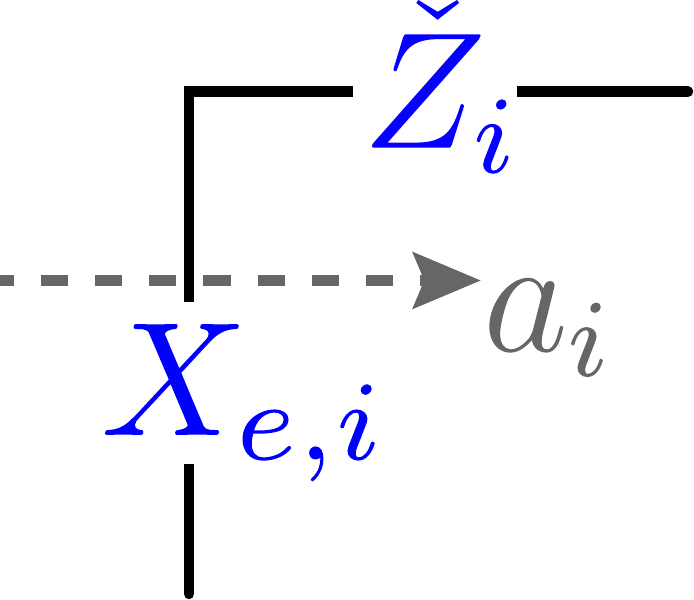}}}~,
    \quad
    \vcenter{\hbox{\includegraphics[scale=0.28]{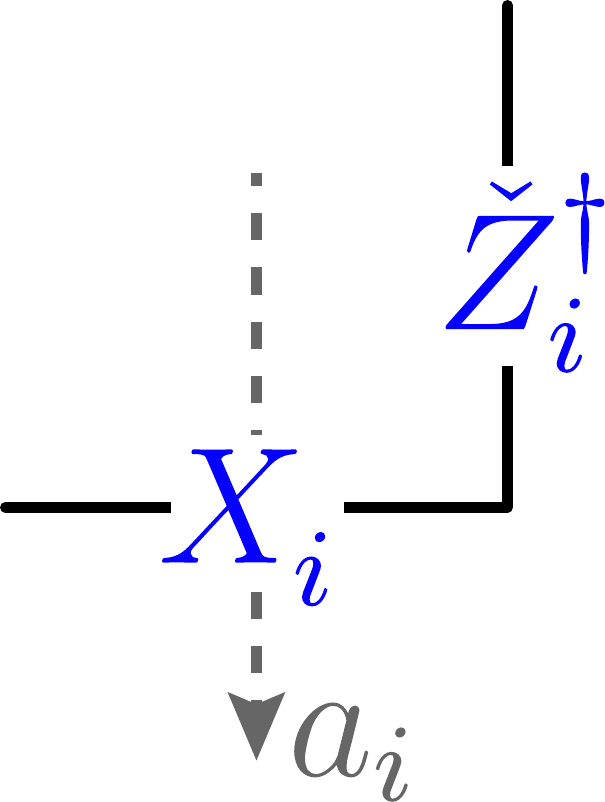}}}.
\end{eqs}
It can be checked that these short string operators indeed produce string operators for the anyon type $a_i$ by using the methods in Section~\ref{sec: anyon theories for subsystem codes} for computing the fusion rules and exchange statistics.

Finally, to verify that we have accounted for all of the stabilizers, gauge operators, and logical operators, we compare the total number of qudits $\mathbf{N}_Q$ to the number of stabilized qudits $\mathbf{N}_S$, gauge qudits $\mathbf{N}_G$, and logical qudits $\mathbf{N}_L$. Similar to the derivation of Eq.~\eqref{eq: counting qubits z41} in Section~\ref{sec: z41 definition}, these satisfy the consistency condition:
\begin{align} \label{eq: counting consistency}
 \mathbf{N}_S + \mathbf{N}_G + \mathbf{N}_L = \mathbf{N}_Q.
\end{align}
We simplify the counting in what follows by considering the $M$ layers independently. We denote the number of qudits, stabilized qudits, gauge qudits, and logical qudits associated to the $i$th layer by $\mathbf{N}_Q^{(i)}$, $\mathbf{N}_S^{(i)}$, $\mathbf{N}_G^{(i)}$, and $\mathbf{N}_L^{(i)}$, respectively. According to Eq.~\eqref{eq: counting consistency}, these must obey the condition:
\begin{align} \label{eq: consistent counting}
\sum_{i=1}^M \left( \mathbf{N}^{(i)}_S + \mathbf{N}^{(i)}_G + \mathbf{N}^{(i)}_L \right) = \sum_{i=1}^M \mathbf{N}^{(i)}_Q.
\end{align}
For the purpose of counting, we assume that the subsystem code is defined on a torus.

We start with the case in which the $i$th layer is characterized by $n_i=0$.
Each~plaquette has two qudits, so the number of $N_i$-dimensional qudits in the $i$th layer is:
\begin{align}
    \mathbf{N}_Q^{(i)} = 2 \mathbf{P},
\end{align}
where $\mathbf{P}$ is the number of plaquettes. 

We next count the number of stabilized qudits $\mathbf{N}_S^{(i)}$ associated to the $i$th layer, starting with the nonlocal stabilizers. To this end, we let $q_i$ be the smallest integer such that the string operator for $a_i^{q_i}$ along a non-contractible path represents a nonlocal stabilizer. The corresponding nonlocal stabilizer stabilizes a qudit of dimension $N_i/q_i$. In terms of $N_i$-dimensional qudits, the number of qudits stabilized by the string operator is:
\begin{align}
\mathbf{W}^{(i)}_S \equiv \log_{N_i}(N_i/q_i) = 1 - \log_{N_i}(q_i).
\end{align}
On a torus, the nonlocal stabilizers are generated by two string operators (see Fig.~\ref{fig: braiding}), so the number of qudits stabilized by nonlocal stabilizers is thus $2 \mathbf{W}^{(i)}_S$. As for the local stabilizers, there is one order $N_i$ local stabilizer generator for each plaquette. However, the product of all of these stabilizers is the identity. Therefore, there are $N_i$ global relations amongst the local generators. 
This means that there are $\mathbf{P} - 1$ stabilized qudits from the local stabilizers, which gives us the total number of stabilized qudits:
\begin{align}
    \mathbf{N}^{(i)}_S = \mathbf{P} - 1 + 2 \mathbf{W}^{(i)}_S.
\end{align} 

Now, we count the number of gauge qudits. There are three independent gauge generators for each plaquette, each of which has order $N_i$. Since the plaquette gauge generators all multiply to the identity, one of them is redundant. This tells us that the number of $N_i$-dimensional gauge qudits is:
\begin{align}
\mathbf{N}_G^{(i)} = \left[(3\mathbf{P}-1) - (\mathbf{P} - 1 + 2 \mathbf{W}^{(i)}_S) \right]/2 = \mathbf{P}-\mathbf{W}^{(i)}_S.
\end{align}
In the brackets, we have computed the number of independent gauge generators minus the number of independent generators of the stabilizer group. We then divide by two, since the gauge generators form the complete operator algebra, i.e., the Pauli $X$ and Pauli $Z$ operators for the gauge qudits. 

Lastly, we count the number of logical qudits. The nontrivial bare logical operators associated to the $i$th layer are represented by string operators that move the anyon type $a_i$ around a non-contractible path. By assumption, the $q_i$th power of this string operator is a (nonlocal) stabilizer. Therefore, the logical operators, must have order $q_i$. This tells us that the number of logical qudits, in terms of $N_i$-dimensional qudits, is:
\begin{align}
\mathbf{N}^{(i)}_L = \log_{N_i}(q_i).
\end{align}
Using that $\mathbf{W}_S^{(i)} + \mathbf{N}_L^{(i)} = 1$, we see that the counting in the $i$th layer satisfies:
\begin{align} \label{eq: general counting ni0}
\mathbf{N}^{(i)}_S + \mathbf{N}^{(i)}_G + \mathbf{N}^{(i)}_L = (\mathbf{P} - 1 + 2 \mathbf{W}_S^{(i)}) + (\mathbf{P} - \mathbf{W}_S^{(i)}) + \mathbf{N}_L^{(i)} = 2 \mathbf{P} = \mathbf{N}^{(i)}_Q.
\end{align}

We next turn to the case in which the $i$th layer is characterized by $n_i = N_i/2$. Although the qudits have dimension $N_i^2$, we find it convenient to count in terms of $N_i$-dimensional qudits.
This means, for example, that a single $N_i^2$-dimensional qudit is counted as two $N_i$-dimensional qudits. With this, there are six $N_i$-dimensional qudits for each plaquette, giving us the total number of qudits: 
\begin{align}
    \mathbf{N}^{(i)}_Q = 6 \mathbf{P}.
\end{align}

For $n_i = N_i/2$, there are no nonlocal stabilizers. This is because, for any $q_i$ that is not a multiple of $N_i$, the anyon type $a_i$ braids nontrivially with $a_i^{q_i}$. This implies that string operators formed by moving $a_i^{q_i}$ fail to commute with those formed by moving $a_i$ along paths with odd intersection number. To see that $a_i$ braids nontrivially with $a_i^{q_i}$, we compute the braiding relations of $a_i$ and $a_i^{q_i}$ using the identity in Eq.~\eqref{eq: braiding identity}. This tells us that the braiding relations are:
\begin{align}
    B_\theta(a_i, a_i^{q_i}) = e^{\frac{2 \pi i}{N_i}q_i (2 \lfloor t_i \rfloor + 1)}.
\end{align}
For the braiding relations to be trivial, $q_i$ needs to satisfy:
\begin{align} \label{eq: braiding exponent}
    q_i (2 \lfloor t_i \rfloor +1) = \alpha N_i,
\end{align}
for some nonnegative integer $\alpha$. Dividing both sides by $q_i$, we find:
\begin{align}
    2 \lfloor t_i \rfloor +1 = \alpha N_i / q_i.
\end{align}
Therefore, the integer $\alpha N_i / q_i$ must be odd. By assumption, $N_i$ is a power of $2$, so $q_i$ must be a multiple of $N_i$. This implies that $a_i^{q_i}$ is the trivial anyon type.

Hence, the stabilizer group is generated by local stabilizers, and we only need to count the number of qudits stabilized by local stabilizers. We find that there are four stabilized qudits of dimension $N_i$ for every plaquette, corresponding to the vertex stabilizer, plaquette stabilizer, and two edge stabilizers. Note that the vertex stabilizer has order $N_i^2$, but its $N_i$th power is a product of plaquette stabilizers and edge stabilizers. There is one order $N_i$ global relation, however, since the product of all of the vertex stabilizers is the identity. This means that the number of stabilized $N_i$-dimensional qudits is: 
\begin{align}
    \mathbf{N}^{(i)}_S = 4 \mathbf{P} - 1.
\end{align}

Next, there are eight order $N_i$ gauge generators for each plaquette -- with one global relation, given by multiplying the $N_i$th power of the plaquette stabilizer on every plaquette. Thus, the number of gauge qudits is:
\begin{align}
\mathbf{N}^{(i)}_G = \left[ (8 \mathbf{P} - 1) - (4 \mathbf{P} -1) \right]/2 = 2 \mathbf{P}. 
\end{align}

Lastly, there are two order $N_i$ logical operators, meaning that there is only one $N_i$-dimensional logical qudit. Therefore, we have $\mathbf{N}_L^{(i)} = 1$, which gives us: 
\begin{align}
\mathbf{N}^{(i)}_S + \mathbf{N}_G^{(i)} + \mathbf{N}_L^{(i)} = (4 \mathbf{P} - 1) + 2 \mathbf{P} + 1 = 6 \mathbf{P} = \mathbf{N}_Q^{(i)}.
\end{align}

We have now shown that the following condition is satisfied for each layer:
\begin{align}
    \mathbf{N}^{(i)}_S + \mathbf{N}_G^{(i)} + \mathbf{N}_L^{(i)} = \mathbf{N}_Q^{(i)}.
\end{align}
After summing over the $M$ layers, as in Eq.~\eqref{eq: consistent counting}, we see that the consistency condition in Eq.~\eqref{eq: counting consistency} is satisfied. 
This suggests that we have correctly counted the nontrivial bare logical operators and nonlocal stabilizers of the topological subsystem code. Consequently, the anyon types that remain after gauging out the anyon types $\{\overline{a}_i\}_{i=1}^M$ are precisely the anyon types forming the subtheory $\mathcal{A}$.

\section{Further examples of topological subsystem codes} \label{sec: more examples}

In this section, we describe examples of topological subsystem codes, derived by following the general construction in the previous section. The examples in Sections~\ref{sec: 1,m subsystem}, \ref{sec: 1 psi subsystem code}, and \ref{sec: 3F subsystem} are defined on qubits and have been discussed in Refs.~\cite{kitaev2006anyons, Bombin2009fermions, Bombin2010subsystem, Bombin2012universal, Bombin2014structure, Suchara2011subsystem, Roberts20203fermion}. The subsystem code in Section~\ref{sec: generalized honeycomb subsystem code} is based on the generalized honeycomb models of Ref.~\cite{Barkeshli2015generalized}. To the best of our knowledge, the subsystem code in Section~\ref{sec: z3 z9 subsystem} is novel and has not yet been discussed in the literature. 

\subsection{$\ZZ_2^{(0)}$ subsystem code} \label{sec: 1,m subsystem}

The $\ZZ_2^{(0)}$ subsystem code is characterized by an anyon theory with $\ZZ_2$ fusion rules and anyon types labeled by $\{1,m\}$. The anyon type $m$ is a boson with trivial mutual braiding relations:
\begin{align}
\theta(m)=1, \quad B_\theta(m,m) =1.
\end{align} 
Therefore, $m$ is transparent, and the $\ZZ_2^{(0)}$ anyon theory is non-modular. As the notation suggests, the $\ZZ_2^{(0)}$ anyon theory is equivalent to the subtheory generated by the $m$ anyon type of the $\ZZ_2$ TC. Indeed, according to Section~\ref{sec: general}, the construction proceeds by gauging out the $m$ anyon type in a $\ZZ_2$ TC.

The $\ZZ_2^{(0)}$ subsystem code is defined on a square lattice with a qubit at each edge. The gauge group is given by:
\begin{align}
\mathcal{G} \equiv \left \langle i, \quad
\vcenter{\hbox{\includegraphics[scale=.25]{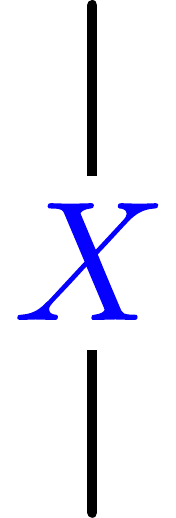}}}, \quad
\vcenter{\hbox{\includegraphics[scale=.25]{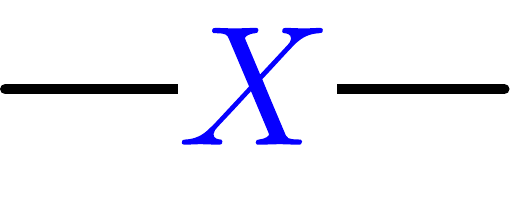}}}, \quad
\vcenter{\hbox{\includegraphics[scale=.25]{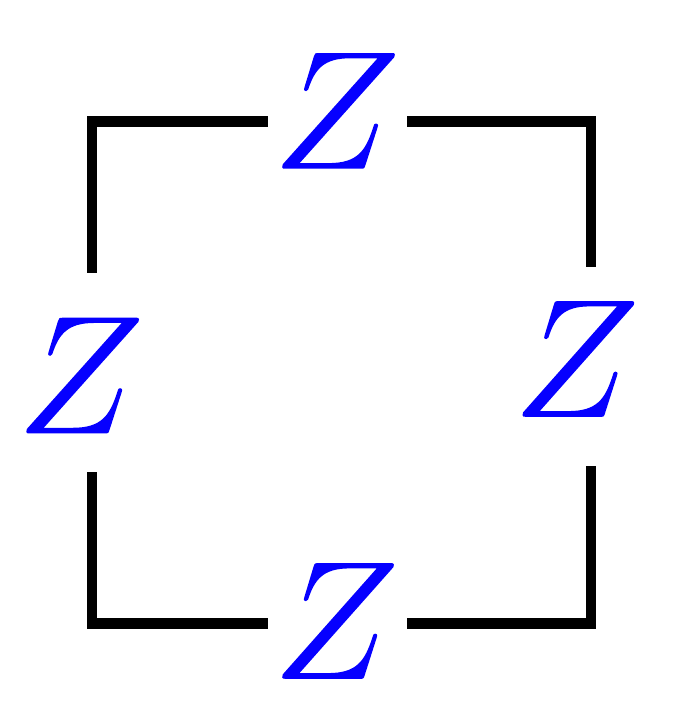}}}
\right \rangle.
\end{align}
The stabilizer group is generated by the string operators for the $m$ anyon types of the $\ZZ_2$ TC, for example:
\begin{align}
\vcenter{\hbox{\includegraphics[scale=.23]{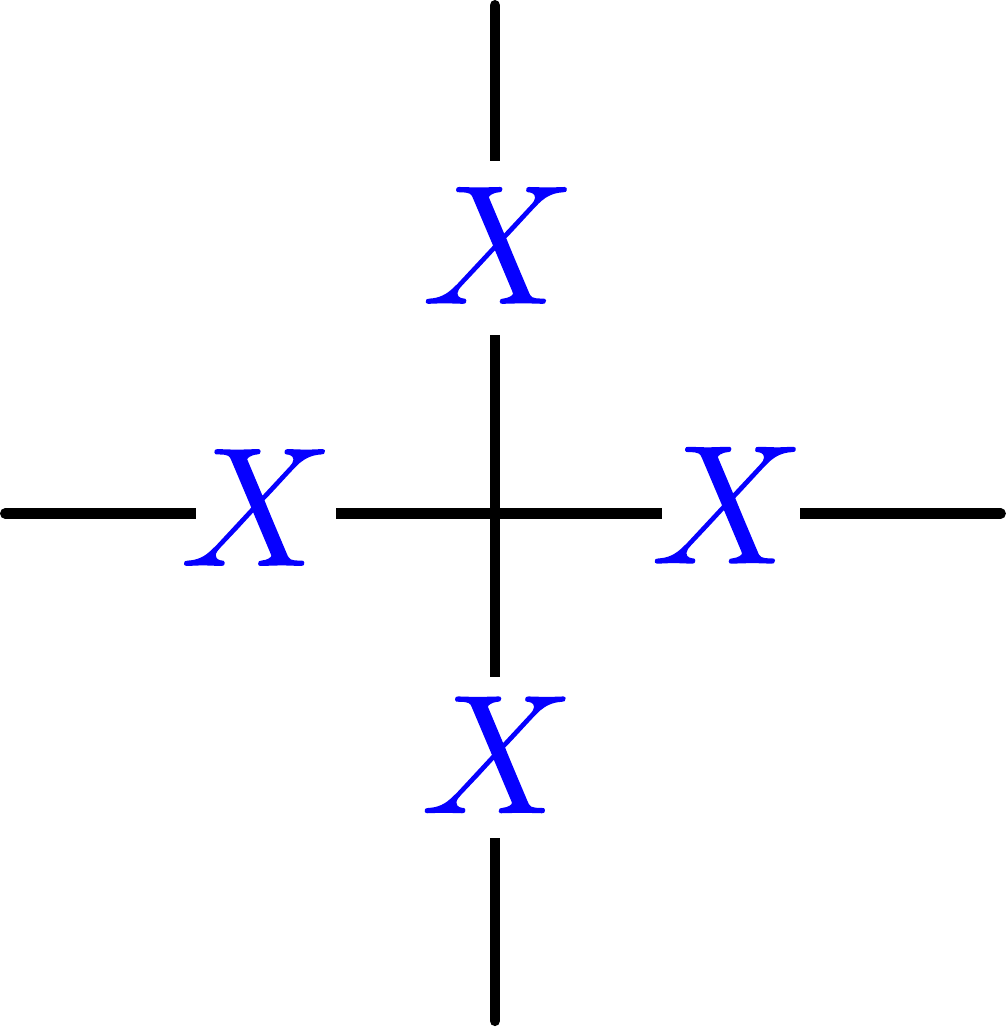}}}, \quad
\vcenter{\hbox{\includegraphics[scale=.22]{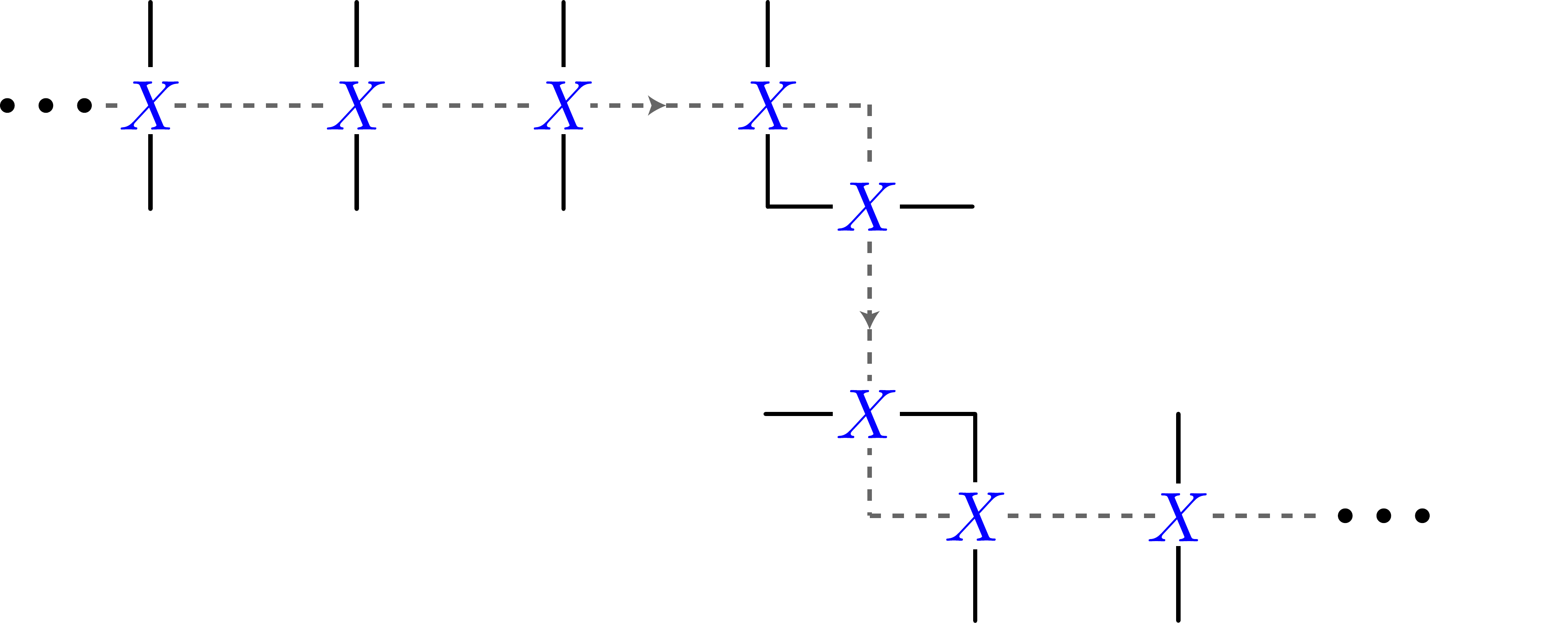}}}.
\end{align}
Since the $m$ anyon type is transparent, this topological subsystem code has nonlocal stabilizers created by moving $m$ around non-contractible paths. There are, however, no opaque anyon types, so the topological subsystem code does not have any logical operators. Consequently, the logical subsystem is trivial, and the subsystem code does not encode any qubits. In Ref.~\cite{Bombin2014structure}, this subsystem code is referred to as the `subsystem toric code'. However, this should not be confused with the two-dimensional subsystem toric code of Ref.~\cite{Bravyi2013Subsystem}.

The parameter space of Hamiltonians defined by the subsystem code contains Hamiltonians of the form:
\begin{align}
H_{\{J_e,J_p\}} \equiv \sum_{e \in E_h} J_e \, \vcenter{\hbox{\includegraphics[scale=.25]{Figures/z2mh2.pdf}}} + \sum_{e \in E_v} J_e \vcenter{\hbox{\includegraphics[scale=.25]{Figures/z2mv2.pdf}}} + \sum_p \vcenter{\hbox{\includegraphics[scale=.25]{Figures/z2plaquette2.pdf}}},
\end{align}
where $E_h$ and $E_v$ denote the sets of horizontal edges and vertical edges, respectively.
Although the anyon theory has a nontrivial anyon type, it is a transparent boson. As such, it does not enforce any anyonic excitations for the Hamiltonians. Indeed, by taking the coefficients of the edge terms to be sufficiently large, we see that the resulting Hamiltonian has trivial topological order. Note that this is the Hamiltonian of a  pure $\ZZ_2$ lattice gauge theory with Gauss's law manifesting as a $1$-form symmetry of the Hamiltonians in the parameter space. 

\subsection{$\ZZ_2^{(1)}$ subsystem code} \label{sec: 1 psi subsystem code}

Similar to the $\ZZ_2^{(0)}$ subsystem code, the $\ZZ_2^{(1)}$ subsystem code is based on an anyon theory with $\ZZ_2$ fusion rules. However, in this case, the nontrivial anyon type, denoted by $\psi$, is a transparent fermion:
\begin{align}
\theta(\psi)=-1, \quad B_\theta(\psi, \psi) = 1.
\end{align}
The $\ZZ_2^{(1)}$ anyon theory can be viewed as the subtheory of the $\ZZ_2$ TC generated by the $em$ anyon type. Accordingly, the general construction in Section~\ref{sec: general} tells us that the subsystem code can be derived from a $\ZZ_2$ TC by gauging out the $em$ anyon type.

In more detail, the $\ZZ_2^{(1)}$ subsystem code is defined on a square lattice with a qubit on each edge. The gauge group is generated by the plaquette term of the $\ZZ_2$ TC as well as short string operators for the $em$ anyon types:
\begin{align}
\mathcal{G}' \equiv \left \langle i, \quad
\vcenter{\hbox{\includegraphics[width=.1\textwidth]{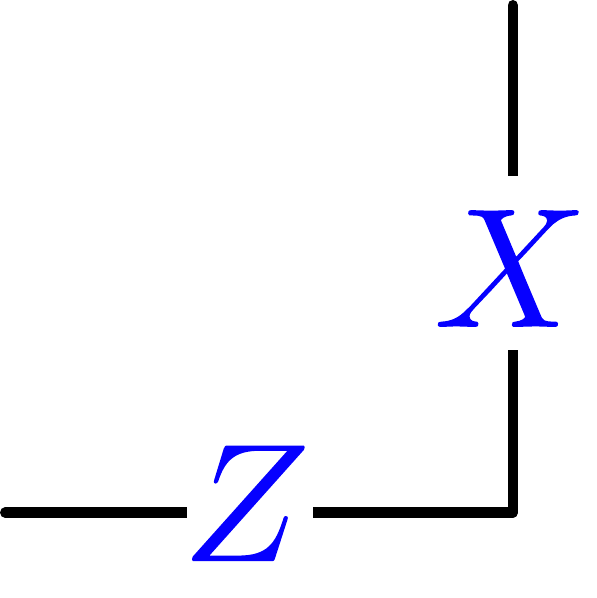}}}, \quad
\vcenter{\hbox{\includegraphics[width=.1\textwidth]{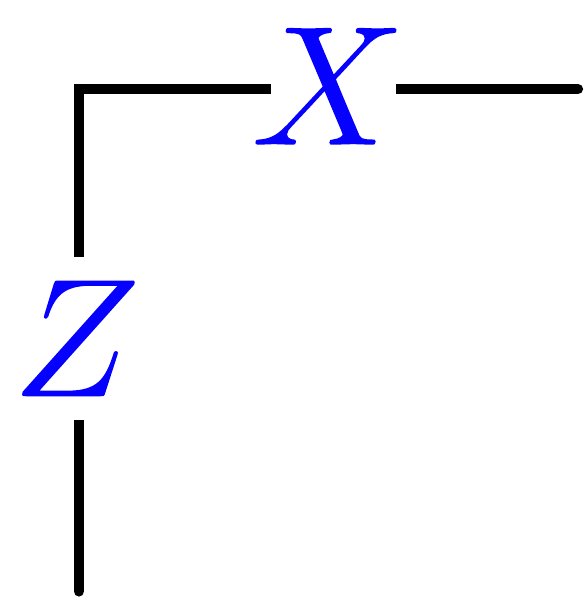}}}, \quad
\vcenter{\hbox{\includegraphics[width=.1\textwidth]{Figures/z2plaquette2.pdf}}}
\right \rangle.
\end{align}
The stabilizer group is then generated by the closed string operators for $\psi$, such as:
\begin{align}
\vcenter{\hbox{\includegraphics[scale=.25]{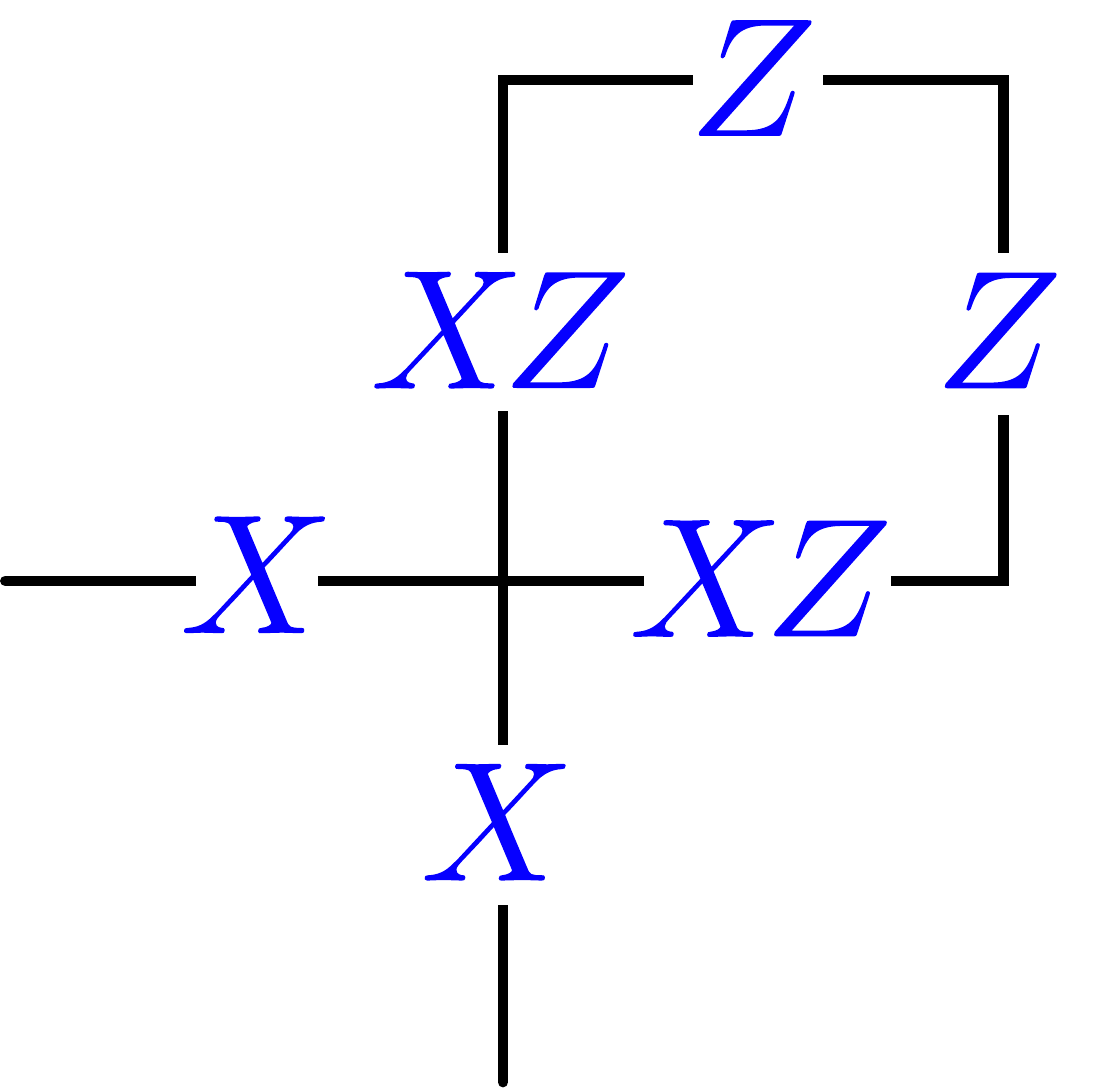}}}, \quad
\vcenter{\hbox{\includegraphics[scale=.25]{Figures/emlongstring2.pdf}}}.
\end{align}
There are no logical operators, because $\psi$ has trivial braiding relations with itself. This implies that the string operators cannot form the Pauli $X$ and Pauli $Z$ operators of a logical subsystem. 

The gauge group $\mathcal{G}'$ can be put into a more familiar form by multiplying the plaquette term by the edge terms and conjugating the gauge generators by a Hadamard gate on all of the horizontal edges. This gives us the gauge group $\mathcal{G}$:
\begin{align}
\mathcal{G} \equiv \left \langle i, \quad 
\vcenter{\hbox{\includegraphics[scale=0.25]{Figures/xedgesquare}}}, \quad
\vcenter{\hbox{\includegraphics[scale=0.25]{Figures/zedgesquare}}}, \quad
\vcenter{\hbox{\includegraphics[scale=0.25]{Figures/yedgesquare}}}
 \right \rangle.
\end{align}
Furthermore, we map the square lattice to a hexagonal lattice by redefining the lattice according to Fig.~\ref{fig: squaretohexagon}. The gauge group can then be written as:
\begin{align}
\mathcal{G} = \left \langle i, \quad
\vcenter{\hbox{\includegraphics[scale=.32]{Figures/xedge2.pdf}}}, \quad \vcenter{\hbox{\includegraphics[scale=.32]{Figures/zedge2.pdf}}}, \quad 
\vcenter{\hbox{\includegraphics[scale=.32]{Figures/yedge2.pdf}}} 
\right \rangle.
\end{align}
We now see that the gauge generators are nothing other than the Hamiltonian terms of the honeycomb model in Ref.~\cite{kitaev2006anyons}! Moreover, the stabilizers are precisely the conserved quantities that were identified in Ref.~\cite{kitaev2006anyons}.

To make the connection to the honeycomb model explicit, we consider the family of Hamiltonians defined by the $\ZZ_2^{(1)}$ subsystem code. For certain choices of parameters, the Hamiltonians belonging to the parameter space take the form:
\begin{align} \label{eq: honeycomb Hamiltonian}
H_{\{J_e\}} = -\sum_{e \in x\text{-edges}} J_e \, \vcenter{\hbox{\includegraphics[scale=.32]{Figures/xedge2.pdf}}} - \sum_{e \in y\text{-edges}} J_e \,  \vcenter{\hbox{\includegraphics[scale=.32]{Figures/yedge2.pdf}}} - \sum_{e \in z\text{-edges}} J_e \,  \vcenter{\hbox{\includegraphics[scale=.32]{Figures/zedge2.pdf}}}.
\end{align}
Here, we have labeled the edges as $x$-edges, $y$-edges, and $z$-edges according to Fig.~\ref{fig: xyzedges}. This family of Hamiltonians includes the honeycomb model of Ref.~\cite{kitaev2006anyons}. We note that all of the gapped Hamiltonians in the parameter space must admit an emergent fermion. This can be checked explicitly, given that the Hamiltonians are exactly solvable and can be mapped to a system of non-interacting fermions, as demonstrated in Ref.~\cite{kitaev2006anyons}. Remarkably, by tuning the coefficients $J_e$ and introducing perturbations that break the time-reversal symmetry, the Hamiltonians realize all of the anyon theories of the $16$-fold way.

\subsection{$\ZZ_N^{(1)}$ subsystem codes} \label{sec: generalized honeycomb subsystem code}

This next example reproduces the generalized honeycomb models of Ref.~\cite{Barkeshli2015generalized}. The subsystem codes are characterized by the $\ZZ_N^{(1)}$ anyon theory, which has anyon types that form a group $\ZZ_N$ under fusion. We label the anyon types as $\{1, a, a^2, \ldots, a^{N-1}\}$. The exchange statistics and braiding relations are given by:
\begin{align}
\theta(a^p) = e^{2 \pi i p^2/N}, \quad B_\theta(a^p,a^q) = e^{4 \pi i pq / N},
\end{align}
for $p,q$ in $\ZZ_N$. This matches the subtheory generated by $em$ in a $\ZZ_N$ TC. According to the prescription in Section~\ref{sec: general}, the corresponding subsystem code is built from a $\ZZ_N$ TC by gauging out the $e^{-1}m$ anyon type.

Explicitly, the subsystem code is defined on a square lattice with an $N$-dimensional qudit on each edge. The gauge group produced by the construction in Section~\ref{sec: general} is:
\begin{align}
\mathcal{G}' \equiv \left \langle 
\vcenter{\hbox{\includegraphics[scale = .23]{Figures/e3mh}}}, \quad
\vcenter{\hbox{\includegraphics[scale = .23]{Figures/z2emv2.pdf}}}, \quad
\vcenter{\hbox{\includegraphics[scale = .23]{Figures/BpTCz41}}}
\right \rangle.
\end{align}
Up to roots of unity, the bare logical group is generated by $em$ string operators supported on closed paths, such as:
\begin{align}
\vcenter{\hbox{\includegraphics[scale = .23]{Figures/emloop}}}, \qquad \vcenter{\hbox{\includegraphics[width=.5\textwidth]{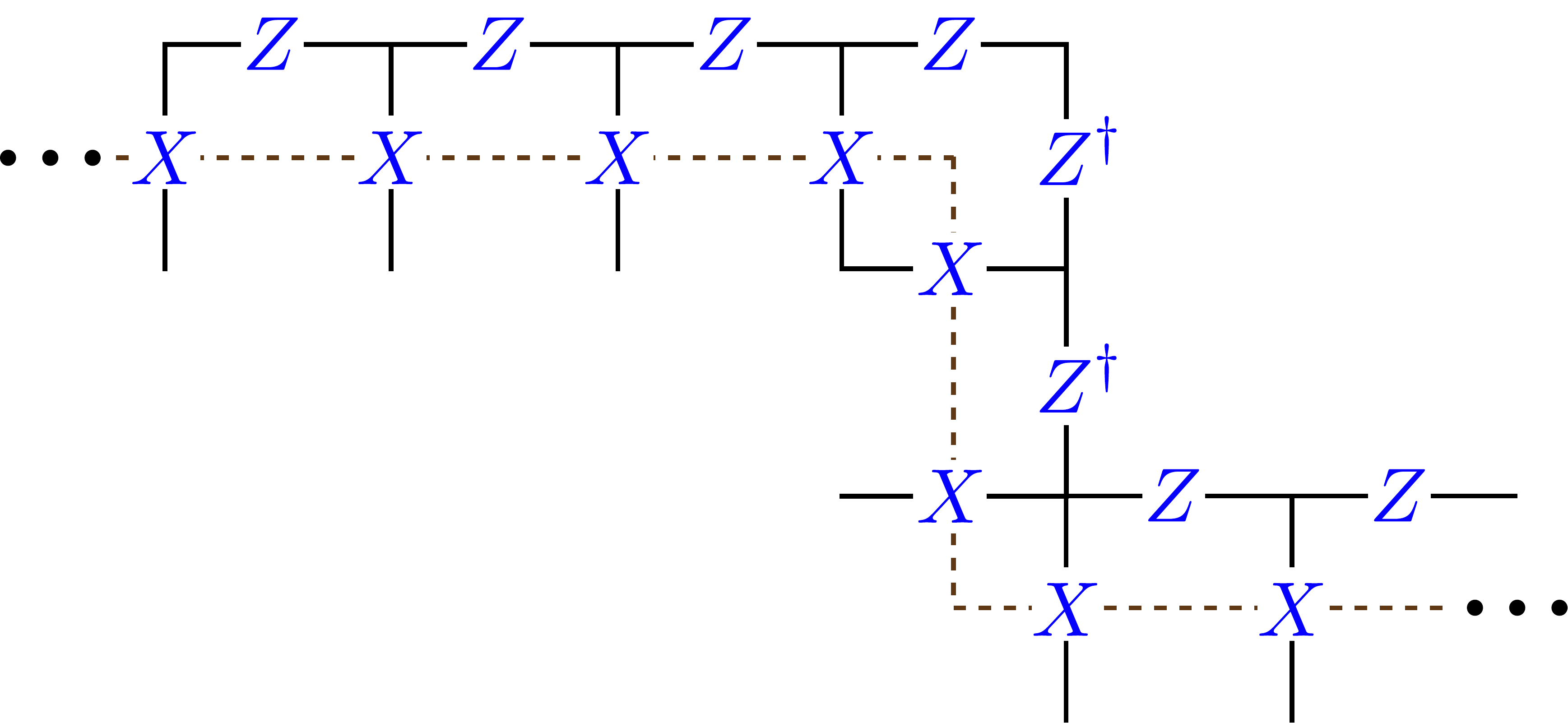}}}.
\end{align}
If $N$ is odd, then all of the anyon types are opaque. Consequently, the subsystem code encodes a single $N$-dimensional qudit on a torus. If $N$ is even, on the other hand, then the $a^{N/2}$ anyon type is transparent. This means that the subsystem code only encodes an $N/2$-dimensional qudit on a torus.

In direct analogy to the construction of the $\ZZ_4^{(1)}$ subsystem code, the gauge group can be simplified by multiplying the plaquette gauge generator by the edge gauge generators. We also apply the following Clifford transformation on the horizontal edges:
\begin{align} \label{eq: F Clifford}
F: X \mapsto Z, \quad Y \mapsto Z^\dagger X, \quad Z \mapsto X^\dagger.
\end{align}
Finally, by mapping the system to a hexagonal lattice (Fig.~\ref{fig: squaretohexagon}), we obtain the gauge group:
\begin{align}
\mathcal{G} \equiv \left \langle 
\vcenter{\hbox{\includegraphics[scale=.32]{Figures/xedge2.pdf}}}, \quad \vcenter{\hbox{\includegraphics[scale=.32]{Figures/zedge2.pdf}}}, \quad 
\vcenter{\hbox{\includegraphics[scale=.32]{Figures/yedge2.pdf}}} 
\right \rangle.
\end{align}
This corresponds to Hamiltonians of the form:
\begin{align} \label{eq: generalized honeycomb Hamiltonian}
H_{\{J_e\}} = -\sum_{e \in x\text{-edges}} J_e \, \vcenter{\hbox{\includegraphics[scale=.32]{Figures/xedge2.pdf}}} - \sum_{e \in y\text{-edges}} J_e \,  \vcenter{\hbox{\includegraphics[scale=.32]{Figures/yedge2.pdf}}} - \sum_{e \in z\text{-edges}} J_e \,  \vcenter{\hbox{\includegraphics[scale=.32]{Figures/zedge2.pdf}}},
\end{align}
which, for certain coefficients, gives the model in Ref.~\cite{Barkeshli2015generalized}.

\subsection{Three-fermion subsystem code} \label{sec: 3F subsystem}

We now describe a topological subsystem code based on the three-fermion ($3$F) anyon theory. Using notation from Ref.~\cite{Roberts20203fermion}, we label the $3$F anyon types as $\{1, \psi_r, \psi_b, \psi_g \}$. These generate a $\ZZ_2 \times \ZZ_2$ group under fusion, where all three nontrivial anyon types have fermionic exchange statistics:
\begin{align}
\theta(1) = 1, \quad \theta(\psi_r) = -1, \quad \theta(\psi_b) = -1, \quad \theta(\psi_g) = -1.
\end{align}
The nontrivial braiding relations are then captured by the following, for any $i,j$ in $\{r,b,g\}$:
\begin{align}
B_\theta(\psi_i, \psi_j) = -1.
\end{align}
This anyon theory appears as a subtheory of two copies of $\ZZ_2$ TC \cite{Bombin2012universal}. To see this, let us label the anyon types of the two $\ZZ_2$ TCs by:
\begin{align}
\{1, e_1, m_1, \psi_1\} \times \{1, e_2, m_2, \psi_2\},
\end{align}
where $\psi_1 =e_1 m_1$ and $\psi_2 = e_2 m_2$ are fermions.
The anyon types can be relabeled as:
\begin{align}
\{1, \, e_1\psi_2, \, m_1\psi_2, \, \psi_1\} \times \{1, \, \psi_1e_2, \, \psi_1m_2, \, \psi_2\},
\end{align}
which is equivalent to two copies of the $3$F anyon theory. Thus, the construction of the $3$F subsystem code proceeds by gauging out the  $\psi_1e_2$ and $\psi_2$ anyon types in two copies of $\ZZ_2$ TC. 

\begin{figure}[t] 
\centering
\hspace{2.5cm}\includegraphics[width=.5\textwidth]{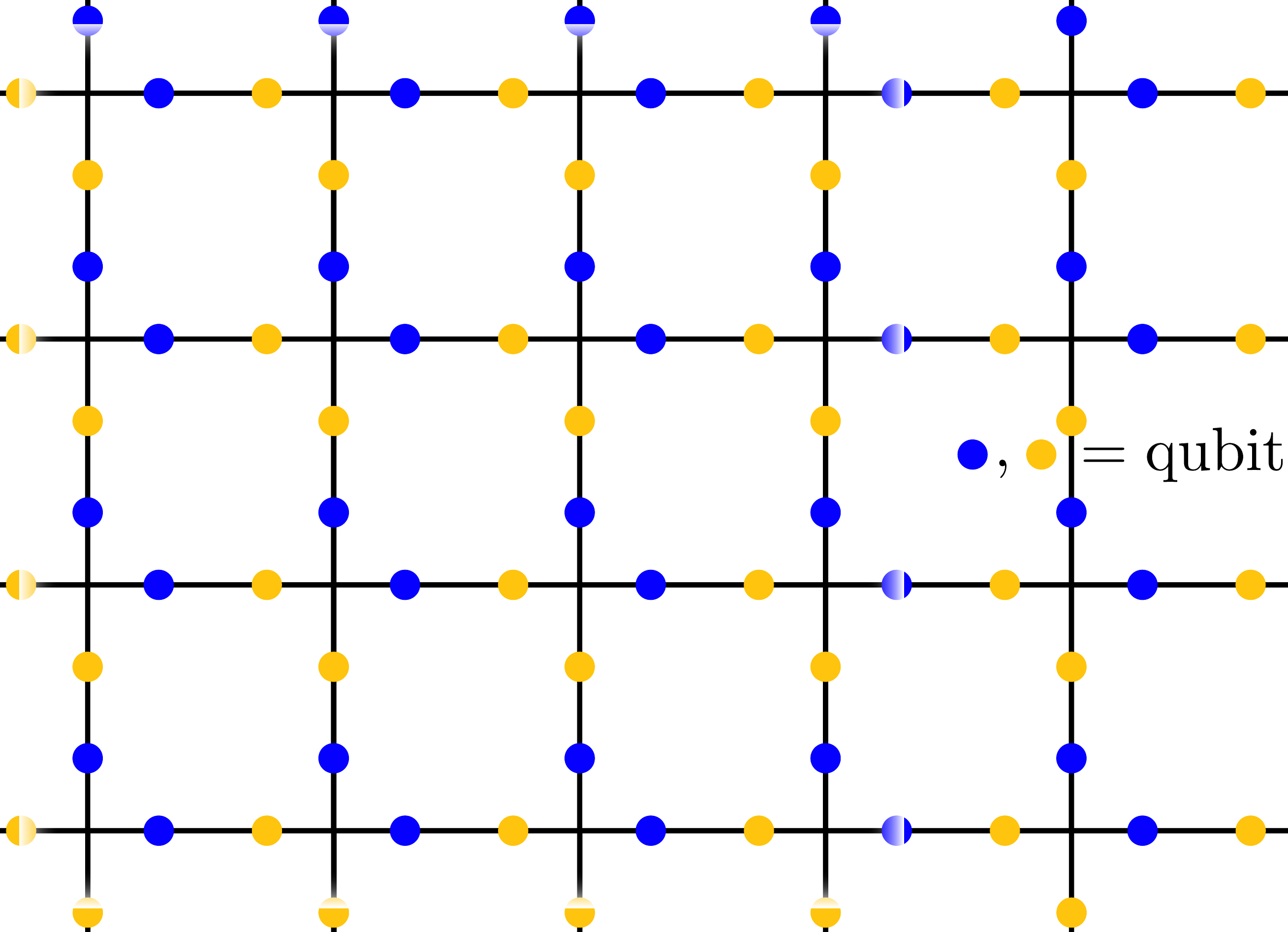}
\caption{The three-fermion subsystem code is defined on a square lattice with two qubits on each edge (blue and yellow).}
\label{fig: 3Fdof}
\end{figure}

The resulting $3$F subsystem code is built on a square lattice with two qubits at each edge, as shown in Fig.~\ref{fig: 3Fdof}. The gauge group is given by:
\begin{align}
\mathcal{G} \equiv \left \langle i, \quad
\vcenter{\hbox{\includegraphics[scale = .23]{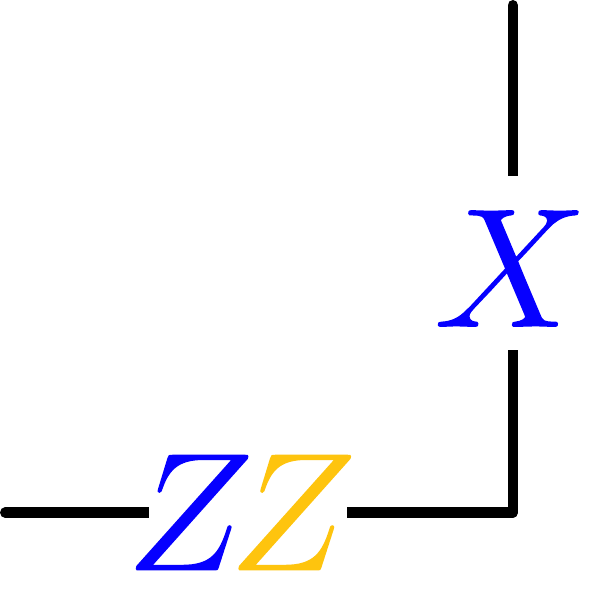}}}, \,\,\,
\vcenter{\hbox{\includegraphics[scale = .23]{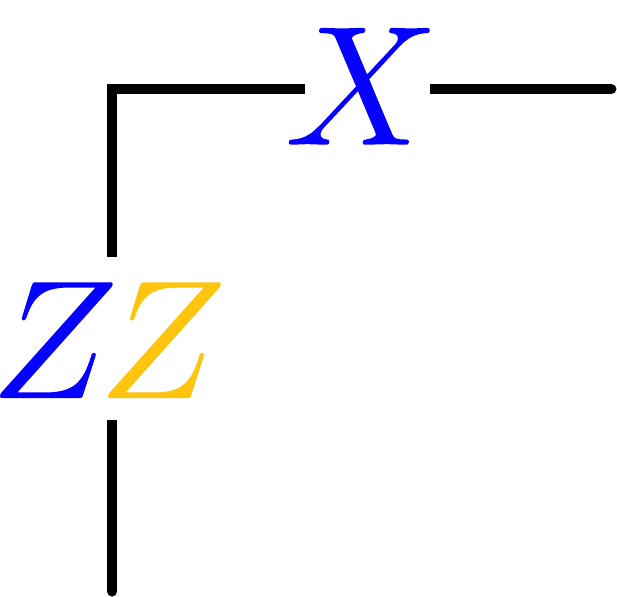}}}, \,\,\,
\vcenter{\hbox{\includegraphics[scale = .23]{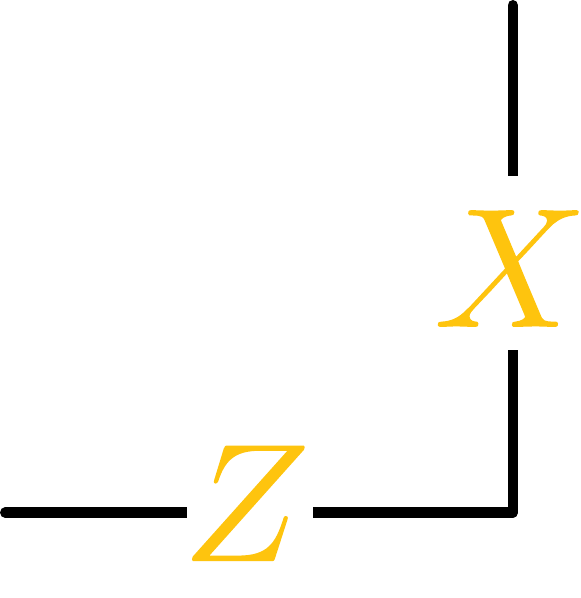}}}, \,\,\,
\vcenter{\hbox{\includegraphics[scale = .23]{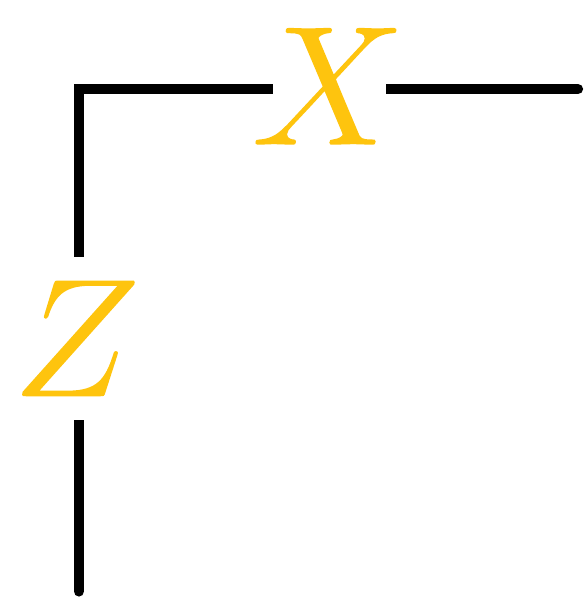}}}, \,\,\,
\vcenter{\hbox{\includegraphics[scale = .23]{Figures/z2plaquette2.pdf}}}, \,\,\,
\vcenter{\hbox{\includegraphics[scale = .23]{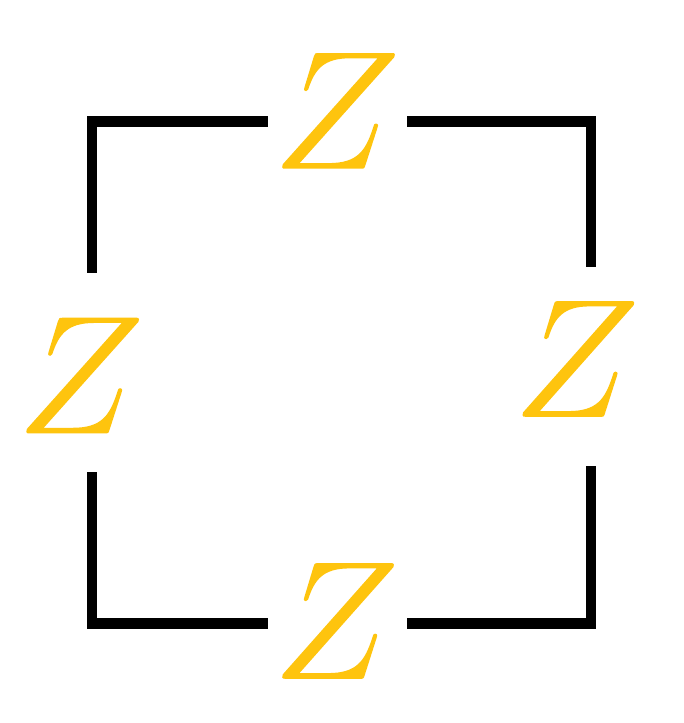}}}
 \right \rangle,
\end{align}
where the blue operators and yellow operators act on the first and second qubit at the edge, respectively. We take the stabilizer group $\mathcal{S}$ to be generated by loops of $e_1 \psi_2$ and $\psi_1$ string operators supported along contractible paths:
\begin{align}
\mathcal{S} \equiv \left \langle 
\vcenter{\hbox{\includegraphics[scale = .23]{Figures/z2emloop2.pdf}}}, \quad
\vcenter{\hbox{\includegraphics[scale = .23]{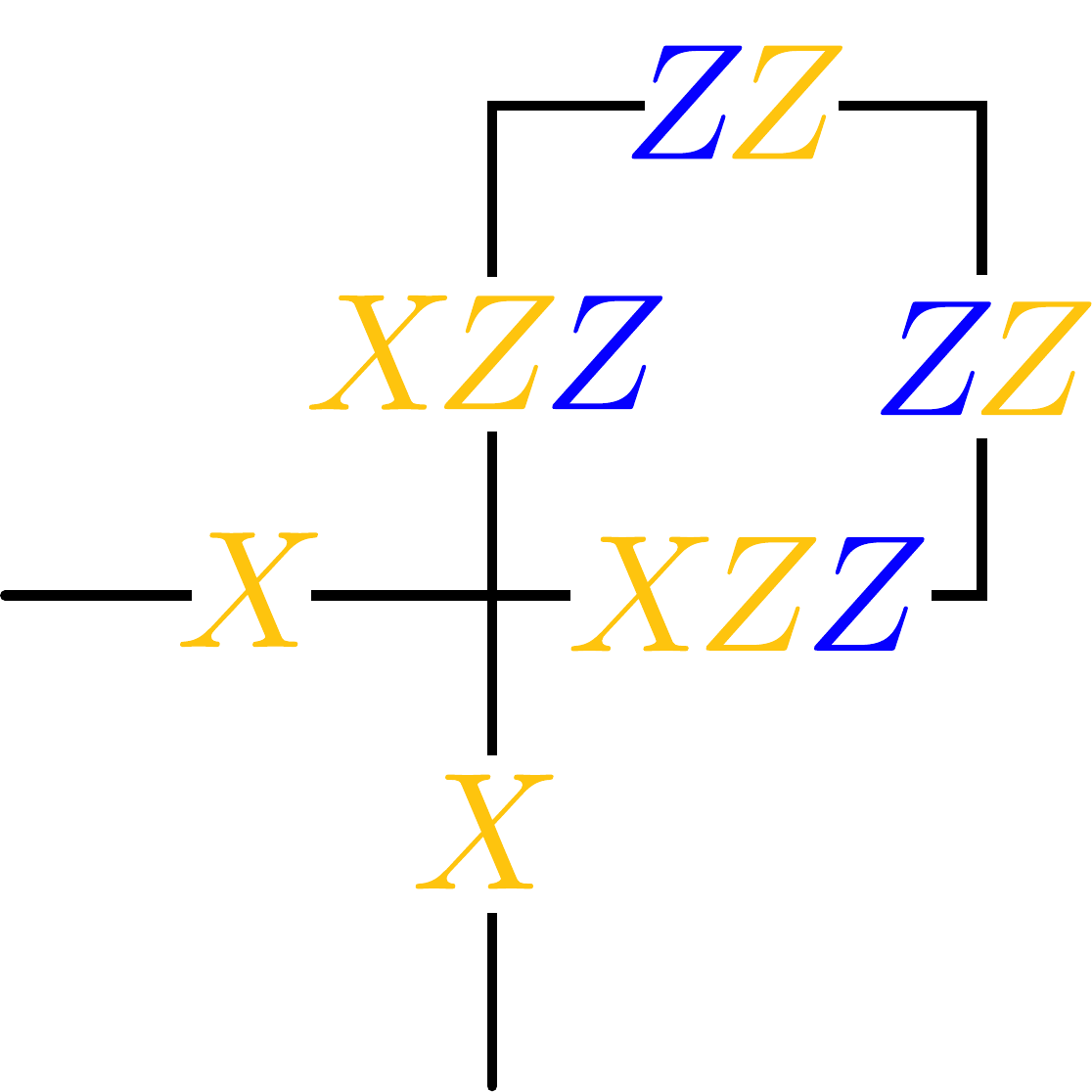}}}
 \right \rangle.
\end{align}
The nontrivial bare logical operators can be represented (up to products of stabilizers) by $e_1 \psi_2$ and $\psi_1$ string operators wrapped around non-contractible paths, such as:
\begin{align}
\vcenter{\hbox{\includegraphics[width=.45\textwidth]{Figures/emlongstring2.pdf}}}, \quad
\vcenter{\hbox{\includegraphics[width=.45\textwidth]{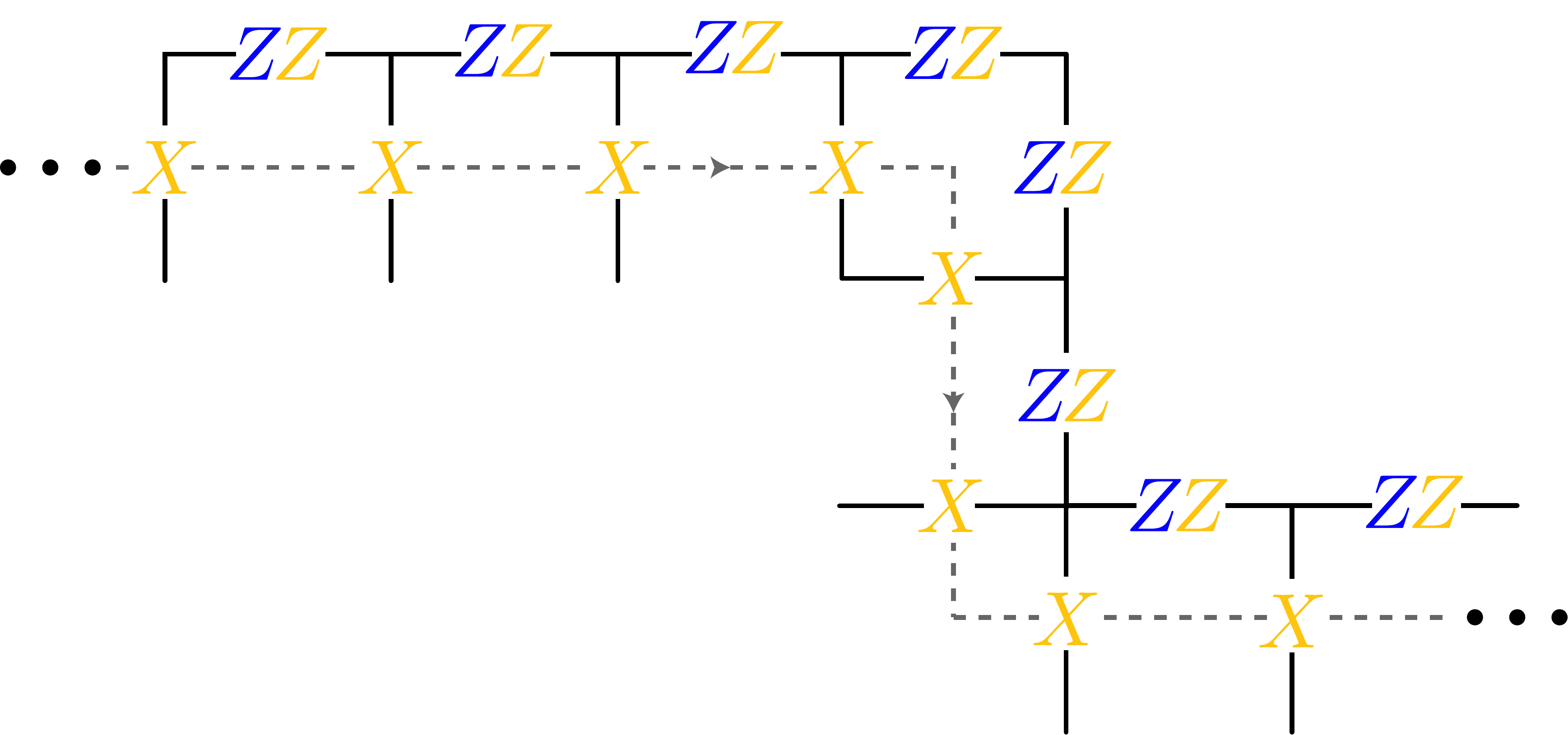}}}.
\end{align}
On a torus, the logical subsystem is composed of two qubits. 

We note that the topological subsystem code of Refs.~\cite{Bombin2009fermions, Bombin2010subsystem} is also characterized by the $3$F anyon theory. We expect that the two topological subsystem codes characterized by the $3$F anyon theory can be mapped to one another using ancillary qubits and constant-depth Clifford unitaries. 
However, we do not work out the explicit mapping here. As observed in Ref.~\cite{Bombin2010subsystem}, subsystem codes characterized by the $3$F anyon theory do not admit surface code implementations with rough and smooth boundaries. 
This is because the $3$F anyon theory is chiral [see Eq.~\eqref{eq: chiral anyons}] and as such, does not admit gapped boundaries (see also Ref.~\cite{Bombin2010subsystem}). On a planar geometry, logical information can instead be stored and manipulated using twist defects, as described in Ref.~\cite{Roberts20203fermion}. 

It is interesting to note that, if the family of Hamiltonians $H_{\{J_G\}}$ for the 3F subsystem code exhibits the 3F anyon theory for some choice of coefficients $\{J_G\}$, then there is a potential route towards constructing a model in the $E_8$ phase. In particular, we could first stack two copies of the system tuned to the 3F phase, which each have chiral central charge $4$~modulo $8$. We could then condense a Lagrangian subgroup (defined in Section~\ref{app: TQD review}), leaving us with a system that has no anyonic excitations and a chiral central charge of $8$. While there are existing models that belong to the 3F phase, the construction of a model in the $E_8$ phase is complicated by string operators that do not have a simple form. In our case, the string operators for the fermions are Pauli operators that are well-localized along a path. 

\subsection{$\ZZ_p \times \ZZ_{p^2}$ subsystem code} \label{sec: z3 z9 subsystem}

The last family of examples that we would like to share is a topological subsystem code characterized by a non-modular anyon theory with $\ZZ_p \times \ZZ_{p^2}$ fusion rules for prime $p$. We refer to this as the $\ZZ_p \times \ZZ_{p^2}$ subsystem code. What makes this anyon theory interesting is that it cannot be decomposed into a stack of simpler anyon theories. This is distinct from modular Abelian anyon theories, which can always be decomposed into a stack of anyon theories whose fusion rules are either $\ZZ_p$ or $\ZZ_{2^n} \times \ZZ_{2^n}$ \cite{Wang2020abelian}. 

We label the generators of the $\ZZ_p \times \ZZ_{p^2}$ fusion group by $a_1$ and $a_2$, where $a_1$ is order $p$, and $a_2$ is order $p^2$. The exchange statistics and braiding relations of $a_1$ and $a_2$ are given by:
\begin{align} \label{eq: z3 z9 statistics}
    \theta(a_1) = \theta(a_2) = 1, \quad B_\theta(a_1,a_2) = e^{2 \pi i / p}.
\end{align}
Intuitively, 
this anyon theory is the subtheory of the $\ZZ_{p^2}$ TC generated by $e^{p}$ and $m$. Alternatively, it can be viewed as a $\ZZ_{p}$ TC in which the diagonal $\ZZ_{p}$ subgroup of the fusion group has been extended by a transparent boson. For the purpose of following the general construction in Section~\ref{sec: general}, however, we interpret the anyon theory as a subtheory belonging to a stack of a $\ZZ_{p}$ TC and a $\ZZ_{p^2}$ TC. The $\ZZ_{p} \times \ZZ_{p^2}$ subtheory is then generated by the anyon types $m_1$ and $e_1m_2$, where $m_1$ and $e_1$ belong to the $\ZZ_{p}$ TC, and $m_2$ belongs to the $\ZZ_{p^2}$ TC. 

Before describing the construction of the subsystem code, we would like to argue that the anyon theory cannot be decomposed into simpler anyon theories. In particular, we show that there is no pair of $\ZZ_{p}$ and $\ZZ_{p^2}$ subgroups with the property that the generators braid trivially with one another and generate the full fusion group. All of the order $p$ anyon types are of the form $a_1^j a_2^{pk}$ for $j,k \in \ZZ_p$,
while all the remaining anyons (i.e., $a_1^la_2^n$ for $l\in \ZZ_p$ and $n \in \ZZ_p^2$ with $n \neq 0 \  (\text{mod}\  p)$ are of order $p^2$.
The braiding between these two anyons is trivial only when $B(a_1^j a_2^{pk},a_1^la_2^n ) = e^{\frac{2\pi i}{p}(jn+pkl)}=1$. Since $n \neq 0 \  (\text{mod}\ p)$, this implies $j=0\  (\text{mod}\ p)$. Thus, the only potential generators of the $\ZZ_p$ subgroup is $a_2^{pk}$ for $k\in \ZZ_p$. However, these anyons are multiples of $a_2^{k}$ and therefore cannot form an independent $\ZZ_p$ subgroup. To conclude, the anyon theory cannot be decomposed into an anyon theory with fusion group $\ZZ_p$ times another with fusion group $\ZZ_{p^2}$.

\begin{figure}[t] 
\centering
\hspace{2.5cm}\includegraphics[width=.5\textwidth]{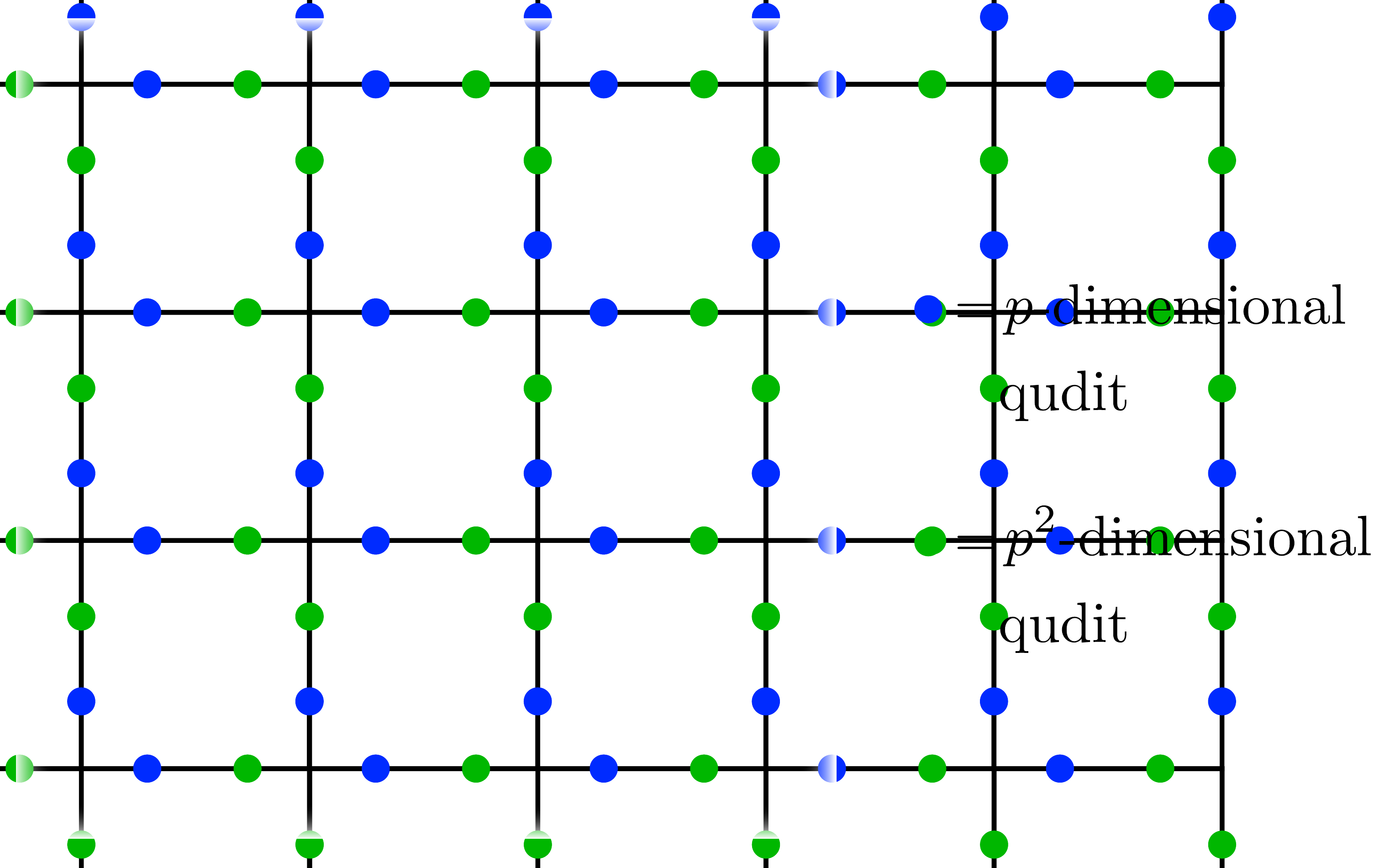}
\caption{The Hilbert space of the $\ZZ_p \times \ZZ_{p^2}$ subsystem code is composed of $p$-dimensional qudit (blue) and a $p^2$-dimensional qudit (green) on each edge of a square lattice.}
\label{fig: z3z9dof}
\end{figure}

We now turn to the construction of the $\ZZ_p \times \ZZ_{p^2}$ subsystem code. The construction starts with a $\ZZ_p$ TC and a decoupled $\ZZ_{p^2}$ TC. These are defined on a square lattice with a $p$-dimensional qudit and a $p^2$-dimensional qudit at each edge (Fig.~\ref{fig: z3z9dof}). In terms of the parameters $t_i$ and $p_{ij}$ in Section~\ref{sec: general}, the $\ZZ_p \times \ZZ_{p^2}$ subtheory of interest is specified by $t_1 = t_2 =0$, and $p_{12}=p_{21}=1$. This tells us that we need to gauge out the anyon types $m_1^{-1}e_2^p$ and $m_2^{-1}$. This gives us the following gauge group:
\begin{align}
\mathcal{G} \equiv \left \langle 
\vcenter{\hbox{\includegraphics[scale=.23]{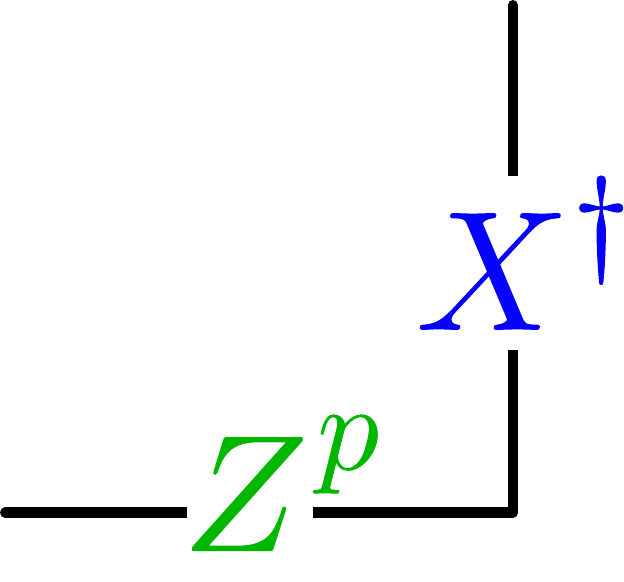}}}, \,\,\,
\vcenter{\hbox{\includegraphics[scale=.23]{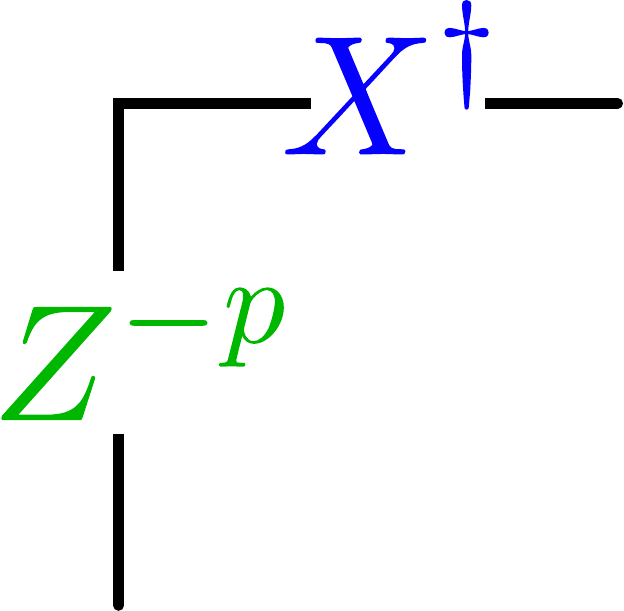}}}, \,\,\,
\vcenter{\hbox{\includegraphics[scale=.23]{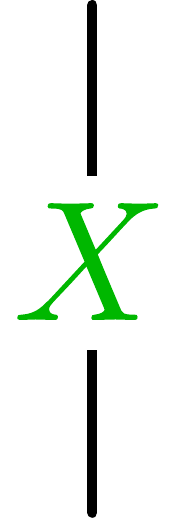}}}, \,\,\,
\vcenter{\hbox{\includegraphics[scale=.23]{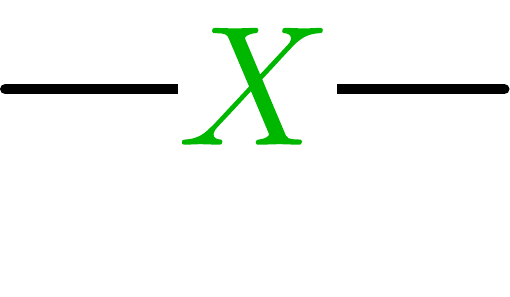}}}, \,\,\,
\vcenter{\hbox{\includegraphics[scale=.23]{Figures/BpTCz41.pdf}}}, \,\,\,
\vcenter{\hbox{\includegraphics[scale=.23]{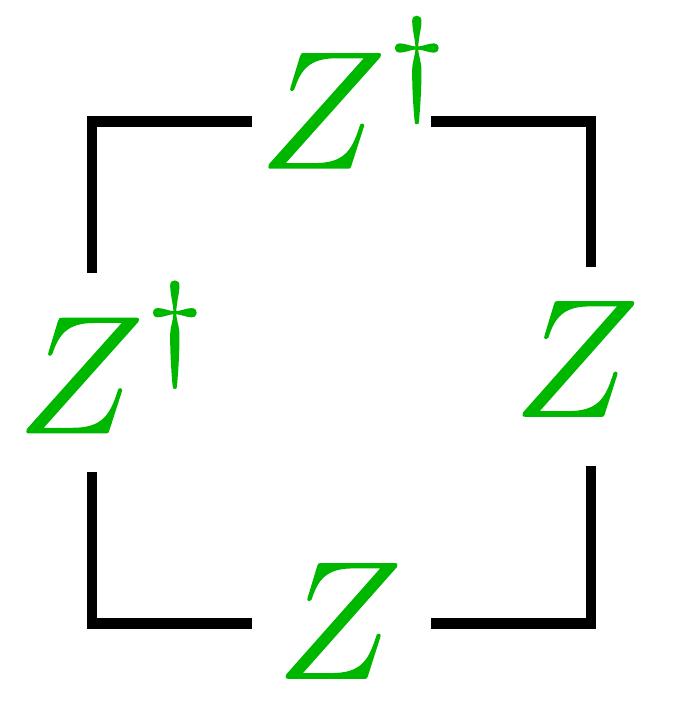}}}
 \right \rangle,
\end{align}
The bare logical group is generated by small loops of string operators for $m_1$ and $e_1m_2$:
\begin{eqs}
    \vcenter{\hbox{\includegraphics[scale=.23]{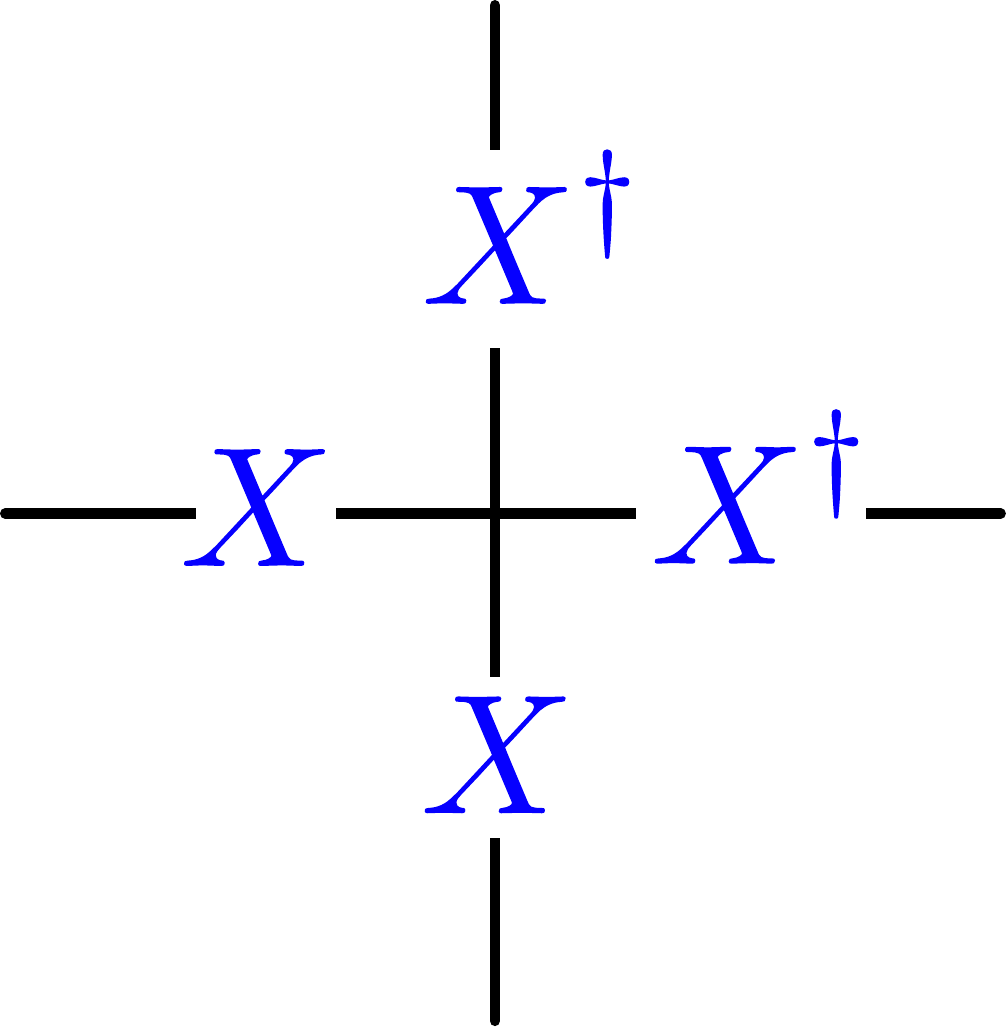}}}, \qquad
\vcenter{\hbox{\includegraphics[scale=.23]{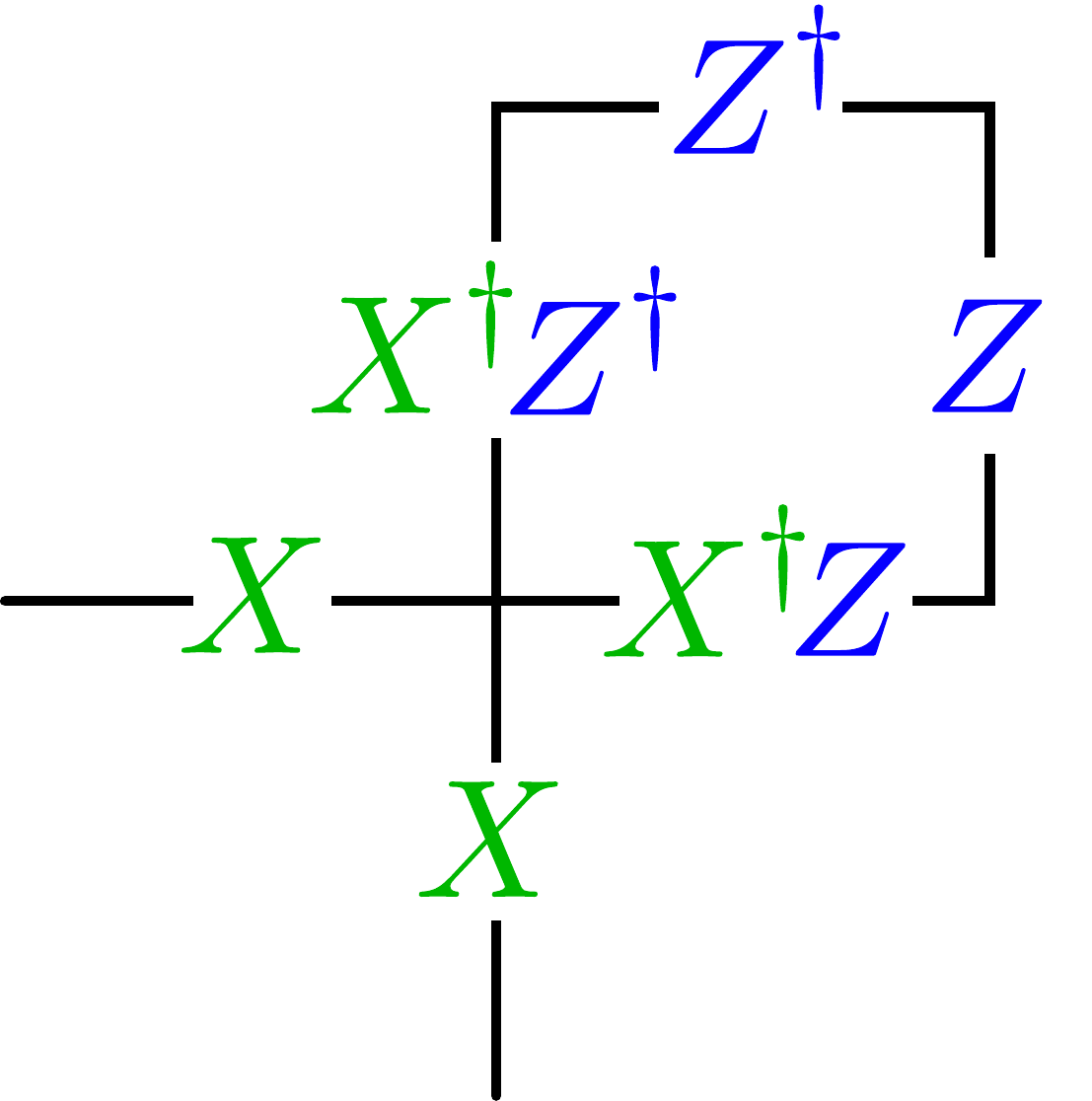}}},
\end{eqs}
as well as the corresponding string operators wrapped along non-contractible paths:
\begin{align}
\vcenter{\hbox{\includegraphics[width=.45\textwidth]{Figures/z2longmstring2.pdf}}}, \quad
\vcenter{\hbox{\includegraphics[width=.45\textwidth]{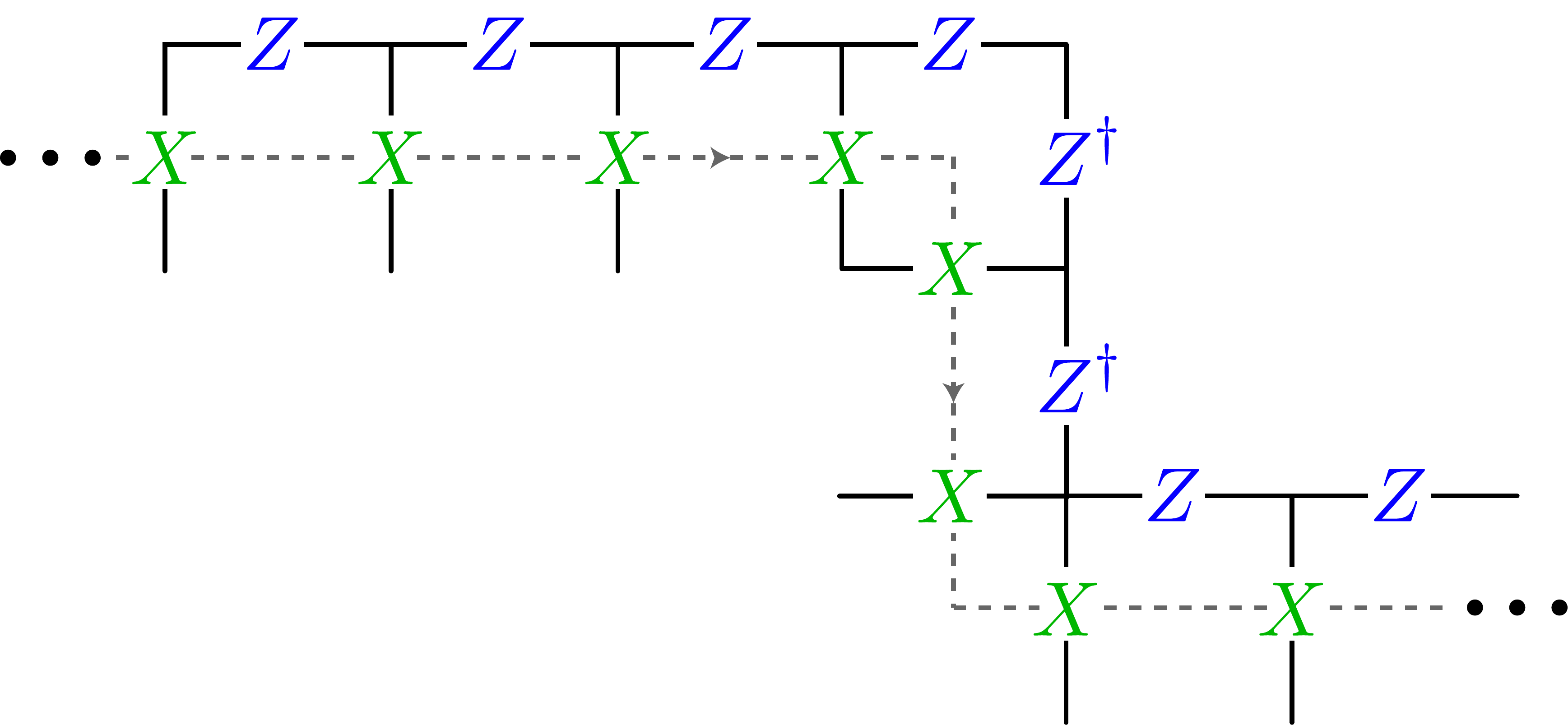}}}.
\end{align}
Since $a_2^p$ is transparent, the subsystem code encodes just two $p$-dimensional qudits on a torus.

\section{Discussion} \label{sec: discussion}

In this work, we constructed a Pauli topological subsystem code for every Abelian anyon theory, which includes non-modular anyon theories and those that do not admit a gapped {boundary} to a product state. Our construction started with a Pauli topological stabilizer code on composite-dimensional qudits, similar to those introduced in Ref.~\cite{Ellison2022Pauli}. We then gauged out a set of anyon types, thereby converting the stabilizer code into a subsystem code characterized by a proper subtheory of the initial anyon theory. We demonstrated that every Abelian anyon theory can be obtained from that of decoupled copies of TCs by condensing and gauging out anyon types (see also Ref.~\cite{Hsin2018Global}). We provided several examples of our construction, including a $\ZZ_4^{(1)}$ subsystem code characterized by a non-modular anyon theory, and a chiral semion subsystem code characterized by an anyon theory without gapped boundaries. To conclude, we comment on some additional structures of Pauli topological subsystem codes and describe a number of avenues for further investigation.

\subsubsection*{Classification of TI Pauli topological subsystem codes}

For Pauli topological stabilizer codes, there is a natural definition of two equivalent codes. We say two Pauli topological stabilizer codes are equivalent if they can~be mapped to one another by a constant-depth Clifford circuit (CDCC) with ancillary qudits~\cite{Bombin2012universal, Haah2018classification}. This is motivated by the definition of topological order~\cite{chen2010local}, in which two gapped Hamiltonians belong to the same topological phase of matter if their ground states can be related to one another by the finite-time evolution of a local Hamiltonian (using ancillary qudits). Since CDCCs capture the piece-wise time evolution of a local Hamiltonian, this means that if two Pauli topological stabilizer codes are equivalent, then their corresponding Hamiltonians belong to the same topological phase of matter.

For TI Pauli topological subsystem codes, however, the choice of equivalence relation is less straightforward. This is because Pauli topological subsystem codes do not necessarily correspond to a single gapped phase of matter, as they are generically associated to a parameter space of Hamiltonians. Therefore, we suggest two possible equivalence relations for classifying topological subsystem codes. The first is to adopt the equivalence relation of topological stabilizer codes and say that two TI Pauli topological subsystem codes are equivalent if they differ from one another by a CDCC with ancillary qudits. This is still natural to consider, given that CDCCs preserve the set of correctable errors of TI Pauli topological subsystem codes, up to a rescaling of the length $\ell_S$. Our work shows that, under this equivalence relation, TI Pauli topological subsystem codes are partially classified by Abelian anyon theories, since CDCCs are unable to change the associated anyon theory. 

The second equivalence relation that we consider is designed to capture the properties of topological subsystem codes that are beyond those of topological stabilizer codes. In particular, we say two TI Pauli topological subsystem codes are equivalent if they can be mapped to one another by stacking with Pauli topological stabilizer codes and applying a CDCC. In this case, all Pauli topological stabilizer codes, including the TQD stabilizer codes of Ref.~\cite{Ellison2022Pauli}, belong to the trivial equivalence class. The subsystem TC in Ref.~\cite{Bravyi2013Subsystem}, for example, also belongs to the trivial class, since it can be mapped via a CDCC to a subsystem code consisting of single-site gauge generators stacked with a $\ZZ_2$ TC. The nontrivial equivalence classes are instead represented by topological subsystem codes that cannot be realized by topological stabilizer codes. This includes the topological subsystem codes that are characterized by non-modular anyon theories or anyon theories that do not admit a gapped boundary to the vacuum.

For this second equivalence relation, we expect that the TI Pauli topological subsystem codes are partially classified by an object that we refer to as the classical Witt monoid. Here, the elements of the classical Witt monoid are equivalence classes of Abelian anyon theories, where two anyon theories are equivalent if they can be related by stacking modular Abelian anyon theories with gapped boundaries. The equivalence classes then form a monoid under the operation of stacking. In general, there are no inverses, since the non-modular anyon theories cannot be converted into modular anyon theories through stacking. For the subset of modular anyon theories, however, the equivalence classes indeed form a group, known as the classical Witt group~\cite{Davydov2013,Davydov2013b,Shirley22}.

Based on Ref.~\cite{Shirley22}, we further conjecture that the equivalence classes of TI Pauli topological subsystem codes are in one-to-one correspondence with the equivalence classes of WW models under constant-depth circuits. This agrees with the recent finding that WW models based on modular Abelian anyon theories with gapped boundaries can be disentangled (to a paramagnet Hamiltonian) by a constant-depth circuit~\cite{Bauer2022WW}, i.e., they belong to the trivial class of WW models. On the other hand, if the anyon theory is modular and does not admit a gapped boundary, then it is expected that the WW model cannot be disentangled by a constant-depth circuit~\cite{HFH18,Haah21Clifford,Shirley22, Haah2022Unitary}, so it belongs to a nontrivial equivalence class. Furthermore, if the anyon theory is non-modular, then the WW model describes a system with intrinsic topological order, so it cannot be disentangled by a constant-depth circuit. This aligns with the expectation that the nontrivial classes correspond to non-modular anyon theories and those without gapped boundaries.

\subsubsection*{Phase diagrams of the subsystem codes}

As described in Section~\ref{sec: definition of topological subsystem codes}, each topological subsystem code defines a family of Hamiltonians, which may realize a variety of phases and phase transitions by tuning the interaction strengths. An interesting direction for future work is thus to explore the phase diagrams of our topological subsystem codes. For example, the phase diagram of the $\ZZ_2^{(1)}$ subsystem code in Section~\ref{sec: 1 psi subsystem code} has been thoroughly studied in Ref.~\cite{kitaev2006anyons}. In this case, the Hamiltonians are exactly solvable, given that they are dual to a system of non-interacting fermions. However, the families of Hamiltonians defined by the more general topological subsystem codes introduced in this work do not admit a mapping to non-interacting fermions. This suggests that more sophisticated methods are needed to probe the phase diagram.

One means of gaining insight into the phase diagram is to consider the anyonic ${1\text{-form}}$ symmetries of the topological subsystem code, as described in Appendix~\ref{app: 1form}. In Section~\ref{sec: anyon theories for subsystem codes}, we argued that an anyonic $1$-form symmetry associated to an Abelian anyon theory $\mathcal{A}$ forces the gapped Hamiltonians in the phase diagram to have anyonic excitations corresponding to the anyon types of $\mathcal{A}$, up to condensing transparent bosons. Under the~widely-accepted assumption that gapped Hamiltonians in two dimensions are characterized by modular anyon theories, we expect that the topological phases of matter realized by the family of Hamiltonians must be modular extensions of the anyon theory $\mathcal{A}$ or anyon theories obtained from condensing transparent bosons. Indeed, for the $\ZZ_2^{(1)}$ subsystem code, the gapped phases in the phase diagram are characterized by modular extensions of the $\ZZ_2^{(1)}$ anyon theory. In fact, the gapped phases of matter realize all of the so-called minimal modular extensions\footnote{Here, we say an extension $\mathcal{C}$ of an anyon theory $\mathcal{A}$ is a minimal modular extension if the only anyon types of $\mathcal{C}$ that braid trivially with all of the anyon types in $\mathcal{A}$ are precisely the transparent anyon types of $\mathcal{A}$. In general, minimal modular extensions are known to exist only in certain cases where an obstruction vanishes. 
However, it was recently proven that a minimal modular extension does in fact exist for all non-modular anyon theories that contain a single transparent fermionic anyon type~\cite{JohnsonFreydReutter21}.} of the $\ZZ_2^{(1)}$ anyon theory given by the $16$-fold way (given a perturbation that breaks the time-reversal symmetry, see \cite{kitaev2006anyons}). We pose the following question: do the gapped phases of matter in the phase diagram for a topological subsystem code based on the Abelian anyon theory $\mathcal{A}$ realize all modular extensions of $\mathcal{A}$?

\subsubsection*{Topological defects}

We next discuss defects in Pauli topological subsystem codes, which are valuable for the implementation of fault-tolerant logical gates~\cite{Bravyi1998boundary, Bombin2010Twist, Fowler2012surface, Brown2017poking, Webster2020defects}. We refer to any redefinition of the gauge generators contained within a $d$-dimensional region as a $d$-dimensional defect. For example, a $0$-dimensional defect might correspond~to flipping the sign of a stabilizer, while a $1$-dimensional defect might be defined by inserting a dislocation into the lattice -- as in the $e$ and $m$ permuting defect of the $\ZZ_2$ TC~\cite{Bombin2010Twist, Kitaev2012}.
We further say the defect is a topological defect if, when defined on a region without boundary, it preserves~locality and does not introduce any local logical qudits. We then impose an equivalence relation~on $d$-dimensional topological defects, such that two topological defects are equivalent if one can be constructed from the other by inserting lower-dimensional topological defects and conjugating the gauge group by a constant-depth circuit whose support is contained in the neighborhood of the $d$-dimensional defect. We call a topological defect trivial if it belongs to the same equivalence class as the trivial defect, i.e., no modification to the gauge generators.
We leave the full classification of topological~defects as an open question and simply comment on some key examples below.  

It is important to note that inserting defects generically breaks the translation invariance of the topological subsystem code. However, we emphasize that topological defects preserve the locality of the gauge group (property (ii) of Definition~\ref{def: topological subsystem code}) and do not introduce any local logical qudits. Moreover, for any contractible region $R$ that is sufficiently far from the defects, we expect that the subsystem code still satisfies a local topological property on $R$, similar to that of Eq.~\eqref{eq: topological property torus} in Lemma~\ref{lem: topological property torus}.

The $0$-dimensional topological defects are partially classified by the anyon types of the Pauli topological subsystem code. This is because we can redefine the gauge generators in a $0$-dimensional region by multiplying by the $U(1)$-valued phases determined by a representative gauge~twist. By construction, the $U(1)$-valued phases cannot be reproduced by conjugation with Pauli operators -- or any unitary operators, for that matter.\footnote{Note that unitary operators that only fail to commute with Pauli operators by roots of unity are, by the definition of the Clifford hierarchy, Pauli operators~\cite{Gottesman1999Universal, Pastawski2015Fault-tolerant}.} Thus, the nontrivial anyon types represent nontrivial $0$-dimensional defects. 

We suggest that a general method for constructing nontrivial $1$-dimensional topological defects is to employ the concept of higher gauging, introduced in Ref.~\cite{Roumpedakis2022Higher}. In particular, in Ref.~\cite{Roumpedakis2022Higher}, it was shown that, at the level of continuum topological quantum field theories, $1$-dimensional defects can be constructed in two-dimensional topological orders by condensing (or proliferating) anyonic excitations along a $1$-dimensional path. They argued that an order $N$ anyon type $a$ can be consistently condensed along a path $\gamma$ if only if $\theta(a)$ is an $N$th root of unity, i.e., it satisfies:
\begin{align} \label{eq: consistent higher gauging}
    \theta(a)^N = 1.
\end{align}
For example, this means that the order four semion in the $\ZZ_4^{(1)}$ anyon theory can be condensed along a path $\gamma$, while the order two semion in the chiral semion anyon theory cannot be consistently condensed along a path.

In Ref.~\cite{Roumpedakis2022Higher}, it was further noted that two $1$-dimensional topological defects can be fused together by defining them along neighboring paths $\gamma$ and $\gamma'$. A defect along $\gamma$ is then called invertible if there exists a defect along a parallel neighboring path $\gamma'$ such that their fusion is a trivial defect. It was argued in Ref.~\cite{Roumpedakis2022Higher} that, for modular Abelian anyon theories, a $1$-dimensional defect created by condensation is invertible if and only if the condensed anyon type has nontrivial exchange statistics. For non-modular anyon theories, 
we expect that, regardless of the exchange statistics, the condensation of a transparent anyon type yields a trivial invertible topological defect. 

For a topological subsystem code, the condensation of an anyon type $a$ along a path $\gamma$ is implemented by measuring (mutually commuting) short string operators for $a$ along $\gamma$. This adds the short string operators to the stabilizer group and thereby proliferates the anyon type $a$ along $\gamma$. For example, the defect that permutes the $e$ and $m$ anyon types in the $\ZZ_2$ TC~\cite{Bombin2010Twist, Kitaev2012} can be constructed by measuring the $em$ short string operators along a path. We expect that this approach to constructing defects yields topological defects, assuming that the condensed anyon type satisfies Eq.~\eqref{eq: consistent higher gauging}.
Based on the arguments of Ref.~\cite{Roumpedakis2022Higher}, we further conjecture that, up to the addition of $0$-dimensional defects and conjugation by a constant-depth circuit along the defect, every $1$-dimensional topological defect in a Pauli topological subsystem code can be constructed by composing defects created by condensation.

Lastly, we comment that a subclass of $2$-dimensional topological defects correspond to holes, created by condensing a bosonic anyon type in a $2$-dimensional patch. This is exemplified by the $e$-condensing and $m$-condensing holes of the $\ZZ_2$ TC~\cite{Kitaev2012, Fowler2012surface, Brown2017poking}. We note that, furthermore, a boundary to a trivial topological subsystem code (i.e., with single-site gauge generators) can be constructed by condensing a set of bosonic anyon types that generate a Lagrangian subgroup (see Appendix~\ref{app: TQD review}). The system with a boundary can be interpreted as a topological subsystem code with a $2$-dimensional topological defect.

\subsubsection*{Utility as quantum error-correcting codes}

Although we have introduced a large class of topological subsystem codes, we have not~assessed their utility as quantum error-correcting codes.~Moving forward, it is important~to compare them with other topological quantum error-correcting codes by, for example, computing their error thresholds and identifying the set of fault-tolerant logical gates that are intrinsic to the code (possibly involving defects). In an upcoming work, we discuss the $\ZZ_3^{(1)}$ subsystem code in more detail. This subsystem code exhibits a number of beneficial properties: it is defined on a low-degree graph, admits two-body checks, allows for a simple fault-tolerant implementation of the Clifford group using topological defects, and has high biased-noise error thresholds.  

\subsubsection*{Three-dimensional topological subsystem codes}

Our focus in this text has been on topological subsystem codes in two dimensions. This leaves the classification of topological subsystem codes in three dimensions as an exciting direction for future work. In analogy to non-modular anyon theories for two-dimensional systems, three-dimensional topological subsystem codes have the potential to host transparent point-like and loop-like excitations as well as point-like excitations with restricted mobility, i.e., fractonic order~\cite{Nandkishore2019Fractons}. We expect that the three-dimensional subsystem toric code~\cite{Kubica2022subsystemTC} and the gauge color code~\cite{Bombin2007branyons,Bombin2015Gaugecolorcodes, Brown2016gaugecolorcode} are two such examples characterized by transparent loop-like excitations.

In three dimensions, there are also topological orders that are anomalous~\cite{JohnsonFreyd2020topologicalZ2,Fidkowski2021Gravitationalanomaly,Chen2021Gravitationalanomaly}. This is to say that they cannot be realized in three dimensions without a four-dimensional bulk. We expect that, nonetheless, three-dimensional topological subsystem codes can be characterized by anomalous topological orders. The simplest example is an anomalous $\ZZ_2$ gauge theory, which features a fermionic point-like excitation and a fermionic loop-like excitation~\cite{JohnsonFreyd2020topologicalZ2,Fidkowski2021Gravitationalanomaly,Chen2021Gravitationalanomaly}. In Ref.~\cite{Fidkowski2021Gravitationalanomaly}, this topological order is labeled as FcFl, to denote the fact that it has a fermionic charge and a fermionic loop. Remarkably, two copies of the FcFl topological order is equivalent, by relabeling the excitations, to two copies of the FcBl topological order, where FcBl topological order is the non-anomalous $\ZZ_2$ gauge theory with a point-like fermion and a bosonic loop-like excitation. This suggests that we can construct a Pauli topological subsystem code characterized by the FcFl topological order by stacking two FcBl topological stabilizer codes and gauging out the excitations corresponding to one copy of FcFl topological order.

\subsubsection*{Dynamics of the subsystem codes}

Recently, subsystem codes have gained interest in the context of non-equilibrium quantum dynamics. In Ref.~\cite{Hastings2021dynamically}, it was found that, by appropriately scheduling the measurements of the gauge generators in the $\ZZ_2^{(1)}$ subsystem code, it is possible to realize a dynamically generated logical qubit. Such codes and generalizations~\cite{Aasen22,Davydova22,Kesselring2022condensation} have since been referred to as Floquet codes, due to the periodic schedule of measurements, and generalizations beyond subsystem codes have been proposed~\cite{Aasen22,Davydova22,Kesselring2022condensation}. 
In future work, we consider Floquet codes built from the subsystem codes described in this work. We find that the dynamically generated logical subspace is that of the parent topological stabilizer code used to construct the topological subsystem code in Section~\ref{sec: general}. Along these lines, two recent works~\cite{Sriram2022measurementKitaev,Lavasani2022monitoredKitaev} studied the measurement-only dynamics of a system evolving under random measurements of the gauge generators of the $\ZZ_2^{(1)}$ subsystem code. It would be interesting to also consider the measurement-only dynamics based on the topological subsystem codes of this work, since they host a more rich set of anyonic $1$-form symmetries. 
We also point out that it may be fruitful to study our topological subsystem codes in the context of quantum many-body scars~\cite{Moudgalya2022fragmentation,Moudgalya2022exhaustive,Moudgalya2022Symmetries} and infinite-temperature quantum memories~\cite{Wildeboer2022infinite}, given that they have an extensive number of conserved quantities. 

\vspace{0.2in}
\noindent{\it Acknowledgements -- } T.D.E. thanks Meng Cheng for valuable discussions about the structure of non-modular anyon theories and the connection between WW models and Abelian TQDs in Appendix~\ref{app: WW}. D.J.W. thanks David Long and Andrew Doherty for discussions about the potential of an $E_8$ lattice Hamiltonian. Y.-A.C.~is supported by the JQI fellowship and by the Laboratory for Physical Sciences through the Condensed Matter Theory Center. 
A.D. is supported by the Simons Foundation through the collaboration
on Ultra-Quantum Matter (651438, AD) and by the Institute for Quantum
Information and Matter, an NSF Physics Frontiers Center (PHY-1733907).
W.S. is supported by the Simons Foundation through
the collaboration on Ultra-Quantum Matter (651444, WS).
N.T. is supported by the Walter Burke Institute for Theoretical Physics at Caltech.

\appendix

\section{Correcting and cleaning topological subsystem codes} \label{app: cleaning lemma} 

In this appendix, we prove the correctability condition in Eq.~\eqref{eq: topological subsystem code ii}. We also prove a cleaning lemma for TI topological subsystem codes, similar to the cleaning lemma for stabilizer codes in Ref.~\cite{bravyi2009no} and for subsystem codes in Refs.~\cite{Bravyi2011local} and~\cite{Haah2012tradeoff}. Importantly, the cleaning lemma described here allows us to `clean' nonlocal stabilizers as well as nontrivial bare logical operators. As a consequence of the cleaning lemma, we show that, up to products of local stabilizers, the nontrivial bare logical operators and nonlocal stabilizers of a topological subsystem code on a torus can be represented by operators supported on the region pictured in Fig.~\ref{fig: toruscleaning}. Furthermore, in Appendix~\ref{app: bare logicals and nonlocal stabilizers}, we use the cleaning lemma to argue that every nontrivial bare logical operator and nonlocal stabilizer can be represented by products of string operators formed by moving nontrivial anyon types along non-contractible paths.

\begin{figure}[tb]
\centering
\includegraphics[width=.35\textwidth]{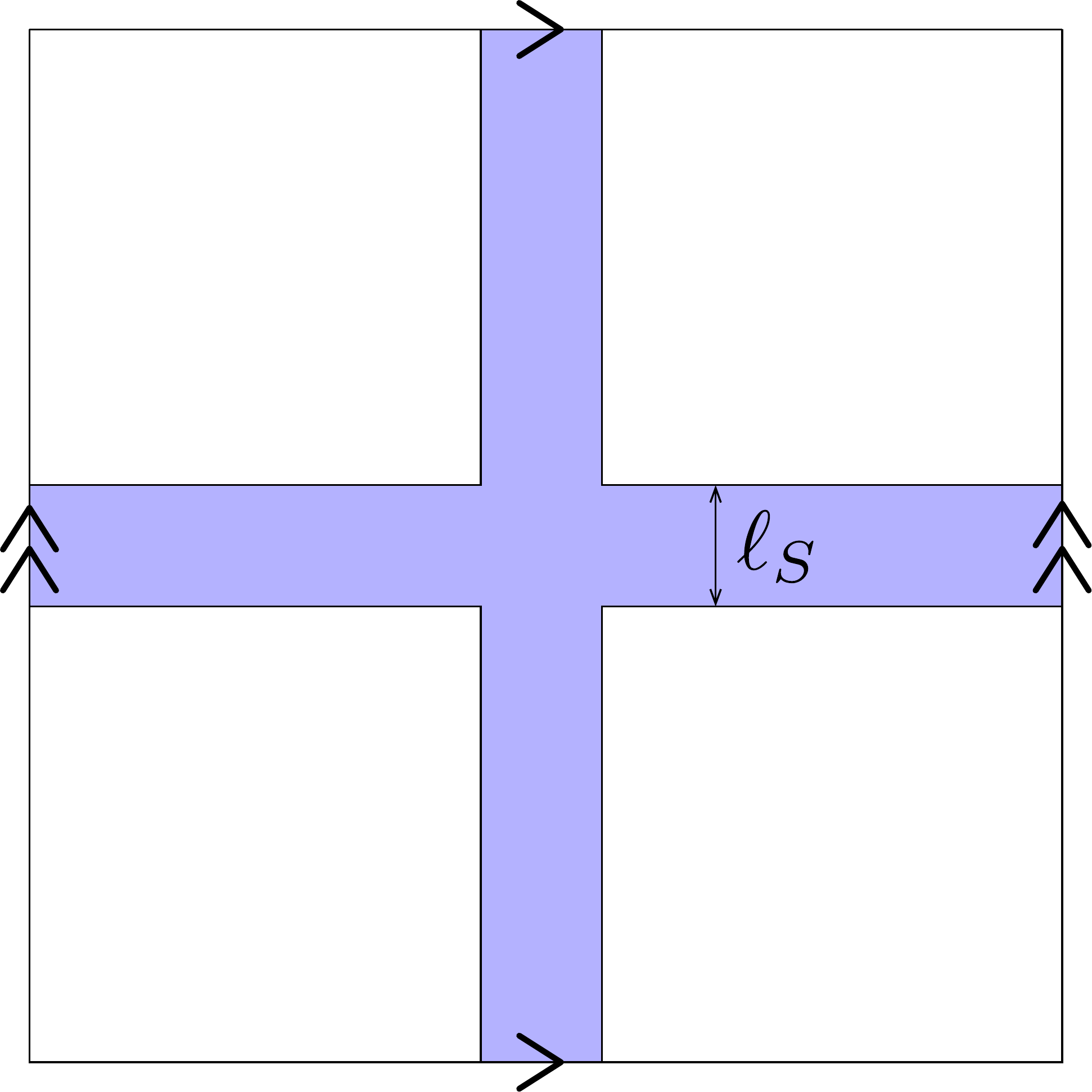}
\caption{In this appendix, we consider TI topological subsystem codes on a torus. The cleaning lemma shows that the nontrivial bare logical operators and nonlocal stabilizers can be represented (i.e. up to locally generated stabilizers) by operators supported entirely on a region of intersecting strips of width $\ell_S$ (blue). 
}
\label{fig: toruscleaning}
\end{figure}

The correctability condition and the cleaning lemma both follow from considering the implications of the topological property of Definition~\ref{def: topological subsystem code} on a torus. For convenience, we recall the topological property for a TI topological subsystem code with a gauge group $\mathcal{G}$ and stabilizer group $\mathcal{S}$:\\
\vspace{-.3cm}
\begin{adjustwidth}{.4cm}{.4cm}
\noindent \emph{On an infinite plane, the stabilizer group $\mathcal{S}$ admits a set of generators whose supports have linear size less than a constant-sized length $\ell_S$, and $\mathcal{Z}_\mathcal{P}(\mathcal{S}) \propto \mathcal{G}$.} \\
\end{adjustwidth}
\vspace{-.3cm}
This implies that TI topological subsystem codes on a torus satisfy the following.
\begin{lemma}[Topological property on a torus] \label{lem: topological property torus}
Let $\mathcal{G}$ and $\mathcal{S}$ denote the gauge group and stabilizer group, respectively, of a TI topological subsystem code on an $L \times L$ torus. For any region $R$ whose linear size is less than $L- \ell_S$, the following condition holds:
\begin{align} \label{eq: topological property torus}
    \mathcal{Z}_\mathcal{P}(\tilde{\mathcal{S}}_{R}) \cap \mathcal{P}(R) \propto \mathcal{G}(R),
\end{align}
where $\tilde{\mathcal{S}}_{R}$ is the subgroup of locally generated stabilizers truncated to $R$, $\mathcal{P}(R)$ is the group of Pauli operators supported entirely on $R$, and $\mathcal{G}(R)$ is defined by $\mathcal{G}(R) = \mathcal{G} \cap \mathcal{P}(R)$.
\end{lemma}

\begin{figure*}[tb]
\centering
\subfloat[\label{fig: toruscutting}]{\raisebox{.9cm}{\hbox{\includegraphics[scale=.3]{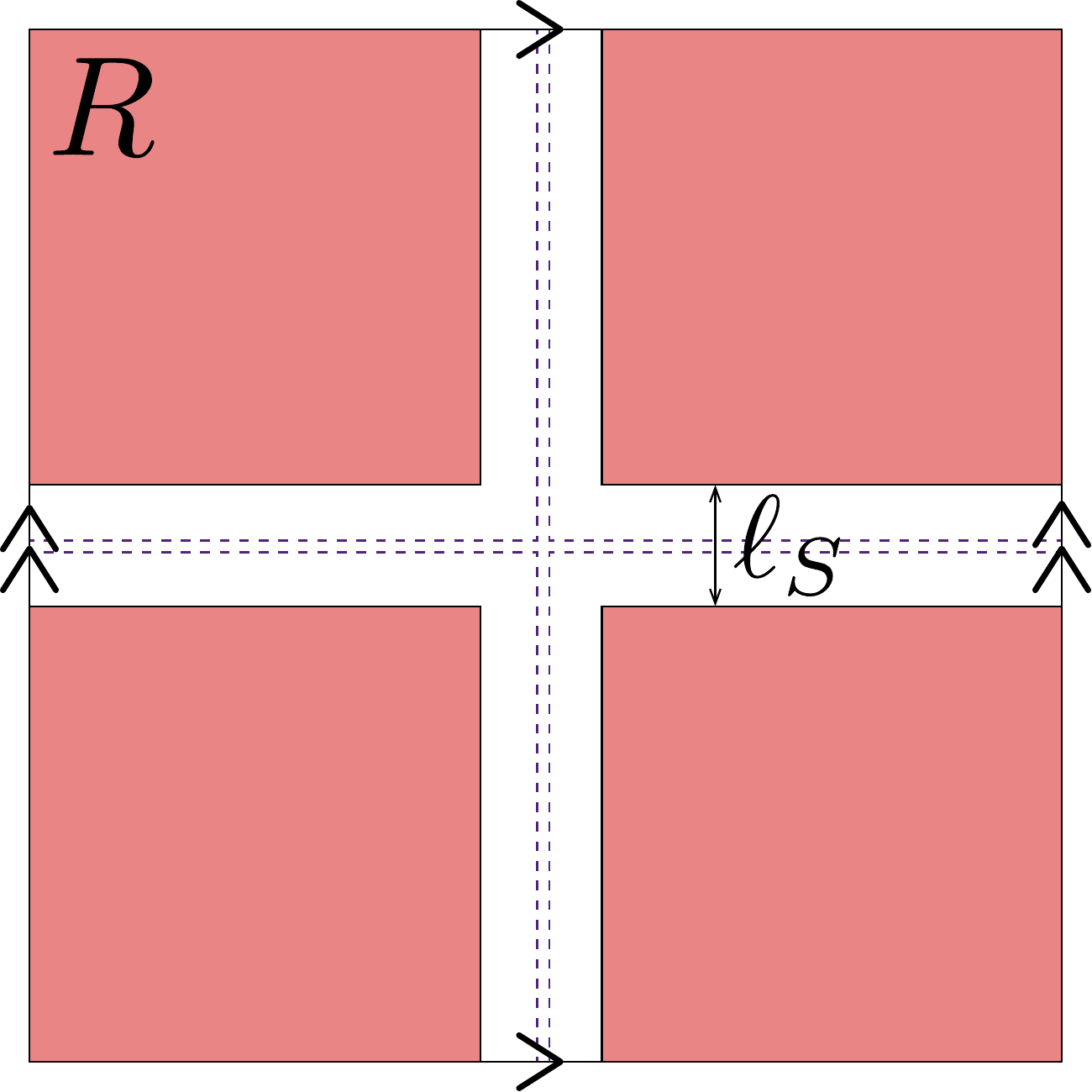}}}} \qquad
\subfloat[\label{fig: torusembedding}]{\includegraphics[scale=.3]{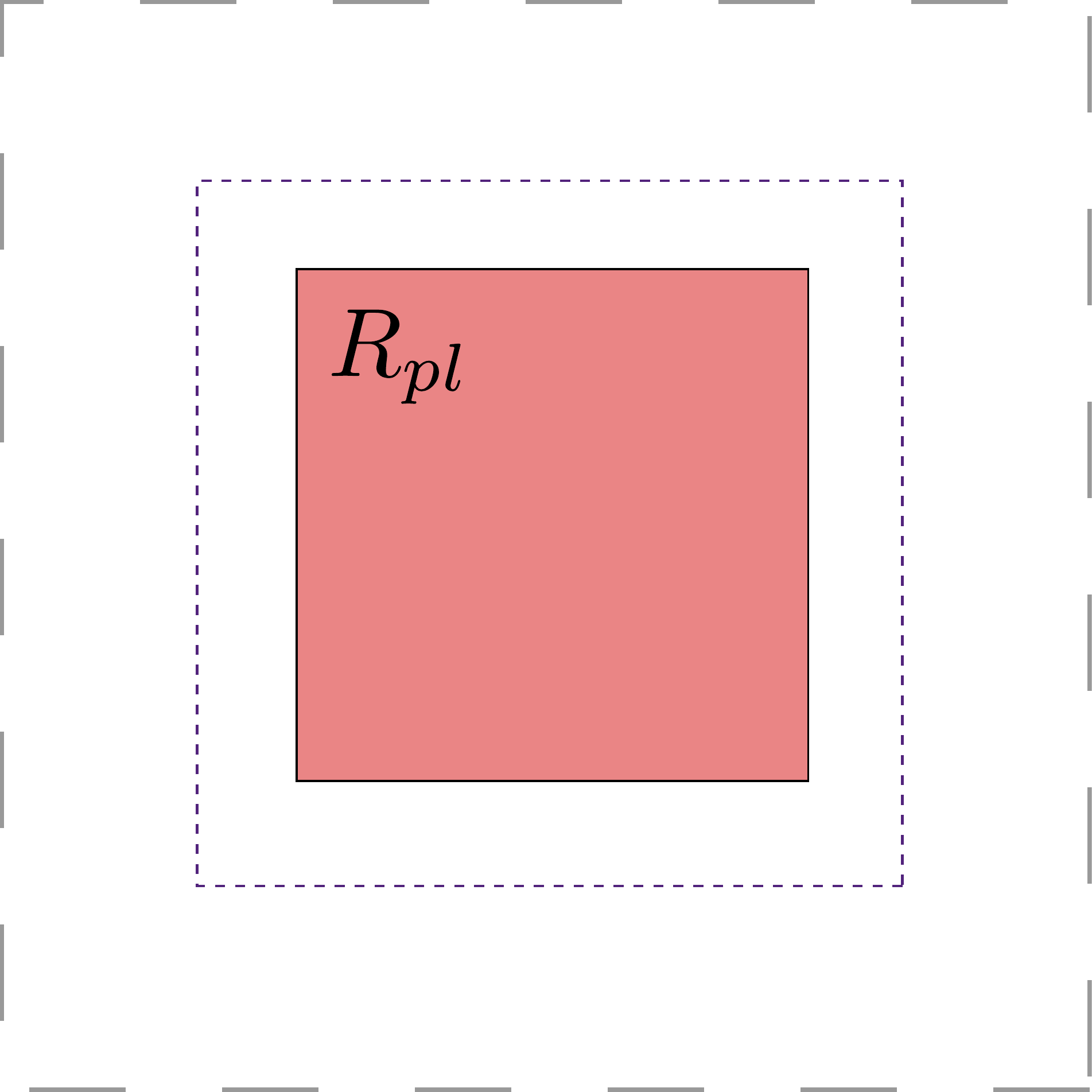}}
\caption{We map the Pauli group on a finite-sized torus to a subgroup of the Pauli group on the infinite plane. (a) We accomplish this by cutting the torus along the meridian and equator (dashed purple). (b) After cutting the torus, it can be unfolded and embedded in the plane. The region $R$ (red, left) is mapped to the region $R_{pl}$ (red, right) of the infinite plane.}
\label{fig: toruscutandembed}
\end{figure*}

\noindent \emph{Proof of Lemma~\ref{lem: topological property torus}:} To begin, we express the condition $\mathcal{Z}_\mathcal{P}(\mathcal{S}) \propto \mathcal{G}$ of the topological~property (Definition~\ref{def: topological subsystem code}) in a more local form. We replace $\mathcal{S}$ with the subgroup of stabilizers $\tilde{\mathcal{S}}$ generated by local stabilizers. This gives us:
\begin{align} \label{eq: topological condition local 1}
    \mathcal{Z}_\mathcal{P}(\mathcal{\tilde{S}}) \propto \mathcal{G}.
\end{align}
Next, we let $R_{pl}$ be an arbitrary connected subregion of the infinite plane, and define $\mathcal{P}(R_{pl})$ to be the subgroup of Pauli operators whose support is contained within $R_{pl}$. We then take the intersection of $\mathcal{P}(R_{pl})$ with both sides of the condition in Eq.~\eqref{eq: topological condition local 1} to find:
\begin{align} \label{eq: local topological condition 1}
    \mathcal{Z}_\mathcal{P}(\mathcal{\tilde{S}}) \cap \mathcal{P}(R_{pl}) \propto \mathcal{G} \cap \mathcal{P}(R_{pl}).
\end{align}
To proceed, we notice that the stabilizer group $\mathcal{\tilde{S}}$ on the left-hand side can be replaced by the subgroup $\mathcal{\tilde{S}}_{R_{pl}}$ of Pauli operators obtained by truncating elements of $\mathcal{\tilde{S}}$ to $R_{pl}$.
Therefore, the expression in Eq.~\eqref{eq: local topological condition 1} can be rewritten as:
\begin{align} \label{eq: local topological condition 2}
    \mathcal{Z}_\mathcal{P}(\mathcal{\tilde{S}}_{R_{pl}}) \cap \mathcal{P}(R_{pl}) \propto \mathcal{G}(R_{pl}),
\end{align}
where $\mathcal{G}(R_{pl})$ is shorthand for the subgroup of  {gauge operators} whose support is fully contained within $R_{pl}$. 

The condition in Eq.~\eqref{eq: local topological condition 2} holds for any connected region $R_{pl}$ in the infinite plane. More importantly, it also holds for certain choices of regions $R$ on an $L \times L$ torus. In particular, it holds for any region $R$ whose linear size is less than $L - \ell_S$. To see this, we map the region $R$ on a torus to a region $R_{pl}$ in the infinite plane. Explicitly, since the linear size of $R$ is less than $L-\ell_S$, we can find paths around the meridian and equator of the torus that do not intersect $R$. We then imagine cutting the torus along these paths and embedding it in the infinite plane (see Fig.~\ref{fig: toruscutandembed}). This allows us to identify $\mathcal{P}(R)$ with $\mathcal{P}(R_{pl})$. Given that the subsystem code is translation invariant, we can also identify $\mathcal{G}(R)$ with the group $\mathcal{G}(R_{pl})$.

\begin{figure*}[tb]
\centering
\subfloat[\label{fig: torusstabtruncation2}]{\raisebox{1.1cm}{\hbox{\includegraphics[scale=.3]{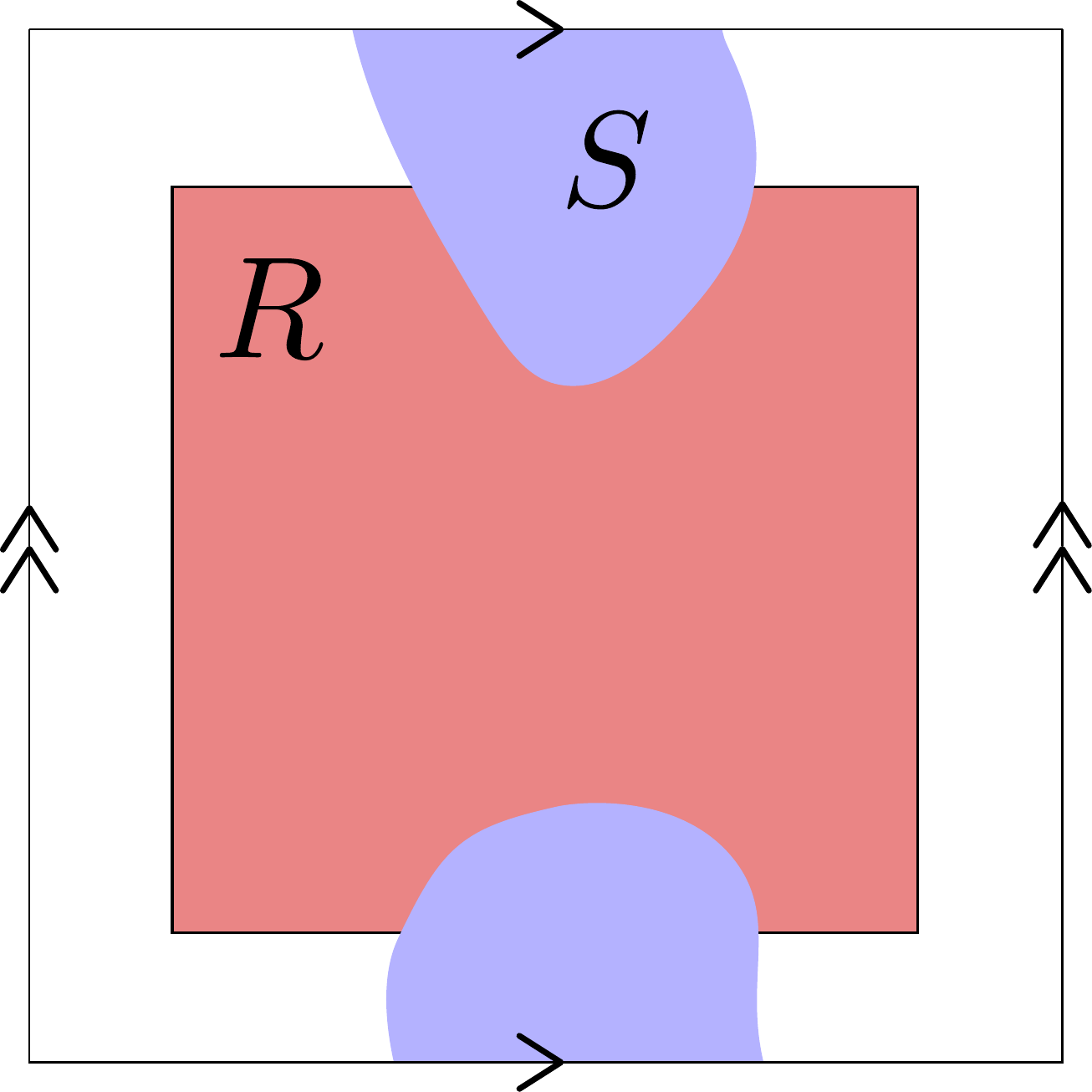}}}} \qquad
\subfloat[\label{fig: tursstabtruncation1}]{\includegraphics[scale=.3]{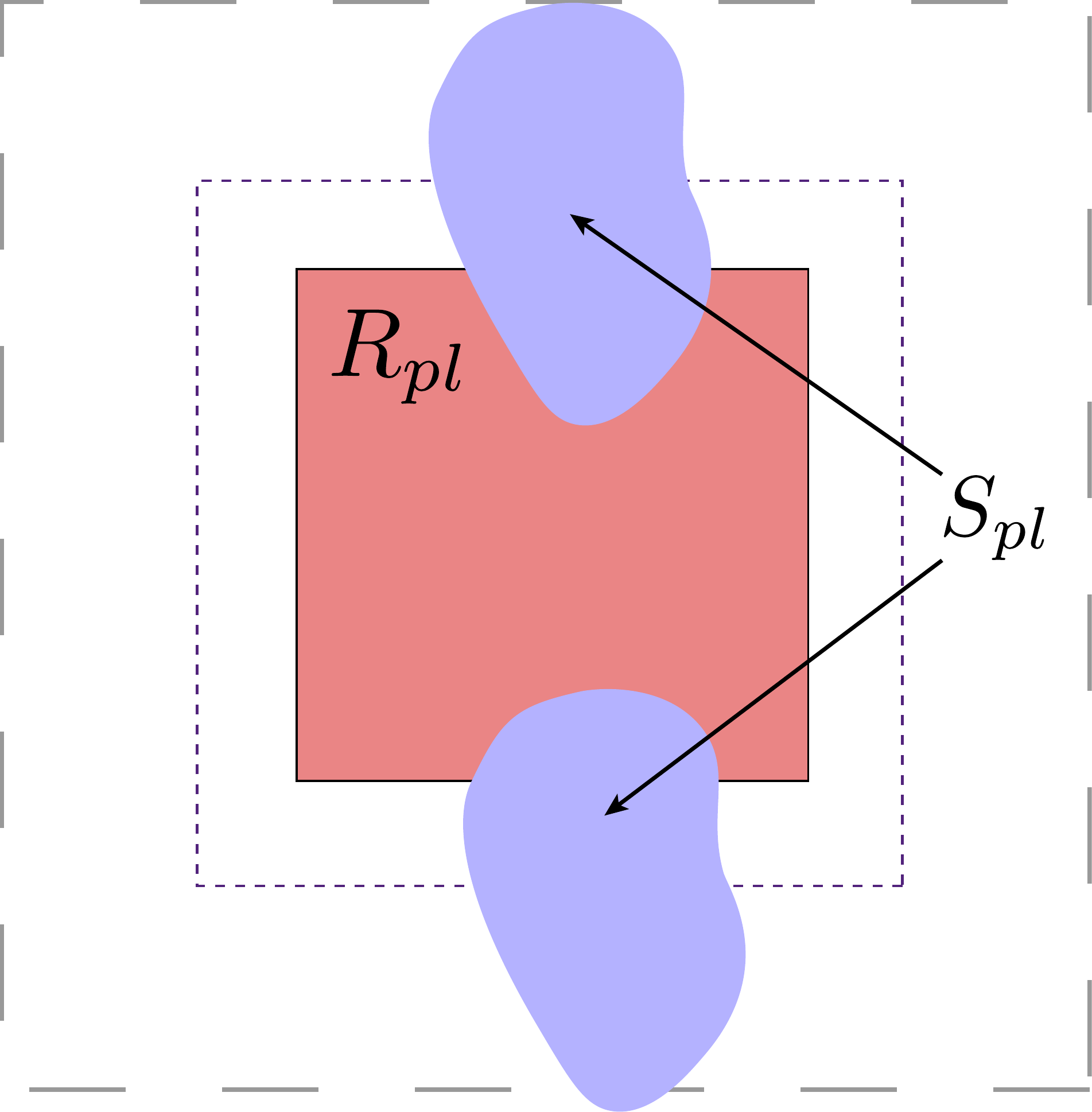}}
\caption{(a) A locally generated stabilizer $S$ (blue) is truncated to the region $R$ (red) on the torus. (b) The truncation of the locally generated stabilizer $S_{pl}$ (blue) is identified with the truncation of $S$ to $R$.}
\label{fig: torusstabtruncation}
\end{figure*}

\begin{figure*}[tb]
\centering
\subfloat[\label{fig: planestabtruncation1}]{\includegraphics[scale=.3]{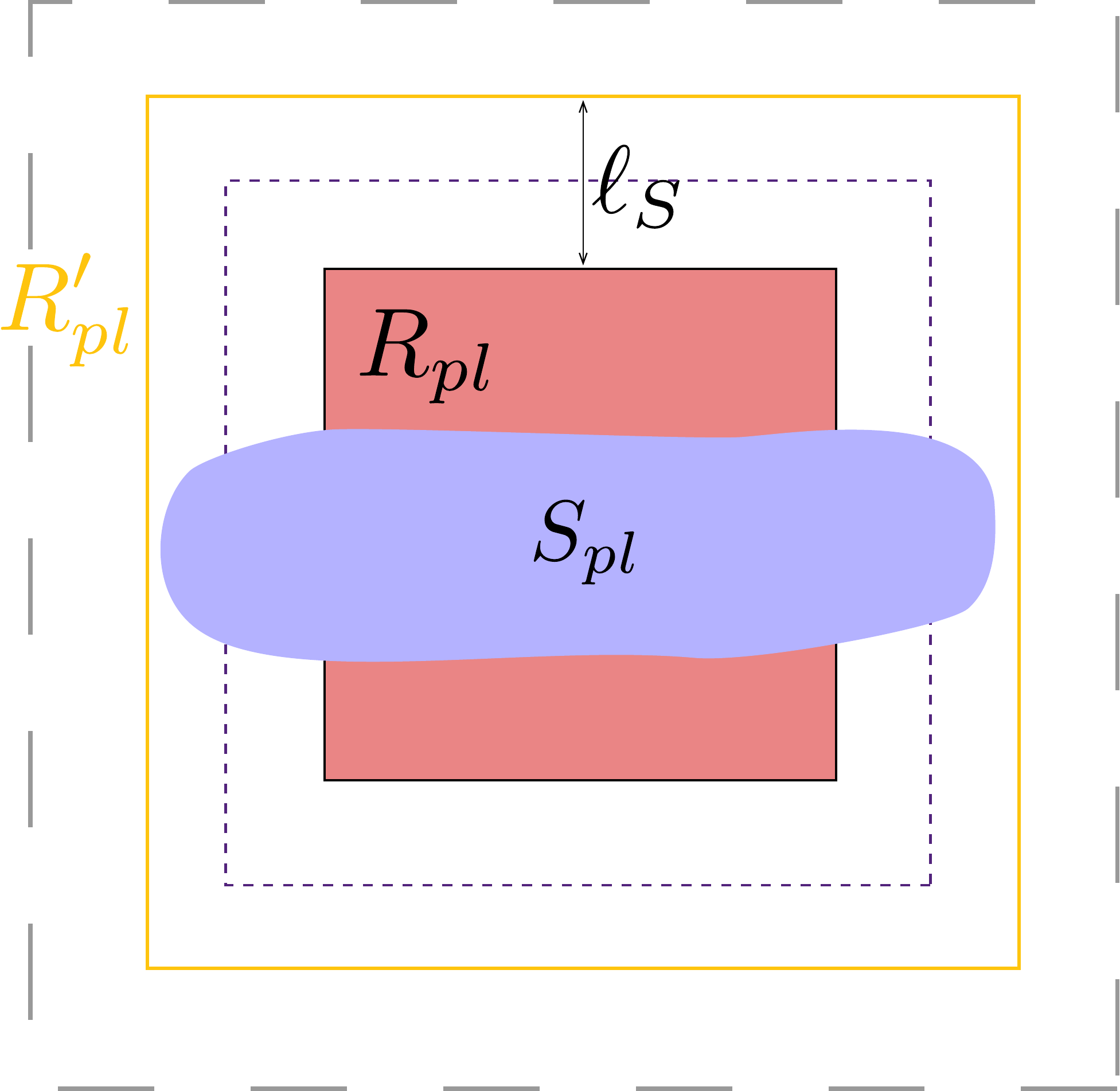}} \qquad
\subfloat[\label{fig: planestabtruncation2}]{\raisebox{.9cm}{\hbox{\includegraphics[scale=.3]{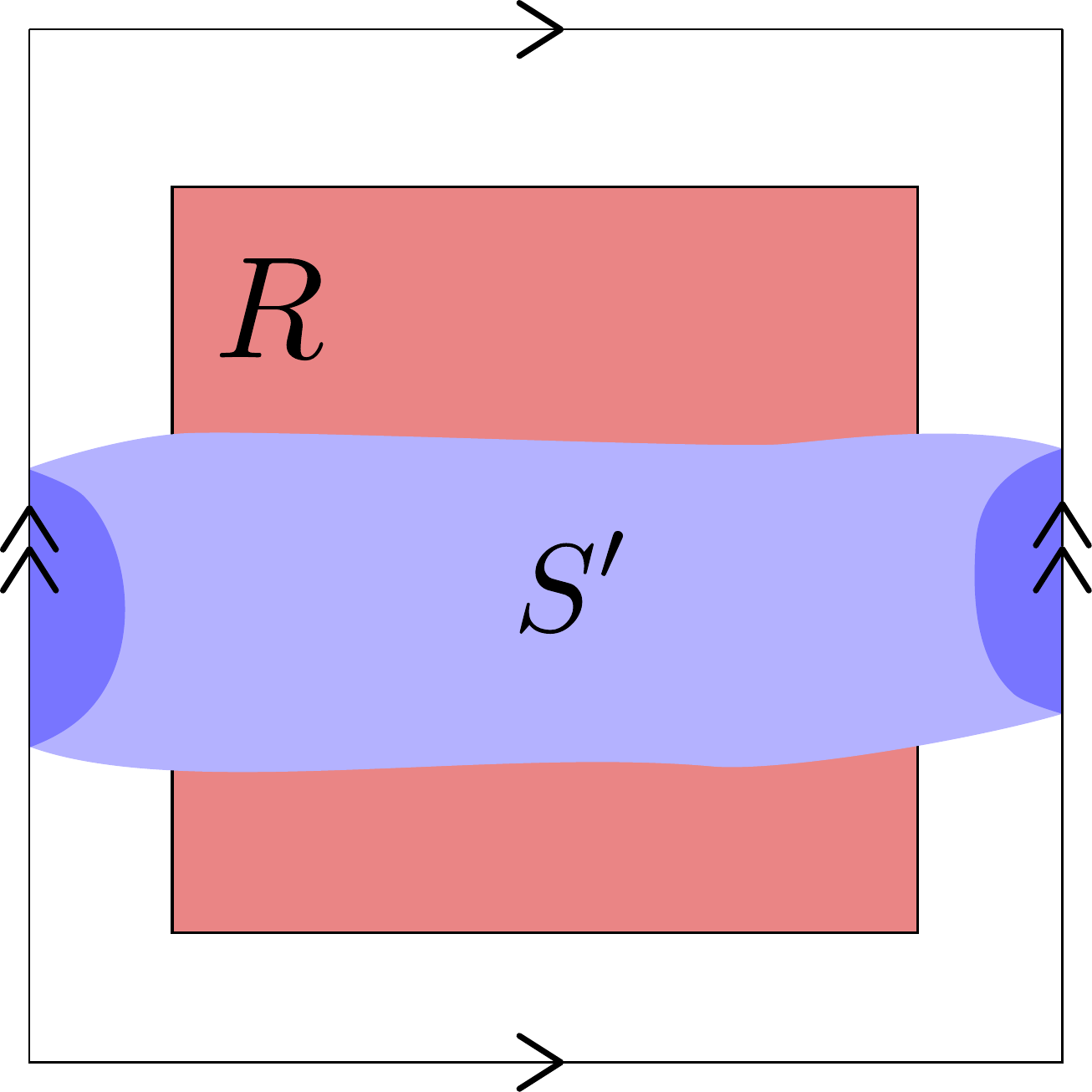}}}}
\caption{(a) The region $R'_{pl}$ (yellow) contains the embedded torus (dashed purple) and is at least a distance $\ell_S$ from the boundary of $R$. A locally generated stabilizer $S_{pl}$ can be truncated to $S'_{pl}$ (blue) on $R'_{pl}$ without affecting its truncation to $R_{pl}$. (b) The operator $S'_{pl}$ can be identified with a locally generated stabilizer $S'$ (blue) on a torus. The truncation of $S'$ to $R$ is identified with the truncation of $S_{pl}$ to $R_{pl}$.}
\label{fig: planestabtruncation}
\end{figure*}

We next identify the subgroup $\tilde{\mathcal{S}}_{R}$ with the subgroup $\tilde{\mathcal{S}}_{R_{pl}}$. First, we show that the elements of $\tilde{\mathcal{S}}_{R}$ can be identified with a subgroup of the elements of $\tilde{\mathcal{S}}_{R_{pl}}$. In particular, suppose we truncate a locally generated stabilizer $S$ on a torus to $R$. We can identify $S$ with an element of $\mathcal{P}(R_{pl})$. Translation invariance then allows us to find a locally generated stabilizer $S_{pl}$ on the infinite plane, such that the truncation of $S$ to $R$ is identified with the truncation of $S_{pl}$ to $R_{pl}$, as illustrated in Fig.~\ref{fig: torusstabtruncation}. Second, we argue that $\tilde{\mathcal{S}}_{R_{pl}}$ can be identified with a subgroup of $\tilde{\mathcal{S}}_{R}$. Let $R_{pl}'$ be the region in the infinite plane such that the boundary of $R_{pl}'$ is a distance $\ell_S$ from the boundary of $R_{pl}$. Given any locally generated stabilizer $S_{pl}$ on the infinite plane, we can find a locally generated stabilizer $S_{pl}'$ contained within $R_{pl}'$ with the property that the truncation of $S_{pl}$ to $R_{pl}$ is the same as the truncation of $S_{pl}'$ to $R_{pl}$. This is because we can truncate $S_{pl}$ to $R_{pl}'$ by throwing out the local stabilizer generators supported outside of $R_{pl}'$. Since the local stabilizer generators can be contained within regions of linear size $\ell_S$, this does not affect the truncation to $R_{pl}$. As shown in Fig.~\ref{fig: planestabtruncation}, the stabilizer $S_{pl}'$ on $R_{pl}'$ defines a locally generated stabilizer $S'$ on the torus such that the truncation of $S'$ to $R$ is identified with the truncation of $S_{pl}$ to $R_{pl}$. Therefore, we have identified $\tilde{\mathcal{S}}_{R}$ with $\tilde{\mathcal{S}}_{R_{pl}}$.

Given that we can identify $\mathcal{P}(R)$, $\mathcal{G}(R)$, and $\tilde{\mathcal{S}}_{R}$ with $\mathcal{P}(R_{pl})$, $\mathcal{G}(R_{pl})$, and $\tilde{\mathcal{S}}_{R_{pl}}$, respectively, the condition in Eq.~\eqref{eq: local topological condition 2} tells us that, for any region $R$ whose linear size is less than $L - \ell_S$, we have:
\begin{align} 
    \mathcal{Z}_\mathcal{P}(\mathcal{\tilde{S}}_{R}) \cap \mathcal{P}(R) \propto \mathcal{G}(R),
\end{align}
which is the condition in Eq.~\eqref{eq: topological property torus}.~$\square$\\

\begin{proposition}[Correctable errors of topological subsystem codes] \label{prop: correctability}
For any TI topological subsystem code on an $L \times L$ torus, the set $\mathsf{E}$ of Pauli operators whose supports have linear size less than $L/2-\ell_S$ is correctable, where $\ell_S$ is the constant-sized length defined in property (iii) of Definition~\ref{def: topological subsystem code}. Moreover, for any errors $E_1, E_2 \in \mathsf{E}$, there is some gauge operator $G$ such that:
\begin{align}
    \Pi_{\tilde{C}}E_1 E_2^\dagger \Pi_{\tilde{C}} \propto G \Pi_{\tilde{C}}.
\end{align}
Here, $\Pi_{\tilde{C}}$ denotes the projector onto the subspace $\mathcal{H}_{\tilde{C}}$ in Eq.~\eqref{eq: local code space}.
\end{proposition}

\noindent \emph{Proof of Proposition~\ref{prop: correctability}:} 
We first argue that any Pauli operator $P_{R}$ supported entirely on a region $R$ with linear size less than $L-\ell_S$ is detectable. If $P_{R}$ fails to commute with an element of $\tilde{\mathcal{S}}_{R}$, then $\Pi_{\tilde{C}}P_{R}\Pi_{\tilde{C}}=0$. Otherwise, if $P_{R}$ commutes with the elements of $\tilde{\mathcal{S}}_{R}$, then $P_{R}$ must belong to $\mathcal{G}(R)$ up to a phase, according to Lemma~\eqref{lem: topological property torus}. Therefore, in either case, we find:
\begin{align} \label{eq: detectability}
    \Pi_{\tilde{C}}P_{R} \Pi_{\tilde{C}} \propto G \Pi_{\tilde{C}},
\end{align}
for some gauge operator $G$. After replacing $\Pi_{\tilde{C}}$ with a projector on to the code space, this is precisely the condition for $P_{R}$ to be detectable. 

\begin{figure}[tb]
\centering
\includegraphics[scale=.3]{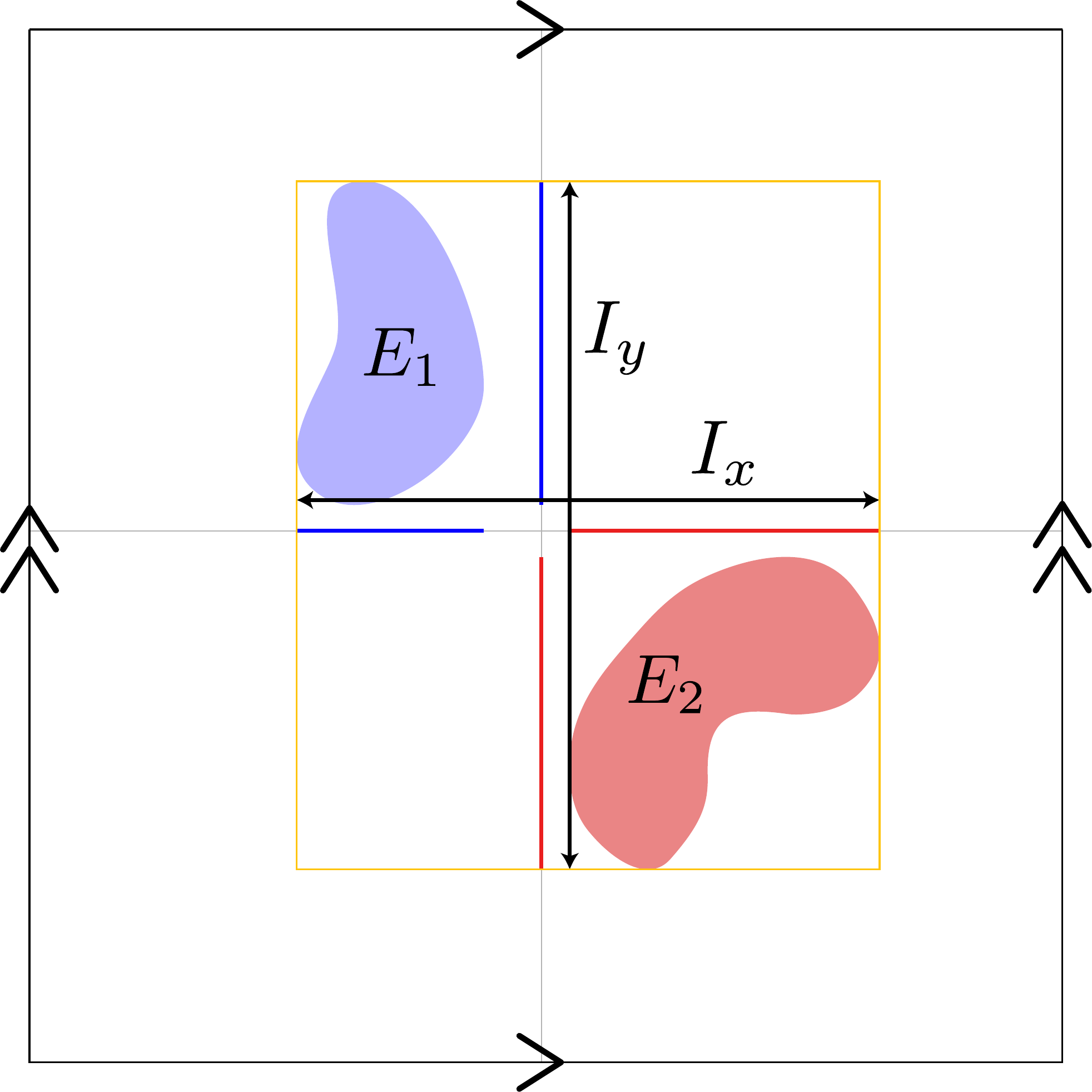}
\caption{The support of the errors $E_1$ and $E_2$ (blue and red, respectively) is projected onto the meridian and equator of the torus. The projection of the supports onto the equator and the meridian fit within the intervals $I_x$ and $I_y$ of the equator and meridian, respectively. The supports of $E_1$ and $E_2$ are contained within the region $I_x \times I_y$, whose linear size is less than $L-\ell_S$.}
\label{fig: supportprojection}
\end{figure}

We now argue that the set $\mathsf{E}$ is correctable. This amounts to showing that any product $E_1 E_2^\dagger$, with $E_1,E_2 \in \mathsf{E}$, is detectable. From the argument above, it suffices to show that the support of the operator $E_1 E_2^\dagger$ can be contained within a region of linear size $L - \ell_S$. To make this explicit, we can project the supports of $E_1$ and $E_2$ onto the equator and the meridian of the torus, as shown in Fig.~\ref{fig: supportprojection}. The projected supports are contained in intervals $I_x$ and $I_y$ with length less than $L-\ell_S$. Therefore, the support of $E_1E_2^\dagger$ can be contained within an $I_x\times I_y$ rectangle. From Eq.~\eqref{eq: detectability}, we have:
\begin{align}
    \Pi_{\tilde{C}}E_1 E_2^\dagger \Pi_{\tilde{C}} \propto G \Pi_{\tilde{C}}.
\end{align}
Thus, the set of errors $\mathsf{E}$ is correctable.~$\square$ \\

\vspace{-.3cm}

We now leverage Lemma~\ref{lem: topological property torus} to prove a cleaning lemma for TI topological subsystem codes.

\begin{lemma}[Cleaning lemma for topological subsystem codes] \label{lem: cleaning lemma}
Let $\mathcal{G}$ and $\mathcal{S}$ be the gauge group and stabilizer group, respectively, of a TI topological subsystem code on an $L \times L$ torus. Also, let $R$ be a connected region whose linear size is less than $L-\ell_S$, where $\ell_S$ is the length defined in Definition~\ref{def: topological subsystem code}. Then for any element $T$ in the bare logical group $\mathcal{L} = \mathcal{Z}_\mathcal{P}(\mathcal{G})$, there exists a product of local stabilizers ${S} \in \tilde{\mathcal{S}}$ such that the support of $T{S}$ is contained within the complement of $R$.
\end{lemma}

\noindent \emph{Proof of Lemma~\ref{lem: cleaning lemma}:} Taking the centralizer of both sides of Eq.~\eqref{eq: topological property torus} and the intersection with $\mathcal{P}(R)$, we obtain:
\begin{align} \label{eq: cleaning stabilizers}
\tilde{\mathcal{S}}_{R} \propto \mathcal{Z}_\mathcal{P}(\mathcal{G}(R)) \cap \mathcal{P}(R).
\end{align}
Here, we have used that, for any subgroup $\mathcal{T}$ of the Pauli group,  $\mathcal{Z}_\mathcal{P}(\mathcal{Z}_\mathcal{P}(\mathcal{T}))=\left \langle \{\omega\}, \mathcal{T} \right \rangle $, for some set of phases $\{ \omega \}$. Given an arbitrary nontrivial bare logical operator or nonlocal stabilizer $T$, the truncation to $R$ commutes with the gauge operators supported entirely within $R$. Therefore, the truncation belongs to $\mathcal{Z}_\mathcal{P}(\mathcal{G}(R))\cap \mathcal{P}(R)$. According to Eq.~\eqref{eq: cleaning stabilizers}, the truncation must be proportional to an element of $\tilde{\mathcal{S}}_{R}$. This means that we can clean $T$, so that it is supported on the complement of $R$ by multiplying by some element of $\tilde{\mathcal{S}}_{R}$. That is, for some ${S} \in \tilde{\mathcal{S}}$, the Pauli operator $T{S}$ is not supported on $R$. ~$\square$ \\

\vspace{-.3cm}

\noindent A direct application of the cleaning lemma shows that, for any TI topological subsystem code defined on a torus, every nontrivial bare logical operator and nonlocal stabilizer can be represented by an operator supported on a pair of intersecting strips of width $\ell_S$ as shown in Fig.~\ref{fig: toruscleaning}. 

\section{Anyonic $1$-form symmetries} \label{app: 1form} 

In Section~\ref{sec: anyon theories for subsystem codes}, we described the anyon types of a TI topological subsystem code in terms of automorphisms of the gauge group, referred to as gauge twists.
In this appendix, we argue that the anyon theories of TI topological subsystem codes can instead be defined according to the conserved quantities shared by the family of Hamiltonians $H_{\{J_G\}}$, i.e., the elements of the bare logical group $\mathcal{L}$. In particular, we interpret loop-like symmetries of the Hamiltonians as string operators that correspond to moving anyon types along closed paths. These loop-like symmetries then define the anyon theory associated to the topological subsystem code. 

To describe the loop-like symmetries of interest, we start by defining $1$-form symmetries  {for two-dimensional systems}, following Ref.~\cite{Qi2021higherform}.
Since the definition of $1$-form symmetries in Ref.~\cite{Qi2021higherform} makes use of concepts from cellular homology, we have reviewed the necessary terminology in Appendix~\ref{app: cellular} for convenience.
We assume throughout this section that the system is defined on an orientable two-dimensional manifold $\mathcal{M}$ with a cellular decomposition defined by a lattice. That is, $\mathcal{M}$ admits a decomposition into $0$-cells, $1$-cells, and $2$-cells, which correspond to the vertices, edges, and plaquettes, respectively, of a lattice embedded in $\mathcal{M}$. 
We let $A$ be an arbitrary finite Abelian group.

 {A $1$-form symmetry is then a unitary projective representation of the group $Z_1(\mathcal{M},A)$, which satisfies certain conditions specified below. Here, $Z_1(\mathcal{M},A)$ is the group of $1$-cycles over $\mathcal{M}$ with coefficients in $A$ (see Appendix~\ref{app:  cellular}). More precisely, a $1$-form symmetry is a map $W$ from $Z_1(\mathcal{M},A)$ to the group of unitary operators such that: 
\begin{enumerate}[label={(\roman*)}]
    \item For any $c$ in $Z_1(\mathcal{M},A)$, the symmetry operator $W(c)$ is a constant-depth quantum circuit of geometrically local gates supported in the vicinity of $c$.\footnote{In particular, $c$ corresponds to a formal sum of edges $e$ weighted by elements $a_e \in A$. The support of $W(c)$ is localized near the edges $e$ for which $a_e \neq 0$.} 
    \item $W$ is a projective representation of $Z_1(\mathcal{M},A)$, i.e., for any $c,c'$ in $Z_1(\mathcal{M},A)$:
        \begin{align}
            W(c)W(c') = \phi\left([c],[c']\right) W(c+c'),
        \end{align}
        for some $U(1)$-valued phase $\phi\left([c],[c']\right)$, which only depends on the homology classes $[c]$ and $[c']$ of $c$ and $c'$.
    \item If $c$ and $c'$ in $Z_1(\mathcal{M},A)$ have a trivial (oriented) intersection number, then:
        \begin{align}
            {\phi\left([c],[c']\right)=1}.
        \end{align}
        This is to say that $W$ forms a linear representation on any subgroup of $Z_1(\mathcal{M},A)$ with $1$-chains that have trivial mutual intersection numbers, such as the group of $1$-boundaries $B_1(\mathcal{M}, A)$.
\end{enumerate}}

 {We recall that $Z_1(\mathcal{M},A)$ can be generated by $1$-cycles $c_{\gamma,a}$ labeled by an oriented loop $\gamma$ and an element $a$ in $A$. We may suggestively write the image of a generator $c_{\gamma,a}$ as $W_\gamma^a$, to make the analogy to anyon string operators more apparent. The condition (i) then implies that, for any generator $c_{\gamma,a}$, the operator $W_\gamma^a$ is a string operator supported along $\gamma$. Intuitively, $W_\gamma^a$ is a string operator that moves an anyon labeled by $a$ along $\gamma$. }

For our purposes, we only require that the $1$-form symmetry forms a projective representation of $Z_1(\mathcal{M},A)$ up to products of local stabilizers supported along the $1$-cycles. More specifically, we consider representations $W$ of $Z_1(\mathcal{M},A)$ such that, for any $c,c'$ in $Z_1(\mathcal{M},A)$, $W(c)$ and $W(c')$ obey the relation:
\begin{align} \label{eq: up to local stabilizers def}
W(c)W(c') \sim \phi\left([c],[c']\right)W(c+c').
\end{align}
Here, $\sim$ denotes that the operators are equal up to products of local stabilizers near the edges of $c+c'$. If $W$ satisfies Eq.~\eqref{eq: up to local stabilizers def}, we say that $W$ is a $1$-form symmetry up to local stabilizers. This is motivated by the fact that anyon string operators do not necessarily generate the same group as the group formed by the anyon types under fusion. For example, in the DS stabilizer model of Ref.~\cite{Ellison2022Pauli}, the string operators of the semions square to products of local stabilizers, despite the semions having order two under fusion.

We further say a representation $W$ of $Z_1(\mathcal{M},A)$ is topologically faithful up to local stabilizers if it satisfies the property:
\begin{align} \label{eq: faithful up to local stabilizers def}
W(c) \sim W(c') \text{ implies } [c]=[c'],
\end{align}
for arbitrary $1$-cycles $c$ and $c'$.\footnote{This is inspired by the definition of a faithful $1$-form symmetry in Ref.~\cite{Qi2021higherform}, in which a $1$-form symmetry $W$ is called faithful if $W(c) = W(c') \text{ implies } c = c' $.} This means that, if $c$ belongs to a nontrivial homology class, then $W(c)$ cannot be a product of local stabilizers. This is important for ensuring that the $1$-form symmetry operators along non-contractible $1$-cycles can be interpreted as string operators for nontrivial anyon types. 

With this, we define an anyonic $1$-form symmetry:
\begin{definition}[Anyonic $1$-form symmetry]
 {An anyonic $1$-form symmetry is a $1$-form symmetry up to local stabilizers that is topologically faithful up to local stabilizers.}
\end{definition}
The anyonic $1$-form symmetries generated by $\mathcal{L}$ define the anyon theory of a topological subsystem code.
To unambiguously define the anyon theory, however, we first impose an equivalence relation on the set of anyonic $1$-form symmetries. We say $W$ and $W'$ are equivalent, denoted $W \sim W'$, if, for every $c$ in  $Z_1(\mathcal{M},A)$, we have
$W(c) \sim W'(c)$.
This is to say, two anyonic $1$-form symmetries $W$ and $W'$ are equivalent if, for any $c$ in $Z_1(\mathcal{M},A)$, the operators $W(c)$ and $W'(c)$ differ by products of local stabilizers supported nearby $c$. 
For TI topological subsystem codes, we assume that, on a torus,\footnote{If $H_1(\mathcal{M},A)$ is trivial, then Eq.~\eqref{eq: faithful up to local stabilizers def} does not impose any constraints on the $1$-form symmetries. As such, the anyonic $1$-form symmetries are highly under-constrained.} there exists an anyonic $1$-form symmetry $W$ generated by $\mathcal{L}$ such that every other anyonic $1$-form symmetry generated by $\mathcal{L}$ is a subgroup.
We refer to this as a maximal anyonic $1$-form symmetry. 
Such a maximal anyonic $1$-form symmetry defines the anyon theory of the TI topological subsystem code. The fusion rules are determined by the group structure of $A$ and the exchange statistics can be computed by truncating the symmetry operators and using the formula in Eq.~\eqref{eq: statistics formula}.

The definition of an anyon theory in terms of a maximal anyonic $1$-form symmetry is closely related to the definition in
Section~\ref{sec: anyon theories for subsystem codes}. As shown in Appendix~\ref{app: bare logicals and nonlocal stabilizers}, the nontrivial bare logical operators and nonlocal stabilizers can be represented by products of string operators formed by moving nontrivial anyon types along non-contractible cycles. This means that anyonic $1$-form symmetry operators can be truncated to yield open string operators that represent anyon types as described in Section~\ref{sec: anyon theories for subsystem codes}. Conversely, for every anyon type defined in Section~\ref{sec: anyon theories for subsystem codes}, the string operators generate an anyonic $1$-form symmetry. By assumption, these generate a subgroup of a maximal anyonic $1$-form symmetry. Therefore, the anyon types defined by a maximal anyonic $1$-form symmetry are the same as those determined by the commutation relations of operators with gauge operators as in Section~\ref{sec: anyon theories for subsystem codes}.

\section{Background on cellular homology} \label{app: cellular} 

To define $1$-form symmetries, as in Ref.~\cite{Qi2021higherform}, it convenient to use the language of cellular homology. In this appendix, we review the terminology from cellular homology necessary to follow the discussion in Appendix~\ref{app: 1form}. Further details and more precise statements about cellular homology can be found in Ref.~\cite{Hatcher2002algebraic}.

\begin{figure}[t]
\centering
\includegraphics[scale=.26]{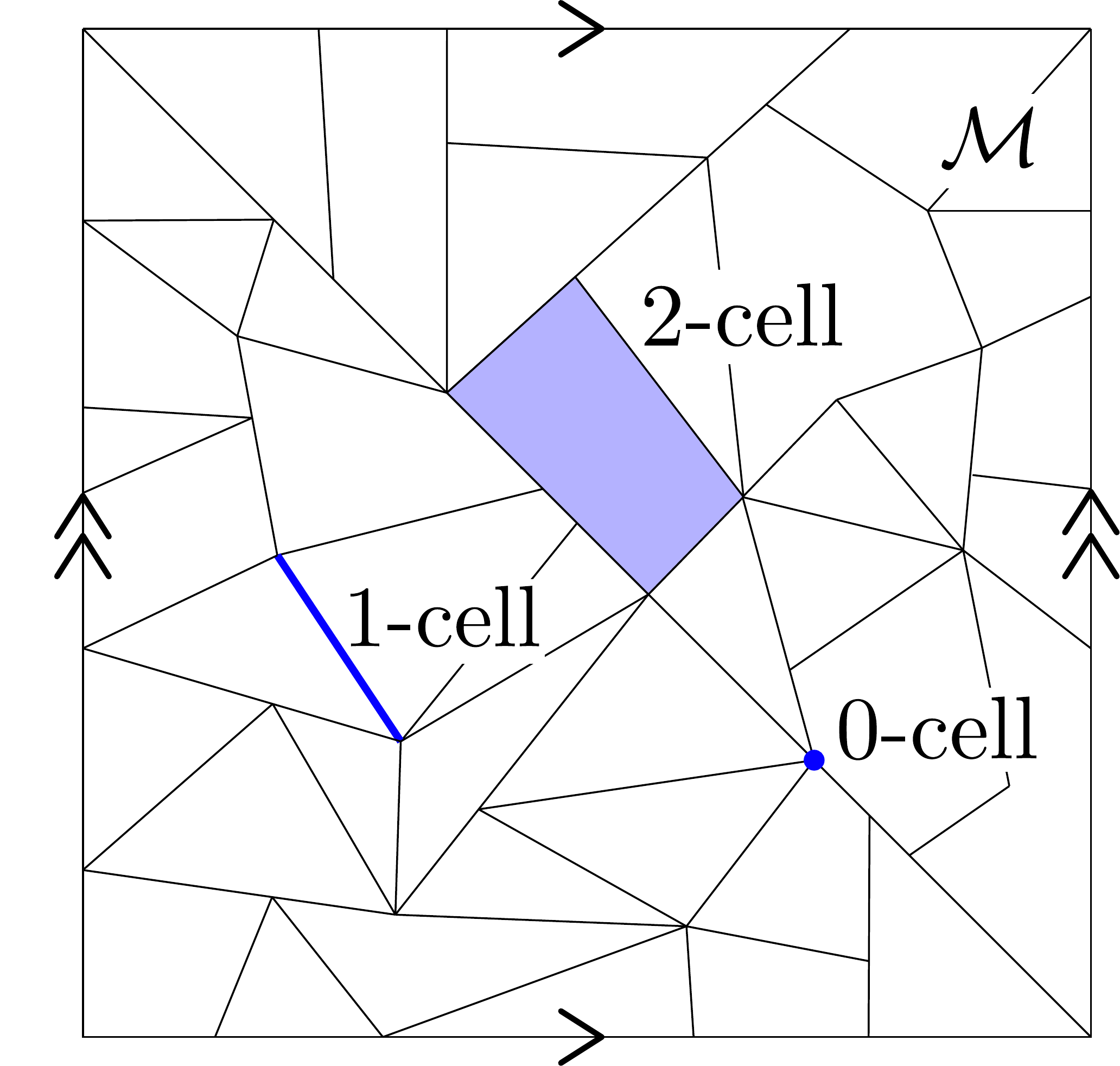}
\caption{The manifold $\mathcal{M}$ (in this case a torus) is divided into $0$-cells, $1$-cells, and $2$-cells.}
\label{fig: cellulation}
\end{figure}

In the discussion below, we use $\mathcal{M}$ to denote an orientable, two-dimensional manifold equipped with a cellular  decomposition into $0$-cells, $1$-cells, and $2$-cells, as pictured in Fig.~\ref{fig: cellulation}. For our purposes, the $0$-cells, $1$-cells, and $2$-cells, respectively, correspond to the vertices, edges, and plaquettes of a two-dimensional lattice embedded in $\mathcal{M}$. 
We use the notation $\la v \ra$ to represent a $0$-cell and write a $1$-cell oriented from $\la v \ra$ to $\la w \ra$ as $\la v,w \ra$. Cellular homology applies to a wider class of cellular decompositions, but for simplicity, we restrict to the case where the cellulation is given by a lattice.

We first define $n$-chains over $\mathcal{M}$ with coefficients in an Abelian group $A$. An $n$-chain is a formal sum of $n$-cells weighted by elements of $A$. For example, a $1$-chain $c_1$ takes the general form:
\begin{align} \label{eq: general 1chain}
c_1 = \sum_{\la v,w \ra} a_{ \la v, w \ra} \la v,w \ra,
\end{align}
where the sum is over (oriented) edges, and $a_{\la v,w \ra}$ belongs to $A$. Reversing the orientation of an $n$-cell is equivalent to inverting the coefficient. Using the example in Eq.~\eqref{eq: general 1chain}, we find that reversing the orientation of each edge gives:
\begin{align}
c_1 =- \sum_{\la v, w \ra} a_{\la v, w \ra} \la w, v \ra.
\end{align}
Two $n$-chains can be added together using the group laws of $A$. For example, the sum of the $1$-chains $c_1$ and $c_1'=\sum_{\la v,w \ra} a'_{\la v, w \ra} 
\la v,w \ra$ is given by the $1$-chain $c_1+c_1'$:
\begin{align}
c_1 + c_1' = \sum_{\la v,w \ra} \left(a_{\la v, w \ra} + a'_{\la v,w \ra}\right)\la v,w \ra.
\end{align}
Hence, the set of $n$-chains forms a group under addition, denoted as $C_n(\mathcal{M},A)$.

\begin{figure*}[tb]
\centering
\includegraphics[scale=.35]{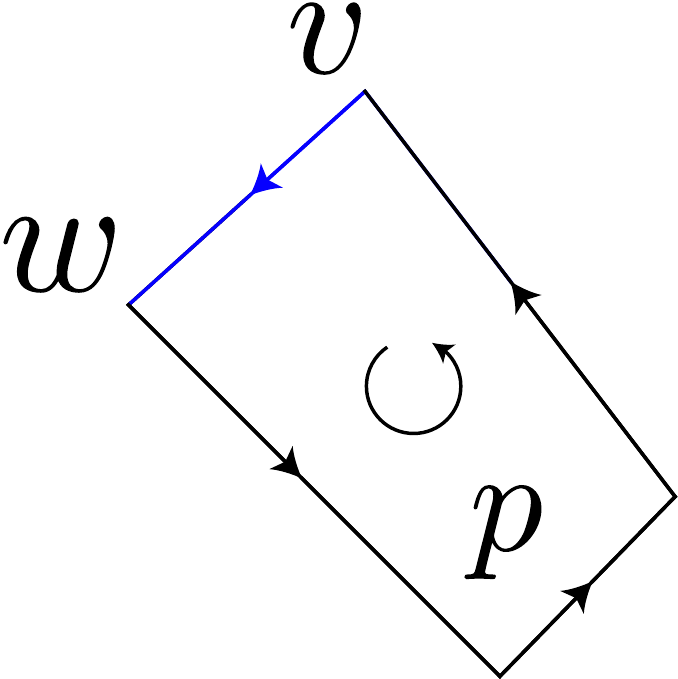} \qquad
\includegraphics[scale=.35]{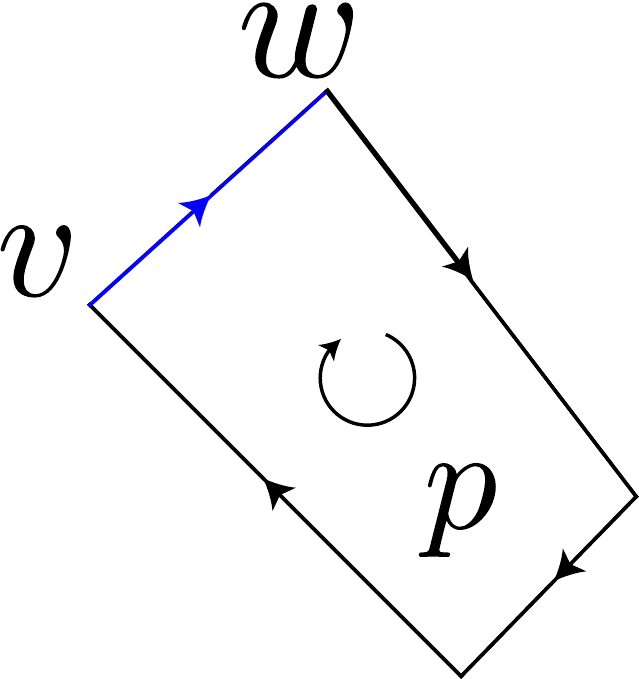}
\caption{The orientation of the edges in Eq.~\eqref{eq: boundary 2cell def} is induced by the orientation of the $2$-cell $p$.}
\label{fig: boundary2cell}
\end{figure*}

Next, we introduce the boundary operator $\partial$, which is a linear map from the group of $n$-chains $C_n(\mathcal{M},A)$ to the group of $(n-1)$-chains $C_{n-1}(\mathcal{M},A)$. We take $\partial c_0 = 0$, for any $0$-chain $c_0$, and we define the boundary operator on $1$-chains by its action on the general $1$-chain $c_1$ in Eq.~\eqref{eq: general 1chain}:
\begin{align}
\partial c_1 = \sum_{\la v,w \ra} a_{\la v, w \ra}\big[\la v \ra - \la w \ra \big].
\end{align}
To define the boundary operator on $2$-chains, we define the action of the boundary operator on a relatively simple $2$-chain $a_pp$, where $a_p$ is in $A$, and $p$ is a $2$-cell. The action of $\partial$ extends to more general $2$-chains by linearity. $\partial (a_pp)$ is given by a particular sum of oriented edges in the boundary of the plaquette $p$ weighted by $a_p$. In particular, $\partial (a_pp)$ is the $1$-chain:
\begin{align} \label{eq: boundary 2cell def}
\partial(a_p p) = \sum_{\la v,w \ra \in \partial p} a_p \la v, w \ra,
\end{align}
where the sum is over edges $\la v, w \ra$ in the boundary of $p$, oriented according to the orientation induced on the boundary by the orientation of $p$ (see Fig.~\ref{fig: boundary2cell}). We note that $\partial \partial =0$.

An $n$-chain $c_n$ is called an $n$-cycle if $\partial c_n =0$. Given that the boundary operator is linear, the $n$-cycles form a group under addition. We use $Z_n(\mathcal{M},A)$ to denote the group of $n$-cycles over $\mathcal{M}$ with coefficients in $A$. We note that, as mentioned in Appendix~\ref{app: 1form}, $Z_1(\mathcal{M},A)$ can be generated by $1$-cycles  labeled by an oriented loop and an element of $A$. More explicitly, the group of $1$-cycles is generated by $1$-cycles of the form:
\begin{align}
 c_{\gamma,a} =  \sum_{\la v,w \ra \in \gamma} a \la v,w \ra, 
\end{align}
where $a$ is an element of $A$, $\gamma$ is an oriented loop of edges, and the sum is over edges in $\gamma$, oriented according to the orientation of $\gamma$.

Finally, we define the cellular homology group $H_1(\mathcal{M},A)$. $H_1(\mathcal{M},A)$ is defined in terms of $Z_1(\mathcal{M},A)$ and the group of $1$-boundaries, denoted as $B_1(\mathcal{M},A)$. The group $B_1(\mathcal{M},A)$ is the group of $1$-chains $c_1$ for which $\partial c_2 = c_1$, for some $2$-chain $c_2$. Note that, since $\partial \partial =0$, $B_1(\mathcal{M},A)$ is a subgroup of $Z_1(\mathcal{M},A)$. With this, $H_1(\mathcal{M},A)$ is the quotient group:
\begin{align}
H_1(\mathcal{M},A) = Z_1(\mathcal{M},A) / B_1(\mathcal{M},A).
\end{align}
In words, the elements of $H_1(\mathcal{M},A)$ are equivalence classes of $1$-cycles with the property that two $1$-cycles belong to the same class if they differ by a $1$-boundary. We write the equivalence class of a $1$-cycle $c_1$ as $[ c_1 ]$. We say two $1$-cycles $c_1$, $c_2$ are homologous if $[ c_1 ]=[ c_2 ]$.

\section{Fluxes in topological subsystem codes} \label{app: fluxes}

In this appendix, we extend the concept of a flux, introduced in Ref.~\cite{Bombin2014structure}, to systems~of composite-dimensional qudits. 
Similar to the role of gauge twists in the definition~of~anyon types, to define fluxes, we begin by introducing the notion of a stabilizer twist.
\begin{definition}[Stabilizer twist] \label{def: stabilizer twist}
Letting $\mathcal{S}$ be the stabilizer group of a TI Pauli topological subsystem code and $\tilde{\mathcal{S}}$ be the subgroup of locally generated stabilizers, a stabilizer twist is a group homomorphism $\Sigma: \mathcal{S} \to U(1)$, such that $\Sigma(S) = 1$ for all but finitely many $S$ in some set of independent local generators of $\tilde{\mathcal{S}}$.
\end{definition}
Note that a stabilizer twist encodes the same information as the stabilizer syndrome for the group $\tilde{\mathcal{S}}$. It is also important to note that, while some stabilizer twists are simply restrictions of gauge twists to the stabilizer group, there are stabilizer twists that are not of this form. For example, a stabilizer twist $\Sigma$ that violates the vertex stabilizer $S$ of the $\ZZ_2^{(0)}$ subsystem code in Section~\ref{sec: 1,m subsystem}, i.e. $\Sigma(S) \neq 1$, does not correspond to the restriction of a gauge twist. The important difference is that stabilizer twists can violate arbitrarily many gauge generators, so long as it violates only finitely many stabilizers.

With this, for a TI Pauli topological subsystem code defined on an infinite plane, we define the fluxes as equivalence classes of stabilizer twists. 
\begin{definition}[Flux]
A flux of a TI Pauli topological subsystem code is an equivalence class of stabilizer twists under the following equivalence relation: two stabilizer twists $\Sigma$ and $\hat{\Sigma}$ are equivalent, or represent the same flux, if there exists a Pauli operator $P$, such that $\Sigma(S) = \Phi_P(S)\hat{\Sigma}(S)$, for all $S$ in the stabilizer group.
\end{definition}
Here, $\Phi_P(S)$ is the commutator of $P$ and $S$ defined in Eq.~\eqref{eq: trivial gauge functional}.
We say a flux is trivial if it belongs to the same equivalence class as the trivial homomorphism. Intuitively, we think of a nontrivial flux as being represented by a stabilizer twist $\Sigma$ given by a semi-infinite product of Pauli operators $W$ as $\Sigma(S) = WSW^\dagger S^\dagger$, for any stabilizer $S$. The string operator $W$ may fail to commute with extensively many gauge generators along its length. We point out that a subset of the fluxes correspond to the anyon types of the topological subsystem code, since the gauge twists that represent anyon types define stabilizer twists when restricted to the stabilizer group. The undetectable anyon types yield trivial fluxes, while the detectable anyon types correspond to nontrivial fluxes.

In general, we expect that the fluxes of a Pauli topological subsystem code are given by the detectable anyon types as well as the so-called tube-like anyon types of the parent topological stabilizer code used to construct the subsystem code (see Appendix~\ref{app: WW}). Indeed, the tube-like anyon types of the parent topological stabilizer code violate the stabilizers formed by loops of string operators for the transparent anyon types. The tube-like anyon types do not correspond to anyon types of the subsystem code, though, because the string operators fail to commute with the short string operators of the transparent anyon types along their length. We note that, since the short string operators of the transparent anyon types are local gauge operators, and thus terms in the Hamiltonians $H_{\{J_G\}}$ in Eq.~\eqref{eq: subsystem Hamiltonian}, the fluxes given by tube-like anyon types may be confined or deconfined excitations depending on the coefficients $\{J_G\}$.

Following the discussion in Section~\ref{sec: anyon theories for subsystem codes}, we can define the fusion of fluxes and create string operators to move fluxes. The exchange statistics and braiding relations of fluxes are not well defined, however. This is because the string operators are only defined up to gauge operators. Indeed, gauge operators do not affect the flux, since they commute with all of the stabilizers. This means that the string operators can be redefined by products of gauge operators along their length. Since gauge operators generically have nontrivial commutation relations with one another, the braiding relations of fluxes can be changed by redefining the string operators. Nonetheless, the braiding relations between the anyon types and the fluxes are well defined. This is because the string operators of the anyon types commute with all of the gauge operators along their length. Therefore, redefining the string operator of the flux by gauge operators does not affect the braiding relations, in this case.

\section{Detectable anyon types are opaque} \label{app: detectable=opaque} 

Here, we prove that detectable anyon types are opaque, i.e., if an anyon type is represented by operators that necessarily fail to commute with stabilizers, then there must be some anyon type with which it braids nontrivially. 
Since every anyon type of a topological stabilizer code is either trivial or detectable, this implies that the anyon theories of topological stabilizer codes must be modular.
We also note that undetectable anyon types are transparent. This is because, for some choice of Pauli operators at the endpoints, the corresponding open string operators commute with all of the stabilizers, including, in particular, the stabilizers formed from contractible loops of anyon string operators. Therefore, they must braid trivially with every anyon type.

In the proof that detectable anyon types are opaque below, we let $\mathcal{G}$ be the gauge group of a topological subsystem code and take $\mathcal{S}$ to be the associated stabilizer group. For simplicity, we assume that the system is defined on an infinite plane, although the argument can be generalized to the torus straightforwardly. 

\begin{figure}[tb]
\centering
\includegraphics[width=.4\textwidth]{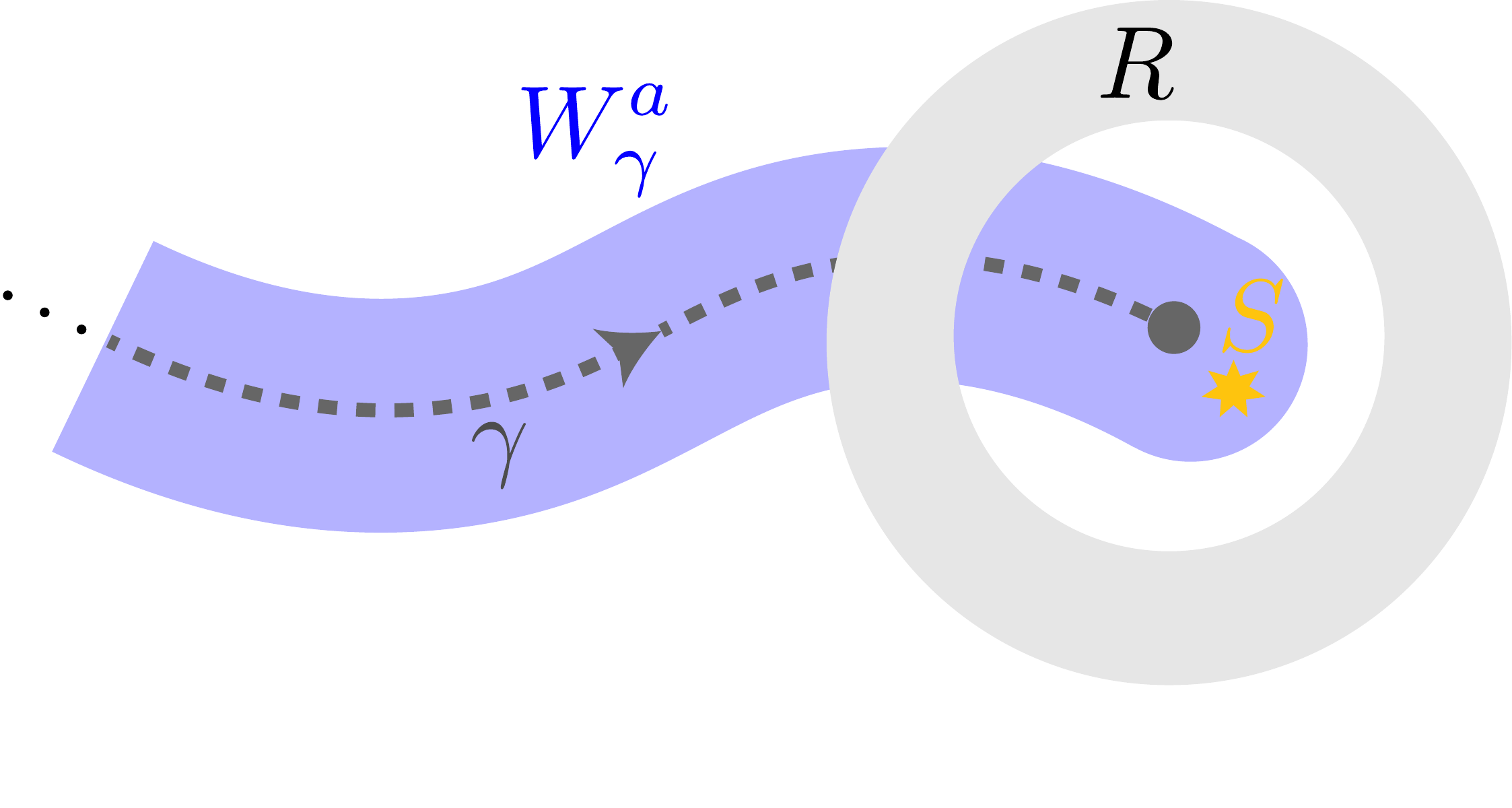}
\caption{The string operator $W_\gamma^a$ (blue) is supported along an open path $\gamma$. We can choose the generators of $\tilde{\mathcal{S}}$ so that $W_\gamma^a$ only fails to commute with the sole generator $S$ (yellow). We argue that $S$ can be chosen so that it is supported fully within an annular region $R$ (gray).}
\label{fig: Wgammaa}
\end{figure}

\begin{proposition} \label{prop: detectable=opaque}
For any two-dimensional TI topological subsystem code, the detectable anyon types are opaque.
\end{proposition}

\noindent \emph{Proof of Proposition~\ref{prop: detectable=opaque}:} Consider a string operator $W_\gamma^a$ for a detectable anyon type $a$, supported along an open path $\gamma$ with one endpoint. Since $a$ is detectable, $W_\gamma^a$ must fail to commute with a generator of $\mathcal{S}$ at its endpoint. The string operator $W_\gamma^a$ may also fail to commute with gauge operators that do not belong to $\mathcal{S}$. By redefining the generators of $\mathcal{S}$, we can ensure that $W_\gamma^a$ fails to commute with only one of the generators of $\mathcal{S}$. Let us call this stabilizer $S$. Furthermore, since the stabilizer group of a topological subsystem code is locally generated when defined on an infinite plane, we can assume that $S$ is local. The string operator $W_\gamma^a$ and the violated stabilizer $S$ are pictured in Fig.~\ref{fig: Wgammaa}. Note that there are no local Pauli operators that violate only the stabilizer $S$. If there existed such a local Pauli operator, then $W_\gamma^a$ could be redefined so that it only fails to commute with elements of $\mathcal{G} - \mathcal{S}$. By assumption, the anyon type $a$ is detectable, so this is not possible.

In what follows, we argue that the violated stabilizer $S$ can be chosen so that it is fully supported on an annular region $R$ around the endpoint of $\gamma$, as shown in Fig.~\ref{fig: Wgammaa}. This shows us that $W_\gamma^a$ fails to commute with a loop-like stabilizer wrapped around its endpoint. Hence, the anyon type $a$ braids nontrivially with the anyon type associated to the loop-like stabilizer. 

We start by defining a new subsystem code in which $S$ no longer belongs to the gauge group. For systems of prime-dimensional qudits, we can always redefine the gauge generators so that $S$ is itself a gauge generator. Subsequently, we can remove $S$ from the set of gauge generators. For systems of composite-dimensional qudits, it might not be possible to choose $S$ as a gauge generator, without introducing redundancy. In general, we can choose the gauge generators so that $S$ is a power of some gauge generator $G$. For example, the gauge group might be generated by $Z$ and $X^2$ for a single four-dimensional qudit. In this case, the stabilizer is $Z^2$. We cannot choose $Z^2$ as an independent gauge generator, since it does not generate $Z$. Therefore, to ensure that $S$ is not part of the gauge group, we remove the generator $G$. We let $\mathcal{G}'$ be the gauge group after removing $G$ as a generator.

Relative to the topological subsystem code defined by $\mathcal{G}$, the subsystem code defined by $\mathcal{G}'$ has an additional logical qudit. Indeed, one of the nontrivial bare logical operators is the Pauli operator $S$. The conjugate logical Pauli operator cannot be local. Otherwise, we would be able to redefine $W_\gamma^a$ so that it commutes with all of stabilizers of the original topological subsystem code. 

To proceed, we use the cleaning lemma for subsystem codes derived in Ref.~\cite{Bravyi2011local}. The cleaning lemma states that, for any region $R$, we have the following equality:
 \begin{align} \label{eq: Bravyi cleaning lemma}
 l_{bare}(R) + l(R^c) = 2k.
 \end{align}
Here, $l_{bare}(R)$ is the number of independent nontrivial bare logical operators supported entirely within $R$, $l(R^c)$ is the number of independent nontrivial dressed logical operators contained within the complement $R^c$, and $k$ is the total number of logical qudits. For example, if there are no nontrivial bare logical operators supported entirely within $R$, then each nontrivial logical operator can be `cleaned' by multiplying by gauge operators so that its support is contained in $R^c$.
 
\begin{figure}[tb]
\centering
\includegraphics[scale=.32]{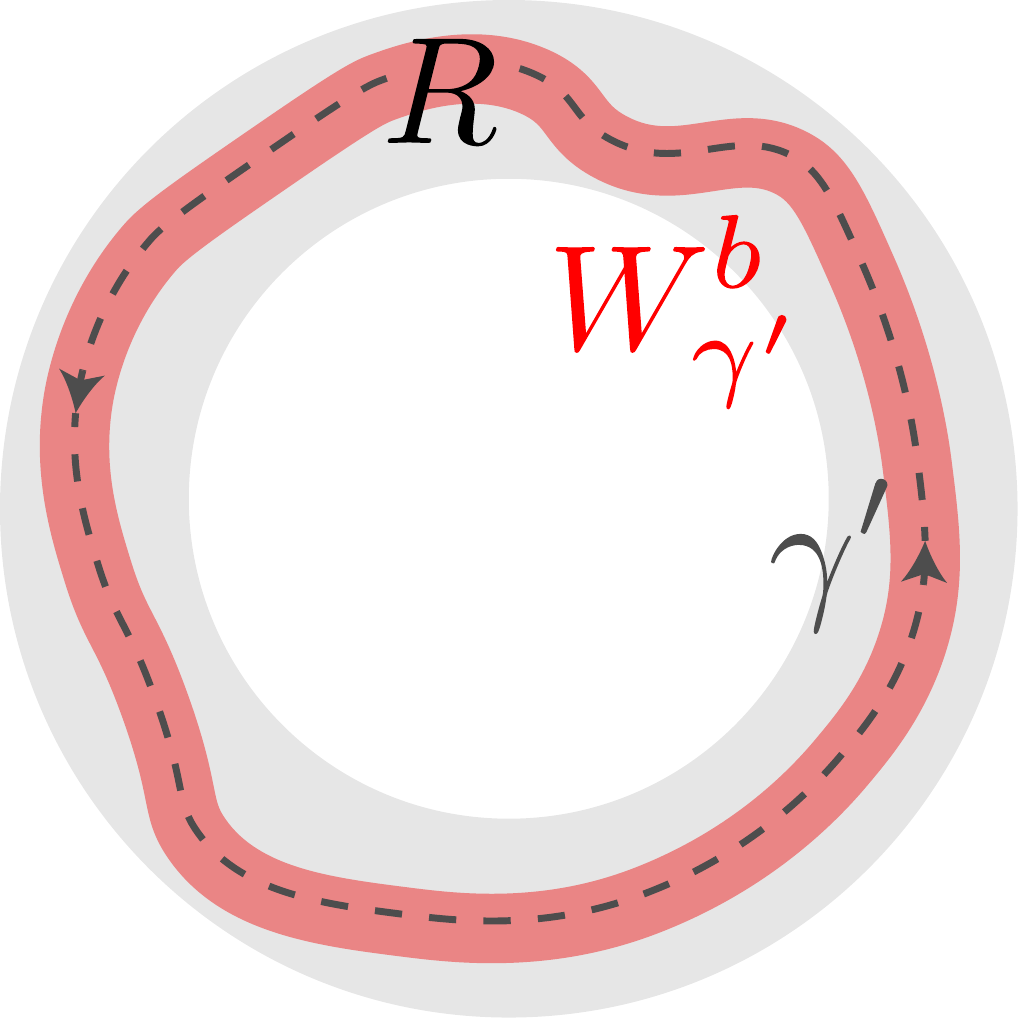}
\caption{The nontrivial bare logical operator $W_{\gamma'}^b$ (red) is supported entirely within $R$ (gray). }
\label{fig: Wgammaprimeb}
\end{figure} 
 
We now apply the cleaning lemma to the modified subsystem code with gauge group $\mathcal{G}'$. We take $R$ to be an annulus around the endpoint of $\gamma$, with a width that is large compared to the supports of the local gauge generators. In particular, we require that $R$ is wide enough such that none of the gauge generators have support in both components of $R^c$. 
We also choose $R$ so that $S$ is not supported in $R$. Assuming that the subsystem code is defined on an infinite plane, removing the generator $S$ of $\mathcal{S}$ introduces one logical qudit. Therefore, we have $k=1$ in Eq.~\eqref{eq: Bravyi cleaning lemma}. The dimension of the qudit is equal to the order of the operator $S$. 

We claim that $S$ is the only independent nontrivial dressed logical operator supported entirely on $R^c$, i.e., we claim that $l(R^c)=1$. As previously mentioned, the conjugate logical operator cannot be supported entirely on the inner disk of $R^c$. Furthermore, Pauli operators on the outer region of $R^c$ do not affect the commutation relations with gauge generators supported on the inner disk of $R^c$. This implies that the conjugate logical operator cannot be fully supported on the two components of $R^c$. Otherwise, we could truncate the operator to the inner disk and derive a contradiction with the fact that the anyon type is detectable. 

Plugging $k=1$ and $l(R^c)=1$ into Eq.~\eqref{eq: Bravyi cleaning lemma}, we find that the number of independent nontrivial bare logical operators supported on $R$ is $l_{bare}(R)=1$. The nontrivial bare logical operator supported on $R$ commutes with $S$. Therefore, it must be equivalent to the logical operator $S$ up to stabilizers. With some foresight, we denote the nontrivial bare logical operator on $R$ by $W^b_{\gamma'}$, as depicted in Fig.~\ref{fig: Wgammaprimeb}. The string operator $W^a_{\gamma}$ has the same commutation relations with $W^b_{\gamma'}$ as with $S$, because $W^a_{\gamma}$ is a dressed logical operator conjugate to $S$. 

The string operator $W^b_{\gamma'}$ is local and commutes with all of the gauge operators in $\mathcal{G}$. Therefore, it must be a stabilizer in the topological subsystem code. This justifies the notation, as we can associate an anyon type $b$ to the loop-like string operator (see Lemma~\ref{lem: string operator anyon types}). We have now found a loop-like string operator $W^b_{\gamma'}$ with nontrivial commutation relations with $W^a_{\gamma}$. Thus, $a$ braids nontrivially with $b$, and the detectable anyon type $a$ must be opaque. $\square$

\section{Nontrivial bare logical operators and nonlocal stabilizers} \label{app: bare logicals and nonlocal stabilizers}

The goal of this appendix is to show that the nontrivial bare logical operators and nonlocal stabilizers of TI topological subsystem codes, i.e., the nonlocal elements of the bare logical group $\mathcal{L}$, can be understood in terms of (Pauli) string operators created by moving anyon types along non-contractible paths. 
We prove that, for TI topological subsystem codes, moving an anyon type $a$ along a non-contractible path produces a nonlocal element of $\mathcal{L}$ if and only if $a$ is a nontrivial anyon type. We then argue that in fact every nontrivial bare logical operator and nonlocal stabilizer can be represented (up to stabilizers or products of local stabilizers, respectively) by a product of string operators created by moving anyon types around non-contractible paths of the torus. This agrees with the examples presented in this text, in which all of the nontrivial bare logical operators and nonlocal stabilizers are indeed represented by string operators wrapped around non-contractible paths. 

To get started, we prove a lemma. In particular, we prove that there is a well-defined anyon type associated to any string operator that is supported on an oriented loop and commutes with all of the gauge operators. 
The associated anyon type is determined by truncating the string operator and analyzing the commutation relations with the gauge operators at the endpoints. We label the string operator by the anyon type $a$, such that the orientation of the truncated string operator points towards $a$.
Thus, we can interpret the string operator as moving the anyon type $a$ around the loop according to its orientation.

\begin{lemma} \label{lem: string operator anyon types}
Any string operator that is supported along an oriented closed path and commutes with all of the gauge operators has a well-defined associated anyon type. 
\end{lemma}

\begin{figure}[tb]
\centering
\raisebox{.5cm}{\hbox{\includegraphics[width=.38\textwidth]{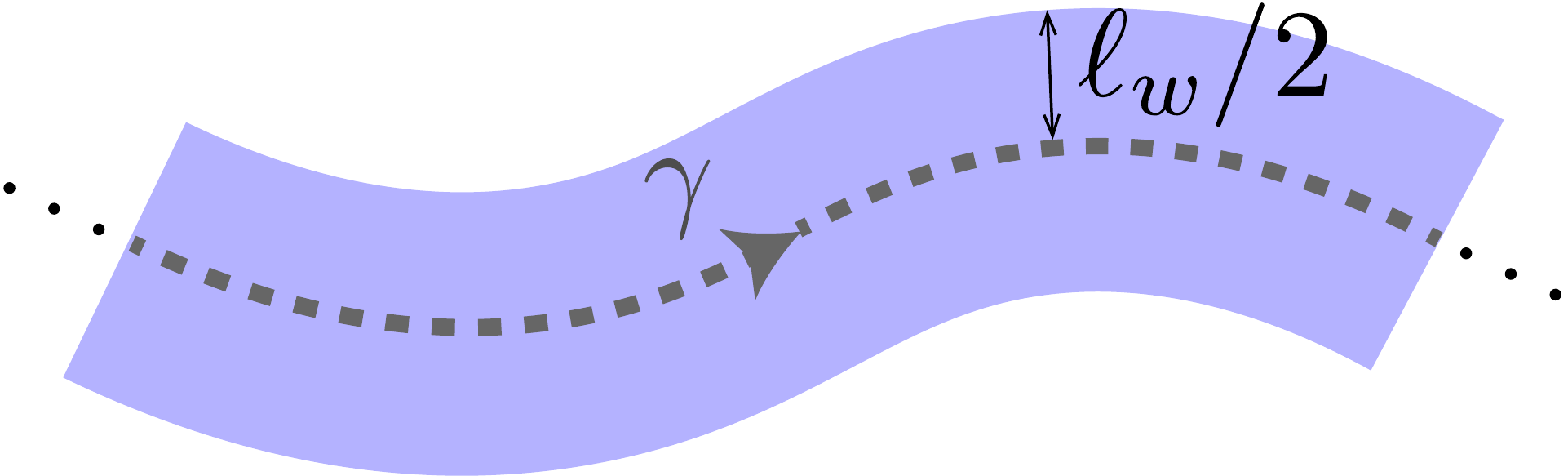}}}
\caption{We assume that the support (blue) of the string operator $W_\gamma$ is supported within a distance $\ell_W/2$ of the path $\gamma$.}
\label{fig: stringwidth}
\end{figure}

\noindent \emph{Proof of Lemma~\ref{lem: string operator anyon types}:} Let $W_\gamma$ be a string operator supported along an arbitrary oriented closed path $\gamma$. We assume that there exists a constant length $\ell_W$ such that the support of the string operator can be contained within a length $\ell_W/2$ of the path $\gamma$ (see Fig.~\ref{fig: stringwidth}). This implies that truncations of $W_\gamma$ are ambiguous only up to Pauli operators supported on disks of diameter $\ell_W$ centered on the endpoints of the truncated path $\gamma'$. Our goal is to show that the anyon type associated to $W_\gamma$ is independent of the choice of truncation, assuming that the two endpoints of the truncated string operator are sufficiently far from one another (specified more precisely below). We start by proving that, for any truncation $W_{\gamma'}$ of $W_\gamma$, the anyon types associated to the endpoints must be inverses of one another. More concretely, let us label the anyon types at the endpoints of $W_{\gamma'}$ by $a$ and $b$ such that the orientation of $\gamma'$ (induced by the orientation of $\gamma$) points from $b$ to $a$. We say that $b$ is at the tail of $\gamma'$ and $a$ is at the head of $\gamma'$. We would like to show that $b = a^{-1}$. 

\begin{figure*}[tb]
\centering
\subfloat[\label{fig: Wgammaprime}]{\includegraphics[scale=.3]{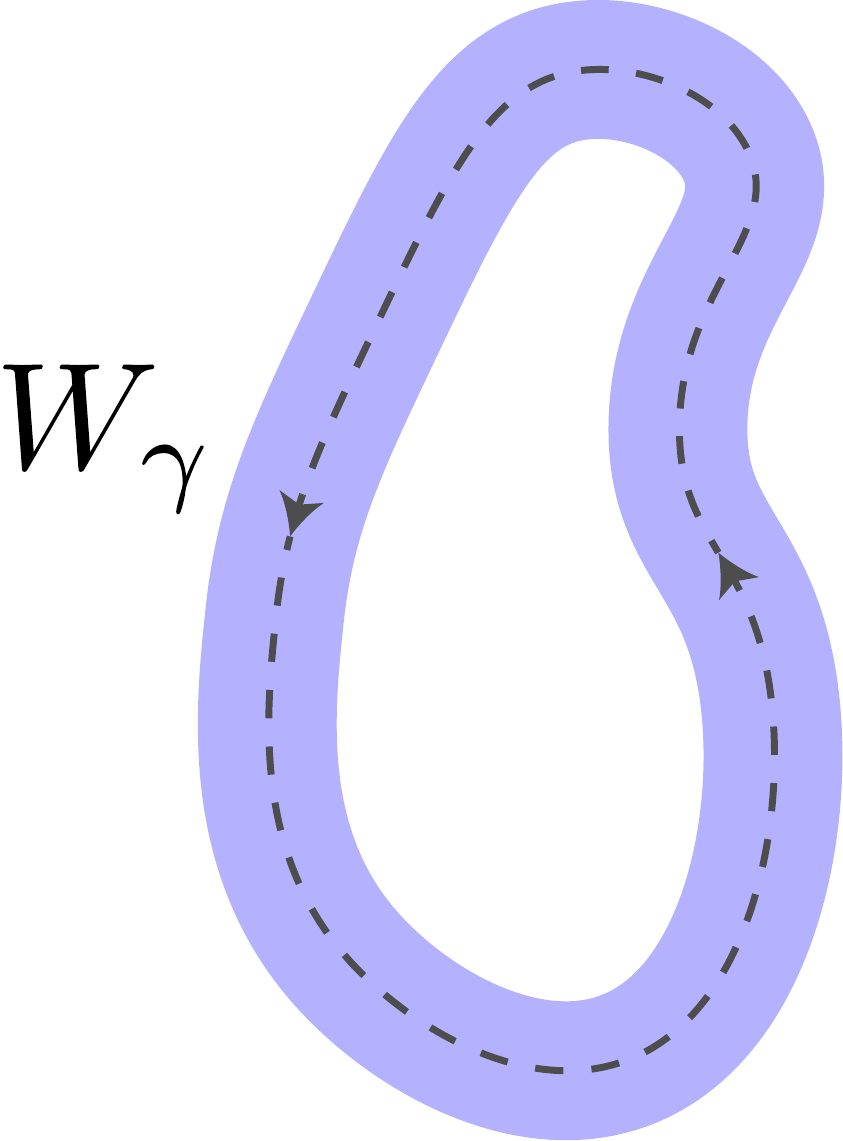}} \qquad
\subfloat[\label{fig: Wgammaprimetr1}]{\includegraphics[scale=.3]{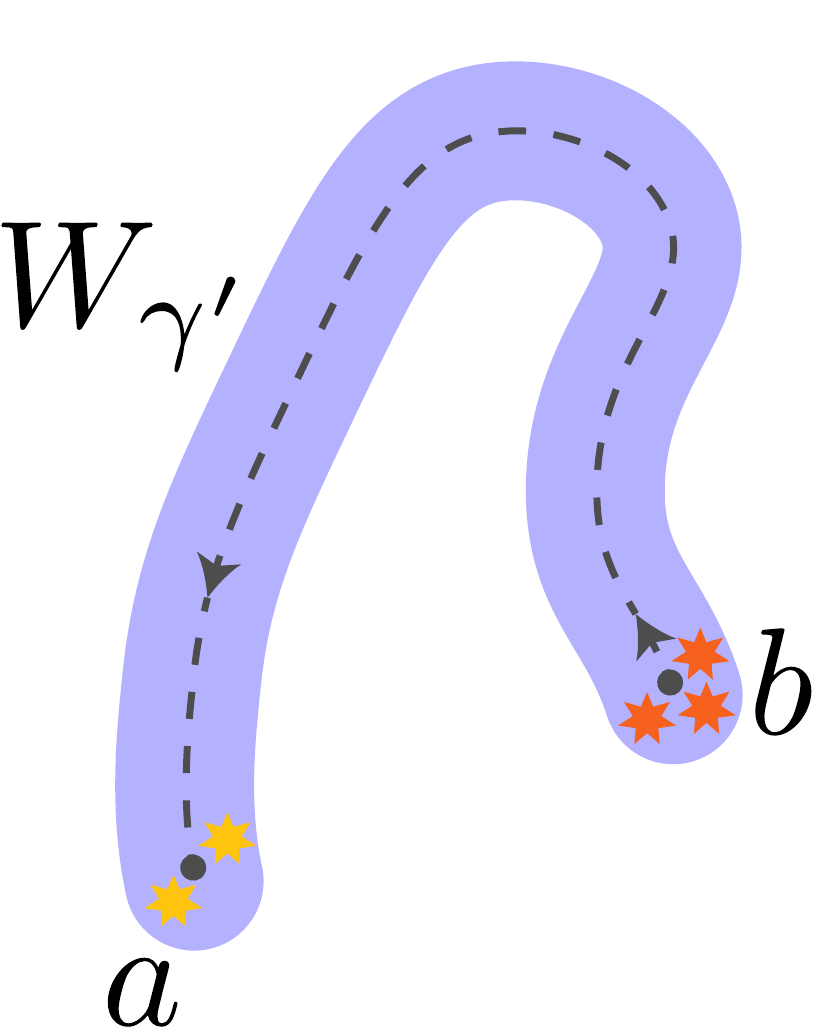}} \qquad
\subfloat[\label{fig: Wgammaprimetr2}]{\includegraphics[scale=.3]{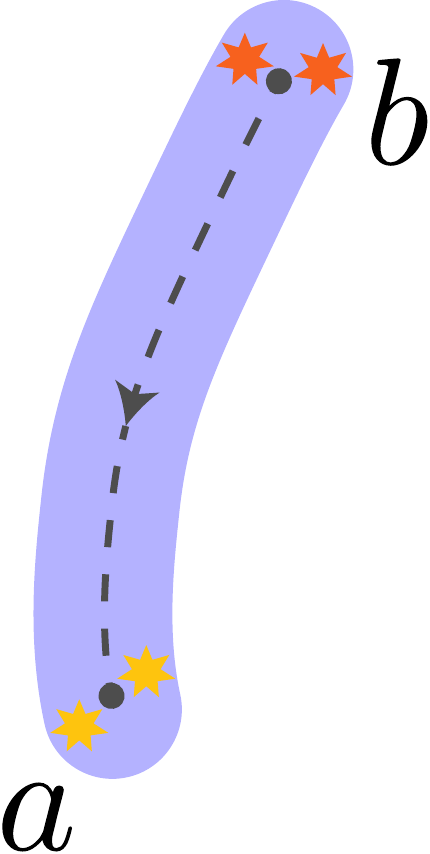}} \qquad
\subfloat[\label{fig: Wgammaprimetr3}]{\includegraphics[scale=.3]{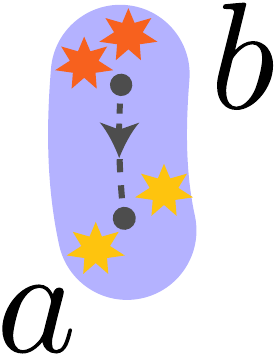}}
\caption{(a),(b) The string operator $W_\gamma$ is truncated to a path $\gamma'$. This defines the string operator $W_{\gamma'}$. (c),(d) The string operator $W_{\gamma'}$ is truncated in constant-sized increments, moving the anyon type $b$ to the anyon type $a$.}
\label{fig: Wgammaprimetruncations}
\end{figure*}

To relate the anyon types $b$ and $a$ at the endpoints of $W_{\gamma'}$, we move $b$ to $a$ by truncating $W_{\gamma'}$ in constant-sized increments, as illustrated in Fig.~\ref{fig: Wgammaprimetruncations}. Each successive truncation is equivalent to applying a product of Pauli operators at the endpoints, so the anyon type $b$ is unchanged by the truncations (by the definition of the anyon type $b$). Once $b$ has been moved to $a$ along $\gamma'$, the pair can be annihilated by using a local Pauli operator. This shows that $b$ is the inverse anyon type of $a$, i.e., $b = a^{-1}$. Note that, in general, topological subsystem codes may have nontrivial defects along closed paths, which permute the anyon types or allow certain anyon types to be condensed. However, assuming that the system is translation invariant, we can block unit cells together, sometimes referred to as ``coarse graining''. After a sufficient amount of coarse graining, we can assume that there are no nontrivial defects.

\begin{figure}[tb]
\centering
\includegraphics[scale=.3]{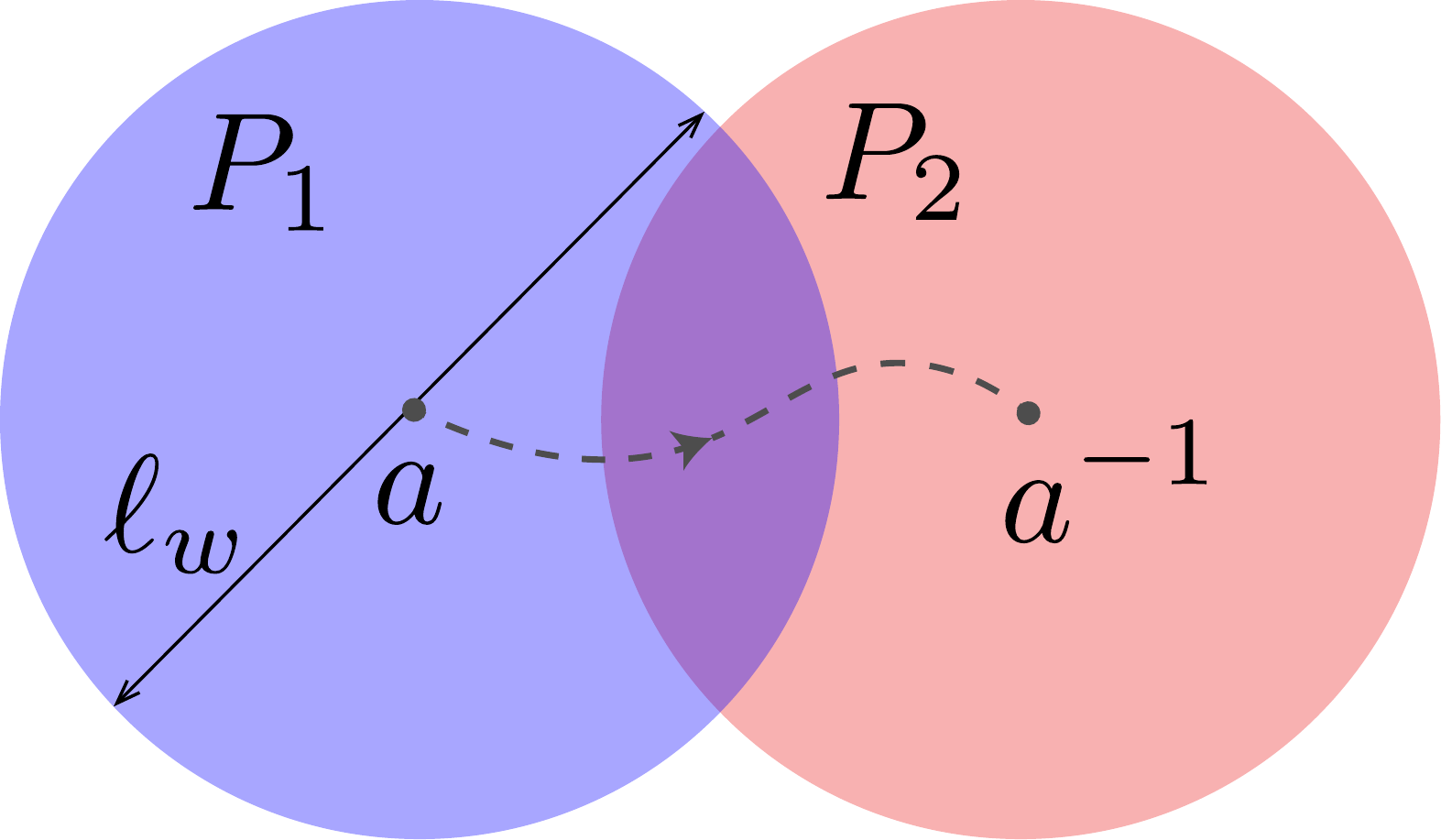}
\caption{The truncation of a string operator of width $\ell_W$ is ambiguous up to Pauli operators $P_1$ and $P_2$ whose supports are contained within a disk of linear size $\ell_W$ centered on the endpoint of the string operator (blue and red, respectively). If the endpoints are further than a distance $\ell_W + \ell_G$ from one another, then the Pauli operators $P_1$ and $P_2$ fail to commute with disjoint sets of gauge generators. Otherwise, the product of $P_1$ and $P_2$ may be a short string operator that moves the anyon type $a$ from one endpoint to the other. In such a case, the truncation of the string operator does not define a unique anyon type for the string operator, because the anyon types at the endpoints can be changed, given the ambiguity of the truncation. }
\label{fig: endpointsclose}
\end{figure}

\begin{figure*}[tb]
\centering
\subfloat[\label{fig: Wgamma}]{\includegraphics[scale=.3]{Figures/Wgammaloop.pdf}} \qquad
\subfloat[\label{fig: Wgamma1}]{\includegraphics[scale=.3]{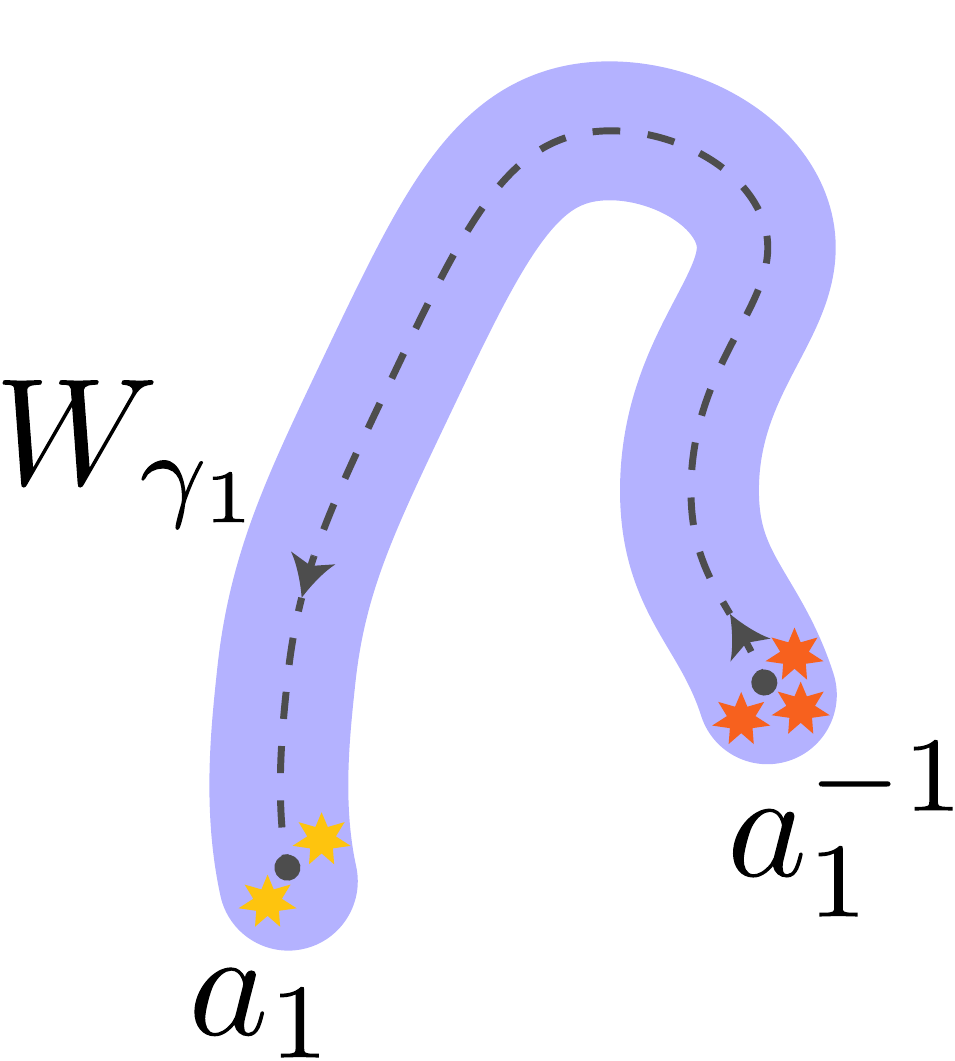}} \qquad
\subfloat[\label{fig: Wgamma2}]{\includegraphics[scale=.3]{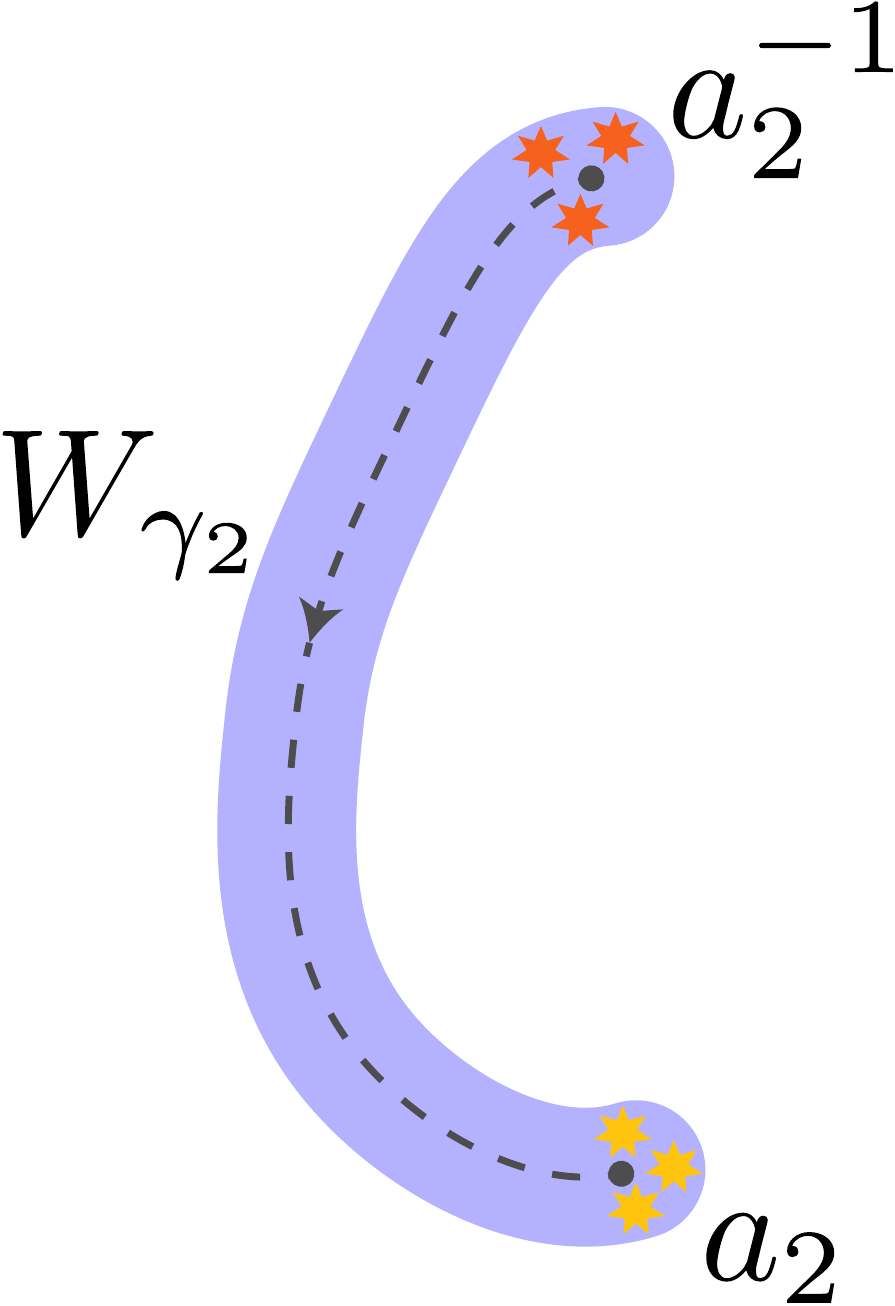}} \qquad
\subfloat[\label{fig: Wgamma3}]{\includegraphics[scale=.3]{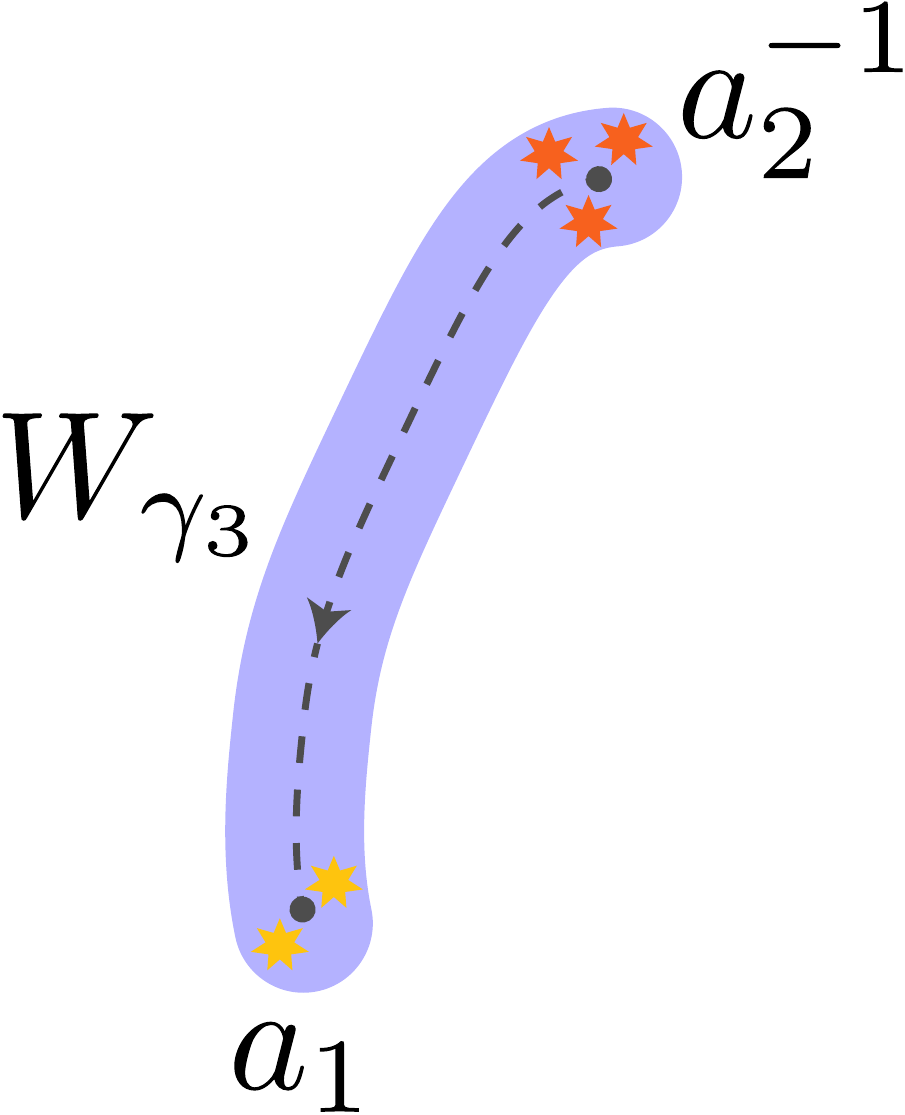}} \qquad
\caption{(a-c) The string operator $W_\gamma$, supported along the loop $\gamma$, is truncated in two different ways -- to the paths $\gamma_1$ and $\gamma_2$. (c) We define a third truncation of $W_\gamma$ to $\gamma_3$, where $\gamma_3$ goes from the tail of $\gamma_2$ to the head of $\gamma_1$.}
\label{fig: Wgammatruncations}
\end{figure*}

With this, we are able to argue that the anyon type associated to $W_\gamma$ is well defined. Let $W_{\gamma_1}$ and $W_{\gamma_2}$ be two truncations of $W_\gamma$, as shown in Figs.~\ref{fig: Wgamma1} and \ref{fig: Wgamma2}. The truncations are arbitrary, except for the requirement that the endpoints of the truncated string operators are well separated. Specifically, we require that the endpoints of the paths are greater than a distance $\ell_W+\ell_G$ from one another. Here, $\ell_G$ is a constant-sized length such that, for some choice of local gauge generators, the support of each gauge generator can be contained within a region of linear size $\ell_G$. This separation of the endpoints ensures that the product of Pauli operators supported on the disks of diameter $\ell_W$ at the endpoints of the paths $\gamma_1$ and $\gamma_2$ are only able to create trivial anyon types at the respective endpoints. For example, if the endpoints are closer than a distance $\ell_W+\ell_G$ from one another, then the truncation could be ambiguous up to a short string operator that changes the anyon types at the endpoints (see Fig.~\ref{fig: endpointsclose}).

We label the anyon types at the endpoints of $W_{\gamma_1}$ by $a_1^{-1}$ and $a_1$, so that the orientation of $\gamma_1$ points from $a_1^{-1}$ to $a_1$. Likewise, we label the anyon types at the endpoints of $W_{\gamma_2}$ by $a_2^{-1}$ and $a_2$. We then consider a third truncation of $W_\gamma$ to a path $\gamma_3$, stretching from the tail of $\gamma_2$ to the head of $\gamma_1$ (see Fig.~\ref{fig: Wgamma3}). We define the truncation $W_{\gamma_3}$ to $\gamma_3$ such that the anyon types at the endpoints are $a_2^{-1}$ and $a_1$, where the orientation of $\gamma_3$ pointing from $a_2^{-1}$ to $a_1$. Since, $W_{\gamma_3}$ is a truncation of $W_\gamma$, the anyon types at its endpoints must be inverses of one another. Therefore, we have that $a_1$ is equal to $a_2$. This shows us that the arbitrary truncations $W_{\gamma_1}$ and $W_{\gamma_2}$ give the same anyon type for the string operator $W_\gamma$.~$\square$ \\

The next two propositions show that the string operator obtained by moving an anyon type $a$ around a non-contractible path is a nonlocal element of $\mathcal{L}$ if and only if $a$ is a nontrivial anyon type. 
For clarity, we have divided the proof into Propositions~\ref{prop: anyon strings trivial} and \ref{prop: anyon strings}. 
As a corollary of Proposition~\ref{prop: anyon strings trivial}, we show that a string operator $W^a_\gamma$ for an anyon type $a$ can be moved by a product of local stabilizers to any other string operator $W^a_{\gamma'}$ for any path $\gamma'$ that is homologous to $\gamma$ (see Appendix~\ref{app: cellular} for a definition of homologous). Intuitively, this means that the string operators can be wiggled around by products of local stabilizers.

\begin{proposition} \label{prop: anyon strings trivial}
Every string operator created by moving a trivial anyon type along a closed path is a product of local stabilizers along that path.
\end{proposition}

\begin{figure*}[t]
\centering
\subfloat[\label{fig: trivialstringblocking}]{\raisebox{1.5cm}{\hbox{\includegraphics[width=.6\textwidth]{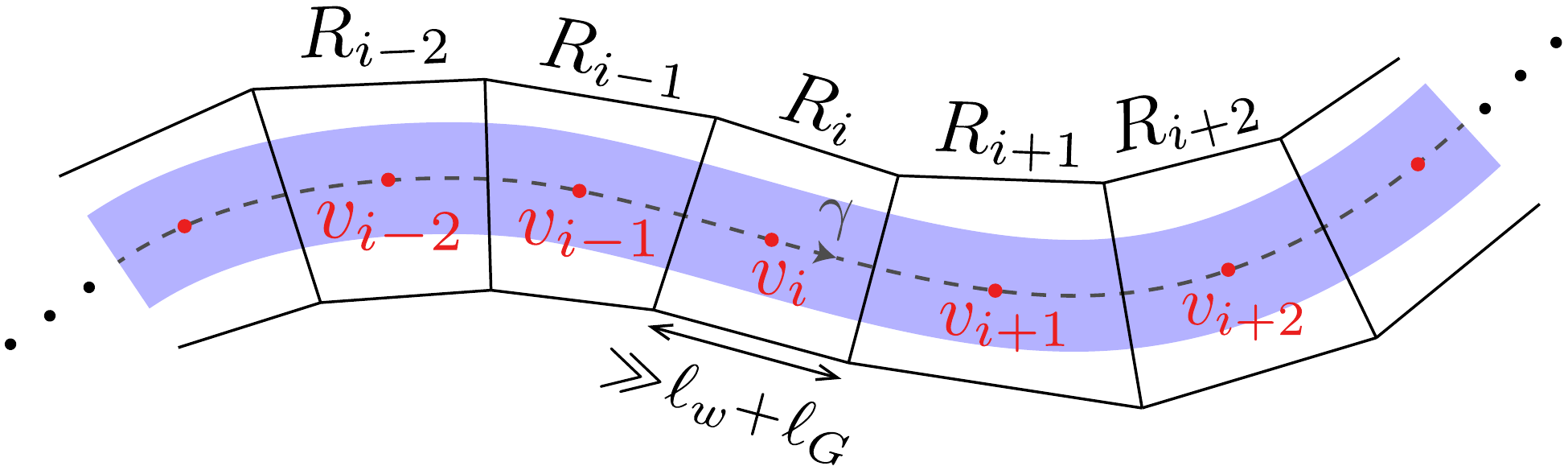}}}} \qquad
\subfloat[\label{fig: trivialshortstring}]{\includegraphics[scale=.5]{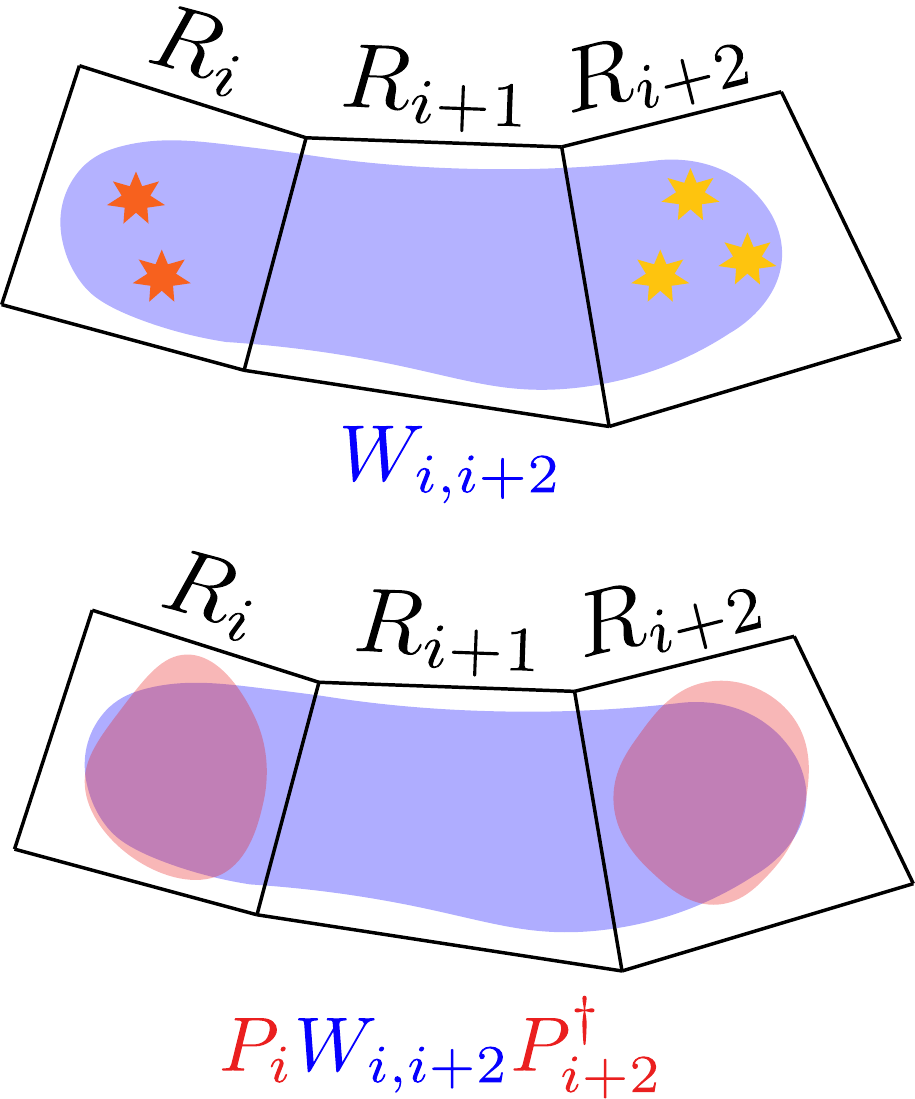}}
\caption{(a) We divide the support of the string operator $W_\gamma$ (blue) into regions $R_i$ of linear size greater than $\ell_W+\ell_G$. We require that each region contains a point $v_i$ (red) along $\gamma$, which is more than a distance $\ell_W+\ell_G$ from the boundary of the region $R_i$. (b) The short string operator $W_{i,i+2}$ (blue) may fail to commute with gauge generators (yellow and orange) in $R_i$ and $R_{i+2}$. The short string operator $P_iW_{i,i+2}P^\dagger_{i+2}$ is modified by Pauli operators $P_i$ and $P_{i+2}$ (red) at its endpoints and commutes with all of the gauge generators.}
\label{fig: trivialstringsprop}
\end{figure*}

\noindent \emph{Proof of Proposition~\ref{prop: anyon strings trivial}:} Let $W_\gamma$ be a string operator formed by moving a trivial anyon type along a closed path $\gamma$. Similar to the proof of Lemma~\ref{lem: string operator anyon types}, we assume that there is a constant-sized length $\ell_W$ such that the string operator $W_\gamma$ is entirely supported within a distance $\ell_W/2$ of $\gamma$ (Fig.~\ref{fig: stringwidth}). We also use the constant-sized length $\ell_G$ defined by the property that every gauge generator can be contained within a region of linear size $\ell_G \times \ell_G$, for some fixed set of local gauge generators.

As shown in Fig.~\ref{fig: trivialstringblocking}, we break the support of $W_\gamma$ into constant-sized regions, whose linear size is at least $\ell_W+\ell_G$. We further require that each region $R_i$ contains a point $v_i$ along $\gamma$, such that $v_i$ is at least a distance $\ell_W+\ell_G$ from the boundary of $R_i$. We then decompose $W_\gamma$ into a product of short string operators $W_{i,i+2}$:
\begin{align} \label{eq: trivial string decomposition}
W_{\gamma} = \prod_i W_{i,i+2},
\end{align}
such that $W_{i, i+2}$ is a truncation from $v_i$ to $v_{i+2}$. Since the endpoints $v_i$ and $v_{i+2}$ of the short string operator are greater than a distance $\ell_W+\ell_G$ from one another, we can apply Lemma~\ref{lem: string operator anyon types}. According to Lemma~\ref{lem: string operator anyon types}, each short string operator $W_{i,i+2}$ creates trivial anyon types localized to $R_i$ and $R_{i+2}$. Therefore, there exists a Pauli operator $P_i$ supported in the vicinity of $v_i$ such that $P_iW_{i,i+2}$ commutes with  the gauge operators in $R_i$ (Fig.~\ref{fig: trivialshortstring}). Since $W_\gamma$ commutes with the gauge operators in $R_i$, the product $W_{i-2,i}P_{i}^\dagger$ must also commute with all of the gauge operators in $R_{i}$.

We now define the operator $S_i$ for each $i$ as:
\begin{align}
S_i \equiv P_i W_{i,i+2} P^\dagger_{i+2}.
\end{align}
By construction, $S_i$ is contained in a constant-sized region and commutes with all of the gauge operators. The topological property of Definition~\ref{def: topological subsystem code} implies that $S_i$ is a locally generated stabilizer. The product of $S_i$ reproduces the expression for $W_\gamma$ in Eq.~\eqref{eq: trivial string decomposition}:
\begin{align}
W_\gamma = \prod_i S_i = \prod_i \left( P_i W_{i,i+2} P^\dagger_{i+2} \right) = \prod_i W_{i,i+2}.
\end{align}
Thus, the string operator $W_\gamma$ is a product of local stabilizers supported along $\gamma$.~$\square$ 

\begin{figure}[tb] 
\centering
\includegraphics[width=.3\textwidth]{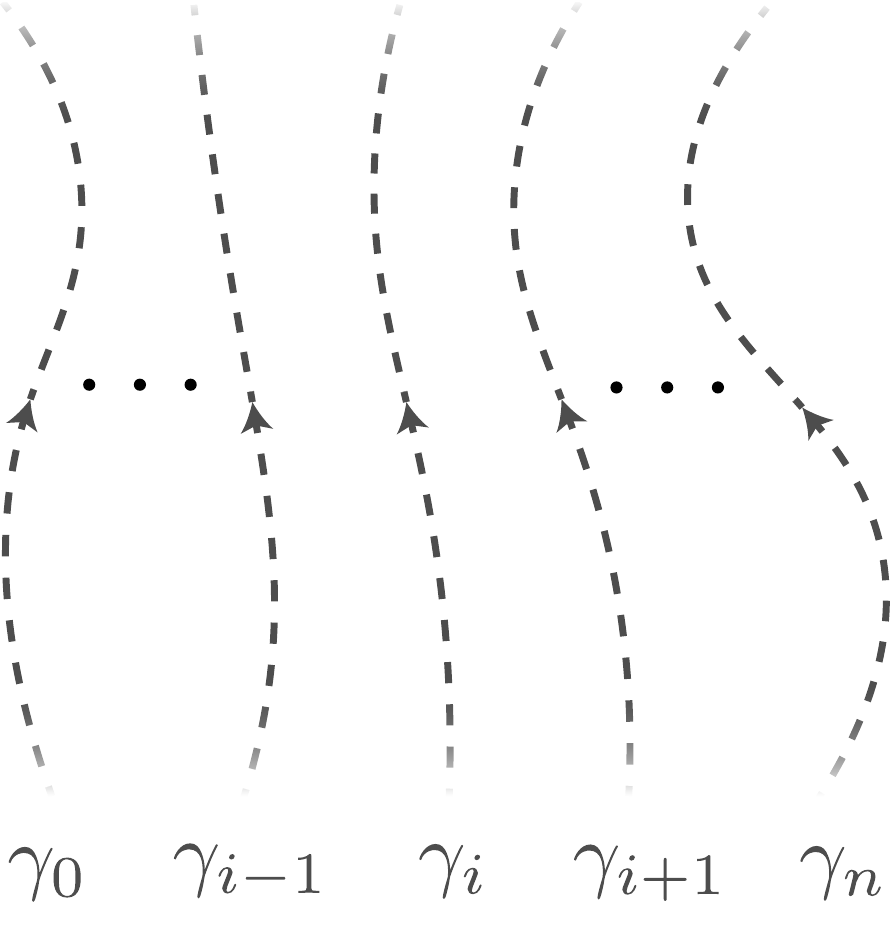}
\caption{Two homologous closed paths $\gamma_0$ and $\gamma_n$ can be interpolated between by a sequence of paths indexed by $i$. }
\label{fig: stringinterpolation}
\end{figure}

\begin{corollary} \label{cor: wiggle strings}
Any two string operators that correspond to the same anyon type and are supported along homologous paths differ by a product of local stabilizers.
\end{corollary}

\noindent \emph{Proof of Corollary~\ref{cor: wiggle strings}:} Consider any two string operators $W^a_{\gamma_0}$ and $W^{a}_{\gamma_n}$ for an anyon type $a$, where the closed paths $\gamma_0$ and $\gamma_n$ are homologous. Given that $\gamma_0$ and $\gamma_n$ are homologous, we can find a sequence of paths, indexed by $i$, that interpolates between the two (see Fig.~\ref{fig: stringinterpolation}). According to Proposition~\ref{prop: anyon strings trivial}, the string operator $(W^a_{\gamma_{i-1}})^{-1}W^{a}_{\gamma_{i}}$ is a product of local stabilizers, since it is a string operator for a trivial anyon type. Therefore, $W^{a}_{\gamma_n}$ can be written as:
\begin{align}
W^{a}_{\gamma_n} = W^a_{\gamma_0} \prod_{i=0}^n \left[ (W^a_{\gamma_{i-1}})^{-1}W^{a}_{\gamma_{i}}\right],
\end{align}
which shows that $W^a_{\gamma_0}$ and $W^{a}_{\gamma_n}$ differ by a product of local stabilizers.~$\square$

\begin{proposition} \label{prop: anyon strings}
Every closed string operator created by moving a nontrivial anyon type along a non-contractible path is either a nontrivial bare logical operator or a nonlocal stabilizer.
\end{proposition}

\begin{figure*}[t]
\centering
\subfloat[\label{fig: R0}]{\includegraphics[width=.3\textwidth]{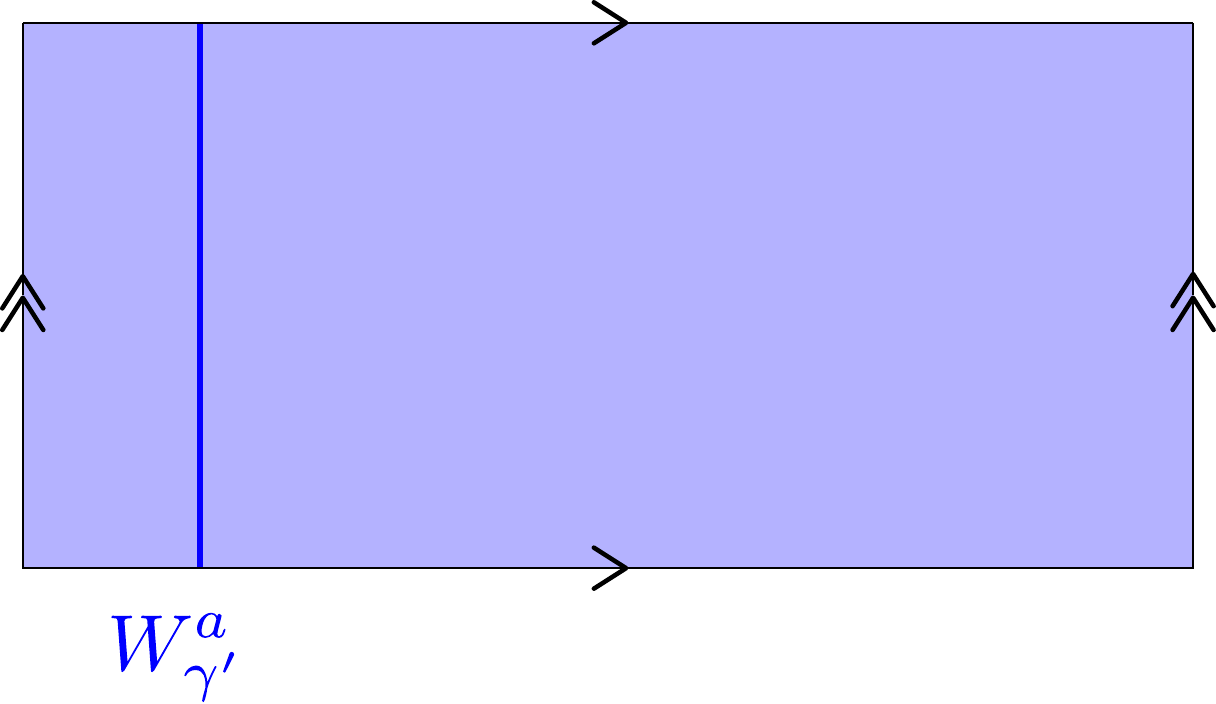}} \quad
\subfloat[\label{fig: R1}]{\includegraphics[width=.3\textwidth]{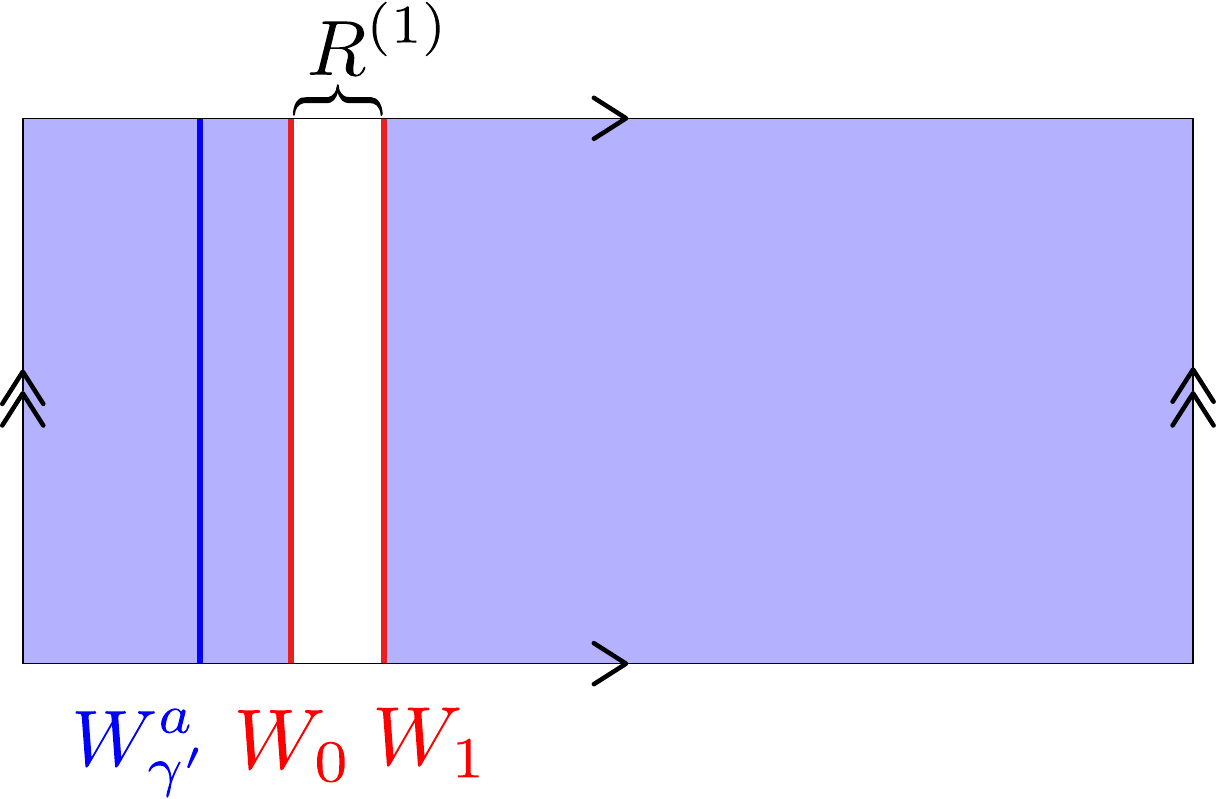}} \quad
\subfloat[\label{fig: R2}]{\includegraphics[width=.3\textwidth]{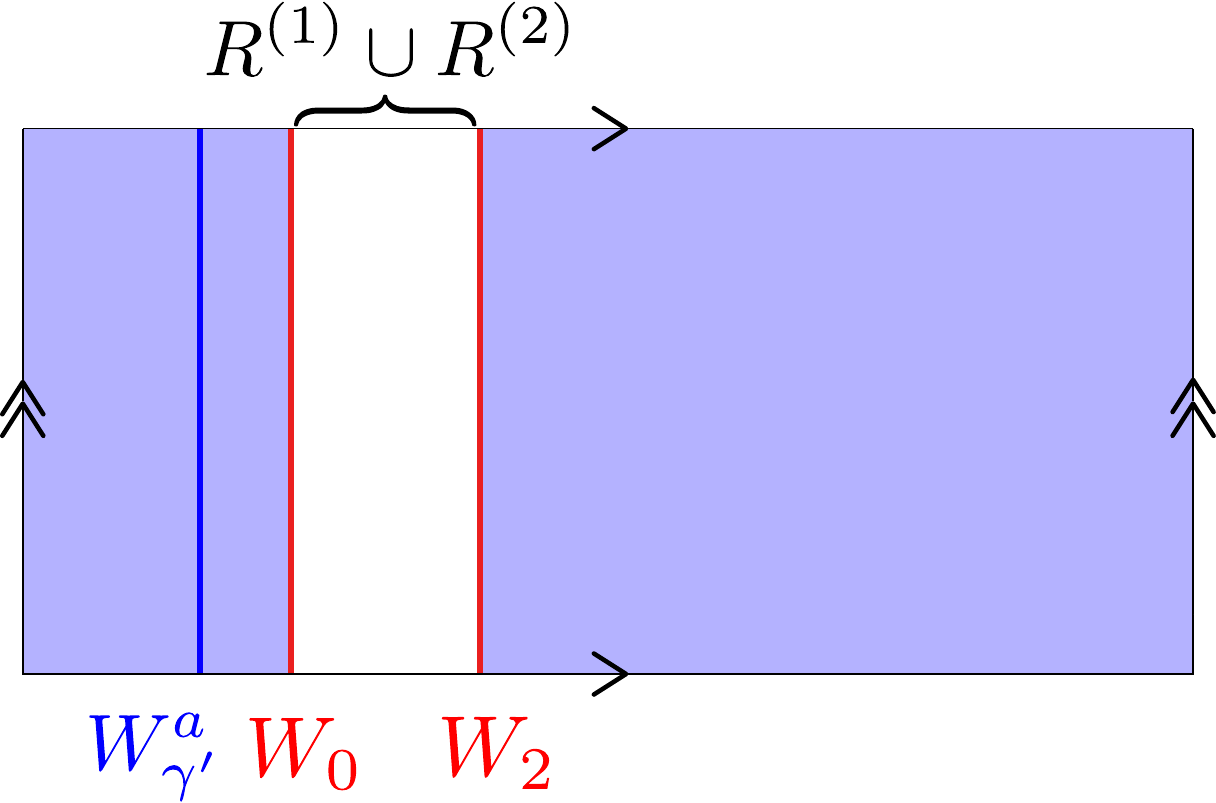}} \\
\subfloat[\label{fig: R3}]{\includegraphics[width=.3\textwidth]{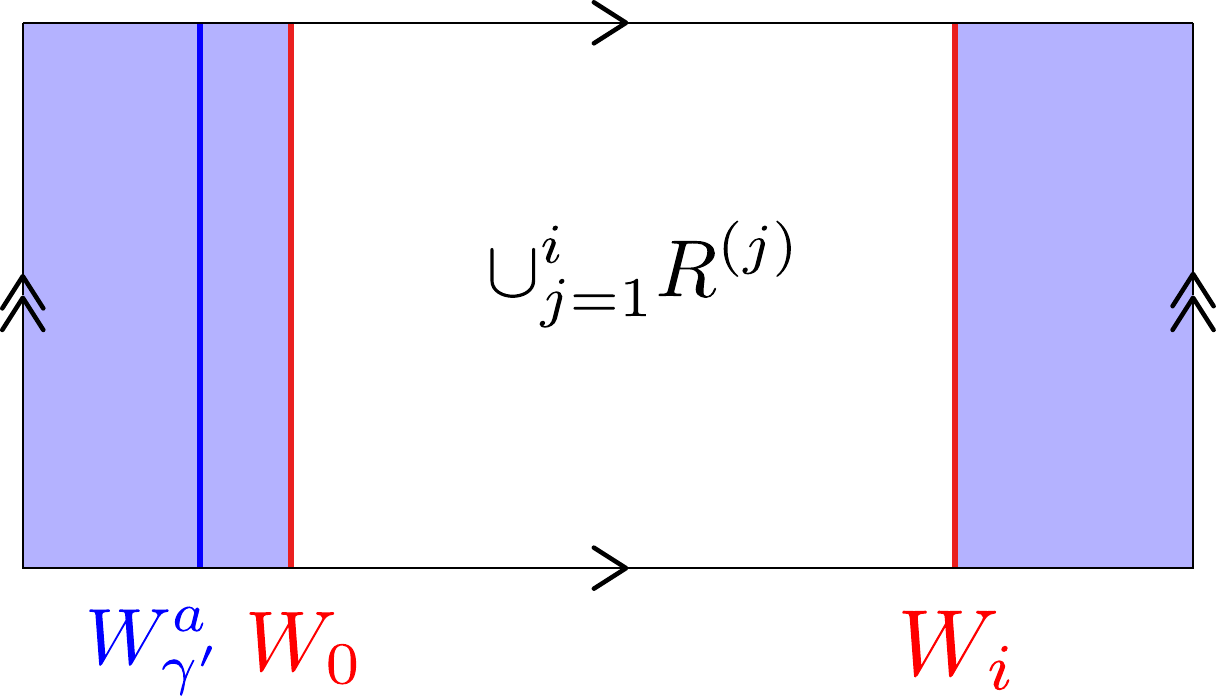}} \quad
\subfloat[\label{fig: R4}]{\includegraphics[width=.3\textwidth]{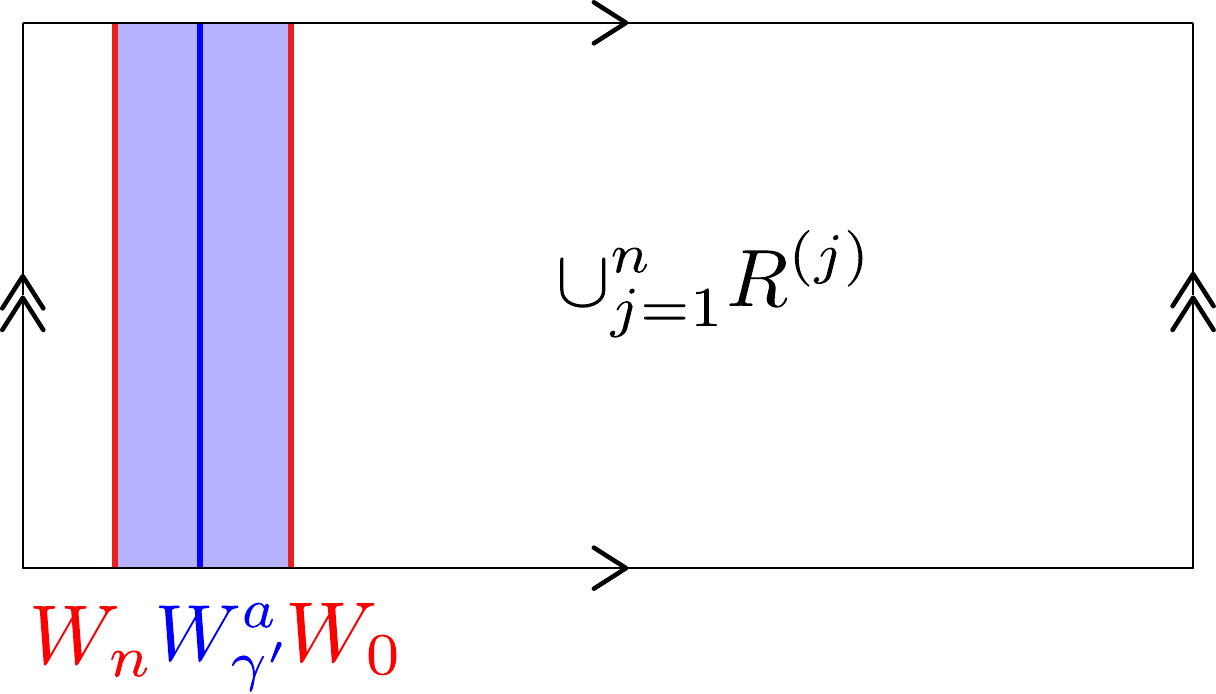}} 
\caption{(a) $W^a_{\gamma'}$ is supported along a non-contractible path $\gamma'$ (dark blue). We assume that $W^a_{\gamma'}$ is generated by local stabilizers supported anywhere on the torus (light blue). (b-e) We successively remove local stabilizers along constant-width strips $R^{(j)}$ from the expansion of $W^a_{\gamma'}$ in terms of local stabilizers. The resulting operator may act non-identically near the boundary of the removed regions (red).}
\label{fig: Rs}
\end{figure*}

\noindent \emph{Proof of Proposition~\ref{prop: anyon strings}:} Let $W^a_{\gamma'}$ be a string operator formed by moving a nontrivial anyon type $a$ around the non-contractible path $\gamma'$ shown in Fig.~\ref{fig: R0}. 
The string operator $W^a_{\gamma'}$ commutes with all of the gauge operators, so it must be either a nontrivial bare logical operator or a stabilizer. Our task is to show that it cannot be generated by products of local stabilizers, implying that it is a nonlocal element of $\mathcal{L}$. To derive a contradiction, suppose that $W^a_{\gamma'}$ is generated by local stabilizers, which in general, may be supported anywhere on the torus -- not necessarily in the vicinity of $W^a_{\gamma'}$. We can then write $W^a_{\gamma'}$ as the following product of local stabilizers (Fig.~\ref{fig: R0}):
\begin{align} \label{eq: stabilizers generating string}
W^a_{\gamma'} = \prod_{S \in \tilde{\mathsf{S}}_\text{loc}} \xi(S) S.
\end{align}
Here, $\tilde{\mathsf{S}}_\text{loc}$ is a set of local generators for the subgroup of locally generated stabilizers $\tilde{S}$, and $\xi(S)$ is a $\{0,1\}$-valued function defined by Eq.~\eqref{eq: stabilizers generating string}.

We now consider removing local stabilizers from the decomposition of $W^a_{\gamma'}$ in Eq.~\eqref{eq: stabilizers generating string}. Specifically, we remove local stabilizers from the strip $R^{(1)}$ of constant width shown in Fig.~\ref{fig: R1}. The width of $R^{(1)}$ is taken to be much larger than the linear size of the support of the local gauge generators. We also assume that the boundaries of $R^{(1)}$ are a constant distance from the support of $W^a_{\gamma'}$. By removing the local stabilizers within $R^{(1)}$, the string operator $W^a_{\gamma'}$ becomes the operator:
\begin{align} \label{eq: W(1) def}
W^{(1)}_{\gamma'} \equiv \prod_{S \in \tilde{\mathsf{S}}_\text{loc}} \xi^{(1)}(S) S,
\end{align}
where $\xi^{(1)}(S)$ is equal to $\xi(S)$ if the support of $S$ is outside of $R^{(1)}$ and $0$ otherwise. The operator $W^{(1)}_{\gamma'}$ acts as $W^a_{\gamma'}$ along $\gamma'$, and may act non-identically along the boundary of $R^{(1)}$. We define $W_0$ and $W_1$ to be the string operators supported near the boundary of $R^{(1)}$ such that:
\begin{align} \label{eq: prod of strings 1}
W^{(1)}_{\gamma'} = W^a_{\gamma'} W_0 W_1.
\end{align}

The string operators $W_0$ and $W_1$ commute with all of the gauge operators. This is because the operator $W_0W_1$ is a product of local stabilizers and the supports of $W_0$ and $W_1$ are well-separated relative to the linear size of the gauge generators. Consequently, $W_0$ and $W_1$ can be interpreted as string operators formed by moving anyon types around a non-contractible path (Lemma~\ref{lem: string operator anyon types}). 
We assume that ${W}_0$ corresponds to an anyon type $b$, in which case ${W}_1$ must correspond to the anyon type $b^{-1}$. This is due to the fact that $W_0W_1$ is a product of local stabilizers within $R^{(1)}$. For any truncation, there exists Pauli operators localized near the endpoints of the truncation such that the product of the Pauli operators and the truncated string operator commute with all of the gauge operators. This implies that the string operator ${W}_0 {W}_1$ corresponds to the trivial anyon type.

We next consider removing a second constant-sized strip of stabilizers from the decomposition of $W^{a}_{\gamma'}$ in Eq.~\eqref{eq: stabilizers generating string}. This time, we remove stabilizers whose supports are contained within the region $R^{(2)}$ shown in Fig.~\ref{fig: R2}. Note that we take the overlap of $R^{(1)}$ and $R^{(2)}$ to be such that every local stabilizer generator in $\tilde{\mathsf{S}}_{loc}$ supported in $R^{(1)}\cup R^{(2)}$ is supported entirely in $R^{(1)}$ or $R^{(2)}$. The operator $W^{(1)}_{\gamma'}$ is mapped to:
\begin{align}
W^{(2)}_{\gamma'} \equiv \prod_{S \in \tilde{\mathsf{S}}_\text{loc}} \xi^{(2)}(S) S,
\end{align}
where $ \xi^{(2)}(S)$ is equal to $ \xi(S)$ if the support of $S$ is outside of $R^{(1)} \cup R^{(2)}$ and $0$ otherwise. The operator $W^{(2)}_{\gamma'}$ acts as $W^{a}_{\gamma'}$ along $\gamma'$ and may act non-identically along the boundary of $R^{(1)} \cup R^{(2)}$. Therefore, it can be written as:
\begin{align}
W^{(2)}_{\gamma'} = W^{a}_{\gamma'} W_0 W_2.
\end{align}
Here, $W_0$ is the string operator in Eq.~\eqref{eq: prod of strings 1}, and $W_2$ is a string operator supported on the boundary of $R^{(2)}$, as depicted in Fig.~\ref{fig: R2}. The string operator $W_2$ commutes with all of the gauge operators and thus can be interpreted as a string operator obtained by moving an anyon type along a non-contractible path. Furthermore, the string operator $W_2$ must correspond the anyon type $b^{-1}$. This follows from translation invariance and the fact that $W_1^{-1}W_2$ is a product of stabilizers supported on $R^{(2)}$. Since $W_1$ corresponds to a string operator for $b^{-1}$, $W_2$ must be a string operator for $b^{-1}$. 

We can continue removing constant-width strips of stabilizers from the decomposition of $W^a_{\gamma'}$ in Eq.~\eqref{eq: stabilizers generating string}. In the $i$th step, we remove the stabilizers in the strip $R^{(i)}$ (Fig.~\ref{fig: R3}), which maps the operator $W^{(i-1)}_{\gamma'}$ to an operator $W^{(i)}_{\gamma'}$. The operator $W^{(i)}_{\gamma'}$ can be expressed as:
\begin{align}
W^{(i)}_{\gamma'} = W^{a}_{\gamma'} W_0 W_i,
\end{align}
where $W_i$ is a string operator supported along the boundary shared by $R^{(i)}$ and $\bigcup_{j=1}^i R^{(j)}$. The operator $W_i$ is a string operator corresponding to the anyon type $b^{-1}$. 

For a sufficiently large $n$, the operator $W^{(n)}_{\gamma'}$ is a product of local stabilizers supported on a constant-width strip localized near $\gamma'$ (Fig.~\ref{fig: R4}). Moreover, $W^{(n)}_{\gamma'}$ is a string operator that can be written as:
\begin{align}
W^{(n)}_{\gamma'} = W^a_{\gamma'} W_0 W_n,
\end{align}
for some string operator $W_n$ supported near $\gamma'$, as shown in Fig.~\ref{fig: R4}. The operator $W_n$ corresponds to a string operator that moves $b^{-1}$ along a non-contractible path, while $W_0$ moves $b$ along the same loop. This means that all together, $W^{(n)}_{\gamma'}$ is a string operator for the anyon type $a$. 

This is a contradiction, however, because $W^{(n)}_{\gamma'}$ is a product of local stabilizers supported near $\gamma'$ even though $a$ is a nontrivial anyon type. 
Therefore, $W^{a}_{\gamma'}$ cannot be written as a product of local stabilizers, as in Eq.~\eqref{eq: stabilizers generating string}. The string operator $W^{a}_{\gamma'}$ formed by moving $a$ along $\gamma'$ must be either a nontrivial bare logical operator or a nonlocal stabilizer.~$\square$ \\

Thus far, we have shown that nontrivial bare logical operators and nonlocal stabilizers can be created by moving nontrivial anyon types around non-contractible loops. According to Appendix~\ref{app: detectable=opaque}, the string operator is a nontrivial bare logical operator if the anyon type is opaque. Otherwise, it is a nonlocal stabilizer and the anyon type is transparent.
We now argue that this construction exhausts all of the nontrivial bare logical operators and nonlocal stabilizers, in the sense that, up to stabilizers, or products of local stabilizers, respectively, all nontrivial bare logical operators and nonlocal stabilizers can be obtained from products of string operators created by moving nontrivial anyon types along non-contractible paths.

\begin{proposition}\label{prop: bare logicals and nonlocal stabilizers}
Every nontrivial bare logical operator and nonlocal stabilizer can be represented by products of string operators that correspond to moving nontrivial anyon types along non-contractible paths. 
\end{proposition}

\begin{figure*}[t]
\centering
\subfloat[\label{fig: Wa}]{\includegraphics[width=.3\textwidth]{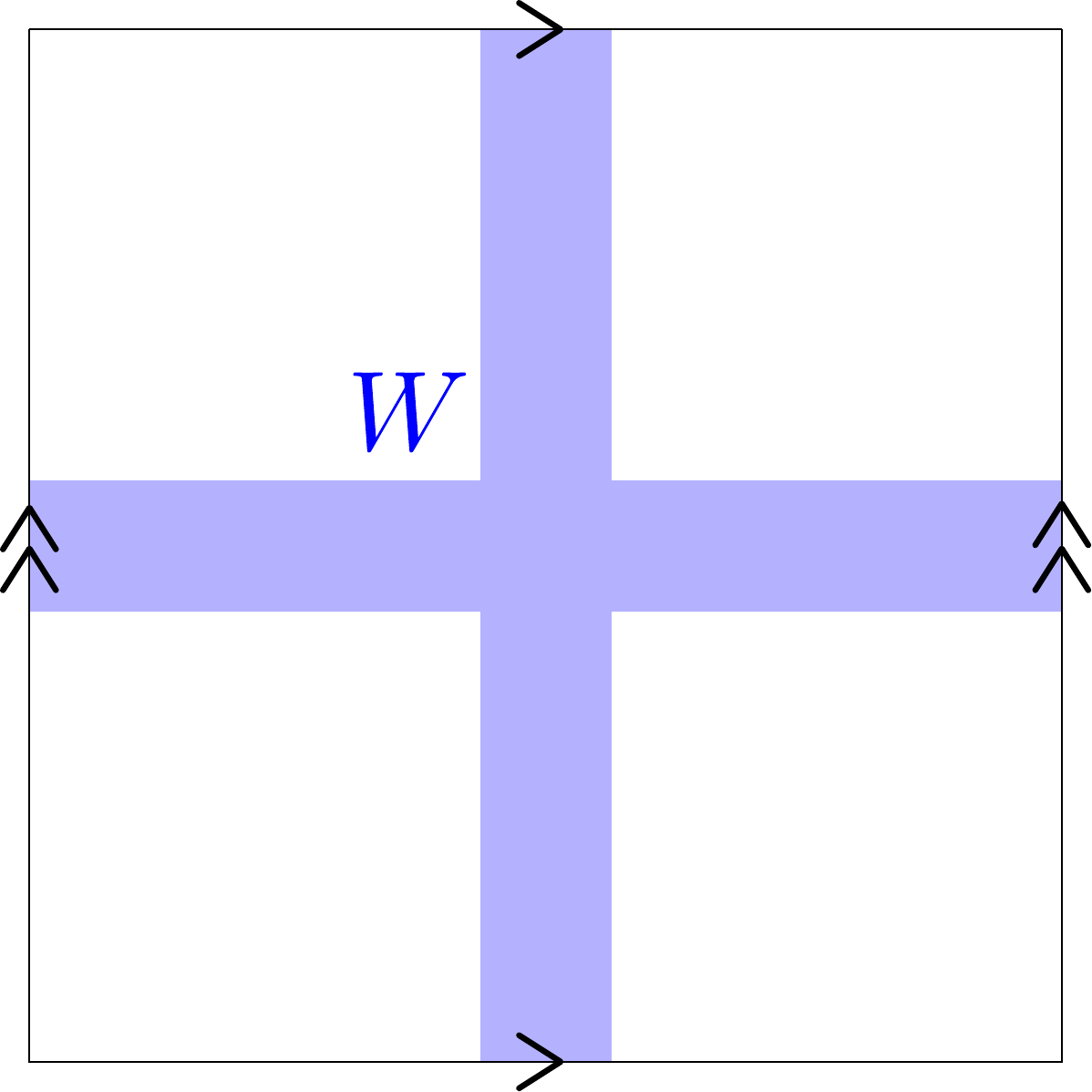}} \quad
\subfloat[\label{fig: Wb}]{\includegraphics[width=.3\textwidth]{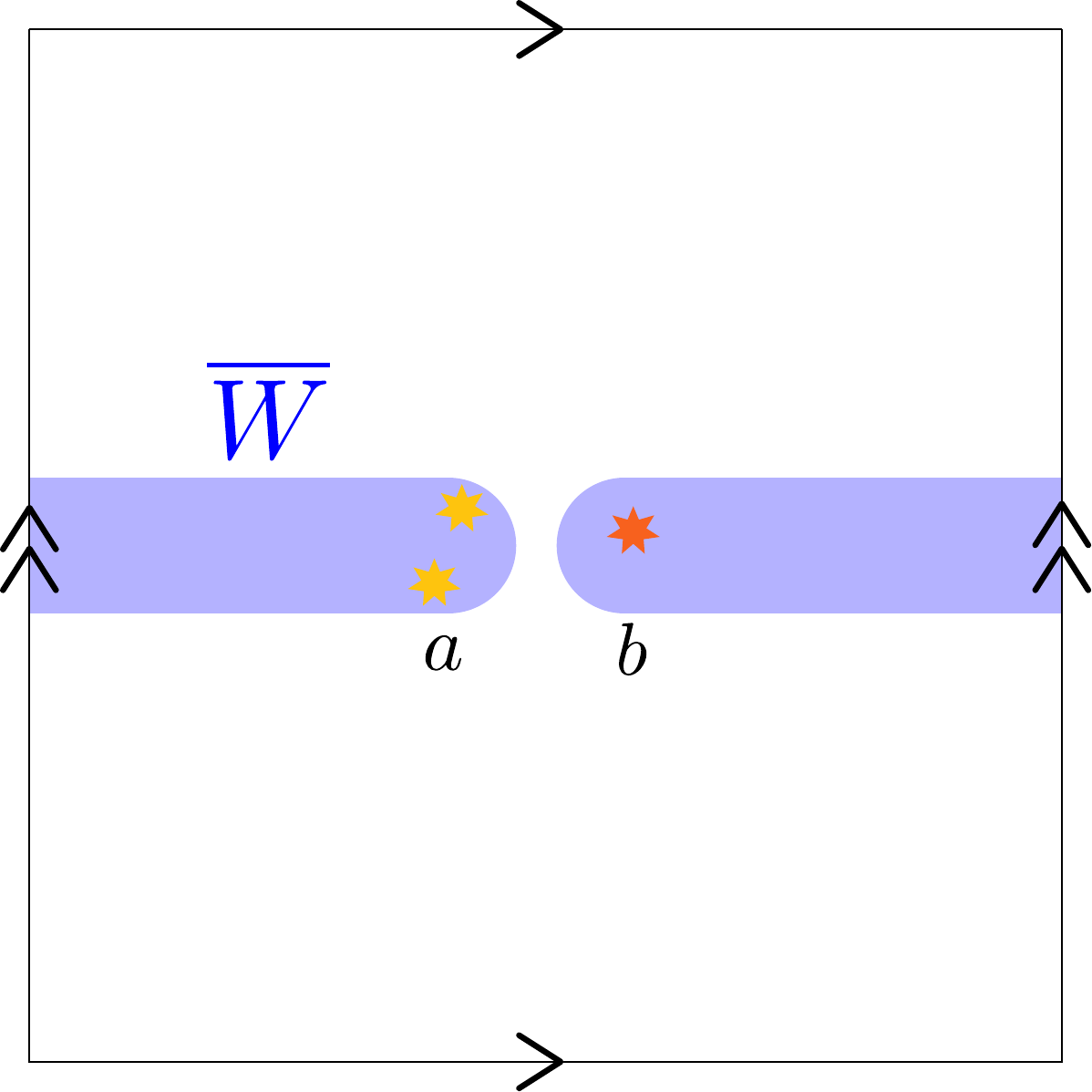}} \quad
\subfloat[\label{fig: Wf}]{\includegraphics[width=.3\textwidth]{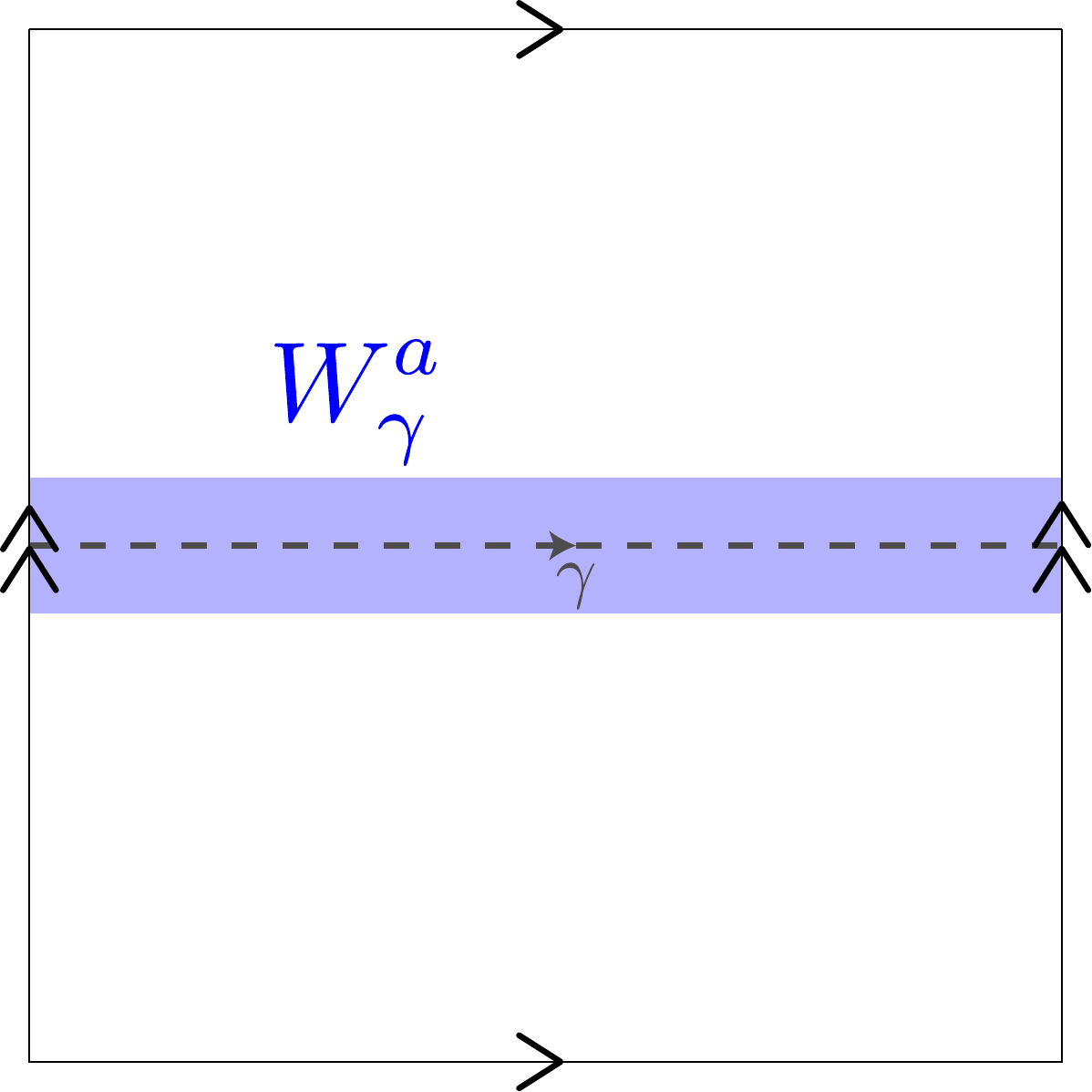}}
\caption{(a) $W$ is an arbitrary nontrivial bare logical operator or nonlocal stabilizer supported on intersecting strips (blue) with constant-sized width. (b) $\overline{W}$ is a truncation of $W$. The anyon types $a$ and $b$ at the endpoints must be inverses of one another. (c) The string operator $W^a_\gamma$ is constructed from $\overline{W}$ by annihilating the inverse anyon types $a$ and $b$.}
\label{fig: Ws}
\end{figure*}

\noindent \emph{Proof of Proposition~\ref{prop: bare logicals and nonlocal stabilizers}:} 
According to the cleaning lemma in Appendix~\ref{app: cleaning lemma}, every nontrivial bare logical operator and nonlocal stabilizer can be represented by an operator whose support is contained within the region shown in Fig.~\ref{fig: Wa}. 
We show further that the nontrivial bare logical operators and nonlocal stabilizers can be generated by string operators wrapped around the non-contractible paths $\gamma$ and $\gamma'$ in Fig.~\ref{fig: braiding}.
To see this, we consider truncating an arbitrary nontrivial bare logical operator or nonlocal stabilizer $W$ supported on the intersecting strips in Fig.~\ref{fig: Wa}. We let $\overline{W}$ denote the truncation of $W$ depicted in Fig.~\ref{fig: Wb}. The truncated operator $\overline{W}$ commutes with the gauge operators away from the endpoints. However, it may fail to commute with the gauge operators near the endpoints. Therefore, the endpoints may correspond to nontrivial anyon types. Regardless of whether they are nontrivial, we label the anyon types at the endpoints by $a$ and $a^{-1}$. The fact that they are inverses of one another follows from the same logic used in the proof of Lemma~\ref{lem: string operator anyon types}. 

With this, we can argue that $W$ decomposes into string operators formed by moving nontrivial anyon types along the paths $\gamma$ and $\gamma'$. Since the anyon types at the endpoints of $\overline{W}$ are inverses of each other, they are annihilated by a Pauli operator supported near the endpoints of $\overline{W}$. This gives us a string operator wrapped along the non-contractible path $\gamma$, as shown in Fig.~\ref{fig: Wf}. We denote the string operator by $W_\gamma^a$. We then define a string operator $W_{\gamma'}^c$ supported along the path $\gamma'$ as $W_{\gamma'}^c \equiv  \left( W_\gamma^a \right)^\dagger W$. Given that $W_{\gamma'}^c$ commutes with all of the gauge operators,  it corresponds to moving some anyon type $c$ around the loop $\gamma'$. Therefore, the operator $W = W_\gamma^a W_{\gamma'}^c$ is generated by the non-contractible loops of string operators $W_\gamma^a$ and $W_{\gamma'}^c$. The nontrivial bare logical operator or nonlocal stabilizer $W$ was arbitrary, so we see that every nontrivial bare logical operator or nonlocal stabilizer can be represented by a product of string operators such as $W_\gamma^a$ and $W_{\gamma'}^c$. 

Furthermore, Proposition~\ref{prop: anyon strings trivial} tells us that the closed string operators for trivial anyon types are equivalent to products of local stabilizers. This means that either $a$ or $c$ must be nontrivial, since $W$ is a nontrivial bare logical operator or a nonlocal stabilizer. If $a$ or $c$ is trivial, then the string operator $W_\gamma^a$ or $W_{\gamma'}^c$, respectively, can be removed by multiplying by products of local stabilizers. We find that the resulting nontrivial bare logical operator or nonlocal stabilizer is represented by products of string operators for nontrivial anyon types, as claimed.~$\square$ \\

\section{Background on twisted quantum doubles} \label{app: TQD review}

A TQD with Abelian anyons is (non-uniquely) specified by three pieces of data \cite{wang2015topological}:
\begin{enumerate}
\item A finite Abelian group $A$, which takes the general form 
$A = \prod_i^M \ZZ_{N_i}$.
\item A set of integers $\{n_i\}_{i=1}^M$, where $n_i$ belongs to $\ZZ_{N_i}$.
\item A set of integers $\{n_{ij}\}_{i,j =1}^M$ for $i \neq j$, where $n_{ij}$ belongs to $\ZZ_{N_{ij}}$ and $N_{ij}$ is the greatest common divisor of $N_i$ and $N_j$.
\end{enumerate}
The anyons of the TQD can be generated by a set of $M$ gauge charges $\{c_i\}_{i=1}^M$ and a choice of $M$ corresponding elementary fluxes $\{\varphi_i\}_{i=1}^M$. The gauge charges are unique up to  
automorphisms of the anyon types -- for example, the gauge charge of a $\ZZ_2$ TC can be chosen as either the $e$ anyon type or the $m$ anyon type. The gauge charges generate the group $A$ under fusion and are bosons with trivial mutual braiding relations:
\begin{align} \label{eq: gauge charge statistics}
\theta(c_i) = 1, \quad B_\theta(c_i,c_j) =1, \quad \forall i,j. 
\end{align}
The elementary fluxes are defined by their braiding relations with the generators of the gauge charges:\begin{align} \label{eq: charge-flux braiding}
B_\theta(\varphi_i, c_j) = \begin{cases}
e^{2\pi i / N_i} & \text{if } i = j,\\
1 & \text{if } i \neq j. 
\end{cases}
\end{align}
This captures the Aharonov-Bohm phase from moving a charge around a flux.
A choice of elementary fluxes is only defined up to fusing with gauge charges, since fusing $\varphi_i$ with gauge charges does not affect the relations in Eq.~\eqref{eq: charge-flux braiding}. Accordingly, the exchange statistics and braiding relations of the elementary fluxes are only determined up to redefinitions by gauge charges. Irrespective of the redefinitions, every choice of elementary fluxes $\{\varphi_i\}_{i=1}^M$ satisfies:
\begin{align}
\theta(\varphi_i)^{N_i} = e^{2 \pi i n_i / N_i}, \quad B_\theta(\varphi_i, \varphi_j)^{N^{ij}} = e^{2 \pi i n_{ij}/N_{ij}},
\end{align}
where $N^{ij}$ is the least common multiple of $N_i$ and $N_j$. Lastly, for each $i$ and some set of integers $m_{ij}$ (associated to redefining the elementary fluxes by gauge charges),
the gauge charges and elementary gauge fluxes obey the following two relations under fusion:\footnote{Note that 
without loss of generality, we may assume that each $N_i$ is a power of a prime. This implies that $N_{ij}$ must be either $N_i$ or $N_j$. If it is equal to $N_j$, then the relation in Eq.~\eqref{eq: TQD fusion group relations} is independent of $m_{ij}$, since the fusion of ${N_j}$ copies of $c_j$ is trivial. On the other hand, if $N_{ij}$ is equal to $N_i$, then the right-hand side has an additional factor of $c_j^{m_{ij}N_{ij}}$. This is equivalent to redefining $\varphi_i$ by $m_{ij}$ copies of $c_j$. Therefore, it is always possible to find a choice of elementary fluxes that satisfies the relation with $m_{ij}=0$, for all $i,j$, as in Ref.~\cite{Ellison2022Pauli}.} 
\begin{align} \label{eq: TQD fusion group relations}
c_i^{N_i} = 1,  \quad \text{and} \quad \varphi_i^{N_i} = c^{2n_i}_i \prod_{j \neq i} c_j^{n_{ij}+m_{ij} N_{ij}}.
\end{align}
We point out that the gauge charges and elementary fluxes generate disjoint groups if and only if $n_i \in \{0, N_i/2 \}$, for all $i$, and $n_{ij} = m_{ij}=0$, for all $i,j$. Otherwise, the elementary fluxes fuse into gauge charges. 

Another important concept 
is that of a Lagrangian subgroup, as it uniquely determines the gauge charges of a TQD (up to automorphisms of the anyon types). We recall that a Lagrangian subgroup $\mathscr{L}$ is a subset of anyon types of an Abelian anyon theory satisfying the following three properties:
\begin{enumerate}[label={(\roman*)}]
\item The anyon types of $\mathscr{L}$ form a subgroup under fusion.
\item The anyon types of $\mathscr{L}$ are bosons with trivial mutual braiding relations, i.e., for any elements $c_i$ and $c_j$ of $\mathscr{L}$, we have $\theta(c_i)=\theta(c_j)=1$ and $B_\theta(c_i,c_j)=1$.
\item For every anyon type $a$ outside of the Lagrangian subgroup, there exists a $c_i$ in $\mathscr{L}$ such that $B_\theta(a,c_i) \neq 1$.
\end{enumerate}
By Eqs.~\eqref{eq: gauge charge statistics}, \eqref{eq: charge-flux braiding}, and \eqref{eq: TQD fusion group relations}, the gauge charges of a TQD generate a Lagrangian subgroup.  
In fact, as argued in Refs.~\cite{Kapustin2011boundary, Kaidi2021higher}, if an Abelian anyon theory has Lagrangian subgroup, then it must be a TQD and the Lagrangian subgroup defines the set of gauge charges up to an automorphism. The high-level explanation is that 
gauging the $1$-form symmetry associated to the anyon types of $\mathscr{L}$ produces a system characterized by the trivial anyon theory. Subsequently, un-gauging the symmetry recovers a TQD corresponding to the initial anyon theory, and by construction, the gauge charges are precisely the elements of $\mathscr{L}$. 

\section{Abelian twisted quantum doubles from Walker-Wang models} \label{app: WW}

In Section~\ref{sec: general}, we identified an Abelian TQD for every Abelian anyon theory $\mathcal{A}$, such that the TQD contains $\mathcal{A}$ as a subtheory. Here, we provide intuition for that particular Abelian TQD by considering the Walker-Wang (WW) model based on $\mathcal{A}$.  

\begin{figure}[t] 
\centering
\includegraphics[width=.7\textwidth]{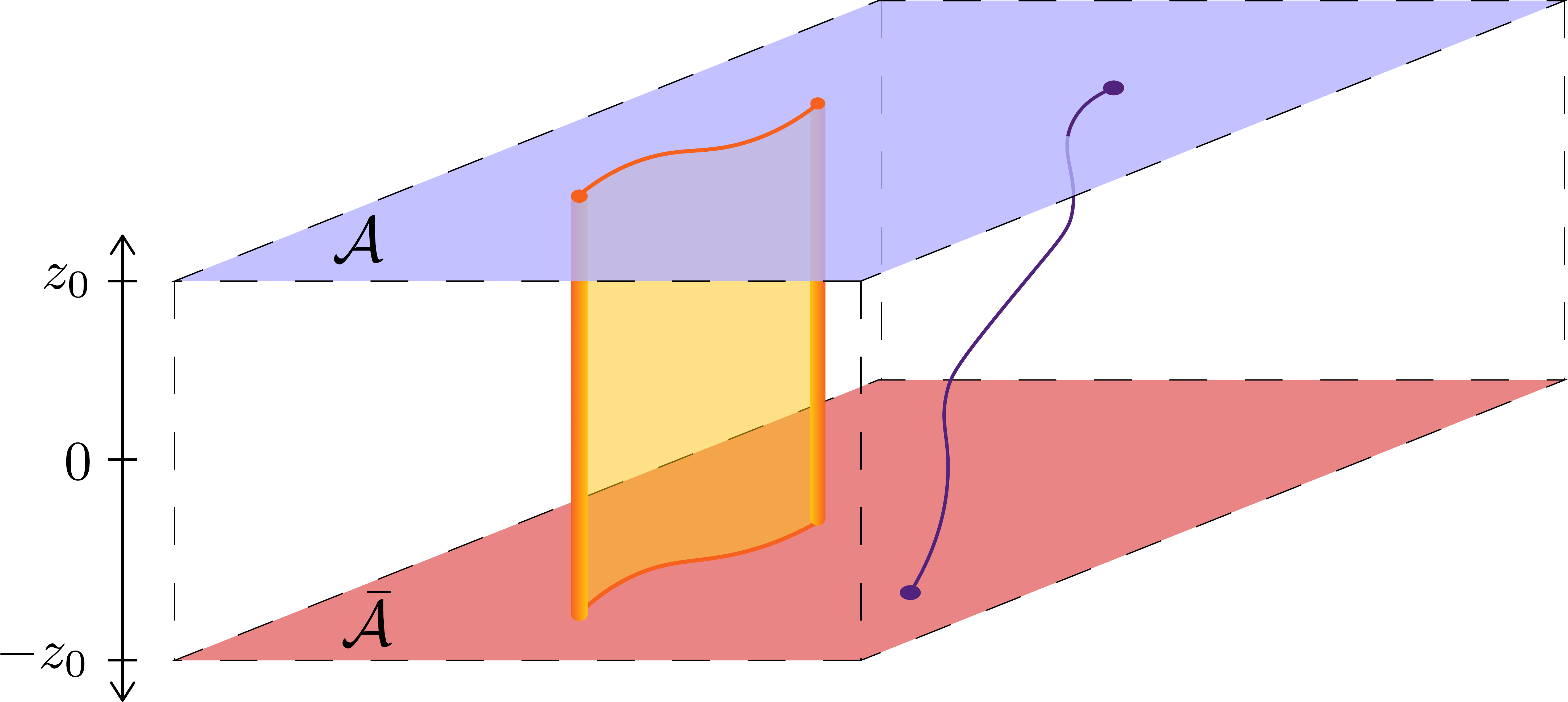}
\caption{We consider a slab of WW model based on the Abelian anyon theory $\mathcal{A}$. The WW model is centered about $z=0$, and its top and bottom boundaries are at $z=z_0$ and $z=-z_0$, respectively. We assume that the WW model has infinite extent in the $xy$-plane. The top surface (blue) hosts the anyon theory $\mathcal{A}$, while the bottom surface (red) hosts the conjugate theory $\bar{\mathcal{A}}$. The transparent anyon types (purple) are free to pass through the three-dimensional bulk to move between the top and bottom surface. In the quasi-2D system, the tube-like anyon types are created by a membrane (yellow) that extends from $z=-z_0$ to $z=z_0$.}
\label{fig: WW}
\end{figure}

We recall that a WW model is a three-dimensional exactly-solvable model defined by a two-dimensional anyon theory  \cite{walker2012}. The precise details of the lattice models in Ref.~\cite{walker2012} are not necessary for the following discussion, but what is important is that the WW model hosts the anyon theory $\mathcal{A}$ on its top surface (see Fig.~\ref{fig: WW}). If the WW model is built on a thin two-dimensional slab, then it gives us a quasi-2D model whose anyon theory includes $\mathcal{A}$. As described below, the anyon theory of the thin WW model is, in fact, equivalent to a TQD with Abelian anyons. Thus, from the anyon theory of the thin WW model based on $\mathcal{A}$, we identify an Abelian TQD whose anyon theory contains $\mathcal{A}$ as a subtheory. 

To make the discussion concrete, we imagine a slab of WW model centered on the $z=0$ plane, so the top surface is at $z=z_0$ and the bottom surface is at $z=-z_0$, as shown in Fig.~\ref{fig: WW}. Although we do not describe the lattice model explicitly, it is important to note that it is symmetric under the transformation $z \to -z$, which inverts the WW model across the $z=0$ plane.

The key feature of the WW model on the slab is that the top surface (at $z=z_0$) hosts the Abelian anyon theory $\mathcal{A}$. This is to say that the anyonic excitations that can be localized to the top surface correspond to the anyon types of $\mathcal{A}$. Likewise, the bottom surface hosts the conjugate anyon theory $\bar{\mathcal{A}}$. 
This is the anyon theory whose anyon types generate the same group as the anyon types of $\mathcal{A}$, but for which the exchange statistics are complex conjugated relative to those of $\mathcal{A}$. That is, for every anyon type $a$ in $\mathcal{A}$, there is a corresponding anyon type $\overline{a}$ in $\bar{\mathcal{A}}$ with the property $\theta(\overline{a}) = \theta(a)^*$. 

The topological excitations in the bulk of the WW model depend on whether the anyon theory $\mathcal{A}$ is modular or non-modular.
If $\mathcal{A}$ is a modular anyon theory, then the WW model has trivial bulk topological order. In this case, the anyon theory takes the form $\mathcal{A} \boxtimes \bar{\mathcal{A}}$, where the operation $\boxtimes$ is physically interpreted as stacking decoupled copies of the anyon theories $\mathcal{A}$ and $\bar{\mathcal{A}}$.
If the anyon theory is non-modular, on the other hand, the bulk has deconfined anyons corresponding to the transparent anyon types of $\mathcal{A}$. This means that the transparent anyon types of $\mathcal{A}$ and $\bar{\mathcal{A}}$ are shared by the two surfaces  -- they are free to move from the top surface to the bottom surface through the three-dimensional bulk. Consequently, the transparent anyon types of $\mathcal{A}$ and $\bar{\mathcal{A}}$ do not correspond to independent anyon types of the WW model. Letting $A$ be the group formed by the anyon types of $\mathcal{A}$ (including the transparent anyon types), there are $|A|$ anyon types associated to the top surface. 
There are then $|A|/|T|$ independent anyon types from the bottom surface, where $T$ is the group generated by the transparent anyon types. 

Furthermore, if the anyon theory $\mathcal{A}$ is non-modular, then the thin WW model has an additional set of anyon types, which we refer to as tube-like anyon types. In the three-dimensional WW model, these correspond to loop-like excitations created by membrane operators. The loop-like excitations are able to condense on the surfaces of the WW model, so the tube-like excitations can be created with a membrane operator stretching from the top surface to the bottom surface, as shown in Fig.~\ref{fig: WW}. In the two-dimensional (i.e., compactified) system, the tube-like excitations become point-like excitations created by string operators. 
Intuitively, the WW models based on non-modular Abelian anyon theories can be obtained from models of 1-form symmetry-protected topological orders by gauging the 1-form symmetry \cite{Delcamp2018}. The tube-like excitations can then be interpreted as the line-like gauge charges, while the transparent anyon types are the point-like gauge fluxes. The tube-like anyon types and the transparent anyon types of $\mathcal{A}$ are in one-to-one correspondence.  

While the tube-like anyon types have bosonic exchange statistics and trivial braiding relations with one another, 
they have nontrivial braiding relations with the transparent anyon types of $\mathcal{A}$. More specifically, for every generator of the transparent anyon types of $\mathcal{A}$, there is a tube-like anyon type with which it braids nontrivially.
This implies that the anyon theory of the thin WW model is modular. All together, there are $|A|$ anyon types from the top surface, $|A|/|T|$ independent anyon types from the bottom surface, and $|T|$ tube-like anyon types. Therefore, the slab of WW model has $|A| \times |A|/|T| \times |T| = |A|^2$ anyon types. 

To argue that the anyon theory of the thin WW model is given by a TQD, we identify
a Lagrangian subgroup of the anyon theory (see Section~\ref{app: TQD review}). 
We claim that a Lagrangian subgroup $\mathscr{L}_{WW}$ is generated by the set of tube-like anyon types and anyon types of the form $a \bar{a}$, where $a$ is  in $\mathcal{A}$ and $\bar{a}$ is the corresponding anyon type in $\bar{\mathcal{A}}$ with $\theta(\bar{a}) = \theta(a)^*$. The first thing to show is that these anyon types form a group under fusion. The tube-like anyon types fuse into anyon types of the form $a\bar{a}$ because of the $z \to -z$ symmetry of the WW model. More specifically, the tube-like anyon types can be created by membrane operators that are invariant under the symmetry, so they must fuse into anyon types that are also invariant. The only possibilities are that the tube-like anyon types fuse into anyon types of the form $a \bar{a}$ or transparent anyon types of $\mathcal{A}$. The anyon types $a \bar{a}$ are invariant under the symmetry, since exchange statistics and braiding relations are complex conjugated upon flipping the orientation of the system. The tube-like anyon types cannot fuse into the transparent anyon types, because the tube-like anyon types have trivial braiding relations with each other, which conflicts with the fact that each tube-like anyon type must braid nontrivially with some transparent anyon type. Therefore, the tube-like anyon types fuse into anyon types of the form $a \bar{a}$. 

Second, the anyon types of $\mathscr{L}_{WW}$ are all bosons with trivial mutual braiding relations. The tube-like anyon types braid trivially with the $a\bar{a}$ anyon types due to the $z \to -z$ symmetry of the WW model. Explicitly, the braiding relation of a tube-like anyon type $c_i$ with $a$ and $\bar{a}$ satisfies:
\begin{align}
B_\theta(c_i, a)^* = B_\theta(c_i, \bar{a}). 
\end{align}
Since $a$ and $\bar{a}$ can be localized to the top and bottom surface, respectively, we then have:
\begin{align}
B_\theta(c_i,a\bar{a}) = B_\theta(c_i, a)B_\theta(c_i, \bar{a}) = |B_\theta(c_i, a)|^2 = 1.
\end{align}
Next, to see that the $a\bar{a}$ anyon types are bosons with trivial mutual braiding relations, we employ the identity in Eq.~\eqref{eq: braiding identity}.
With this, the exchange statistics $\theta(a \bar{a})$ can be computed as:
\begin{align} \label{eq: abara theta}
\theta(a\bar{a}) = B_\theta(a,\bar{a}) \theta(a) \theta(\bar{a}) = |\theta(a)|^2 = 1.
\end{align}
Here, we used that $B_\theta(a,\bar{a})=1$, since $a$ and $\bar{a}$ can be localized to the top and bottom surface, respectively. Similarly, it can be checked that the anyon types of the form $a \bar{a}$ have trivial mutual braiding relations. For example, the braiding relation of $a_1 \bar{a}_1$ and $a_2 \bar{a}_2$ is [using Eq.~\eqref{eq: abara theta}]:
\begin{eqs}
B_\theta(a_1 \bar{a}_1, a_2 \bar{a}_2) = \frac{\theta(a_1 a_2 \bar{a}_1 \bar{a}_2)}{\theta(a_1\bar{a}_1)\theta(a_2 \bar{a}_2)} 
= 1.
\end{eqs} 

Finally, for any anyon type that does not belong to $\mathscr{L}_{WW}$, there is an anyon type in $\mathscr{L}_{WW}$ with which it braids nontrivially. To see this, we note that all of the anyon types of the thin WW model are generated by the anyon types of $\mathcal{A}$ and the anyon types of $\mathscr{L}_{WW}$. Therefore, it suffices to check that each anyon type in $\mathcal{A}$ braids nontrivially with some anyon type in $\mathscr{L}_{WW}$. The transparent anyon types in $\mathcal{A}$ braid nontrivially with the tube-like anyon types of $\mathscr{L}_{WW}$. All of the other anyon types of $\mathcal{A}$ braid nontrivially with some anyon type of the form $a\bar{a}$. This is because, besides the transparent anyon types, every anyon type $a$ in $\mathcal{A}$ braids nontrivially with some anyon type $b$ in $\mathcal{A}$. Thus, $a$ braids nontrivially with $b\bar{b}$, which is in $\mathscr{L}_{WW}$. This confirms that $\mathscr{L}_{WW}$ is a Lagrangian subgroup for the anyon theory of the thin WW model based on $\mathcal{A}$. According to Section~\ref{app: TQD review}, the anyon theory of the thin WW model must be a TQD, and $\mathscr{L}_{WW}$ defines a choice of the gauge charges.

To specify the class of Abelian TQDs described by thin WW models, we first notice that the anyon types of $\mathcal{A}$ correspond to a choice of elementary fluxes of the TQD.
This is because the anyon types of $\mathcal{A}$ along with the gauge charges, defined by $\mathscr{L}_{WW}$, generate all of the anyon types. Likewise, the anyon types of $\bar{\mathcal{A}}$ correspond to a choice of elementary fluxes. Since the anyon types of $\mathcal{A}$ or $\bar{\mathcal{A}}$, respectively, do not fuse into elements of $\mathscr{L}_{WW}$, we see that the gauge charges and the elementary fluxes must form disjoint groups. 
According to the group relations in Eq.~\eqref{eq: TQD fusion group relations}, this means that the TQD associated to the thin WW model is defined by a group $A$ of the general form:
\begin{align}
A = \prod_{i=1}^M \ZZ_{N_i},
\end{align}
and a set of integers $n_i \in \{0, N_i/2 \}$, for all $i$, and $n_{ij}=m_{ij}=0$, for all $i,j$. 
Here, we take $n_i$ to be $0$, if $N_i$ is odd. Such a TQD decomposes into a stack of $M$ decoupled TQDs, where the TQD in the $i$th layer is characterized by the group $\ZZ_{N_i}$ and an integer $n_i \in \{0, N_i/2\}$. This is precisely the TQD described in Section~\ref{sec: Construction of twisted quantum doubles}!

The final step in constructing topological subsystem codes is to gauge out the anyon types belonging to $\bar{\mathcal{A}}$.
The anyon types in $\bar{\mathcal{A}}$ braid nontrivially with the tube-like anyon types and the opaque anyon types of $\bar{\mathcal{A}}$. This leaves only the anyon types in $\mathcal{A}$, as desired. We note that to find the particular form for the generators of the anyon types of $\bar{\mathcal{A}}$ in Eq.~\eqref{eq: anyons to be gauged out}, we considered gauging out an arbitrary set of elementary fluxes of the Abelian TQD. We then identified the anyon types of the topological subsystem code and worked backwards to specify the anyon types that need to be gauged out in order to produce the anyon theory $\mathcal{A}$.

\bibliographystyle{quantum} 

\bibliography{bib}

\end{document}